\pdfoutput=1
\documentclass[12pt]{article}

\font\grande=cmr9.5 scaled \magstep4
\font\medio=cmr9.5 scaled \magstep2
\outer\def\beginsection#1\par{\medbreak\bigskip
      \message{#1}\leftline{\bf#1}\nobreak\medskip
\vskip-\parskip
      \noindent}
\usepackage{graphicx} 
\begin{document}
\bibliographystyle {unsrt}

\titlepage

\begin{flushright}
CERN-TH-2018-093
\end{flushright}
\vspace{2cm}
\begin{center}
{\grande Probing large-scale magnetism}\\
\vspace{0.8cm}
{\grande with the Cosmic Microwave Background}\\
\vspace{1.5cm}
 Massimo Giovannini 
 \footnote{Electronic address: massimo.giovannini@cern.ch} \\
\vspace{1cm}
{{\sl Department of Physics, 
Theory Division, CERN, 1211 Geneva 23, Switzerland }}\\
\vspace{0.5cm}
{{\sl INFN, Section of Milan-Bicocca, 20126 Milan, Italy}}
\vspace*{1.5cm}

\end{center}

\centerline{\medio  Abstract}
Prior to photon decoupling magnetic random fields of comoving intensity 
in the nano-Gauss range distort the temperature and the polarization 
anisotropies of the microwave background, potentially induce a peculiar 
$B$-mode power spectrum and may even generate a frequency-dependent 
circularly polarized $V$-mode. We critically analyze the theoretical 
foundations and the recent achievements of an interesting trialogue 
involving plasma physics, general relativity and astrophysics.
\noindent

\vspace{5mm}

\vfill
\newpage

\pagenumbering{arabic}

\tableofcontents

\newpage

\renewcommand{\theequation}{1.\arabic{equation}}
\setcounter{equation}{0}
\section{A magnetized Universe}
\label{sec1}

\subsection{History, orders of magnitude and units}
At the dawn of the seventeenth century William Gilbert 
published a celebrated treatise entitled 
{\it De Magnete, Magneticisque Corporibus, et de Magno Magnete Tellure} 
\cite{gilbert} where the quest for a coherent presentation 
of electric and magnetic phenomena 
anticipated the spirit, if not the letter, of the Maxwellian unification. 
In his systematic effort, Gilbert even conjectured that
large-scale magnets (like the earth itself) could share 
the same physical properties of magnetic phenomena over much 
shorter distance-scales: a similar kind of extrapolation is
at the heart of modern astrophysical applications 
from planetary sciences to black-holes.  
More than two hundred years later Michael Faraday introduced 
the expression {\em magnetic field}, a wording 
coined by Faraday himself while summarizing an amazing series 
of observations in his {\em Experimental Researches in Electricity} 
\cite{faraday}. Since then, the synergic evolution of physics, 
astronomy and astrophysics has been guided in many cases by 
the study of magnetic fields over different 
length-scales so that today nearly all astrophysical objects, 
from planets to clusters of galaxies, appear to be magnetized, 
at least to a certain degree. 
Through the years the quantity and quality of the answerable questions 
became larger and now we are allowed to ask sensible questions 
on the origin of large-scale magnetism with the hope of receiving 
reasonably definite answers. The present discussion, with its own 
limitations, aims at summarizing in a theoretical perspective 
the various interesting attempts involving the interplay 
between large-scale magnetism and the physics of the 
microwave background. 
\begin{table}[!ht]
\begin{center}
\begin{tabular}{||c|c|c||}
\hline
\hline
\rule{0pt}{4ex} Physical system & Magnetic field intensity & Typical scale of variation\\
\hline
 earth &  ${\mathcal O}(1)$ G & ${\mathcal O}(10^{4}\,\mathrm{km})$ \\
 Jupiter & ${\mathcal O}(10)$ G& ${\mathcal O}(10^{5} \mathrm{km})$\\
 LHC dipoles & ${\mathcal O}(10^{5})$ G  & ${\mathcal O}(15 \,\mathrm{m})$ \\
 neutron stars& ${\mathcal O}(10^{13})$ G & ${\mathcal O}(10\, \mathrm{km})$ \\
 spiral galaxies & ${\mathcal O}(10^{-6})$ G& ${\mathcal O}(30\, \mathrm{kpc})$ \\
regular (Abell) clusters & $ {\mathcal O}(10^{-7})$ G& $ < {\mathcal O}(\mathrm{Mpc})$\\
\hline
\end{tabular}
\caption{The magnetic field intensities of different physical systems are compared in terms 
of their associated scales of variation.}
\label{SEC1TABLE1}
\end{center}
\end{table}

The magnetic fields of physical systems characterized by very different scales of variation are 
compared in Tab. \ref{SEC1TABLE1}. The scale of variation roughly measures
 the distance over which there is an appreciable correlation 
between the values of the field at two spatially separated points. 
The dipoles of the Large Hadron Collider (LHC) are of the order of $10^{5}$ G,
 that is to say almost a million times larger that the earth's magnetic 
field which is roughly $0.3$ G. From Tab. \ref{SEC1TABLE1} we also see that the
 geomagnetic field (as well as the magnetic fields of other planets of the solar system) 
is a million times more intense than the magnetic fields of the galaxies 
and of the intergalactic medium.  One of the largest magnetic field intensities 
we can plausibly imagine in the framework of quantum electrodynamics comes from the
 Schwinger threshold for the production of electron-positron pairs
demanding, at least, a field $m_{e}^2/e$ (where $m_{e}$ is the electron mass and $e$ 
the corresponding charge).  The Schwinger limit implies an intensity
of the order of $10^{13}$ G that is comparable, according to Tab. \ref{SEC1TABLE1} 
with the magnetic fields possibly present at the surface of a neutron star.
The rationale for the huge magnetic fields of neutron stars may be what we call
compressional amplification: since at high conductivity the magnetic flux is frozen 
into the plasma element, as the gravitational collapse takes place the magnetic 
field increases.  We shall preferentially measure magnetic fields in Gauss within the
natural system of units\footnote{In this system we have, in particular, that  $\hbar c= 197.327 
\, \mathrm{MeV}\, \mathrm{fm}$ is equal to $1$ so that 
energies are measured as inverse lengths and vice-versa. The relation between K degrees and eV 
is given by $\mathrm{K} = 8.617\times 10^{-5} \,\mathrm{eV}$.
The conversion between centimetres and seconds follows from the speed of light 
$c= 2.99792\times 10^{10} \,\,\mathrm{cm}/\mathrm{sec}$. The conversion between mbarn and $\mathrm{GeV}^2$ can 
be deduced from $(\hbar c)^2 = 0.389\, \mathrm{GeV}^2 \,\mathrm{mbarn}$. Finally, 
since $e^2/(\hbar c) = 1/137$ the electric charge in natural units will be given by $1/\sqrt{137}$.}
(i.e. $\hbar= c= k_{B}= 1$, where $k_{B}$ is the Boltzmann constant).  In these units the Bohr magneton equals $ 5.788\times 10^{-11} 
{\rm MeV}/{\rm Tesla}$ and the relation between Tesla, Gauss and GeV is given by the following equations
\begin{equation}
1 \, {\rm Tesla} = 10^{4} \, {\rm Gauss},\qquad 1\, \mathrm{Gauss} = 6.9241 \times 10^{-20}\, \mathrm{GeV}^2.
\label{OM}
\end{equation}
W shall often employ the well known metric prefixes to indicate 
the multiples (or the fractions) of a given unit; so for 
instance, $\mu \mathrm{G} = 10^{-6}\, \mathrm{G}$, $\mathrm{n G} = 10^{-9} \mathrm{G}$ and so on.
When needed the typical length-scales will be often expressed in parsec and their multiples: 
recall, in this respect, that $1\, \mathrm{kpc} = 3.085\times 10^{21} \mathrm{cm} $. 
The present value of the Hubble radius is  $H_{0}^{-1} = 4282.7 \, (h_{0}/0.7) \, \mathrm{Mpc}$.
Magnetic fields whose correlation length is larger than the astronomical unit 
( $1\,{\rm AU}  = 1.49 \times 10^{13} {\rm cm}$) will be referred to as  large-scale magnetic fields. 
While this choice is largely conventional, magnetic fields  with approximate correlation 
scale comparable with the earth-sun distance are 
not observed (on the contrary, both the magnetic field of the sun and the one of the earth 
have a clearly distinguishable localized structure). Furthermore simple magnetohydrodynamical estimates 
seem to suggest that the magnetic diffusivity scale (i.e. the scale below which magnetic fields are diffused because 
of the finite value of the conductivity of the corresponding medium) of the order of the AU. For a definition of the magnetic diffusivity 
scale in weakly interacting plasmas see, for instance, section \ref{sec21} and discussion therein.

The central theme of this paper deals with two apparently unrelated phenomena, namely the 
large-scale magnetism and the Cosmic Microwave Background radiation (CMB in what follows) 
originally discovered by Penzias and Wilson \cite{CMB1} and subsequently confirmed by the 
COBE\footnote{The Cosmic Background Explorer (for short COBE) was a satellite which operated from 1989 to 1993 
and provided the best limits on the spectral distortions of the microwave background spectrum and 
the first solid evidence of its temperature anisotropies.} 
satellite mission \cite{CMB2,CMB3,CMB4} which also gave the first solid evidence 
of the large-scale temperature anisotropies. The CMB temperature is given by \cite{CMB5}:
\begin{equation}
T_{\gamma 0} = (2.72548 \pm 0.00057) \, \mathrm{K}.
\label{OM0}
\end{equation}
The energy density of the CMB turns out to be of the same order of the energy density 
of the magnetic energy density stored in the galactic field. More specifically we could say 
\begin{eqnarray}
&& \rho_{\gamma 0} = \frac{\pi^2}{15} T_{\gamma 0}^4 = 2.001\times 10^{-51} \,\,\biggl(\frac{T_{\gamma 0}}{2.72548}\biggr)^4\,\, 
\mathrm{GeV}^4,
\label{OM02}\\
&& \rho_{\mathrm{B}} = \frac{B^2}{8\pi} = 2.002\times10^{-51} \biggl(\frac{B}{3.24 \mu \mathrm{G}}\biggr)^2\,\,\mathrm{GeV}^4,
\label{OM03}
\end{eqnarray}
where Eq. (\ref{OM}) has been used together with the conversion between K degrees and eV.
Equations (\ref{OM02}) and (\ref{OM03}) just account for an interesting numerical coincidence. Needless to say that the galactic 
magnetic field is not exactly $3.24\, \mu \mathrm{G}$: the magnetic field in the Solar neighbourhood has regular component 
and a random contribution so that estimates of the total magnetic field, depending on the way 
we count, range between $2$ and $6 \mu \mathrm{G}$ \cite{haverkorn}. It is 
however relevant to stress that the energy density 
of the CMB, the energy density of the galactic magnetic field and the energy density of the cosmic rays 
are all comparable within one order of magnitude. Two excellent background monographs on large-scale 
magnetism are listed in Refs. \cite{parker,zeldovich}.

\subsection{Magnetic fields in galaxies}
While it is probably true that large-scale magnetism is the birthright of radio-astronomy, 
the very first evidence of galactic and interstellar magnetic fields came from the 
isotropy of the galactic cosmic ray spectrum in the Milky Way and from the polarization of starlight. 
The lack of detection of appreciable anisotropies in cosmic ray spectrum led Fermi \cite{fermi} 
to suggest the existence of a magnetic field of approximate $\mu$G strength 
scrambling the trajectories of the charged particles and making the spectrum 
fairly isotropic. Even if concrete evidences of large-scale magnetic fields in the interstellar
media were still lacking,  magnetic fields were known to be stable in 
highly conducting plasmas thanks to the seminal contributions of Alfv\'en
\footnote{Alfv\'en \cite{alv2} and others \cite{alv3} vocally criticised the suggestion of Fermi and 
claimed that cosmic rays can only be in equilibrium with stars.
Today we do know that this is the case for low-energy cosmic rays but not for the more energetic ones 
around, and beyond, the knee in the cosmic ray spectrum.} \cite{alv1}. 
Few months after Fermi's proposal Hiltner \cite{hiltner} and, independently, 
Hall \cite{hall} observed the polarization of starlight which was later on interpreted 
by Davis and Greenstein \cite{davis} as an effect of galactic magnetic field 
aligning the dust grains. 

After more than three score years of radio-astronomical observations, spiral galaxies are known to 
have magnetic fields in the same range of the Milky Way 
(i.e. ${\mathcal O}(\mu \mathrm{G})$) while elliptical galaxies have similar intensities 
but shorter correlation scales. As already alluded to in connection with Eqs. (\ref{OM02})--(\ref{OM03}), 
galaxies have a regular magnetic field but they also possess a random component:
magnetized domains with typical correlation scales from $100$ pc to few kpc are observed 
in the galactic halo of the Milky Way. Two excellent background monographs 
 on galactic magnetism can be found in Refs. \cite{heiles,ruzmaikin}. In the last decade or so it has been established that planets and stars 
are formed in an environment which is already magnetized \cite{gal1,gal2}
so that, as lucidly argued in a comprehensive review on large-scale magnetism \cite{gal3}, 
the true question before us today does not concern the existence of these fields 
but rather their origin. The measurements of galactic magnetic fields in the Milky Way and 
in external galaxies are reviewed in various papers (see e.g. \cite{gal3,gal4,gal5} and \cite{gal6} for an introduction 
to the main observational techniques). It is often difficult to disentangle the large-scale (ordered) fields from other 
components with smaller correlations scales. In this respect newly developed spectropolarimetric techniques \cite{gal7} 
 for wide-band polarization observations might not only improve the sensitivity but also give 
 synthesized maps of Faraday rotation measure. 

It is at the moment not yet clear if the observed galactic fields are the 
consequence of a strong dynamo action (see e.g. \cite{parker,zeldovich,ruzmaikin}) 
or if their existence somehow precedes the formation of galaxies. 
According to some intriguing suggestions, if the magnetic fields do not flip their sign 
from one spiral arm to the other, then a strong dynamo action 
can be suspected \cite{gal4} (see also \cite{gal5}). In the opposite case the magnetic field of galaxies should (or could)
be primordial (i.e. present already at the onset of gravitational collapse). 
In this perespective a further indication that would support the primordial nature of the magnetic field 
of galaxies would be, for instance, the evidence that not only spirals but also 
elliptical galaxies are magnetized with a correlation scale shorter than in the case of spirals. 
Since elliptical galaxies have a much less efficient rotation, it seems difficult to postulate 
a strong dynamo action as the common origin of the two corresponding magnetic fields.

\subsection{Magnetic fields in clusters}

Magnetic fields are not only associated with galaxies 
but also with clusters which are gravitationally bound systems 
of galaxies. The Milky Way is part of the local group which is our own cluster and 
other members of the local group (e.g. Andromeda and Magellanic clouds) 
have magnetic fields between few and $10$ $\mu\mathrm{G}$. 
While the local group contains fewer members than other 
rich clusters (and it is sometimes referred to as an irregular cluster), 
regular clusters (like the Coma cluster) are magnetized at a level of $0.5 \, \mu\mathrm{G}$  for typical correlation 
scales between $500$ kpc and the Mpc. Magnetic fields of single clusters have been extensively 
analyzed but in the last decade or so remarkable analyses of multi-cluster measurements 
became available \cite{cluster1} (see also \cite{gal3,cluster2} for review articles on these specific themes). 
In the past it was shown that regular clusters have  cores with
a detectable component of Faraday rotation measure. There is now mounting evidence 
that $\mu\mathrm{G}$ magnetic fields are indeed detected 
inside regular Abell clusters \cite{cluster3,cluster4}, as originally 
suggested in \cite{cluster2}.  

Weakly bound systems of clusters (i.e. superclusters) have been also claimed to be magnetized at the $\mu\mathrm{G}$  level:
 this is the case for the local supercluster (formed by the local group and by the Virgo cluster) 
and for the Coma supercluster \cite{supercl1}.  The current indications 
 seem to be encouraging even if crucial ambiguities persist on the way the magnetic field strengths are inferred from 
 the Faraday rotation measurements of superclusters magnetic fields. It is not excluded 
 that the recent progress in spectropolarimetric techniques \cite{gal7} could be used also in the case of superclusters.
In this connection we can mention that the intergalactic magnetic field in cosmic voids can be indirectly probed through its effect on electromagnetic cascades initiated by a source of TeV gamma-rays, such as active galactic nuclei. The original idea
of Plaga \cite{TEV0,TEV1} suggested the possibility of deriving lower limits on the magnetic fields in voids even if 
reasonable statistical analyses seem to cast doubts on the claimed lower limits \cite{TEV2}.

The hope for the near future is connected with the possibility of a 
next generation radio-telescope like the Square Kilometer Array (for short SKA \cite{SKA}). 
The unprecedented collecting area of the instrument and the frequency range 
(hopefully between $0.1$--$25$ GHz) will allow full sky surveys of Faraday Rotation which may be combined 
with the most recent advances spectro-polarimetry \cite{gal7}.
This instrument might not only be directly beneficial 
for the microwave background physics but it might also 
have an amazing impact in pulsar searches \cite{pulsar} which 
are essential for sound determinations of magnetic fields 
from Faraday rotation \cite{gal3,gal4,gal5,gal6}.

To close the circle we can go back to the isotropy 
of the cosmic ray spectrum and remind that nearly 
ten years ago the Auger collaboration advertized a correlation 
between the arrival directions of cosmic rays with energy above 
$6 \times 10^{19}$ eV and the positions of active galactic nuclei within 75 Mpc \cite{auger1}. In the same context 
concurrent analyses demonstrated \cite{auger2} that overdensities on windows of 5 degree radius (and for energies $10^{17.9} eV < E < 10^{18.5}$ eV) were compatible with an isotropic distribution. In a nutshell the claim was that in the highest 
energy domain (i.e. energies larger than $60$ EeV) cosmic rays were not appreciably deflected: within a cocoon of 70 Mpc the intensity of the (uniform) component of the putative magnetic field should be smaller than ${\mathcal O}(\mathrm{nG})$. The evidence of this claim got worse and worse so that the recent analyses suggest \cite{auger3} that no deviation from isotropy is observed on any angular scale in the energy range between $4$ and $8$ EeV. Above $8$ EeV a weak indication for a dipole moment is claimed; no other deviation from isotropy is observed for other moments. While the claimed departure from isotropy 
is still at the level of indication, if cosmic rays would also be roughly isotropic in the EeV range, it would be tempting to conclude for the existence of potentially large magnetic fields (in the $10$ or $100$ nG range) for typical correlation scales larger than $10$ Mpc. It is amusing to note that the speculations of a single source accounting for a nearly isotropic high-energy cosmic ray spectrum in the presence of strong magnetic fields \cite{auger4} now are becoming more plausible.

\subsection{Magnetic fields at the largest scales}
In spite of the remarkable progresses of the last decade, 
as we probe larger and larger distance scales 
the techniques used in the case of galaxies and clusters 
(i.e. Faraday rotation measures or synchrotron emission) become ambiguous.
Therefore if we aim at scrutinizing the magnetization of the whole Universe 
we need to investigate directly the microwave background and its anisotropies.

The idea of employing microwave background physics as a magnetometer has a relatively long history 
which should be traced back to the seminal contributions of Hoyle \cite{hoyle} and Zeldovich 
\cite{zeldovich1}. Less than ten years after the debate that confirmed the existence of a magnetic field 
associated with the galaxy \cite{fermi,alv2,alv3}, Hoyle speculated in favour of a cosmological 
origin for the galactic magnetic fields and mentioned
 CMB physics as a crucial test of his idea. In a contribution entirely devoted to the steady state theory \cite{hoyle}, Hoyle 
discussed at length the origin of galactic magnetism (not really central to the steady state theory) 
and lucidly concluded for a cosmological relevance of the problem: 
if the galactic magnetic fields would result from processes within the galaxy 
(e.g. ejecta of magnetic flux due to finite conductivity effects) the correlation scale 
would be inexplicable and, besides that, the field should be maintained against the magnetic diffusivity.
The same problem actually occurs in geomagnetism where the maintenance of the field 
is insured by some kind of dynamo action. The origin of the magnetic field of the galaxies 
should therefore be understood from the past history of the Universe. Moreover, since the magnetic energy 
density increases faster than the energy density of non-relativistic matter, the role of the magnetic 
fields has to become more prominent as the curvature and the energy density 
of the Universe increase. 

Few years later Zeldovich \cite{zeldovich1} (see also \cite{zeldovich2,zeldovich3}) 
even argued that magnetic fields should primarily account for 
the temperature anisotropies of the microwave background, an idea now ruled out 
by direct tests on the isotropy of the angular power spectra. The analysis of Zeldovich, later discussed and refined by 
many authors along slightly different perspectives, was formulated in the simplest general relativistic 
framework allowing for a uniform (i.e. homogeneous) magnetic field
in a homogeneous (but anisotropic) space-time metric. 
These Bianchi models \cite{ryan} can host magnetic and electric fields 
in various situations more complicated then the one originally
considered in \cite{zeldovich1}. 

More than fifty years after the pioneering speculations of Hoyle and Zeldovich 
the current formulation of the standard cosmological paradigm implies that 
the temperature and the polarization anisotropies observed in the CMB 
are not caused by a large-scale magnetic field but rather by 
curvature inhomogeneities which are Gaussian and (at least predominatly) adiabatic.
 This possibility, originally intuited by Lifshitz \cite{liff} has been subsequently 
analyzed by various authors including Peebles \cite{pee1,pee2,pee3}, Silk \cite{silk},  
Harrison \cite{harrisons}, Novikov and Zeldovich \cite{zeldovichnovikov}.  
Around the same time Rees \cite{rees} showed that the repeated Thomson scattering 
of the primeval radiation during the early 
phases of an anisotropic Universe would modify the black-body spectrum and produce 
linear polarization. Today we know that the polarization anisotropies have an entirely
different spectrum from the temperature fluctuations but their initial conditions 
is common and it comes from the large-scale inhomogeneities 
in the spatial curvature. The modern way of implementing the suggestions 
of Hoyle and Zeldovich is therefore to embed the presence 
of the magnetic random fields in the concordance paradigm and 
to analyze carefully their impact on the CMB observables.

\subsection{Magnetic random fields and CMB observables}

In spite of the efforts both from the theoretical and from the experimental 
sides, our knowledge of pre-decoupling magnetic fields is 
still not satisfactory in many respects and one of the purposes 
of the present article is to contribute to the ongoing debate.
During the past decade there have been specific attempts 
to rule in (or out) the presence of a large-scale 
magnetic field potentially present after neutrino decoupling but prior 
to recombination \cite{magnetized}. In what follows we shall 
simply outline the guiding logic of the present discussion 
and briefly mention the summary of the forthcoming sections. 

Because the current concordance paradigm is consistent with 
the assumption that the extrinsic curvature (i.e. the Hubble rate) dominates 
against the intrinsic (spatial) curvature, the background geometry prior to 
photon decoupling is conformally flat to a very good approximation 
and characterized by a metric tensor:
\begin{equation}
\overline{g}_{\mu\nu}(\tau) = a^2(\tau) \eta_{\mu\nu} , \qquad \eta_{\mu\nu} = \mathrm{diag}(1,\, -1,\, -1,\, -1),
\label{OM1}
\end{equation}
where $\eta_{\mu\nu}$ is the Minkowski metric, $a(\tau)$ is the scale factor and $\tau$ will denote 
throughout the conformal time coordinate. Since large-scale magnetic fields must not break the spatial 
isotropy of the background geometry their form is constrained by rotational invariance, 
by gauge-invariance and by the invariance under infinitesimal coordinate 
transformations on the background geometry (\ref{OM1}). 
The most general two-point function of magnetic random fields respecting these 
requirements is given by:
\begin{equation}
{\mathcal C}_{ij}(r, \tau) = M_{T}(r,\tau)\biggl( \delta_{ij} - \frac{r_{i} r_{j}}{r^2}\biggr) + M_{L}(r,\tau) \frac{r_{i} r_{j}}{r^2} + M_{G}(r,\tau) \epsilon_{i j \ell} \frac{r^{\ell}}{r},
\label{OM2}
\end{equation}
where $M_{T}(r, \tau)$, $M_{L}(r,\tau)$ and $M_{G}(r,\tau)$ denote, respectively, the transverse, the longitudinal and the gyrotropic 
component of the two-point function. Note that $M_{T}(r,\tau)$ and $M_{L}(r,\tau)$ are not independent since the 
two-point function must be divergenceless (see, in particular, Eq. (\ref{IRFVEC2})). If
$M_{G} \neq 0 $ the two-point function is rotationally invariant but not parity-invariant: this term 
arises when the magnetic field $\vec{B}$ has a non-vanishing magnetic gyrotropy 
(i.e. $\vec{B} \cdot \vec{\nabla} \times \vec{B} \neq 0$). 
More detailed discussions on Eq. (\ref{OM2}) and on the theory 
of isotropic random fields 
of different spin can be found in appendix \ref{APPA}.

\begin{figure}[!ht]
\centering
\includegraphics[height=8cm]{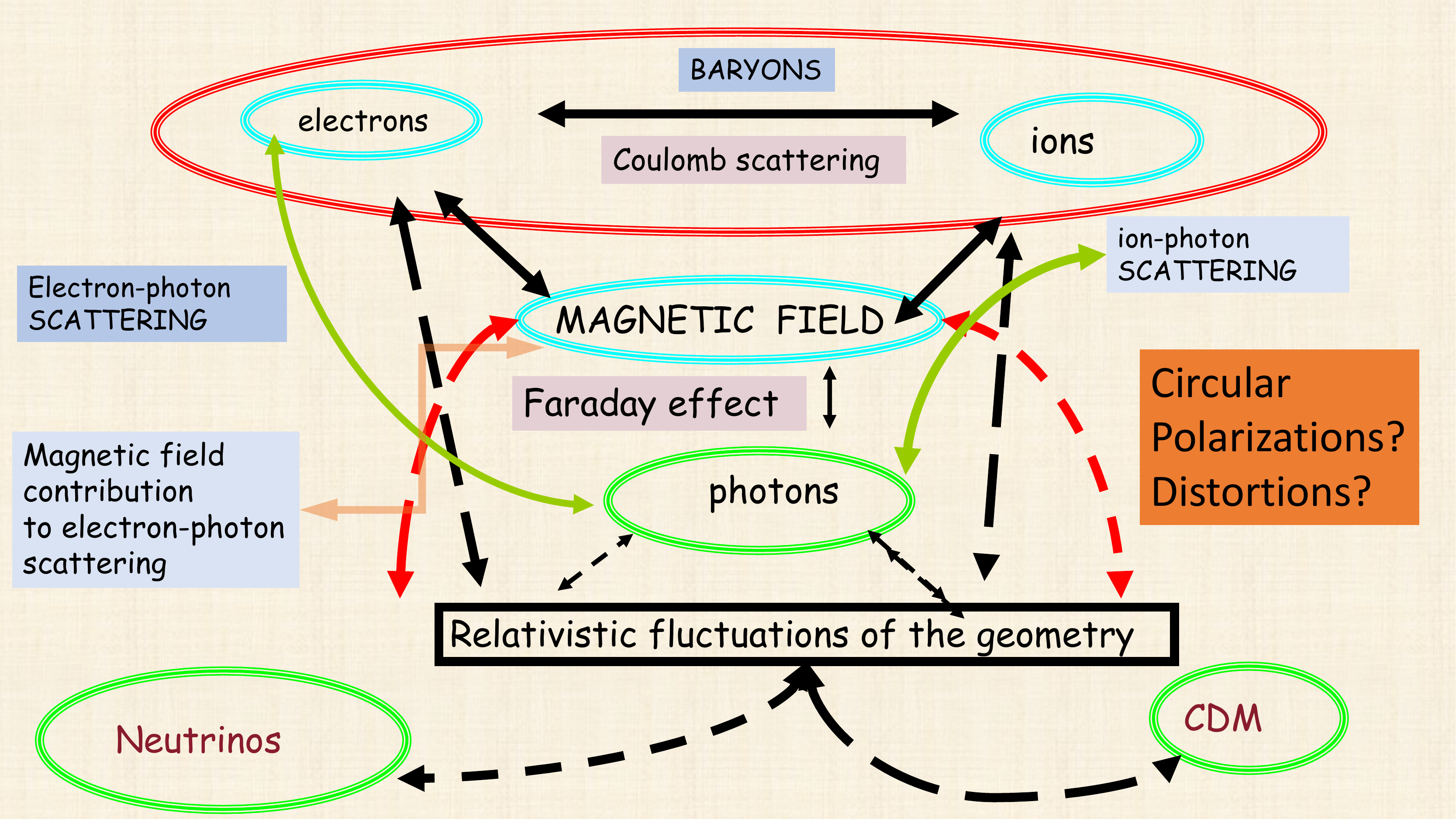}
\caption[a]{The effects of magnetic random fields on the pre-decoupling plasma for temperatures 
much smaller than the 
electron mass.}
\label{figure1}      
\end{figure}

The dynamical effects of the (isotropic) magnetic random fields 
on CMB physics are summarized in Fig. \ref{figure1}. The first and most obvious 
consequence is a modification of the evolution equations of the charged species (i.e. 
electrons and ions) prior to photon decoupling. For this reason the modifications of the
electron-photon and photon-ion scatterings affect 
 the collisional terms of the corresponding radiative transfer equations
 for the temperature and polarization brightness perturbations.
Furthermore the Faraday effect on the linear polarization of the CMB
may induce a $B$-mode polarization\footnote{See, in this respect, the discussion in the first part of section \ref{sec5} 
and References therein.}. There is also the possibility of 
an inverse Faraday effect, namely the rotation of an initial $B$-mode polarization 
of tensor origin. Last but not least magnetic random fields may affect 
the CMB spectrum itself and produce circular polarizations
which have been for long time an observational challenge. 

Since the temperature of the plasma before photon decoupling is 
much smaller than the mass of the lightest 
charge carrier (i.e. the electrons), some of the most notable 
direct effects of the magnetic random fields 
are pictorially illustrated in Fig. \ref{figure1} in a  
qualitative manner suitable for those who might want to avoid 
the more technical aspects of the forthcoming discussions. 
In Fig. \ref{figure1} the {\em direct} effects of the magnetic fields have been indicated with full (double) 
arrows while the {\em indirect} effects have been denoted by dashed arrows. 
The ellipse at the top of the picture reminds that, prior to decoupling, the electrons and ions are interacting 
 strongly via Coulomb scattering. In some cases this 
observation justifies the treatment of the electron-ion fluid as a 
single effective species often dubbed as the baryon fluid. One of the 
most notable exceptions to the previous statement is represented by the 
Faraday effect (i.e. the rotation of the polarization plane of the CMB) 
and more generally by all the phenomena where the magnetic field directly affects 
the propagation of high-frequency electromagnetic disturbances in the plasma.

Even if photons are not electrically charged, magnetic random fields 
have a direct effect on their evolution (as indicated in Fig. \ref{figure1}). 
This apparently counterintuitive phenomenon occurs since, prior 
to decoupling, photons scatter electrons and ions (or, for short, baryons).
The electron-ion-photon system is, to some extent, a unique physical 
entity whose evolution equations are directly modified by the magnetic 
random fields, at least in the low-frequency branch of the spectrum of 
plasma excitations.    

As the dashed lines of Fig. \ref{figure1} suggest, the magnetic random fields interact  indirectly
with all the neutral species of the plasma (i.e. neutrinos, cold dark matter particles and of course photons). 
The neutral species actually appear in the evolution equations of the 
cosmological perturbations which are also affected by the presence 
of random magnetic fields. In particular, through the Hamiltonian and momentum 
constraints, the magnetic random fields
modify the dynamics of the relativistic fluctuations of the geometry 
without affecting the evolution of the background. These two constraints 
determine, respectively, the initial conditions of the density contrasts 
and of the peculiar velocities for all species of the plasma (both 
charged and neutral). As a consequence the normal modes of the system are 
also modified and this occurrence entails, ultimately,  different sets of 
large-scale initial conditions of the Einstein-Boltzmann hierarchy.
All together the direct and indirect effects of the magnetic random fields 
will then determine the final values of the temperature and polarization 
anisotropies of the microwave background. 

The layout of the paper is, in short the following. In section \ref{sec2}  the 
physical scales of the pre-decoupling plasma will be introduced together with
the main ingredients of the concordance paradigm. 
The contribution of the magnetic fields to the electron-photon scattering 
will be analyzed in the second part of the section which will be concluded 
by a discussion of the distortions of the microwave background spectrum.

The evolution equations 
of the various species that interact strongly with the 
magnetic field (i.e. electrons and ions and, ultimately, baryons) will be discussed 
in section \ref{sec3}. After addressing the evolution of the weakly interacting species
(i.e. neutrinos and cold dark matter particles), we shall tackle  
the magnetized scalar, vector and tensor modes of the geometry.  
By introducing the distinction between regular and divergent modes, 
the initial conditions of the Einstein-Boltzmann hierarchy will be specifically studied 
in terms of the normal mode of the system. 

In section \ref{sec4} we shall discuss some of the bounds on the magnetic fields 
derived in the past decade or so from the observed temperature and 
polarization anisotropies. We shall also illustrate the 
main distortions produced by magnetic random fields.
The corresponding shapes of the magnetized 
temperature and polarization anisotropies will be 
briefly described by focussing on the current 
observables (i.e. the temperature  autocorrelations, 
the polarization autocorrelations 
and the temperature-polarization cross-correlations). 

Section \ref{sec5}  will be devoted to the analysis of the Faraday 
effect under different approximations and to the scaling properties 
of the polarization anisotropies. In section \ref{sec6} we shall 
discuss the effects of the magnetic fields on the circular 
polarization (the so-called $V$-mode polarization).
Section \ref{sec7} contains some concluding remarks and 
some perspectives for the incoming decade. 
With the purpose of making this paper 
self-contained, some relevant technical aspects have been 
relegated to the appendix which could be useful 
for those who are also interested 
in the quantitative aspects of the problem.  

\renewcommand{\theequation}{2.\arabic{equation}}
\setcounter{equation}{0}
\section{The pre-decoupling plasma}
\label{sec2}
Prior to photon decoupling the  evolution of the space-time curvature and the relativistic fluctuations 
of the geometry cannot be neglected but the plasma parameter itself is at most of the order of $10^{-7}$, 
a value often encountered in diverse terrestrial plasmas from glow discharges to tokamaks.
Since the Weyl invariance is broken by the masses of the charge carriers the evolution of the 
system cannot simply be reduced to its flat space-time analog. 
In this framework, the magnetic random fields affect the electron-photon scattering
and, ultimately, the explicit form  of the radiative transfer equations 
for the brightness perturbations. 

\subsection{Plasma parameters} 
\label{sec21}
The global neutrality of the plasma for redshifts $ 10^{3} < z < 10^{6} $  implies that the concentration 
of the electrons and of the ions coincides, i.e. $n_{\mathrm{e}}=n_{\mathrm{i}}=n_{0}$ where 
$n_{0} =\eta_{\mathrm{b0}} n_{\gamma 0}$ and  $n_{\gamma 0}$ denotes 
the present concentration of photons while, as usual, $\eta_{\mathrm{b}0}$ is the ratio between photon concentration and baryon concentration. Neglecting, for the moment, the expansion of the background 
geometry the plasma parameter \cite{spitzer,stix,krall} is given by:
\begin{equation}
g_{\mathrm{plasma}} = \frac{1}{V_{\mathrm{D}} n_{0} x_{\mathrm{e}}} = 24 e^{3}  \sqrt{\frac{\overline{\zeta}(3)}{\pi}} 
\sqrt{x_{\mathrm{e}} \,\eta_{\mathrm{b}0}}
= 2.308 \times 10^{-7} \sqrt{x_{\mathrm{e}}} \biggl(\frac{h_{0}^2\Omega_{\mathrm{b0}}}{0.02273}\biggr)^{1/2},
\label{gplasma}
\end{equation}
where $e$ is the electric charge and $\overline{\zeta}(3) = 1.202\,...$;  $x_{\mathrm{e}}$ is the ionization fraction and 
$\Omega_{\mathrm{b}0}$ is present critical fraction of baryons. In Eq. (\ref{gplasma}) 
$\lambda_{\mathrm{D}}$ is the Debye length and $V_{\mathrm{D}}$ is the volume of the Debye sphere: 
\begin{equation}
\lambda_{\mathrm{D}} =\sqrt{\frac{T}{8\pi e^2 n_{0} x_{\mathrm{e}}}},\qquad V_{\mathrm{D}} =\frac{4}{3}\pi \lambda_{\mathrm{D}}^3,
\label{DEBYE}
\end{equation}
where $T$ denotes the temperature of the plasma. 
Both $g_{\mathrm{plasma}}$ and its inverse (measuring
the number of charge carriers within the Debye sphere)
determine all the physically relevant hierarchies between the plasma  parameters. 
Indeed, the Debye length (i.e. $\lambda_{\mathrm{D}}$) is parametrically smaller than 
the Coulomb mean free path (i.e. $\lambda_{\mathrm{Coul}}$) because of one power of $g_{\mathrm{plasma}}$:
\begin{equation}
\frac{\lambda_{\mathrm{D}}}{\lambda_{\mathrm{Coul}}} = \frac{g_{\mathrm{plasma}}}{48\pi}
\ln{\Lambda_{\mathrm{C}}}, \qquad \Lambda_{\mathrm{C}} = \frac{18 \sqrt{2}}{g_{\mathrm{plasma}}},
\label{hier1}
\end{equation}
where $\ln{\Lambda}_{\mathrm{C}}$ defines the Coulomb logarithm\footnote{ In the case of a proton (or of an electron) impinging 
on an electron (or on a proton) the Rutherford cross section is logarithmically 
divergent at large impact parameters when the particles are free. Prior to decoupling the logarithmic 
divergence is avoided  because of the Debye screening length: the cross section is then 
known as Coulomb cross section and the logarithmic divergence is replaced by the so-called Coulomb logarithm
\cite{stix,krall}.}.  In similar terms the plasma frequency of the electrons (i.e. $\omega_{\mathrm{pe}}$) is much larger than the collision frequency that is related, in its turn, to the Coulomb rate of interactions (i.e. $\Gamma_{\mathrm{Coul}}$):
\begin{equation}
\frac{\Gamma_{\mathrm{Coul}}}{\omega_{\mathrm{pe}}} = \frac{\ln{\Lambda_{\mathrm{C}}}}{24 \sqrt{2} \pi } 
g_{\mathrm{plasma}}, \qquad \omega_{\mathrm{pe}} = \sqrt{\frac{4\pi n_{0} x_{\mathrm{e}}}{m_{\mathrm{e}}}}.
\label{hier2}
\end{equation}
Since the conductivity $\sigma$ depends both on the plasma frequency and on the Coulomb rate \cite{spitzer,krall}
we can use Eq. (\ref{hier2}) and express $\sigma$ in terms of the plasma parameter:
\begin{equation}
\sigma = \frac{\omega_{\mathrm{pe}}^2}{4 \pi \Gamma_{\mathrm{Coul}}} = \frac{6 \sqrt{2}}{\ln{\Lambda_{\mathrm{C}}}} \frac{\omega_{\mathrm{pe}}}{g_{\mathrm{plasma}}}.
\label{hier3}
\end{equation}

The three hierarchies discussed in Eqs. (\ref{hier1}), (\ref{hier2}) and (\ref{hier3}) 
should be supplemented by a fourth one, not directly related to the plasma parameter. 
The Hubble radius $r_{\mathrm{H}}= H^{-1}$ 
around equality exceeds (roughly by 20 orders of magnitude) the Debye length
at the corresponding  epoch.  At the same reference time, magnetic 
fields can be present only over sufficiently large scales $L > L_{\sigma}$ where $L_{\sigma}$ 
is the magnetic diffusivity scale\footnote{For typical values of the cosmological parameters, around equality, 
$L_{\sigma} \simeq 10^{-17} r_{\mathrm{H}}$. Magnetic fields 
over typical length-scales $L \simeq {\mathcal O}(r_{\mathrm{H}})$ 
(and possibly larger) can be present without suffering appreciable diffusion.}
 \begin{equation}
L_{\sigma} \simeq (4\pi \sigma H_{\mathrm{eq}})^{-1},\qquad 
\sigma = \frac{T}{e^2  \ln{\Lambda_{\mathrm{C}}}} 
\biggl(\frac{T}{m_{\mathrm{e}}}\biggr)^{1/2}.
\label{Lsigma}
\end{equation}
In these conditions the Larmor radius prior to matter-radiation equality is always 
much smaller than the range of variation of the magnetic field, i.e. 
\begin{equation}
r_{\mathrm{Be}} \ll  L \simeq r_{\mathrm{H}},
 \qquad r_{\mathrm{Be}} = \frac{v_{\perp}}{\omega_{\mathrm{Be}}},\qquad v_{\perp} \simeq v_{\mathrm{th}},
\label{hier4}
\end{equation}
where $v_{\mathrm{th}} \simeq \sqrt{3 T/m_{\mathrm{e}}}$ and $\omega_{\mathrm{Be}}$ is 
the Larmor frequency.  Equation (\ref{hier4}) is the starting point for the so-called guiding center 
approximation \cite{guide1,guide2} which accounts for the motion of charged particles 
in the magnetized plasma and will be relevant when discussing the effects of 
magnetic random fields on the electron-photon scattering. 

\subsection{Gravitating plasmas}
Denoting by $\ell_{P} = \sqrt{8 \pi G}$ the Planck length and by $T_{\mu}^{\nu}$ 
the (covariantly conserved) total energy-momentum tensor of the 
plasma, the Einstein equations shall be written as:
\begin{equation}
R_{\mu}^{\nu} - \frac{1}{2} \delta_{\mu}^{\nu}\, R = \ell_{P}^2 \,T_{\mu}^{\nu}, \qquad \nabla_{\mu} T^{\mu\nu}=0,
\label{COV1}
\end{equation}
where $R_{\mu\nu}$ is the Ricci tensor, $R$ is the Ricci scalar\footnote{As already 
mentioned in connection with Eq. (\ref{OM1}), 
the signature of the metric is mostly minus i.e.  $(+,\,-,\, -,\, -)$; the Ricci tensor is derived 
from the Riemann tensor by contracting the first and third indices, i.e.  
$R_{\mu\nu} = R^{\alpha}_{\,\,\,\mu\, \alpha\, \nu}$. In Eq. (\ref{COV1}) and in the 
remaining part of the paper $\nabla_{\mu}$ denotes a covariant 
derivation.}.  The total energy-momentum tensor $T^{\mu\nu}$ is the sum of all the individual energy-momentum 
tensors of the various species of the plasma:
\begin{equation}
T^{\mu\nu} = T^{\mu\nu}_{(\mathrm{e})} + T^{\mu\nu}_{(\mathrm{i})} + 
T^{\mu\nu}_{(\nu)} + T^{\mu\nu}_{(\gamma)} + T^{\mu\nu}_{(\mathrm{c})} 
+ T^{\mu\nu}_{(\Lambda)} + T^{\mu\nu}_{(\mathrm{EM})}.
\label{COV4}
\end{equation}
In Eq. (\ref{COV4}) the subscripts denote, respectively, the contributions of electrons, ions, neutrinos, photons, cold dark matter (CDM) particles, dark energy and electromagnetic fields. Since  the electrons, the ions and the cold dark matter 
particles are all pressureless, their associated energy-momentum tensor becomes:
\begin{equation}
T^{\alpha\beta}_{(\mathrm{e})} = \rho_{\mathrm{e}} \,u_{(\mathrm{e})}^{\alpha} u_{(\mathrm{e})}^{\beta}, \qquad T^{\alpha\beta}_{(\mathrm{i})} = \rho_{\mathrm{i}}\, u_{(\mathrm{i})}^{\alpha} u_{(\mathrm{i})}^{\beta},\qquad 
T^{\alpha\beta}_{(\mathrm{c})} = \rho_{\mathrm{c}} \,u_{(\mathrm{c})}^{\alpha} u_{(\mathrm{c})}^{\beta}.
\label{COV5}
\end{equation}
The neutrinos are massless in the concordance paradigm and
energy-momentum tensor will have exactly the same form as the one of the photons:
\begin{equation}
T^{\alpha\beta}_{(\nu)} = \frac{4}{3} \rho_{\nu}\, u^{\alpha}_{(\nu)} u^{(\beta)}_{(\nu)} - \frac{\rho_{\nu}}{3} g^{\alpha\beta},\qquad
T^{\alpha\beta}_{(\gamma)} = \frac{4}{3} \rho_{\gamma} \,u^{\alpha}_{(\gamma)} u^{(\beta)}_{(\gamma)} - \frac{\rho_{\gamma}}{3} g^{\alpha\beta}.
\label{COV6}
\end{equation}
In  Eq. (\ref{COV6}) the energy-momentum tensor of the neutrinos should also contain a 
contribution from the anisotropic stress which is fully inhomogeneous and only affects 
the evolution of the relativistic fluctuations of the geometry. 
Finally the energy-momentum tensors of the electromagnetic field and of the dark energy component is given by: 
\begin{equation}
T_{(\mathrm{EM})}^{\alpha\beta} = 
\frac{1}{4\pi} \biggl[ - F^{\alpha\mu} F^{\beta}_{\mu} + \frac{1}{4} 
g^{\alpha\beta} F_{\mu\nu} F^{\mu\nu}\biggr], \qquad T^{\alpha\beta}_{\Lambda} = \rho_{\Lambda} g^{\alpha\beta},
\label{COV7}
\end{equation}
where $F_{\mu\nu}$ is the Maxwell field strength. In Eq. (\ref{COV7}) the dark energy component is parametrized in terms 
of a cosmological constant, as it happens in the context of the concordance paradigm. Thus
 the relativistic fluctuations of the dark energy component are absent.
As soon as we deviate from this choice the dark energy supports 
its own fluctuations. 

The evolution equations of the background follow directly 
by writing Eq. (\ref{COV1}) in the metric of  Eqs. (\ref{OM1}) and they are:
\begin{eqnarray}
&& 3 {\mathcal H}^2 = \ell_{P}^2 a^2 \rho_{\mathrm{t}}, \qquad 2({\mathcal H}^2 - {\mathcal H}' )= \ell_{P}^2 a^2 (p_{\mathrm{t}} + \rho_{\mathrm{t}}),
\label{COV11}\\
&& \rho_{\mathrm{t}}' + 3 {\mathcal H} (\rho_{\mathrm{t}} + p_{\mathrm{t}}) =0,\qquad 
{\mathcal H} = \frac{a^{\prime}}{a},
\label{COV12}
\end{eqnarray}
where the prime denotes a derivation with respect to the conformal time 
coordinate $\tau$; as usual the relation of ${\mathcal H}$ to the standard Hubble rate is given by 
 ${\mathcal H}= a H$ where $H= \dot{a}/a$; note that the overdot denotes a derivation with respect to the cosmic time coordinate $t$. 
 Moreover, by definition of cosmic time coordinate, we also have
 $dt = a(\tau)\, d\tau$. In the paper the derivation with 
 respect to $\tau$ has been also denoted by $\partial_{\tau}$ in all the situations 
 where the use of the prime would lead to potential ambiguities. 
 The total energy density and the total pressure appearing in Eqs. (\ref{COV11}) and (\ref{COV12}) are  
\begin{equation}
\rho_{\mathrm{t}} = \rho_{\mathrm{e}} + \rho_{\mathrm{i}} + \rho_{\gamma} + \rho_{\nu} + \rho_{\mathrm{c}}+  \rho_{\Lambda}, \qquad p_{\mathrm{t}} = \frac{\rho_{\gamma}}{3} + \frac{\rho_{\nu}}{3} -  \rho_{\Lambda}. 
\label{F10a}
\end{equation}
The energy density of the magnetic fields is negligible in comparison with the energy density of the plasma 
but it is not negligible in comparison with the plasma inhomogeneities.
By definition, the isotropic random fields of Eq. (\ref{OM2}) have vanishing mean 
(see appendix \ref{APPA}).  

\subsubsection{Evolution of the electromagnetic fields in curved space}
The evolution of the electromagnetic fields can be summarized in terms 
of the following pair of generally covariant equations: 
\begin{equation}
\nabla_{\mu} F^{\mu\nu} = 4 \pi j^{\nu}, \qquad \nabla_{\mu} \widetilde{F}^{\mu\nu} =0,\qquad \nabla_{\mu} j^{\mu} =0,
\label{COV2COV3}
\end{equation}
where $\widetilde{F}^{\mu\nu}$ is the dual field strength, 
$j^{\nu}$ denotes the (covariantly conserved) total current of the plasma. Equations (\ref{COV2COV3}) 
do not change their form under a Weyl rescaling either when the total current vanishes or whenever the sources 
transform in an appropriate manner. We remind here that a Weyl rescaling
of the four-dimensional metric corresponds to the transformation $G_{\mu\nu} \to g_{\mu\nu} = q(x)\, G_{\mu\nu}$ 
(where $x=(\vec{x},\tau)$ is a generic space-time coordinate). Under a Weyl rescaling the field strength and its dual transform, respectively, as $F^{\mu\nu} = {\mathcal F}^{\mu\nu}/q^2(x)$ and as $\widetilde{F}^{\mu\nu} = \widetilde{{\mathcal F}}^{\mu\nu}/q^2(x)$.  
For instance, in the case of an Ohmic conductor with massless charge carriers
 the total current can be written as $j^{\nu} = \sigma(x) F^{\nu\alpha}\, u_{\alpha}$ where $\sigma(x)$ is the conductivity and Eq. (\ref{COV2COV3}) becomes\footnote{Eq. (\ref{COV9}) is invariant under Weyl rescaling provided the conductivity transforms as 
 $\sigma(x) \to \overline{\sigma}(x) = \sqrt{q(x)} \sigma(x)$ and 
$u_{\alpha}(x) \to \overline{u}_{\alpha}(x) = u_{\alpha}(x)/\sqrt{q(x)}$. Incidentally Eq. (\ref{COV9}) follows 
from the classic Lichnerowicz approach to relativistic magnetohydrodynamics \cite{lich}.}
\begin{equation} 
\partial_{\mu} \biggl[ \sqrt{ - g}\,  F^{\mu\nu} \biggl] = 4 \pi \sqrt{- g} \, \sigma(x) \, F^{\nu\alpha}\, u_{\alpha}, \qquad 
\partial_{\mu} ( \sqrt{ - g}\,  \tilde{F}^{\mu\nu} )=0,
\label{COV9}
\end{equation}
together with the supplementary condition $g^{\alpha\beta} \, u_{\alpha} \, u_{\beta} =1$.  
The gravitating plasma for temperatures smaller than the MeV is not
an Ohmic conductor. Since the masses of the charge carriers dominate against the (approximate) 
temperature of the plasma, the Weyl invariance is not preserved by the total current which
 is due to electrons and ions\footnote{This point is also relevant in an apparently different 
 context, namely the conducting initial conditions of the gauge fields during a quasi-de Sitter stage of 
 expansion \cite{weyl}.} 
\begin{eqnarray}
 j^{\mu} = e \,\widetilde{n}_{\mathrm{i}} u^{\mu}_{(\mathrm{i})} - 
e\, \widetilde{n}_{\mathrm{e}} u^{\mu}_{(\mathrm{e})}, \qquad g_{\mu\nu}\,u^{\mu}_{(\mathrm{e})}\,u^{\nu}_{(\mathrm{e})} = 1, 
\qquad g_{\mu\nu}\,u^{\mu}_{(\mathrm{i})}\,u^{\nu}_{(\mathrm{i})} = 1,
\label{COV8}
\end{eqnarray}
where $e$ denotes the electric charge while $\widetilde{n}_{\mathrm{e}}$ and 
$\widetilde{n}_{\mathrm{i}}$ are the physical concentrations of the electrons and of the ions. 

In a conformally flat background geometry the components of the electromagnetic field strengths expressed
 in terms of the physical electric and magnetic fields are given by $F_{0 i} = - a^2 \, {\mathcal E}_{i}$ and 
 $F_{ij} = - a^2 \epsilon_{ijk} {\mathcal B}^{k}$. Equations (\ref{COV2COV3}) then imply the following form 
of the Maxwell equations:
\begin{eqnarray}
&& \vec{\nabla} \cdot \vec{E} = 4 \pi e (n_{\mathrm{i}} - n_{\mathrm{e}}),\qquad \vec{\nabla} \cdot \vec{B} =0,
\label{S1}\\
&& \vec{\nabla}\times \vec{B} = 4 \pi e (n_{\mathrm{i}}\, \vec{v}_{\mathrm{i}} - 
n_{\mathrm{e}}\, \vec{v}_{\mathrm{e}} ) + \partial_{\tau}\vec{E},\qquad \vec{\nabla} \times \vec{E} = - \partial_{\tau}\vec{B},
\label{S4}
\end{eqnarray}
where the comoving concentrations and the comoving electromagnetic fields are defined as:
\begin{equation}
n_{\mathrm{i}} = a^3 \widetilde{n}_{\mathrm{i}},\qquad 
n_{\mathrm{e}} = a^3 \widetilde{n}_{\mathrm{e}},\qquad \vec{E} = a^2 \vec{{\mathcal E}}, \qquad  
\vec{B} = a^2 \vec{{\mathcal B}}.
\label{S4a}
\end{equation}
In Eq. (\ref{S4}) the peculiar velocities of the electrons and ions are defined as
$u^{k}_{(\mathrm{e})} = u^{0}_{(\mathrm{e})} v_{\mathrm{e}}^{k}$ and by $u^{k}_{(\mathrm{i})} = u^{0}_{(\mathrm{i})} v_{\mathrm{i}}^{k}$ as it follows from the general expression of the four-velocity\footnote{We remind that, by definition,  $u^{i} = dx^{i}/d\lambda = u^{0} v^{i}$ where $\lambda$ is the affine parameter 
and  $u^{0} = d\tau/d\lambda$.}. The peculiar velocity can also be expressed as
$ v^{i} = P^{i}/P^{0}$ where the {\em physical momenta} are often replaced by the {\em comoving three-momenta} $\vec{q}$
defined as:
\begin{eqnarray}
P^{0} = \frac{1}{a^2} \sqrt{m^2 a^2 + q^2}, \qquad  P_{0} = \sqrt{m^2 a^2 + q^2},\qquad q^{i} = a^2 P^{i},
\label{S4c}
\end{eqnarray}
which also implies that $\vec{v} = \vec{q}/\sqrt{q^2 + m^2 a^2}$. 
In the ultrarelativistic limit $\vec{v} = \vec{q}/|\vec{q}|$ (and 
the evolution equations would have the same flat-space-time 
form). Conversely, in the non-relativistic limit, $\vec{v}= \vec{q}/(m a)$ and the 
Weyl invariance is  broken\footnote{The relation of Eq. (\ref{S4c}) 
between physical momenta and comoving momenta neglects the metric 
fluctuations; the inclusion of the metric fluctuations in the relation between 
physical and comoving momenta is crucial for the correct derivation 
of the evolution of the brightness perturbations.}.

\subsubsection{Comoving and physical descriptions}

Since Weyl invariance is broken the plasma descriptions in curved and flat space-time are in general 
not the same. Recalling Eqs. (\ref{hier1}), (\ref{hier2}) and (\ref{hier3}), the plasma 
parameter and the Debye length can be written in terms of the physical concentration $\widetilde{n}_{0}$:
\begin{equation}
\widetilde{g}_{\mathrm{plasma}} = \frac{3}{4 \pi  \widetilde{n}_{0} \widetilde{\lambda}_{\mathrm{D}}^3 x_{\mathrm{e}}}, \qquad 
\widetilde{\lambda}_{\mathrm{D}} =\sqrt{\frac{\widetilde{T}}{8\pi e^2 \widetilde{n}_{0} x_{\mathrm{e}}}},
\label{DEBYE2}
\end{equation}
where the tilde denotes the corresponding physical variable.
For instance the physical temperature and the physical 
concentration are, respectively, $\widetilde{T}= T (a_{0}/a)$ and $\widetilde{n}_{0} = n_{0} (a_{0}/a)^3$; $T$ and $n_{0}$ denote instead the comoving variables\footnote{Note that 
we shall always normalize the scale factor as $a_{0} =1$. This is implies that, at the present time, the comoving 
and the physical values of a given quantity coincide.}. It follows from Eq. (\ref{DEBYE2}) that 
the plasma parameter has the same value in the comoving and in the physical descriptions and it is therefore invariant:
\begin{eqnarray}
 \widetilde{g}_{\mathrm{plasma}} = g_{\mathrm{plasma}} = \frac{3}{4 \pi  n_{0} \lambda_{\mathrm{D}}^3 x_{\mathrm{e}}},\qquad  \widetilde{\lambda}_{\mathrm{D}} = \lambda_{\mathrm{D}} (a/a_{0}), \qquad \lambda_{\mathrm{D}} =\sqrt{\frac{T}{8\pi e^2 n_{0} x_{\mathrm{e}}}},
\label{DEBYE3}
\end{eqnarray} 
where $\lambda_{\mathrm{D}}$ and $g_{\mathrm{plasma}}$ are, respectively, 
the comoving Debye scale and the comoving plasma parameter. 

Because of the difference between comoving and physical three-momenta in the massive 
limit, the plasma  frequencies for electrons and ions can be expressed either 
in comoving or in physical terms:
\begin{equation}
\overline{\omega}_{\mathrm{p}X} = \sqrt{\frac{4\pi n_{\mathrm{i}} e^2}{m_{X} a}}  
\equiv \omega_{\mathrm{p}X} a,
\qquad \omega_{\mathrm{p}X} = \sqrt{\frac{4\pi \tilde{n}_{X} e^2}{m_{X}}},
\label{freq1}
\end{equation}
where $X = \mathrm{i},\, \mathrm{e}$ corresponds either to the electrons or to the ions; moreover 
$\overline{\omega}_{\mathrm{p}X}$ and  $\omega_{\mathrm{p}X}$ denote respectively 
the comoving and the physical frequencies. 
With the same notations the comoving Larmor frequencies for the electrons and for
the ions are instead given by:
\begin{equation}
\overline{\omega}_{\mathrm{B}X} = \frac{e \vec{B}\cdot\hat{n}}{m_{X} a} = \omega_{\mathrm{B}X} a,\qquad  
\omega_{\mathrm{B}X} = \frac{ e \vec{{\mathcal B}}\cdot \hat{n}}{m_{X}},
\label{freq4}
\end{equation}
where the relation between the comoving and the physical magnetic fields is given in Eq. (\ref{S4a}) and $\hat{n}$ denotes 
the magnetic field orientation. A direct consequence of Eqs. (\ref{freq1}) and (\ref{freq4}) is that the explicit expression of 
the comoving Larmor and plasma frequencies depend on the redshift:
\begin{eqnarray}
&& \overline{\omega}_{\mathrm{Be}} = 1.7 \times 10^{-2} \biggl(\frac{\hat{n}\cdot\vec{B}}{\mathrm{nG}}\biggr) (z+1) \, \mathrm{Hz}, \,\,\,
\overline{\omega}_{\mathrm{pe}} = 0.3 \, \sqrt{x_{\mathrm{e}}} \, \biggl( \frac{h_{0}^2 \Omega_{\mathrm{b}0}}{0.022}\biggr)^{1/2} \sqrt{z+1} \,\mathrm{MHz},
\label{FR6}\\
&&\overline{\omega}_{\mathrm{Bi}} = 9.5\times 10^{-6} \biggl(\frac{\hat{n}\cdot\vec{B}}{\mathrm{nG}}\biggr) (z+1) \,\mathrm{Hz}, \,\,\,
\overline{\omega}_{\mathrm{pi}} = 6.6\, \sqrt{x_{\mathrm{e}}} \, \biggl( \frac{h_{0}^2 \Omega_{\mathrm{b}0}}{0.022}\biggr)^{1/2} \sqrt{z+1}\,\mathrm{kHz}.
\label{FR7}
\end{eqnarray}
The use of comoving or physical descriptions depends on the convenience. For instance 
the bounds on the magnetic field intensity are often compiled by using a comoving description.
Conversely  the values of the magnetic field and of the other plasma parameter in the bottom line of
\begin{table}[!ht]
\begin{center}
\begin{tabular}{||l|c|c|c|c|c|c|c|c||||}
\hline
\hline
\rule{0pt}{4ex} Plasma & $\widetilde{n}_{0}[m^{-3}]$ &  $\widetilde{T}$ [keV]& ${\mathcal B}$[G] & $\omega_{\mathrm{pe}}$[Hz]& $\widetilde{\lambda}_{\mathrm{D}}$[m]& $\widetilde{n}_{0} x_{e} \widetilde{\lambda}_{D}^3$& $\widetilde{\Gamma}_{\mathrm{Coul}}$[Hz]\\
\hline
tokamak& $10^{20}$& $10$ &$10^{5}$&$ 10^{11}$ &$ 10^{-5}$& $ 10^{7}$& $10^{4}$  \\
glow discharge& $10^{20}$ & $10^{-3}$ & $10^{3}$ & $6 \times 10^{11}$ & $10^{-7}$& $100$ & $10^{10}$\\
solar corona & $10^{12}$ & $10^{-1}$& $10$ & $10^{7}$ & $10^{-2}$ & $10^{8}$ & $10^{-1}$\\
pre-decoupling & $10^{9}$ & $10^{-3}$ & $< 10^{-2}$& $10^{5}\, \sqrt{x_{e}}$ & $10^{-2}$ & $10^{6}/\sqrt{x_{e}}$ & $10^{-2} x_{e}$\\
\hline
\end{tabular}
\caption{Plasma parameters of some common physical system compared with the pre-decoupling plasma; $x_{e}$ denotes the ionization fraction.}
\label{TABLEA}
\end{center}
\end{table}
 Tab. \ref{TABLEA} are computed for the typical reference temperature of the eV roughly corresponding to the equality between 
matter and radiation occurring at a redshift:
\begin{equation}
1 + z_{\mathrm{eq}} = \frac{a_{0}}{a_{\mathrm{eq}}} = \frac{h_{0}^2 \Omega_{\mathrm{M}0}}{h_{0}^2 
\Omega_{\mathrm{R}0}}= 3228.91 \biggl(\frac{h_{0}^2 \Omega_{\mathrm{M}0}}{0.134}\biggr).
\label{zeq}
\end{equation}
For comparison photon decoupling takes place at a typical redshift $z_{*} = {\mathcal O}(1100)$ (i.e. between $1080$ and $1110$).
In Tab. \ref{TABLEA} we also illustrate the same plasma parameters for other  
examples of highly ionized plasmas. Note that the number of charged carriers
within the Debye sphere is grossly the same for the pre-decoupling plasma, for the solar corona 
and for a tokamak (see second column from the right in Tab. \ref{TABLEA}). Similar 
comparisons can be developed in the case of the other plasma parameters by always 
reminding, as emphasized in Eqs. (\ref{hier1})--(\ref{hier3}) that the various hierarchies are 
controlled either by $g_{\mathrm{plasma}}$ or by its inverse.

\subsubsection{The approximate temperature of the plasma}
The evolution of the approximate temperature of the plasma depends on $g_{\mathrm{plasma}}$. 
Indeed when the plasma contains an equal number of positively and negatively charged species in 
a radiation background its total energy density and pressure are:
\begin{equation}
\overline{\rho}_{\mathrm{tot}} = \rho_{+} + \rho_{-} + \rho_{r}, \qquad 
\overline{p}_{\mathrm{tot}} = p_{+} + p_{-} + p_{r}.
\label{cond1}
\end{equation}
As long as the physical temperatures of the charged species exceed the corresponding masses 
(i.e. $ \widetilde{T}_{\pm} \gg m_{\pm}$), the temperatures 
$\widetilde{T}_{+}$ and $\widetilde{T}_{-}$ approximately coincide with $\widetilde{T}_{r}$ which is, by definition, the temperature of the radiation, i.e. $\widetilde{T}_{+} \simeq \widetilde{T}_{-} \simeq \widetilde{T}_{r}$. 
In the opposite case (i.e. for $\widetilde{T}_{\pm} < m_{\pm}$) the evolution of the various temperatures 
depends on $g_{\mathrm{plasma}}$. From Eqs. (\ref{cond1}) the first principle of the thermodynamics and the adiabaticity of the evolution imply\footnote{The different pressures and energy densities of the charged species are, respectively, $p_{\pm} = \widetilde{n}_{\pm} \widetilde{T}_{\pm}$ and 
$  \rho_{\pm} = m_{\pm} \widetilde{n}_{\pm} + 3\widetilde{n}_{\pm} \, \widetilde{T}_{\pm}/2$. 
For the radiation, assuming ${\mathcal N}_{\mathrm{th}}$  species in approximate thermal equilibrium,
 we have instead $\rho_{r} = \pi^2\, {\mathcal N}_{\mathrm{th}}\, \widetilde{T}_{r}^4/30$ and $p_{r} = \rho_{r}/3$. }:
\begin{equation} 
d \biggl\{ V_{H} \,\biggl[( \widetilde{n}_{+} m_{+} + \widetilde{n}_{-} m_{-}) + \frac{3}{2} \biggl( \widetilde{n}_{+} \widetilde{T}_{+} + 
\widetilde{n}_{-} \widetilde{T}_{-} \biggr) + \rho_{r} \biggr]\biggr\} + \biggl( \widetilde{n}_{+} \widetilde{T}_{+} + 
\widetilde{n}_{-} \widetilde{T}_{-} + p_{r} \biggr) d V_{H} =0,
\label{cond4}
\end{equation}
where $V_{H}(a) = (4 \pi/3) H_{*}^{-3} a^3$ is the fiducial Hubble volume. 
Since the plasma is globally neutral (i.e.  
$\widetilde{n}_{+} = \widetilde{n}_{-} = \widetilde{n}_{0}$),  Eq. (\ref{cond4}) can also be expressed as:
\begin{equation}
d[ a^2 ( \widetilde{T}_{+} + \widetilde{T}_{-})] + a\, \gamma\, d (a \widetilde{T}_{r} ) =0, \qquad \gamma = \frac{2 s}{n_{0}},
\label{cond5}
\end{equation}
where, besides the comoving concentration (i.e. $n_{0} = a^3 \, \widetilde{n}_{0}$) we introduced 
the comoving entropy density  $s =a^3 \, \widetilde{s}$ (with  $\widetilde{s} = 2 \pi^2{\mathcal N}_{\mathrm{th}} \widetilde{T}_{r}^3/45$). The physical initial conditions stipulate that  $\widetilde{T}_{+} \simeq \widetilde{T}_{-} \simeq \widetilde{T}_{r}$ with the result that, thanks to Eq. (\ref{cond5}), 
the common temperature of the different species scales as:
\begin{equation}
\widetilde{T} \simeq a^{- \frac{4 + \gamma}{2 + \gamma}},\qquad \gamma = \frac{4 \pi^4}{45\, \overline{\zeta}(3)} \, \biggl(\frac{n_{r}}{n_{0}}\biggr) = \frac{16}{5} (2\pi)^3\, {\mathcal N}_{\mathrm{th}} \, \biggl(\frac{e^3}{g_{\mathrm{plasma}}}\biggr)^2.
\label{cond6}
\end{equation}
where $\overline{\zeta}(3)$  has been already introduced after Eq. (\ref{gplasma}) and $n_{r} = a^3 \widetilde{n}_{r}$; note that $\widetilde{n}_{r} = {\mathcal N}_{\mathrm{th}} \, \widetilde{T}_{r}^3 \,\overline{\zeta}(3)/\pi^2$.
If $n_{r} \ll n_{0}$, the temperature scales, approximately, as $a^{-2}$ in the opposite case (i.e. $n_{r} \gg n_{0}$) 
the effective temperature evolves, to first order in $1/\gamma$, as $a^{-1}$. Since prior to decoupling $g_{\mathrm{plasma}} \ll 1$ and $\gamma \propto g_{\mathrm{plasma}}^{-2}$, we are exactly in the limit $\gamma \gg 1$.
 
\subsection{Relativistic fluctuations of the geometry}
The relativistic fluctuations of the conformally flat background of Eq. (\ref{OM1}) (i.e.  
$g_{\mu\nu}(\vec{x},\tau) = \overline{g}_{\mu\nu}(\tau) + \delta g_{\mu\nu}(\vec{x},\tau)$) 
can be separated into scalar, vector and tensor modes as originally suggested by Lifshitz \cite{liff,liff2}: 
\begin{equation}
\delta g_{\mu\nu}(\vec{x},\tau) = \delta_{\mathrm{s}} g_{\mu\nu}(\vec{x},\tau) 
+ \delta_{\mathrm{v}} g_{\mu\nu}(\vec{x},\tau) + 
 \delta_{\mathrm{t}} g_{\mu\nu}(\vec{x},\tau),
\label{FLUC1}
\end{equation}
where $\delta_{\mathrm{s}}$, $\delta_{\mathrm{v}}$ and $\delta_{\mathrm{t}}$ denote 
the inhomogeneity preserving, separately, the scalar, vector and tensor nature of the corresponding fluctuations. 
Magnetic random fields affect the evolution of the relativistic 
fluctuations of the geometry and, in particular, of the large-scale curvature 
inhomogeneities. Some relevant aspects of this well known problem will now be swiftly outlined.

\subsubsection{Scalar, vector and tensor modes}
The scalar modes of the geometry are parametrized in terms of four independent functions 
$\psi(\vec{x},\tau)$, $\phi(\vec{x},\tau)$, $F(\vec{x},\tau)$ and $G(\vec{x},\tau)$: 
\begin{eqnarray}
&& \delta_{\mathrm{s}} g_{00}(\vec{x},\tau) = 2 a^2(\tau) \phi(\vec{x},\tau), \qquad \delta_{\mathrm{s}} g_{0i}(\vec{x},\tau) = - a^2(\tau) \partial_{i} F(\vec{x},\tau),
\nonumber\\
&& \delta_{\mathrm{s}} g_{ij}(\vec{x},\tau) = 2 a^2(\tau) [\psi(\vec{x},\tau) \delta_{ij} - \partial_{i}\partial_{j}G(\vec{x},\tau)],
\label{FLUC2}
\end{eqnarray}
The vector modes are described by two independent vectors $Q_{i}(\vec{x},\tau)$ and $W_{i}(\vec{x},\tau)$: 
\begin{equation}
 \delta_{\mathrm{v}} g_{0i}(\vec{x},\tau) = - a^2 Q_{i}(\vec{x},\tau),\qquad \delta_{\mathrm{v}} g_{ij}(\vec{x},\tau) = a^2 \biggl[\partial_{i} W_{j}(\vec{x},\tau) + \partial_{j}W_{i}(\vec{x},\tau)\biggr],
\label{FLUC3}
\end{equation}
subjected to the conditions $\partial_{i} Q^{i} =0$ and $\partial_{i} W^{i} =0$. 
Finally the tensor  modes of the geometry are parametrized in terms of a rank-two tensor in three spatial dimensions, i.e.
\begin{equation}
\delta_{t} g_{ij}(\vec{x},\tau) = - a^2 h_{ij}(\vec{x},\tau), \qquad \partial_{i} h^{i}_{j}(\vec{x},\tau) = h_{i}^{i}(\vec{x},\tau) = 0.
\label{FLUC4}
\end{equation}

For an infinitesimal coordinate shift $x^{\mu} \to \widetilde{x}^{\mu} = x^{\mu} + \epsilon^{\mu}$
the scalar and vector modes of Eqs. (\ref{FLUC2}) and (\ref{FLUC3}) transform according to the Lie derivative 
in the direction $\epsilon^{\mu} = (\epsilon^{0}, \, \epsilon^{i})$. The scalar fluctuations in the tilded 
coordinate system read\footnote{Recalling $\epsilon_{\mu} = a^2(\tau)(\epsilon_{0}, - \epsilon_{i})$, the gauge parameters  $\epsilon_{i}$ can be 
written as the sum of an irrotational part supplemented by a solenoidal contribution (i.e. $\epsilon_{i} = \partial_{i} \epsilon + \zeta_{i}$
where $ \partial_{i} \zeta^{i} =0$) affecting, respectively, the scalar and the vector modes. }
\begin{eqnarray}
&& \phi \to \widetilde{\phi} = \phi - {\cal H} \epsilon_0 - \epsilon_{0}^{\prime} ,\qquad  \psi \to \widetilde{\psi} = \psi + {\cal H} \epsilon_{0},
\label{phipsi}\\
&& F \to \widetilde{F} = F +\epsilon_{0} - \epsilon^{\prime}, \qquad  G \to \widetilde{G} = G - \epsilon.
\label{FE}
\end{eqnarray}
For the sake of simplicity in Eq. (\ref{FE}) the arguments of the various functions have been 
neglected and will be omitted hereunder unless strictly necessary. Following the same notations, the vector modes transform as:
\begin{equation}
 Q_{i} \to \widetilde{Q}_{i} = Q_{i} - \zeta_{i}^{\prime},\qquad W_{i} \to \widetilde{W}_{i}= W_{i} + \zeta_{i}.
\label{QW}
\end{equation}
In the case of the vector modes the gauge choices are extremely limited and while a convenient gauge is $Q_{i}=0$, 
there are two unambiguous gauge-invariant variables that 
arise when combining the fluctuations of the metric with the vector fluctuations of the sources 
(see Eq. (\ref{VEC3}) and discussion thereafter).  In the scalar case the possible gauge choices are more numerous than for the vector modes. 
For instance if $G$ and $F$ are set to zero in Eq. (\ref{FLUC2})  the gauge freedom is 
completely fixed (see Eq. (\ref{FE})) and this choice pins down the conformally Newtonian gauge \cite{bard1} where 
the longitudinal fluctuations of the metric read, in Fourier space,
\begin{equation}
\delta_{\mathrm{s}}\, g_{00}(k,\tau) = 2 a^2 \,\phi(k,\tau),\qquad \delta_{\mathrm{s}} g_{ij} = 2 a^2 \psi(k,\tau) \delta_{ij}.
\label{STR4}
\end{equation}  
By instead setting $\phi$ and $F$ to zero we recover the standard choice of the synchronous 
coordinate system \cite{syn1,bertschingerma} where the metric fluctuations can be written, in Fourier space, as\footnote{In the parametrization of Eq. (\ref{FLUC2})  the fluctuations of the metric are given by $\delta_{\mathrm{s}} g_{ij}(k,\tau) = 2 a^2 (\psi_{\mathrm{S}} \delta_{ij} + k_{i} k_{j} G_{\mathrm{S}})$ implying that 
$\psi_{\mathrm{S}} = - \xi$ and $G_{\mathrm{S}} = (h + 6 \xi)/(2 k^2)$. The parametrization of Eq. (\ref{SYN2}) is more 
standard and this is why we shall stick to it. }  
\cite{bertschingerma}
\begin{equation}
\delta_{\rm s} g_{i j}(k,\tau) = 
a^2(\tau)\biggl[ \hat{k}_{i} \hat{k}_{j} h(k,\tau) + 6 \xi(k,\tau)\biggl(\hat{k}_{i} \hat{k}_{j} - \frac{1}{3} \delta_{ij}\biggr)\biggr], 
\label{SYN2}
\end{equation}
where, as usual, $\hat{k}_{i} = k_{i}/|\vec{k}|$. Finally the third convenient choice for the analyses of large-scale magnetism is the off-diagonal (or uniform curvature) gauge demanding that $\psi = F = 0$ in Eq. (\ref{FLUC2}) \cite{hw1,hw2,NoM1}. 

\subsubsection{Gauge-invariant normal modes of the system}

In the tensor case the normal modes coincide, up to a trivial field redefinition involving the 
scale factor, with the metric fluctuation introduced in Eq. (\ref{FLUC4}). 
The evolution of the tensor modes breaks Weyl invariance and it has been derived 
well before the formulation of the inflationary scenario \cite{grishchuk1,grishchuk2}:
\begin{equation}
h_{ij}^{\prime\prime} + 2 {\mathcal H} h_{ij}^{\prime} - \nabla^2 h_{ij} =0.
\label{NM1}
\end{equation}
In the scalar case the normal modes are the 
curvature perturbations on comoving orthogonal hypersurfaces
\footnote{This gauge is comoving since the velocity 
fluctuation vanishes and it is also orthogonal (i.e. $F=0$) since the off-diagonal fluctuation 
of the metric vanishes.}, conventionally denoted by ${\mathcal R}$. In the comoving orthogonal 
gauge the fluctuations of the spatial curvature correspond to ${\mathcal R}$, i.e. 
$\delta_{\mathrm{s}} ^{(3)}R = - (4/a^2) \nabla^2 {\mathcal R}$.
When the background is  dominated by an irrotational relativistic fluid the evolution of ${\mathcal R}$ is:
\begin{equation}
{\mathcal R}^{\prime\prime} + 2 \frac{z_{\mathrm{t}}^{\prime}}{z_{\mathrm{t}}} {\mathcal R}^{\prime} - c_{\mathrm{st}}^2 \nabla^2 {\mathcal R}=0, \qquad z_{\mathrm{t}} = \frac{a^2 \sqrt{\rho_{\mathrm{t}} + p_{\mathrm{t}}}}{{\mathcal H} \, c_{\mathrm{st}}},
\label{NM2}
\end{equation}
where $c_{\mathrm{st}}^2 = p_{\mathrm{t}}^{\prime}/\rho_{\mathrm{t}}^{\prime}$; note that $\rho_{\mathrm{t}}$ and $p_{\mathrm{t}}$ enter directly the background equations (\ref{COV11}) and (\ref{COV12}). The variable of Eqs. (\ref{NM2}) and (\ref{NMR4}) has been first discussed by Lukash \cite{lukash} (see also \cite{strokov,luk2}) when analyzing the
quantum excitations of an irrotational and relativistic fluid.  The canonical normal mode identified in Ref. \cite{lukash}  is invariant under infinitesimal coordinate transformations as required in the context of the  Bardeen formalism \cite{bard1}. 
If the background is instead dominated by a single scalar field $\varphi$ the analog 
of Eq. (\ref{NM2}) can be written as:
\begin{equation}
{\mathcal R}^{\prime\prime} + 2 \frac{z_{\varphi}^{\prime}}{z_{\varphi}} {\mathcal R}^{\prime} - \nabla^2 {\mathcal R}=0, \qquad
 z_{\varphi}= \frac{a \varphi^{\prime}}{\mathcal H}.
\label{NM3}
\end{equation}
Equation (\ref{NM3}) has been derived in the case of scalar field matter in Refs.  \cite{KS} and  \cite{chibisov}. These analyses follow the same logic of \cite{lukash}  (see Eq. (\ref{NM2})). The normal modes of Eqs. (\ref{NM2}) and (\ref{NM3})  coincide with the (rescaled) curvature perturbations on comoving orthogonal hypersurfaces \cite{br1,bard2}.
Once the curvature perturbations are computed (either from Eq. (\ref{NM2}) or from Eq. (\ref{NM3})) the 
metric fluctuations can be easily derived in a specific gauge. Since ${\mathcal R}$ is gauge-invariant, 
its value is, by definition, the same in any coordinate system even if its expression changes from one gauge 
to the other. For instance, in the synchronous (i.e. Eq. (\ref{SYN2})) and longitudinal (i.e. Eq. (\ref{STR4})) gauges  
the expression of ${\mathcal R}$ is, respectively\begin{equation} 
{\mathcal R}^{(S)} = \frac{{\mathcal H}}{{\mathcal H}^2 - {\mathcal H}'} \xi' + \xi, \qquad {\mathcal R}^{(L)} = - \psi - \frac{{\mathcal H} ( {\mathcal H} \phi + \psi^{\prime})}{{\mathcal H}^2 - {\mathcal H}^{\prime}}.
\label{NMR4}
\end{equation}
Even if the expressions ${\mathcal R}^{(L)}$ and ${\mathcal R}^{(S)}$ of Eq. (\ref{NMR4}) are formally different, the invariance 
under infinitesimal coordinate transformations implies that the values of ${\mathcal R}$ 
computed in different gauges must coincide, i.e. ${\mathcal R}^{(S)} = {\mathcal R}^{(L)} = {\mathcal R}$.

\subsection{The concordance paradigm} 
The $\Lambda$CDM paradigm\footnote{$\Lambda$ stands for the dark energy component and CDM 
refers to the cold dark matter component. A peculiar property of the scenario is 
that the dark energy component does not fluctuate.} is just a useful compromise between 
the available data, the standard cosmological model and the number of ascertainable parameters. 
The turning point shaping the present form of the $\Lambda$CDM scenario has been
the WMAP program with the first analysis\footnote{The first observational evidence of large-scale polarization of the CMB 
has been actually obtained by the DASI (Degree Angular Scale Interferometer) 
experiment \cite{dasi1}.}  of the cross-correlation between the temperature and the 
polarization anisotropies \cite{chone28,chone29}. The position of the first Doppler peak 
in the temperature autocorrelations and the location of the first anti-correlation 
peak of the polarization implied that the source of large-scale inhomogeneities 
accounting for the CMB anisotropies had to be adiabatic and Gaussian fluctuations of the spatial 
curvature \cite{chone28,chone29}. This evidence, subsequently confirmed by the following 
data releases of the WMAP experiment \cite{chone30,chone31,chone32} and by the 
Planck collaboration \cite{chone33a,chone33b,chone34a,chone34am,chone34b}, 
justifies and motivates the current formulation of the 
concordance paradigm where the dominant source of 
large-scale inhomogeneity are the adiabatic curvature perturbations. 
  
\subsubsection{The pivotal parameters}
 The $\Lambda$CDM paradigm is formulated in terms of six 
 pivotal parameters\footnote{The critical fractions are sometimes assigned as 
$\omega_{X0} = h_{0}^2 \Omega_{X0}$ where $X = \gamma,\,\nu,\,b,\,c,\, \Lambda$. 
Even if this way of presenting the parameters is conceptually more sound, we shall avoid such 
a notation which might be confused with the angular frequencies.} that are customarily chosen as follows:
{\it i)} the present critical fraction of baryonic matter  [i.e. $\Omega_{b0}= \rho_{b0}/\rho_{crit}= {\mathcal O}(0.048)$];  
{\it ii)} the present critical fraction of CDM particles, [i.e. $\Omega_{c0}= \rho_{c0}/\rho_{crit}={\mathcal O}(0.26)$];
{\it iii)} the present critical fraction of dark energy, [i.e. $\Omega_{\Lambda}= \rho_{\Lambda}/\rho_{crit}={\mathcal O}(0.7)$];
{\it iv)} the indetermination on the Hubble rate\footnote{In units of $100 \,\mathrm{km} \, \mathrm{Hz}/\mathrm{Mpc}$ the Hubble rate is given by $H_{0} = 100 \, h_{0} \, \mathrm{km} \, 
 \mathrm{Hz}/\mathrm{Mpc}$.} [i.e.  $h_{0}= {\mathcal O}(0.7)$]; {\it v)} the spectral index scalar inhomogeneities [i.e. $n_{\mathrm{s}}= {\mathcal O}(0.967)$];
{\it vi)} the optical depth at reionization [i.e.  $\epsilon_{re} ={\mathcal O}(0.07)$].

The parameters of the $\Lambda$CDM paradigm can be inferred either by considering 
a single class of data (e.g. microwave background observations) or by 
requiring the consistency of the scenario with the three observational data sets represented, generally speaking, 
by the temperature and polarization anisotropies of the microwave background,
 by the extended galaxy surveys (see e.g. \cite{chone35,chone36}) and by the supernova observations (see e.g. \cite{chone37,chone38}). There have been 5 different releases of the WMAP 
 data \cite{chone28,chone29,chone30,chone31,chone32} corresponding to one, 
 three, five, seven and nine years of integrated observations. The various releases led to compatible (but slightly different) determinations of the pivotal parameters of the $\Lambda$CDM paradigm.
A similar comment holds for the two releases of the Planck collaboration \cite{chone33a,chone33b,chone34a,chone34am,chone34b}.
Various terrestrial observations of the temperature and polarization anisotropies have been reported (see 
e.g. \cite{chone39,chone40,chone41,chone41a,chone42,chone42a,chone43,chone44}) and they have been sometimes used 
to infer specific limits on magnetic random fields. 
\begin{table}[!ht]
\begin{center}
\begin{tabular}{||l|c|c|c|c|c|c||}
\hline
\hline
\rule{0pt}{4ex} Data & WMAP5  & WMAP7  & WMAP9 & PLANCK\\
\hline
$\Omega_{b0}$& $0.0441\pm 0.0030$ & $ 0.0449\pm 0.0028$& $0.0463 \pm 0.0024$ &$0.0486 \pm 0.0010$ \\
$\Omega_{c0}$& $0.214\pm 0.027$& $ 0.222\pm0.026$& $0.233\pm0.023 $& $0.2589 \pm 0.0057$\\
$\Omega_{\Lambda}$& $0.742\pm 0.030 $ &$ 0.734\pm 0.029$&$0.721\pm0.025$&$0.6911\pm 0.0062$\\
$H_{0}$ & $71.9^{+2.6}_{-2.7} $ & $71.0\pm 2.5$&$ 70.0\pm 2.2 $ &$67.74 \pm 0.46$\\
$n_{\mathrm{s}}$ & $0.963^{+0.014}_{-0.015}$ & $0.963\pm 0.014$&$ 0.972\pm 0.013$&$0.9667\pm 0.0040$\\
$\epsilon_{re}$ & $0.087\pm 0.017$ & $0.088\pm 0.015$&$0.089\pm0.014$&$0.066 \pm0.012$\\
$r_{T}$ & $< 0.43$& $< 0.36$ & $< 0.34$& $<0.1$\\
\hline
\end{tabular}
\caption{The six pivotal parameters of $\Lambda$CDM scenario. The best fits of the WMAP5, WMAP7 and WMAP9  data alone are compared with the Planck fiducial set of parameters.}
\label{TABLEAA}
\end{center}
\end{table}
In Tab. \ref{TABLEAA} the determination of the various $\Lambda$CDM parameters 
of the last three data releases of the WMAP experiment is compared with the Planck 
fiducial set of parameters. In the last line, for reference,  the limits on the 
tensor-to-scalar ratio have been illustrated.

\subsubsection{Neutrinos, photons and baryons}
The minimal $\Lambda$CDM paradigm (sometimes referred to as the vanilla $\Lambda$CDM scenario) 
assumes that the neutrinos are strictly massless and the tensor modes of the geometry are absent.
The radiation component can therefore be directly computed from the photon and 
from the neutrino temperatures. The energy density of a massless neutrino background is today 
\begin{equation}
\rho_{\nu0} = \frac{21}{8} \biggl(\frac{4}{11}\biggr)^{4/3} \rho_{\gamma0},\qquad h_{0}^2 \Omega_{\nu0} = 
1.68\times 10^{-5},
\label{OMnu}
\end{equation}
while the contribution of the photons is given by  $h_{0}^2 \Omega_{\gamma 0} = 2.47 \times10^{-5}$.
The factor  $(4/11)^{4/3}$ stems 
from the relative reduction of the neutrino (kinetic) temperature (in comparison 
with the photon temperature) after weak interactions fall out of thermal 
equilibrium (see e.g. \cite{naselskyb,weinbergb} and also appendix A of Ref. \cite{primer}).
The photon fraction in the radiation plasma  (i.e $R_{\gamma} = \rho_{\gamma}/\rho_{R}$) and the corresponding neutrino fraction (i.e. $R_{\nu} = \rho_{\nu}/\rho_{R}$) obey, by definition,  $R_{\gamma} = 1 - R_{\nu}$ and $R_{\nu}$ is:
\begin{equation}
R_{\nu} = \frac{\rho_{\nu}}{\rho_{\gamma} + \rho_{\nu}} = \frac{3 \times (7/8)\times (4/11)^{4/3}}{ 1 + 3 \times (7/8)\times (4/11)^{4/3}} = 0.4052,
\label{L6}
\end{equation}
where $3$ counts the degrees of freedom associated with the 
massless neutrino families, $(7/8)$ arises because neutrinos follow 
the Fermi-Dirac statistics. The energy density of radiation in critical units is given today by:
\begin{equation}
h_{0}^2 \Omega_{R0} = h_{0}^2 \Omega_{\gamma0} + h_{0}^2 \Omega_{\nu0} = 4.15\times 10^{-5}.
\label{omegaR1}
\end{equation}
To close the circle, the relative weight of photons and baryons depends on 
the redshift and it is parametrized in terms of $R_{b}(z)$:
\begin{equation}
R_{b}(z) = \frac{3}{4} \frac{\rho_{\mathrm{b}}}{\rho_{\gamma}}  = 
0.664 \biggl(\frac{ h_{0}^2\Omega_{b0}}{0.023}\biggr) \biggl(\frac{1051}{z + 1}\biggr),
\label{Rbdef}
\end{equation}
where $z$ denotes the redshift. Note that the baryon to photon ratio determines the sound speed of the baryon-photon system and the sound horizon, namely:
\begin{equation}
c_{\mathrm{sb}}(\tau) = \frac{1}{\sqrt{3[1 + R_{b}(\tau)]}}, \qquad r_{\mathrm{s}}(\tau_{*})= \int_{0}^{\tau_{*}} d\tau c_{\mathrm{sb}}(\tau).
\label{soundhor}
\end{equation}
The value of the sound speed affects then the relative positions of the Doppler peak in the temperature autocorrelations (in the jargon the $TT$ correlations) and of the first anti-correlation peak of the temperature-polarization power spectrum (customarily referred to as the $TE$ correlations).

\subsubsection{Large-scale inhomogeneities} 
The adiabatic scalar fluctuations are the dominant source of large-scale
inhomogeneties in the $\Lambda$CDM scenario and they are customarily 
introduced in terms of the gauge-invariant curvature perturbation on 
comoving orthogonal hypersurfaces already mentioned in  Eqs. (\ref{NM2}), 
(\ref{NM3}) and (\ref{NMR4}). The scalar random field ${\mathcal R}(\vec{x},\tau)$  
corresponding to the curvature perturbation is described, in Fourier space, by its associated
power spectrum $P_{{\mathcal R}}(k,\tau)$ (see Eqs. (\ref{IRFSC1}), 
(\ref{IRFSC2}) and (\ref{IRFSC3}) of appendix \ref{APPA} for a specific 
discussion of the underlying notations).

The scalar power spectrum is customarily assigned as a power-law 
for typical length-scales larger than the Hubble 
radius at the corresponding time  and well before matter-radiation equality
\footnote{In Fourier space these two requirements translate, respectively, 
into $k\tau <1$  and $ \tau < \tau_{\mathrm{eq}}$ (or equivalently
$z > z_{\mathrm{eq}} = {\mathcal O}(3200)$)}:
\begin{equation}
P_{{\mathcal R}}(k,\tau) = {\mathcal A}_{{\mathcal R}}   \biggl(\frac{k}{k_{\mathrm{p}}}\biggr)^{n_{\mathrm{s}}-1}, \qquad {\mathcal A}_{{\mathcal R}} = {\mathcal O}(2.4)\times 10^{-9}
 \qquad k_{\mathrm{p}} = 0.002 \,\, \mathrm{Mpc}^{-1}.
\label{SPS}
\end{equation}
where $n_{\mathrm{s}}$ is the scalar spectral index already introduced above and tabulated in 
Tab. \ref{TABLEAA} (see the fifth line from the top); ${\mathcal A}_{{\mathcal R}}$ measures 
the amplitude of curvature perturbations at the pivot scale 
$k_{\mathrm{p}}$ and its value is deduced from the overall normalization 
of the temperature autocorrelations. While the value of $k_{\mathrm{p}}$ is largely conventional,
 the choice of Eq. (\ref{SPS}) corresponds to an effective harmonic $\ell_{\mathrm{eff}} \simeq 30$.
In Eq. (\ref{SPS}) the scale invariant limit (sometimes dubbed Harrison-Zeldovich \cite{harrison,zeldovichnovikov} limit) occurs for $n_{\mathrm{s}} \to 1$ (or $(n_{\mathrm{s}} -1) \to 0$). 

For adiabatic initial conditions of the Einstein-Boltzmann hierarchy the $TT$ angular power spectrum has the celebrated first acoustic peak for 
$\ell_{\mathrm{d}}\simeq 220$. The first (anticorrelation) peak of the $TE$ spectra occurs instead for $\ell_{\mathrm{ac}} \simeq 3 \,\ell_{\rm d}/4 < \ell_{\mathrm{d}} \simeq 150$.  As correctly argued already in the first data release of the WMAP experiment \cite{chone28} this is the best 
evidence of the adiabatic nature of large-scale curvature perturbations. 
Recalling Eqs. (\ref{Rbdef}) and (\ref{soundhor}) we have that 
the large-scale contribution to the temperature $TT$ correlation goes, in Fourier space, as $\cos{(k c_{\mathrm{sb}} \tau_{\mathrm{dec}} )}$ where
 $\tau_{\mathrm{dec}}$ denotes approximately the decoupling time (see e.g. \cite{chone28,hus} and also \cite{naselskyb,weinbergb,primer}).
For the same set of adiabatic Cauchy data the $TE$ correlation oscillates, always in Fourier space, as $\sin{( 2 k c_{\mathrm{sb}} \tau_{\mathrm{dec}}) }$.  
Consequently the first (compressional) peak of the temperature autocorrelation corresponds to 
$k c_{\mathrm{sb}} \tau_{\rm dec}\sim \pi$, while  the first peak of the cross-correlation will arise for $k c_{\mathrm{sb}} \tau_{\rm dec}\sim 3\pi/4$ i.e., as anticipated, $\ell_{\mathrm{ac}} \simeq 3\, \ell_{\mathrm{d}}/4$. These analytic results can be obtained by working to first-order in the tight-coupling expansion \cite{pee3,chone28,hus,TC2,TC3}.

From the position of the first anticorrelation peak 
it is possible to derive limits on the contribution of the magnetic random fields 
to the initial conditions of the Einstein-Boltzmann hierarchy \cite{MOD1}. 
The adiabatic nature of large-scale curvature inhomogeneities implies that the fluctuations 
of the specific entropy are either absent or strongly constrained. 
Non-adiabatic (or entropic) fluctuations are easily generated both in the early and in the late Universe and 
have been scrutinized in a number of different context \cite{nad1,nad2}. 
Magnetic random fields have been also studied in connection with the entropic 
fluctuations of the plasma \cite{nad3} (see also section \ref{sec3}). 
Depending on the nature of the entropic solution, one of the most notable results of the 
Planck experiment implies the possibility of a non-adiabatic component smaller than about $2\%$ of the 
adiabatic component \cite{chone34b}.

While the tensor modes of the geometry obeying Eq. (\ref{NM1}) do 
not appear in the minimal $\Lambda$CDM paradigm, 
their power spectrum is customarily assigned as:
\begin{equation}
P_{\mathrm{T}}(k) = {\mathcal A}_{\mathrm{T}} \biggl(\frac{k}{k_{\mathrm{p}}}\biggr)^{n_{\mathrm{T}}}, \qquad 
{\mathcal A}_{\mathrm{T}} = r_{\mathrm{T}} {\mathcal A}_{{\mathcal R}},
\label{int1}
\end{equation}
where $n_{T}$ is the tensor spectral index and $r_{T}$ is often referred to as the tensor-to-scalar ratio 
(see also Eq. (\ref{IRFTEN4}) of appendix \ref{APPA}).
Equation (\ref{int1})  requires the addition of two supplementary parameters: the spectral index and the 
amplitude. If the inflationary phase is driven by a single scalar degree of freedom 
and if the radiation dominance kicks in almost suddenly after inflation, 
the whole tensor contribution can be solely parametrized  in terms of $r_{\mathrm{T}}$. 
The rationale for the latter statement is that  $r_{\mathrm{T}}$ not only determines the tensor amplitude but also, thanks to the algebra obeyed by the slow-roll parameters, the  slope of the tensor power 
spectrum\footnote{ To lowest order in the slow-roll 
expansion, therefore, the tensor spectral index is slightly red and it is related to $r_{\mathrm{T}}$ (and to the slow-roll parameter) as $n_{\mathrm{T}} \simeq - r_{\mathrm{T}}/8 \simeq  - 2 \epsilon$ where, by definition, the slow-roll parameter is $\epsilon = -\dot{H}/H^2$ and it measures the rate of decrease of the Hubble 
parameter during the inflationary epoch.}.
In Tab. \ref{TABLEAA} the upper limits on $r_{T}$ have been tabulated in the seventh line from the top.
The direct measurement of $r_{T}$ can be achieved if the $B$-mode polarization is directly observed.
The first detection of a $B$-mode polarization, coming from the lensing of the CMB anisotropies, has been published by the South Pole Telescope \cite{chone43}. The $B$-mode polarization induced by the lensing of the CMB anisotropies is, however, qualitatively different from the $B$-mode induced by the tensor modes of the geometry\footnote{ 
The Bicep2 experiment \cite{chone43} reported the detection of a primordial $B$-mode component compatible with a tensor to scalar ratio $r_{T}=0.2^{+0.07}_{-0.05}$.  Unfortunately the signal turned out to be affected by a serious contamination of a polarized foreground.}.

\subsection{Magnetized radiative transfer equations}
The electron-photon scattering is customarily discussed without any 
magnetic field \cite{chandra,pera}. However, prior to decoupling the photons 
scatter electrons (an ions) in a magnetized environment. 
Magnetic random fields affect the Stokes parameters as well as 
the evolution of the scalar, vector and tensor 
brightness perturbations \cite{mg2010} (see also \cite{TS1,TS2}).
 
\subsubsection{Jones and Mueller calculus}
\begin{figure}[!ht]
\centering
\includegraphics[height=7cm]{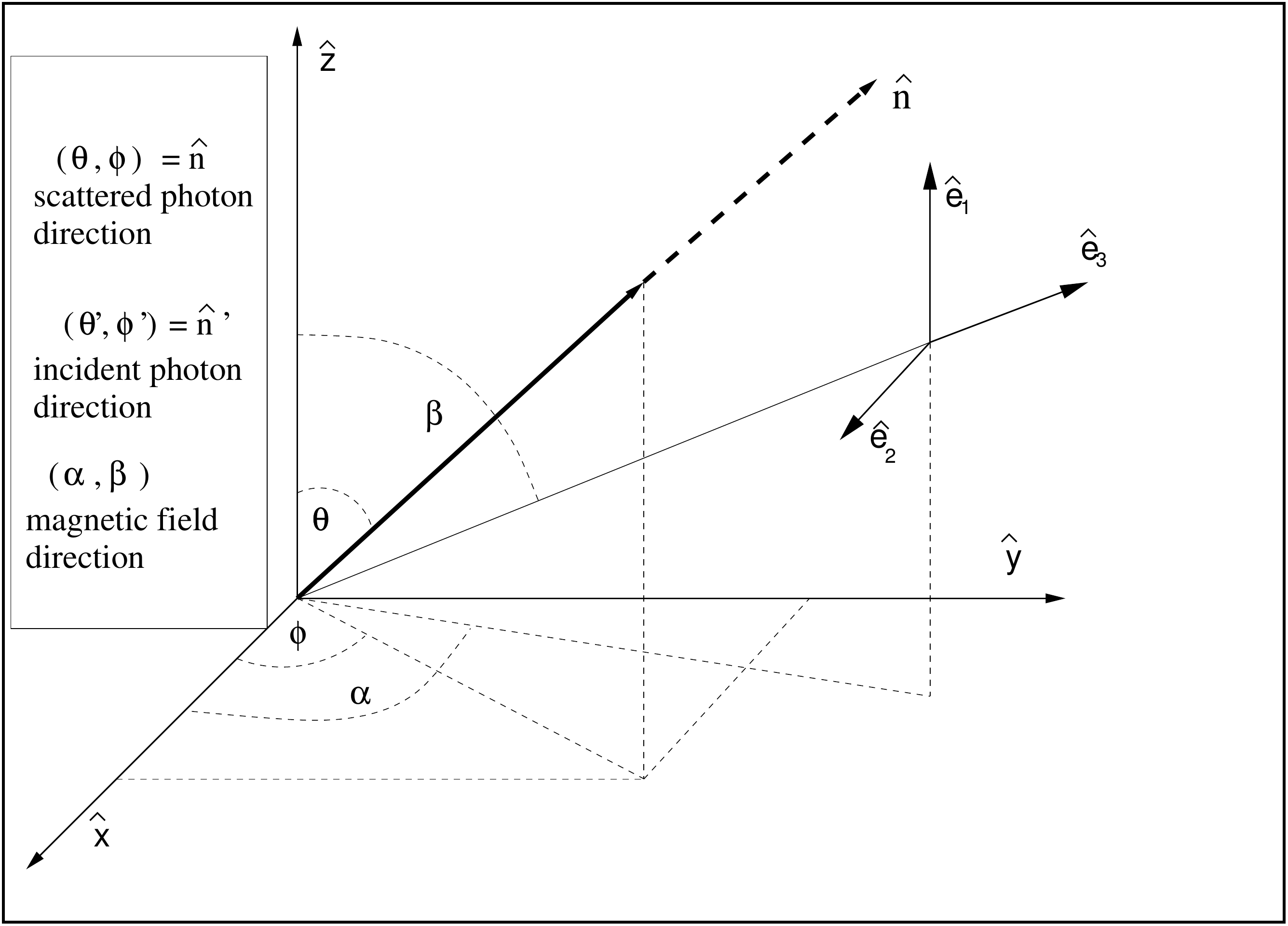}
\caption[a]{Schematic view of the relation between the coordinate system defining the scattered 
radiation field and the reference frame where the magnetic field is oriented along the third axis.}
\label{figure2}      
\end{figure}

The four Stokes parameters can either be organized in a $2\times 2$ matrix (as in Jones calculus) 
or in a column vector with four entries (as in the case of Mueller calculus). 
See Ref. \cite{robson} for an introduction to the Jones and Mueller approaches
 to the polarized radiative transfer equations. In what follows the polarization tensor  
$ {\mathcal P}_{ij}=  {\mathcal P}_{ji}= E_{i}\, E_{j}^{*}$ shall be organized first in a $2\times2$ 
matrix whose explicit form is:
\begin{equation}
{\mathcal P}= \left(\matrix{ I + Q
& U - i V &\cr
U + i V & I - Q &\cr}\right) = \left( I\,{\bf 1} +U\, \sigma_{1} + V\, \sigma_2 +Q\, \sigma_3 \right),
\label{PM}
\end{equation}
where  ${\bf 1}$ denotes the identity matrix while $\sigma_{1}$, $\sigma_{2}$ and 
$\sigma_{3}$ are the three Pauli matrices. The problem depends on $6$ angles: 
{\it i)} $\Omega= (\vartheta,\varphi)$ defines the directions of the scattered photons, 
{\it ii)} $\Omega^{\prime} = (\vartheta',\varphi')$ accounts for the directions of the incident 
photons and {\it iii)} $\Omega^{\prime\prime}=(\alpha,\beta)$ denotes the magnetic field direction\footnote{
Hereunder $\Omega$ and $\Omega^{\prime}$ will denote the angular variables and must not be confused with the energy density in critical units.}. 
The radial, azimuthal and polar directions of the scattered radiation are defined as:
\begin{eqnarray}
&& \hat{r} = (\cos{\varphi} \sin{\vartheta},\, \sin{\varphi} \sin{\vartheta},\, \cos{\vartheta}),
\nonumber\\
&& \hat{\vartheta} = (\cos{\varphi} \cos{\vartheta},\, \sin{\varphi} \cos{\vartheta},\, -\sin{\vartheta}),\qquad  \hat{\varphi} = ( - \sin{\varphi},\, \cos{\varphi},\, 0).
\label{CS1}
\end{eqnarray}
As it can be checked the orientation of the unit vectors is such that $\hat{r} \times \hat{\vartheta} = \hat{\varphi}$.
The spherical polar coordinates of the incident photons are assigned as in Eq. (\ref{CS1}) but the angles $(\vartheta,\varphi)$ are replaced 
by $(\vartheta^{\prime},\varphi^{\prime})$. In Fig. \ref{figure2} the (thick) dashed line denotes the direction of $\hat{n} = (\vartheta,\varphi)$.  
The direction defined by $\hat{e}_{1}$, $\hat{e}_{2}$ and $\hat{e}_{3}$ is determined 
by the angles $\alpha$ and $\beta$ illustrated in Fig. \ref{figure2} and defined as:  
\begin{eqnarray}
&&\hat{e}_{1} = (\cos{\alpha} \cos{\beta},\, \sin{\alpha} \cos{\beta}, \, - \sin{\beta}),
\nonumber\\
&&\hat{e}_{2}= (- \sin{\alpha},\, \cos{\alpha},\, 0),\qquad \hat{e}_{3} 
= (\cos{\alpha} \sin{\beta}, \sin{\alpha} \sin{\beta}, \cos{\beta}).
\label{CS4}
\end{eqnarray}

With these specifications, the evolution of the matrix ${\mathcal P}$ can be formally written as
\begin{equation}
 \frac{d {\mathcal P}}{d\tau} + \epsilon' {\mathcal P} =  \frac{3 \epsilon^{\prime}}{8 \pi} \int\, d\Omega^{\prime}\,  M(\Omega,\Omega',\Omega^{\prime\prime})\, {\mathcal P}(\Omega') \,M^{\dagger}(\Omega,\Omega',\Omega^{\prime\prime}),
\label{EV1}
\end{equation}
where $d\Omega' = \sin{\vartheta^{\prime}}\, d\vartheta^{\prime} \,d\varphi^{\prime}$  and the dagger in Eq. (\ref{EV1}) defines, as usual, the complex conjugate of the transposed matrix. Defining the rate of electron-photon scattering $\Gamma_{\gamma\mathrm{e}}$ we have that the differential optical depth, the Thomson cross section and the classical 
radius of the electron are given, respectively, as
\begin{equation}
\epsilon' =a \Gamma_{\gamma\mathrm{e}} = a \tilde{n}_{\mathrm{0}} x_{\mathrm{e}} \sigma_{\mathrm{e}\gamma}, \qquad \sigma_{\gamma\mathrm{e}} = \frac{8}{3} \pi r_{\mathrm{e}}^2, \qquad r_{\mathrm{e}}= \frac{e^2}{m_{\mathrm{e}}}.
\label{diffop}
\end{equation}
In the differential optical depth  the contribution of the ions is neglected since 
the mass of the ions is much larger than the mass of the electrons; we shall also follow 
this practice even if it is not strictly necessary. 
For reference the explicit components of the matrix $M(\Omega,\Omega',\Omega^{\prime\prime})$ 
are separately reported in Eqs. (\ref{DIP4})--(\ref{DIP5}) and (\ref{DIP6})--(\ref{DIP7}).

\subsubsection{Magnetized electron-photon scattering}
When the photons impinge the electrons or ions in a magnetized environment, the magnetic field can be treated 
in the guiding centre approximation  \cite{guide1,guide2} already mentioned in connection 
with Eq. (\ref{hier4}). Denoting with $\vec{B}(\vec{x},\tau)$ the comoving magnetic field intensity, 
the guiding centre approximation stipulates that
\begin{equation}
B_{i}(\vec{x},\tau) \simeq B_{i}(\vec{x}_{0}, \tau)  + (x^{j} - x_{0}^{j}) \partial_{j} B_{i} +...
\label{CS2}
\end{equation}
where the ellipses stand for the higher orders in the gradients leading, both, to curvature and drift corrections. 
The wavelength of the incident radiation at the decoupling epoch (i.e. $\lambda^{(\mathrm{rec})}_{\gamma}= {\mathcal O}(\mu\mathrm{m})$) is much smaller than $L=|\vec{x} - \vec{x}_{0}|$ and $L_{0}=|\vec{x}_{0}|$ and it is also much smaller than the Hubble radius (see Eq. (\ref{hier4})). 
The perturbative expansion of  Eq. (\ref{CS2}) holds provided the scale of variation of the magnetic field is much larger than the Larmor radius. If this is the case the spatial gradients can be approximately neglected. 

The scattered electric field are then computed as in the standard case 
by superimposing the scattered electric fields due to the electrons and to the ions:
\begin{equation} 
\vec{E}^{(\mathrm{out})}_{(\mathrm{e})} = - e \frac{\vec{r} \times [\vec{r} \times \vec{a}_{(\mathrm{e})}]}{r^3},\qquad 
\vec{E}^{\mathrm{out}}_{(\mathrm{i})} =  e \frac{\vec{r} \times[\vec{r} \times \vec{a}_{(\mathrm{i})}]}{r^3},
\label{sc1}
\end{equation}
where $\vec{a}_{(\mathrm{e})}$ and $\vec{a}_{(\mathrm{i})}$ are the corresponding accelerations. 
Thus the outgoing electric field can also be written as:
\begin{equation}
\vec{E}^{(\mathrm{out})} = \vec{E}^{(\mathrm{out})}_{(\mathrm{e})}  + \vec{E}^{\mathrm{out}}_{(\mathrm{i})}=- e \frac{\vec{r} \times [\vec{r} \times \vec{A}]}{r^3} \equiv - \frac{e}{r} \biggl[ \hat{r} (\vec{A}\cdot\hat{r}) - \vec{A} \biggr],
\label{sc1a}
\end{equation}
where  the vector $\vec{A} = (\vec{a}_{(\mathrm{e})}- 
\vec{a}_{(\mathrm{i})})$ in the frame of Eq. (\ref{CS4}) can be decomposed as 
$\vec{A} = (A_{1} \hat{e}_{1} + A_{2} \hat{e}_{2} + A_{3} \hat{e}_{3})$. 
Denoting by $E_{1}= (\vec{E}\cdot\hat{e}_{1})$, $E_{2} = (\vec{E} \cdot\hat{e}_{2})$ and $E_{3} =(\vec{E} \cdot\hat{e}_{2})$ the components 
of the electric fields of the incident radiation in the local frame, we have from the geodesics of electrons and 
ions\footnote{For the calculation of the scattering matrix the magnetic field can be aligned along $\hat{e}_{3}$ but since 
$\hat{e}_{3}$ has arbitrary orientation with respect to the fixed coordinate system, the magnetic field itself 
will have an arbitrary orientation parametrized by $\alpha$ and $\beta$.}: 
\begin{eqnarray}
&& A_{1} =  
\frac{\overline{\omega}_{\mathrm{pe}}^2}{4 \pi  n_{0}} \, \zeta(\overline{\omega}) \biggl[ \Lambda_{1}(\overline{\omega}) E_{1} - i f_{\mathrm{e}}(\overline{\omega}) \Lambda_{2}(\overline{\omega}) E_{2} \biggr],
\label{sc2}\\
&& A_{2} = \frac{\overline{\omega}_{\mathrm{pe}}^2}{4 \pi  n_{0}} \, 
\zeta(\overline{\omega}) \biggl[ \Lambda_{1}(\overline{\omega}) E_{2} + i f_{\mathrm{e}}(\overline{\omega}) \Lambda_{2}(\overline{\omega})\,E_{1} \biggr],\qquad  A_{3} =  - \frac{\overline{\omega}_{\mathrm{pe}}^2}{4 \pi  n_{0}} \Lambda_{3}(\overline{\omega}) E_{3},
\label{sc4}
\end{eqnarray}
where, as previously remarked (see Eqs. (\ref{freq1}) and (\ref{freq4})), 
 $\overline{\omega}_{\mathrm{Be,\,i}}$ and $\overline{\omega}_{\mathrm{pe,\,i}}$ 
denote, respectively, the comoving Larmor and plasma frequencies for electrons (and ions). 
In Eqs. (\ref{sc2}) and (\ref{sc4}) the functions $\Lambda_{i}(\overline{\omega})$ (with $i = 1, 2, 3$) 
as well as $\zeta(\overline{\omega})$  all depend upon the comoving angular frequency of the photon:
\begin{eqnarray}
&&\Lambda_{1}(\overline{\omega}) = 1 + 
\biggl(\frac{\overline{\omega}^2_{\mathrm{pi}}}{\overline{\omega}^2_{\mathrm{pe}}}\biggr) \biggl( 
\frac{ \overline{\omega}^2 - \overline{\omega}^2_{\mathrm{Be}}}{\overline{\omega}^2 - \omega^2_{\mathrm{Bi}}}\biggr),
\qquad \Lambda_{2}(\overline{\omega}) = 1 - \biggl(\frac{\overline{\omega}^2_{\mathrm{pi}}}{\overline{\omega}^2_{\mathrm{pe}}}\biggr)
\biggl(\frac{\overline{\omega}_{\mathrm{Bi}}}{\overline{\omega}_{\mathrm{Be}}}\biggr) \biggl( 
\frac{ \overline{\omega}^2 - \overline{\omega}_{\mathrm{Be}}^2}{\overline{\omega}^2 - \overline{\omega}^2_{\mathrm{Bi}}}\biggr),
\nonumber\\
&& \Lambda_{3}(\overline{\omega}) = 1 + \biggl(\frac{\overline{\omega}^2_{\mathrm{pi}}}{\overline{\omega}^2_{\mathrm{pe}}}\biggr), \qquad  \zeta(\overline{\omega}) = \frac{\overline{\omega}^2}{\overline{\omega}_{\mathrm{Be}}^2 - \overline{\omega}^2} = \frac{1}{f_{\mathrm{e}}^2(\overline{\omega}) -1}.
\label{LAM3}
\end{eqnarray}
Recalling the discussion of the comoving plasma and Larmor frequencies of Eqs. (\ref{FR6})--(\ref{FR7})  
the numerical value of $f_{\mathrm{e}}(\overline{\omega})$ for typical cosmological parameters is given by 
\begin{equation}
f_{\mathrm{e}}(\overline{\omega}) = \biggl(\frac{\overline{\omega}_{\mathrm{Be}}}{\overline{\omega}}\biggr)
 = 2.79 \times 10^{-12} \biggl(\frac{B}{\mathrm{nG}}\biggr)
 \biggl(\frac{\mathrm{GHz}}{\overline{\nu}}\biggr) (z+1),
 \label{FEa}
 \end{equation}
where  $B= |\hat{e}_{3}\cdot\vec{B}|$. Using Eq. (\ref{sc1a}) and thanks to Eqs. (\ref{sc2})--(\ref{sc4}) the relation between the outgoing and the ingoing electric fields is given by: 
\begin{eqnarray}
&& E^{(\mathrm{out})}_{\vartheta}(\Omega, \Omega^{\prime},\Omega^{\prime\prime}) = \frac{r_{\mathrm{e}}}{r} \biggl[ M_{\vartheta\vartheta}(\Omega, \Omega^{\prime},\Omega^{\prime\prime})  E^{(\mathrm{in})}_{\vartheta}(\Omega^{\prime}) + 
M_{\vartheta\varphi}(\Omega, \Omega^{\prime},\Omega^{\prime\prime}) E^{(\mathrm{in})}_{\varphi}(\Omega^{\prime})  \biggr],
\nonumber\\
&& E^{(\mathrm{out})}_{\varphi}(\Omega, \Omega^{\prime},\Omega^{\prime\prime})  = \frac{r_{\mathrm{e}}}{r} \biggl[ M_{\varphi\vartheta}(\Omega, \Omega^{\prime},\Omega^{\prime\prime}) E^{(\mathrm{in})}_{\vartheta}(\Omega^{\prime}) + M_{\varphi\varphi}(\Omega, \Omega^{\prime},\Omega^{\prime\prime}) 
E^{(\mathrm{in})}_{\varphi}(\Omega^{\prime})  \biggr],
\label{electric}
\end{eqnarray}
where, $M_{ij}(\Omega,\Omega^{\prime},\Omega^{\prime\prime})$ are the components of the matrix $M(\Omega,\Omega^{\prime},\Omega^{\prime\prime})$ appearing in Eq. (\ref{EV1})
 and defined as:
\begin{equation}
M(\Omega,\Omega^{\prime},\Omega^{\prime\prime}) = \left(\matrix{ M_{\vartheta\varphi}(\Omega,\Omega^{\prime},\Omega^{\prime\prime})
&M_{\vartheta\vartheta}(\Omega,\Omega^{\prime},\Omega^{\prime\prime}) &\cr
M_{\varphi\vartheta}(\Omega,\Omega^{\prime},\Omega^{\prime\prime}) &M_{\varphi\varphi}(\Omega,\Omega^{\prime},\Omega^{\prime\prime})&\cr}\right).
\label{LIMM}
\end{equation}
The explicit expressions of $M_{ij}(\Omega,\Omega^{\prime},\Omega^{\prime\prime})$ (see Eqs. (\ref{DIP4})--(\ref{DIP6})) in the limit of vanishing magnetic field  imply\footnote{ In the limit of vanishing magnetic field we have 
$f_{e} \to 0$; moreover, from Eq. (\ref{LAM3}),  $\Lambda_{1} \to 1$,  $\Lambda_{2} \to 1$, $\Lambda_{3} \to 1$ and $\zeta \to -1$.}:
\begin{eqnarray}
&& M_{\vartheta\vartheta}(\mu,\varphi,\nu,\varphi') = - \sqrt{ 1 - \mu^2} \sqrt{1 - \nu^2} - \mu \nu \cos{(\varphi'- \varphi)},
\nonumber\\
&& M_{\vartheta\varphi}(\mu,\varphi,\nu,\varphi') = \mu \sin{(\varphi' - \varphi)}, \qquad  M_{\varphi\vartheta}(\mu,\varphi,\nu,\varphi')  = - \nu \sin{(\varphi' - \varphi)},
\nonumber\\
&& M_{\varphi\varphi}(\mu,\varphi,\nu,\varphi') = - \cos{(\varphi' - \varphi)},
\label{LIM2}
\end{eqnarray}
where $\mu = \cos{\vartheta}$ and $\nu = \cos{\vartheta'}$. Equation (\ref{LIM2}), modulo the different 
conventions, coincides exactly with the standard result (see e.g. \cite{chandra}). 

\subsubsection{Magnetized brightness perturbations}
From Eq. (\ref{EV1}), using the explicit form of the matrix elements 
of Eqs. (\ref{DIP4})--(\ref{DIP5}) and (\ref{DIP6})--(\ref{DIP7}) 
it is straightforward to obtain the evolution 
column matrix ${\mathcal I}$ whose entries are
the four Stokes parameters (i.e. respectively, $I$, $Q$, $U$ and $V$) \cite{robson}:
 \begin{equation}
 \frac{d {\mathcal I}}{d\tau} + \epsilon' {\mathcal I} = \frac{3 \epsilon^{\prime}}{16\pi} \int d\Omega' \,\, {\mathcal T}(\Omega, \Omega',\Omega^{\prime\prime}) {\mathcal I}(\Omega'),
 \label{EV2}
 \end{equation}
where ${\mathcal T}(\Omega,\Omega',\alpha,\beta)$ is a  $4\times4$ matrix derived from the matrix elements of  Eqs. (\ref{DIP4})--(\ref{DIP5}) and (\ref{DIP6})--(\ref{DIP7}). A different form of the evolution equations (\ref{EV2}) involves a further average over the local magnetic field direction:
 \begin{equation}
 \frac{d \overline{{\mathcal I}}}{d\tau} + \epsilon' \overline{{\mathcal I}} = \frac{3 \epsilon^{\prime}}{16\pi} \int d\Omega' \,d\Omega^{\prime\prime}\, {\mathcal T}(\Omega, \Omega',\Omega^{\prime\prime}) {\mathcal I}(\Omega').
 \label{EV2a}
 \end{equation}
The scalar, vector and tensor fluctuations of the geometry of Eqs. (\ref{FLUC2}), (\ref{FLUC3}) and (\ref{FLUC4}) 
contribute to the evolution of the total brightness perturbation defined as:
\begin{equation}
 \Delta_{X}(\vec{x},\tau) = \Delta^{(\mathrm{s})}_{X}(\vec{x},\tau) + \Delta^{(\mathrm{v})}_{X}(\vec{x},\tau) + \Delta^{(\mathrm{t})}_{X}(\vec{x},\tau),
\label{BRDEC1}
\end{equation}
where $X=I,\,Q,\,U,\,V$ denotes, generically, one of the four Stokes parameters and where the superscripts refer, respectively, to the scalar, vector and tensor modes of the geometry. In the case of the intensity the relation between the Stokes parameters and the brightness perturbations is
\begin{equation}
I(\vec{x},\tau, q, \hat{n})= f_{0}(q) \biggl[ 1 - \frac{\partial \ln{f_{0}}}{\partial \ln{q}} \Delta_{\mathrm{I}}(\vec{x},\tau, \hat{n})\biggr],
\label{add1}
\end{equation}
where $f_{0}(q)$ is the (unperturbed) Bose-Einstein distribution, $q$ is the modulus of the comoving three-momentum (see Eq. (\ref{S4c})) and $\hat{n}$ denotes, as usual, the direction of the photon. Note that $\Delta_{\mathrm{I}}$ does not depend on $q$ and this is the advantage of using $\Delta_{\mathrm{I}}$ 
instead of  using directly the fluctuation of the intensity in the form  $I(\vec{x},\tau, q, \hat{n})= f_{0}(q)[ 1 +f^{(1)}(\vec{x},\tau, q, \hat{n})]$. Similar notations will be used for the remaining Stokes parameters. 

The collisionless and the collisional contributions can be separately treated.
As an example, in the case of the intensity, the collisionless contribution is\footnote{To avoid notational confusions, the partial derivations with respect to $\tau$ (customarily denoted with  a prime in the other sections) will be denoted by $\partial_{\tau}$; the partial derivations with respect to the spatial coordinates 
will be instead denoted by $\partial_{i}$ with $i= 1,\,2,\,3$.}:
\begin{eqnarray}
&& {\mathcal L}_{I}^{(\mathrm{s})}(\hat{n},\vec{x},\tau) = \partial_{\tau} \Delta^{(\mathrm{s})}_{\mathrm{I}}  + \hat{n}^{i} \partial_{i} \Delta^{(\mathrm{s})}_{\mathrm{I}} + \epsilon' \Delta^{(\mathrm{s})}_{\mathrm{I}} + \frac{1}{q} \biggl(\frac{d q}{d\tau}\biggr)_{\mathrm{s}},
\label{BRDEC6}\\
&& {\mathcal L}_{I}^{(\mathrm{v})}(\hat{n}, \vec{x},\tau) = \partial_{\tau} \Delta^{(\mathrm{v})}_{\mathrm{I}} + \hat{n}^{i} \partial_{i} \Delta^{(\mathrm{v})}_{\mathrm{I}} + \epsilon' \Delta^{(\mathrm{v})}_{\mathrm{I}} + \frac{1}{q}\biggl(\frac{d q}{d\tau}\biggr)_{\mathrm{v}},
\label{BRDEC7}\\
&&  {\mathcal L}_{I}^{(\mathrm{t})}(\hat{n},\vec{x},\tau) = \partial_{\tau} \Delta^{(\mathrm{t})}_{\mathrm{I}} + \hat{n}^{i} \partial_{i} \Delta^{(\mathrm{t})}_{\mathrm{I}} + \epsilon' \Delta^{(\mathrm{t})}_{\mathrm{I}} + \frac{1}{q} \biggl(\frac{d q}{d\tau}\biggr)_{\mathrm{t}},
\label{BRDEC8}
\end{eqnarray}
where $q = \hat{n}_{i} q^{i}$ is the modulus of the comoving three-momentum 
whose derivative with respect to $\tau$ depends on the scalar, vector and tensor fluctuations of the metric:
\begin{eqnarray}
&& \biggl(\frac{d q}{d\tau}\biggr)_{\mathrm{s}} = - q \partial_{\tau} \psi + q \hat{n}^{i} \partial_{i} \phi, 
\label{BRDEC9a}\\
&& \biggl(\frac{d q}{d\tau}\biggr)_{\mathrm{v}} = \frac{q}{2} \hat{n}^{i} \hat{n}^{j} (\partial_{i} \partial_{\tau}W_{j} + \partial_{\tau}\partial_{j} W_{i}),\qquad  \biggl(\frac{d q}{d\tau}\biggr)_{\mathrm{t}} = - \frac{q}{2} \, \hat{n}^{i}\, \hat{n}^{j} \, \partial_{\tau}h_{ij}.
\label{BRDEC10}
\end{eqnarray}
In Fourier space the evolution of the brightness perturbations can be expressed, in the scalar case, as an expansion in $f_{\mathrm{e}}(\overline{\omega})$:
\begin{eqnarray}
&& \partial_{\tau} \Delta^{(\mathrm{s})}_{\mathrm{I}}
+ ( i k\mu + \epsilon') \Delta^{(\mathrm{s})}_{\mathrm{I}} = \partial_{\tau} \psi - i k \mu \phi + \epsilon' {\mathcal A}_{I} + \epsilon' \, f_{\mathrm{e}}(\overline{\omega}) \, 
{\mathcal B}_{I} + \epsilon' \, f_{\mathrm{e}}^2(\overline{\omega})  \, {\mathcal C}_{I}, 
\label{BRI}\\
&&\partial_{\tau} \Delta^{(\mathrm{s})}_{\mathrm{Q}} + ( i k\mu + \epsilon') \Delta^{(\mathrm{s})}_{\mathrm{Q}} =  \epsilon' {\mathcal A}_{Q} + \epsilon' \, f_{\mathrm{e}}(\overline{\omega}) \, 
{\mathcal B}_{Q} + \epsilon' \, f_{\mathrm{e}}^2(\overline{\omega})  \, {\mathcal C}_{Q},
\label{BRQ}\\
&&   \partial_{\tau} \Delta^{(\mathrm{s})}_{\mathrm{U}} + ( i k\mu + \epsilon') \Delta^{(\mathrm{s})}_{\mathrm{U}} =  \epsilon' {\mathcal A}_{U} + \epsilon' \, f_{\mathrm{e}}(\overline{\omega})  \, 
{\mathcal B}_{U} + \epsilon' \, f_{\mathrm{e}}^2(\overline{\omega})  \, {\mathcal C}_{U},
\label{BRU}\\
&&  \partial_{\tau} \Delta^{(\mathrm{s})}_{\mathrm{V}}+ ( i k\mu + \epsilon') \Delta^{(\mathrm{s})}_{\mathrm{V}} =  \epsilon' {\mathcal A}_{V} + \epsilon' \, f_{\mathrm{e}}(\overline{\omega})  \, 
{\mathcal B}_{V} + \epsilon' \, f_{\mathrm{e}}^2(\overline{\omega})  \, {\mathcal C}_{V},
\label{BRV}
\end{eqnarray}
where, for $X= I,\,Q,\, U,\, V$. Note that ${\mathcal A}_{X}$ denotes the leading order result of the expansion, ${\mathcal B}_{X}$ denotes the next-to-leading order (NLO) correction while 
${\mathcal C}_{X}$ denotes the next-to-next-to-leading (NNLO) term \cite{mg2010}.

The terms at the left hand sides of Eqs. (\ref{BRI})--(\ref{BRQ}) and (\ref{BRU})--(\ref{BRV}) 
are obtained by integrating the collisional contributions over  $\nu = \cos{\vartheta^{\prime}}$ and over $\varphi^{\prime}$.
Following the standard practice, to facilitate the integration over $\nu$ of the collisional terms 
the four brightness perturbations have been expanded in a series of Legendre polynomials $P_{\ell}(\nu)$ as
\begin{equation}
\Delta_{X}(\nu,k,\tau) = \sum_{\ell=0}^{\infty} (-i)^{\ell} (2\ell + 1) \, P_{\ell}(\nu)\, \Delta_{X\,\ell}(k,\tau).
\label{r1a}
\end{equation}
Denoting by $S_{\mathrm{P}} =  (\Delta_{I 2} + \Delta_{Q 0} + \Delta_{Q 2})$ the usual combination of the quadrupole of the intensity and of the monopole and quadrupole of the linear polarization,
the  leading order contribution for the for brightness perturbations appearing in  Eqs. (\ref{BRI})--(\ref{BRQ}) and (\ref{BRU})--(\ref{BRV})  is:
\begin{eqnarray}
{\mathcal A}_{I} &=& \Delta_{I 0} + \mu v_{\mathrm{b}} - \frac{P_{2}(\mu)}{2} S_{\mathrm{P}},\qquad  {\mathcal A}_{V} = - \frac{3}{2} \, i \,\mu\, \Delta_{V 1},
\label{AI}\\
{\mathcal A}_{Q} &=& \frac{3}{4} (1 - \mu^2) S_{\mathrm{P}}, \qquad {\mathcal A}_{U} = 0,
\label{AVAU}
\end{eqnarray}
where the notation $\vec{v}^{(\mathrm{s})}= \vec{k} v_{\mathrm{b}}$ has been 
employed for the scalar component of the Doppler term. The NLO and the NNLO are rather lengthy and can be found in \cite{mg2010}. 

Another useful way of presenting the results for the magnetized perturbations is to average 
the source terms over the magnetic field directions by integrating over $\alpha$ and $\beta$ the collisional terms, as suggested in Eq. (\ref{EV2a}). The 
result of this further integration is:
\begin{eqnarray}
 \partial_{\tau} \overline{\Delta}^{(\mathrm{s})}_{\mathrm{I}}
+ ( i k\mu + \epsilon') \overline{\Delta}^{(\mathrm{s})}_{\mathrm{I}} &=& \partial_{\tau} \psi - i k \mu \phi + \epsilon' \biggl[ \overline{\Delta}_{I 0} + \mu v_{\mathrm{b}} - \frac{P_{2}(\mu)}{2} \overline{S}_{\mathrm{P}} 
\nonumber\\
&+& f_{\mathrm{e}}^2 \biggl(\frac{2}{3} \overline{\Delta}_{I 0} + \frac{P_{2}(\mu)}{6} \overline{S}_{\mathrm{P}}\biggr)\biggr]
\label{AVI}\\
\partial_{\tau} \overline{\Delta}^{(\mathrm{s})}_{\mathrm{Q}} + ( i k\mu + \epsilon') \overline{\Delta}^{(\mathrm{s})}_{\mathrm{Q}} &=&  \epsilon'  \frac{(f_{\mathrm{e}}^2 - 3) (\mu^2 -1)}{4} \overline{S}_{\mathrm{P}},
\label{AVQ}\\
\partial_{\tau} \overline{\Delta}^{(\mathrm{s})}_{\mathrm{U}} + ( i k\mu + \epsilon') \overline{\Delta}^{(\mathrm{s})}_{\mathrm{U}} &=&  0,
\label{AVU}\\
\partial_{\tau} \overline{\Delta}^{(\mathrm{s})}_{\mathrm{V}}+ ( i k\mu + \epsilon') \overline{\Delta}^{(\mathrm{s})}_{\mathrm{V}} &=& - \frac{i \, \epsilon'}{2}( 3 + f_{\mathrm{e}}^2) \overline{\Delta}_{V 1}.
\label{AVV}
\end{eqnarray}

The same discussion presented in the scalar case can also be carried on in the vector and tensor cases \cite{mg2010}.
Consider first the case where the propagation of the long-wavelength gravitational wave is parallel to the direction of the magnetic field intensity 
(i.e.  $\alpha = \beta =0$ and $\hat{k} \parallel \hat{e}_{3}$). The azimuthal dependence can be decoupled from the radial 
dependence and the brightness perturbations will be: 
\begin{eqnarray}
\Delta_{\mathrm{I}}^{(\mathrm{t})}(\varphi,\mu,k,\tau) &=& (1 - \mu^2) \biggl[ \cos{2 \varphi} \,{\mathcal Z}_{\oplus}(\mu,k,\tau) + \sin{2 \varphi} \,{\mathcal Z}_{\otimes}(\mu, k, \tau)\biggr],
\label{T7}\\
\Delta_{\mathrm{Q}}^{(\mathrm{t})}(\varphi,\mu,k,\tau) &=& (1 + \mu^2) \biggl[ \cos{2 \varphi}\, {\mathcal T}_{\oplus}(\mu,k,\tau) + \sin{2 \varphi}\, {\mathcal T}_{\otimes}(\mu,k,\tau)\biggr],
\label{T8}\\
\Delta_{\mathrm{U}}^{(\mathrm{t})}(\varphi,\mu,k,\tau) &=& 2 \mu \biggl[ -\sin{2 \varphi} \, {\mathcal T}_{\oplus}(\mu,k,\tau) + \cos{2 \varphi}\, {\mathcal T}_{\otimes}(\mu,k,\tau)\biggr],
\label{T9}\\
\Delta_{\mathrm{V}}^{(\mathrm{t})}(\varphi,\mu,k,\tau) &=& 2 \mu \biggl[ \cos{2 \varphi} \, {\mathcal S}_{\oplus}(\mu,k,\tau) + \sin{2 \varphi}\, {\mathcal S}_{\otimes}(\mu,k,\tau)\biggr].
\label{T10}
\end{eqnarray}
The particular angular dependence of the brightness perturbations 
is fixed by the contribution of the tensor mode of the geometry to the collisionless part of the Boltzmann equation. According to Eq. (\ref{BRDEC10}),
this contribution is proportional, in Fourier space, to $\hat{n}^{i} \hat{n}^{j} \partial_{\tau} h_{i j}(\vec{k},\tau)$ where $\hat{n}$, as usual, denotes 
the outgoing photon direction. By recalling the explicit form of the two polarizations of the gravitational wave (see Eq. (\ref{IRFTEN8}))
we have, without much effort, that 
\begin{equation}
\hat{n}^{i} \hat{n}^{j} \partial_{\tau} h_{i j}(\vec{k},\tau) =  [ (\hat{n}\cdot\hat{a})^2 
- (\hat{n}\cdot\hat{b})^2] \partial_{\tau} h_{\oplus}(\vec{k},\tau) + 2 (\hat{n}\cdot\hat{a}) (\hat{n} \cdot \hat{b})
\partial_{\tau} h_{\otimes}(\vec{k},\tau)
\label{T10a}
\end{equation} 
where $\hat{a}$ and $\hat{b}$ are two unit vectors orthogonal to $\hat{k}$ and mutually orthogonal. 
Equation (\ref{T10a}) has the same azimuthal dependence of Eq. (\ref{T7}) which is just written in more explicit terms.
 The other expressions of Eqs. (\ref{T8}), (\ref{T9}) and (\ref{T10}) directly follow by consistency with the other evolution equations 
for the brightness perturbations. After some algebra, the evolution of ${\mathcal Z}$, ${\mathcal T}$ and 
${\mathcal S}$ becomes:
\begin{eqnarray}
&&\partial_{\tau} {\mathcal Z} + (i k\mu + \epsilon') {\mathcal Z} - \frac{1}{2}\partial_{\tau} h   =  
\epsilon' \zeta^2(\overline{\omega}) [\Lambda_{1}^2(\overline{\omega}) - f_{\mathrm{e}}^2(\overline{\omega}) \Lambda_{2}^2(\overline{\omega})] \Sigma^{(\mathrm{t})},
\label{T11}\\
&& \partial_{\tau}{\mathcal T} +  (i k\mu + \epsilon') {\mathcal T} + \epsilon' {\mathcal T} = -  \epsilon' \zeta^2(\overline{\omega}) [\Lambda_{1}^2(\overline{\omega}) - f_{\mathrm{e}}^2(\overline{\omega}) \Lambda_{2}^2(\overline{\omega})] \Sigma^{(\mathrm{t})},
\label{T12}\\
&& \partial_{\tau}{\mathcal S} + (i k\mu + \epsilon') {\mathcal S} =0,
\label{T13}
\end{eqnarray}
where ${\mathcal Z}$, ${\mathcal T}$, ${\mathcal S}$ and $h$ denote either the $\oplus$ or the $\otimes$ polarization. 
By expanding ${\mathcal Z}$, ${\mathcal T}$  in series of Legendre polynomials: 
 \begin{eqnarray}
&& {\mathcal Z}(\nu,k,\tau)  = \sum_{\ell} (-i)^{\ell} (2\ell + 1) \, P_{\ell}(\nu)\, {\mathcal Z}_{\ell}(k,\tau).
\label{ZZ1}\\
&& {\mathcal T}(\nu,k,\tau)  = \sum_{\ell} (-i)^{\ell} (2\ell + 1) \, P_{\ell}(\nu)\, {\mathcal T}_{\ell}(k,\tau),
\label{TT1a}
\end{eqnarray}
the source term $\Sigma^{(\mathrm{t})}$ can also be expressed as:
\begin{eqnarray}
\Sigma^{(\mathrm{t})} &=& \frac{3}{32} \int_{-1}^{1} d \nu 
[  (1 - \nu^2)^2 {\mathcal Z}(\nu)
- ( 1 + \nu^2)^2 {\mathcal T}(\nu) - 4 \nu^2 {\mathcal T}(\nu)] 
\nonumber\\
&=& \frac{3}{70}{\mathcal  Z}_{4} + \frac{{\mathcal Z}_{2}}{7} - \frac{{\mathcal Z}_{0}}{10}- \frac{3}{70} {\mathcal T}_{4} + \frac{6}{7} {\mathcal T}_{2} - \frac{3}{5} {\mathcal T}_{0},
\label{T14}
\end{eqnarray}
where, as usual, ${\mathcal Z}_{\ell}$ and ${\mathcal T}_{\ell}$ 
denote the $\ell$-th mulipoles of the corresponding functions. When the relic tensor propagates orthogonally to the magnetic field direction, the two tensor polarizations will obey different equations. 
We can also compute the evolution equations by averaging the source 
functions over the directions of the magnetic field as suggested in Eq. (\ref{EV2a}).
In this case the, on top of the integrations over $\nu = \cos{\vartheta^{\prime}}$ and $\varphiÕ$ we have to perform also the integrals 
over $\alpha$ and $\beta$. According to Eq. (\ref{EV2a}) the evolution equations of the tensor polarizations with averaged collisional terms is then given by
\begin{eqnarray}
&& \partial_{\tau} \overline{{\mathcal Z}} + ( i k \mu + \epsilon') \overline{{\mathcal Z}} - \frac{1}{2} \partial_{\tau} h = \frac{\epsilon'}{15} [ \zeta^2 ( 7 \Lambda_{1}^2 - 5 f_{\mathrm{e}}^2 \Lambda_{2}^2) - 6 \zeta \Lambda_{1} \Lambda_{3} + 2 \Lambda_{3}^2] \overline{\Sigma}^{(\mathrm{t})},
\label{T5a}\\
&& \partial_{\tau} \overline{{\mathcal T}} + ( i k \mu + \epsilon') \overline{{\mathcal T}} = - \frac{\epsilon'}{15} [ \zeta^2 ( 7 \Lambda_{1}^2 - 5 f_{\mathrm{e}}^2 \Lambda_{2}^2) - 6 \zeta \Lambda_{1} \Lambda_{3} + 2 \Lambda_{3}^2] \overline{\Sigma}^{(\mathrm{t})},
\label{T6a}\\
&& \partial_{\tau} \overline{{\mathcal S}} + ( i k \mu + \epsilon') \overline{{\mathcal S}}=0.
\label{T7a}
\end{eqnarray}
The same analysis discussed in the scalar and tensor case has been completed for 
the vector modes; these results will not be discussed here but can be found in Ref. \cite{mg2010}.

\subsection{Spectral distortions?}
Non-interacting Planckian distributions are preserved by the adiabatic evolution so that the 
ratio of the Planckian temperatures at two different redshifts is given by the ratio of the redshifts.
Since the baryon to photon ratio is about $10^{-10}$ (see e.g. Eq. (\ref{gplasma}) and discussion therein)
once the thermal equilibrium is established (for instance at the epoch of nucleosynthesis)
 the transition from the ionized primordial plasma to neutral atoms at recombination  
 does not significantly alter the microwave background spectrum. 
 The remarkable precision with which the CMB spectrum is  fitted by a 
 Planckian distribution provides limits on possible energy releases 
 for redshifts $z < 10^{7}$. There are three important classes 
 of spectral distortions corresponding to energy releases at different 
 epochs: {\it i)} Compton distortions (due to late 
 energy releases for $z< 10^{5}$); {\it ii)} Bose-Einstein (or chemical potential) distortions 
 (due to early energy releases $10^{5} < z < 10^{7}$); 
 {\it iii)} Free-free distortions (due to very late energy releases much after decoupling, i.e. $z\ll 10^{3}$).
 The free-free distortions  occur for $z\ll 10^{3}$ and are therefore not directly relevant for magnetic random fields present prior to photon decoupling. There have been recently various interesting review articles stressing the importance and the challenges for improved measurements 
of the CMB intensity spectrum \cite{sironi,dezotti}. 
Spectral distortions of the CMB are in principle a unique source of precious informations up to reshifts $z\simeq 10^{7}$. Unfortunately all the attempts so far carried out for detecting distortions failed and all of them were based on comparisons among absolute measurements of the CMB temperature at different frequencies \cite{sironi}.

The spectral distortions of the CMB are parametrized in terms of two parameters: the dimensionless chemical potential\footnote{ By dimensionless chemical potential we mean the Bose-Einstein occupation number is expressed as $\overline{n} = 1/(e^{x+ \mu} -1)$ where $x= k/T$ and $\mu$ is the dimensionless chemical potential. } and the comptonization parameter denoted, respectively, by $\mu$ and $y$. The COBE/FIRAS limits were obtained by
measuring the difference between the cosmic microwave background and a precise blackbody spectrum \cite{COBEFIRAS}
and they imply the two well known bounds $|\mu| < 9\times 10^{-5}$ and $|y| < 15 \times 10^{-6}$. 
 
Before the accurate determinations of the temperature and the polarization anisotropies, 
one of the hopes to constrain large-scale magnetic fields came from the spectral distortions 
of the CMB and probably the first attempt in this direction is due to Ref. \cite{peter}.
If magnetic fields exist before the decoupling epoch, they create both  $\mu$- and $y$-type distortions. 
If magnetic random fields are present in the plasma they radiate. By computing 
the associated Poynting vector, it is possible to obtain the analog of the Larmor 
formula for a stochastic velocity field.  The bounds on the magnetic fields 
depend on the redshift and by studying the Bose-Einstein distortion 
 a rather strong bound of $10^{-10}$ G has been originally obtained
  on the uniform component of the magnetic field. 
 From the $y$ distortion the limit was less demanding  implying an intensity 
smaller than $3.4 \times 10^{-8}$ \cite{peter}. The perspective of Ref. \cite{peter} has been subsequently criticized in Ref. \cite{kar} (see, in particular, the second paper) by suggesting that the cyclotron effect does 
not lead to large $\mu$-distortions at small redshifts. According to \cite{kar} the damping 
of pre-decoupling magnetic fields does lead to $\mu$ and $y$ distortions. By imposing 
the COBE/FIRAS limit we do get a limit on the magnetic field intensity 
of the order of $3\times 10^{-8}$ between comoving coherence length $400$ pc and  $0.6$ Mpc
(see also \cite{oth0,oth1}). More recently the hopes 
for new CMB experiments with improved sensitivities to distortions stimulated 
various reprises of these themes \cite{oth2}. 

All the attempts so far envisaged for detecting CMB distortions failed probably because they were based 
on comparisons among absolute measurements of the CMB temperature at different frequencies.
It is however not excluded that new experimental ideas could succeed like, for instance 
measurements of the frequency derivative of the CMB temperature over large frequency intervals, as suggested 
in \cite{sironi}. So far the limits on magnetic fields distorting the microwave background spectrum are, as expected, weaker than the ones obtainable from the analysis of the temperature and polarization anisotropies. 

\renewcommand{\theequation}{3.\arabic{equation}}
\setcounter{equation}{0}
\section{Magnetized $\Lambda$CDM scenario}
\label{sec3}
While in the conventional case the large-scale solutions 
of the Einstein-Boltzmann hierarchy are discussed 
in various textbooks \cite{naselskyb,weinbergb,primer}, 
in what follows the discussion will be focussed on 
the modifications caused by the magnetic random fields. 
Since the curvature inhomogeneities and the fluctuations 
of the plasma mix with the evolution of the magnetic 
random fields, the large-scale solutions 
of the  Einstein-Boltzmann hierarchy 
differ both from the adiabatic solution and from the 
conventional entropic modes.
\begin{figure}[!ht]
\centering
\includegraphics[height=7cm]{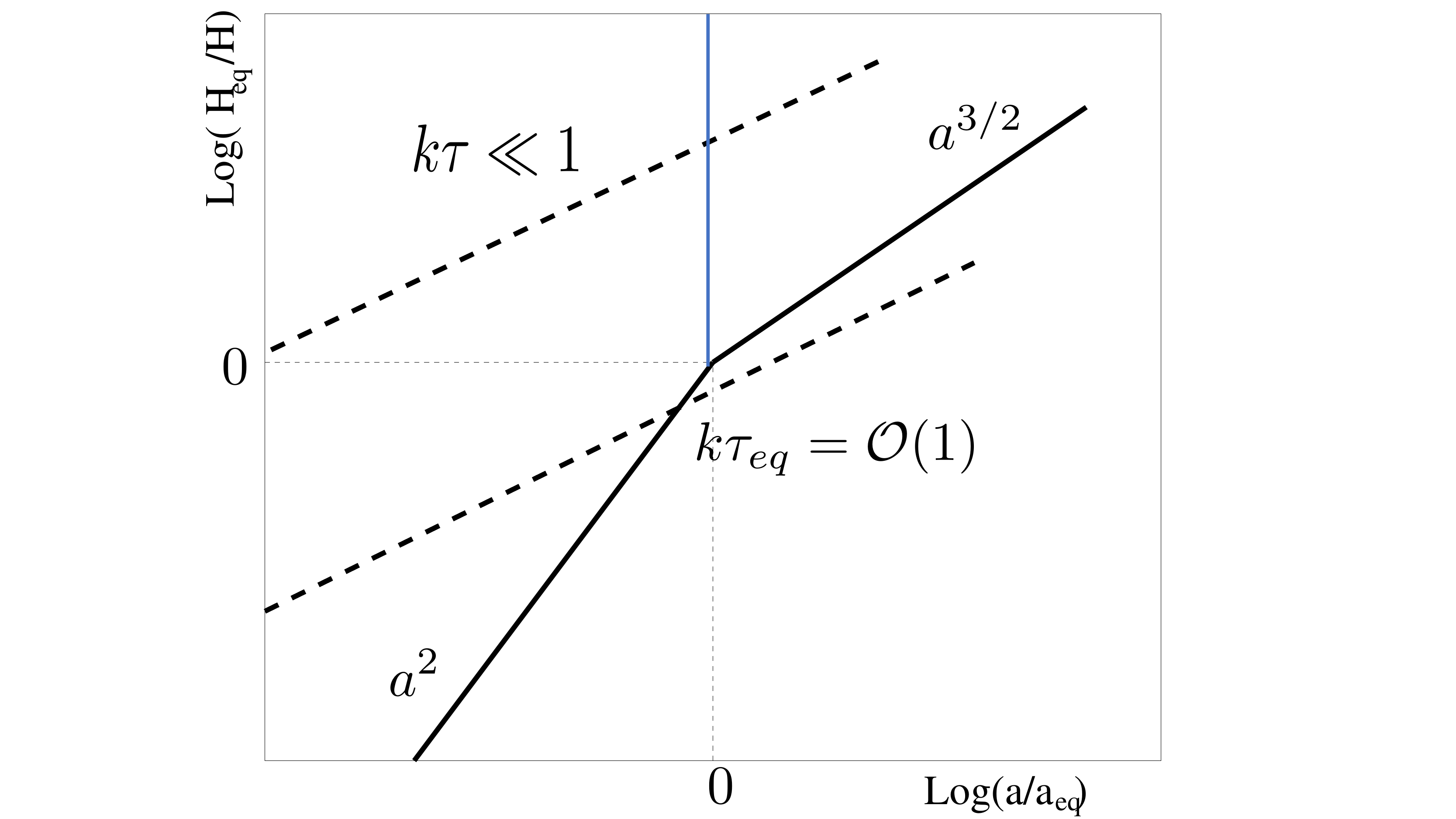}
\caption[a]{The  evolution of the particle horizon 
is illustrated together with  the physical scales determining the initial 
conditions of the Einstein-Boltzmann hierarchy.}
\label{figurescale}      
\end{figure}
By schematically illustrating the evolution of the Hubble radius across 
matter-radiation equality, Fig. \ref{figurescale} summarizes 
the main scales determining the 
Cauchy data of the Einstein-Boltzmann hierarchy: 
on the vertical axis the common logarithm of the particle horizon 
is reported as a function of common logarithm of the scale factor. 
While the dashed lines in Fig. \ref{figurescale} denote two
wavelengths larger than the particle horizon, the scale which is 
about to cross the Hubble radius corresponds to 
$k \tau_{\mathrm{eq}} = {\mathcal O}(1)$.
Prior to equality and when the relevant wavelengths 
of the large-scale fluctuations exceed the particle horizon, 
the initial conditions of the Einstein-Boltzmann hierarchy are then set:
in Fig. \ref{figurescale} this regime corresponds to 
$\tau < \tau_{\mathrm{eq}}$ (or $a< a_{\mathrm{eq}}$) 
and $k \tau \ll 1$. The (qualitative) description of large-scale 
cosmological perturbations \cite{naselskyb,weinbergb, primer} 
stipulates that a given wavelength exits the Hubble radius at some typical conformal time  
$\tau_{\mathrm{ex}}$ during an inflationary stage of expansion and approximately reenters 
at $\tau_{\mathrm{re}}$, when the Universe still expands but in a decelerated manner. 
By a mode being beyond the horizon we only mean that the physical wavenumber 
is much less than the expansion rate:  this does not necessarily 
have anything to do with causality \cite{weinberg}.

\subsection{Magnetized scalar modes} 
The scalar fluctuations of the geometry introduced in Eq. (\ref{FLUC1})
are subjected to the Hamiltonian constraint (imposing a relation between 
the density contrasts of the various species of the plasma) 
and to the momentum constraint (determining a specific relation
among the peculiar velocities of the different species). 
Because of the simultaneous presence of two constraints the analysis 
is comparatively more challenging than in the case of the 
vector modes (where only the momentum constraint survives) 
and of the tensor inhomogeneities (where there are no constraints). 
The notion of the magnetized initial conditions for the 
scalar modes of the Einstein-Boltzmann hierarchy has 
been firstly discussed and pursued Ref. \cite{MOD1} by using the 
complementary descriptions provided by the 
longitudinal and synchronous gauges introduced in Eqs. (\ref{STR4}) and (\ref{SYN2}). 
While the conformally Newtonian description is free from spurious gauge modes, 
the synchronous description is more suitable for 
the numerical treatment of the problem\footnote{Since the prototypical versions of
Cosmics and Cmbfast \cite{MOD2a,MOD2c,MOD2d} the Boltzmann codes 
are entirely based on the synchronous description or on its variations.}.

During the past decade the analysis of the magnetized scalar modes 
converged to the same standard employed when constraining more standard sets of initial data like, for instance, the four non-adiabatic modes \cite{nad1,nad2}. The non-Gaussian effects 
associated with the scalar modes have been firstly scrutinized in \cite{MOD2}.
The first calculations of the temperature and polarization anisotropies induced 
by the magnetized adiabatic mode can be found in \cite{MOD3} and other explicit calculations have been reported in Refs. \cite{MOD4,MOD4a}.  Beside the magnetized adiabatic mode \cite{MOD1} 
also the entropic solutions have been generalized to accommodate the presence of magnetic random fields. While the latter solutions have been comparatively less studied than their adiabatic counterpart, the non-adiabatic modes in combinations with the magnetic contribution may interfere either 
constructively or destructively but so far no explicit bounds on these solutions 
have been discussed besides the ones reported in \cite{nad3}. 
 
\subsubsection{Strongly interacting species}
In the Vlasov-Landau approach (appropriately extended to 
curved backgrounds)  the evolution equations for the distribution functions for 
electrons and ions can be written as\footnote{In Eq. (\ref{VL1}) the subscripts refer, respectively either to 
the case of the electrons and to the case of the ions; the plus sign at the left hand side refers to the ions while the minus 
refers to the electrons.}:
\begin{equation}
  \partial_{\tau} f_{\mathrm{e,i}}  + \vec{v} \cdot \vec{\nabla}_{\vec{x}} f_{\mathrm{e,i}} \mp e \bigl[ \vec{E} + \vec{v} \times \vec{B}\bigr]\cdot\vec{\nabla}_{\vec{q}}  f_{\mathrm{e,i}} = \bigl[\partial_{\tau} f_{\mathrm{e,i}} \bigr]_{\mathrm{coll}},
\label{VL1}
\end{equation}
where $\vec{B}$, $\vec{E}$ and $\vec{v}$ are, respectively, the comoving electromagnetic fields and the peculiar velocity already defined in Eqs. (\ref{S4a}) and (\ref{S4c}); in the non-relativistic 
limit (which is the relevant one in the case of Eq. (\ref{VL1}) the comoving three-momentum is $\vec{q} = m a \vec{v}$ and $m$ is the mass of the charge carrier (i.e. either electron or ion). 
The collisional terms at the right hand side of Eq. (\ref{VL1}) are different for electrons and ions.
By perturbing Eq. (\ref{VL1}) around a solution describing an approximate kinetic equilibrium, 
the evolution of the various moments of the (perturbed) distribution functions can be 
derived and they will lead, respectively, to  the equations for charge concentration (from the zeroth-order moment),
to the  equations for the velocities (from the first-order moment) and so on. In the flat space-time 
case this analysis is well known and can be found, for instance, in \cite{spitzer,krall}. 
Defining the charge concentrations 
and the velocities as appropriate moments of the distribution function
\begin{eqnarray}
&&n_{\mathrm{e}}(\vec{x},\tau) = n_{0} \int d^{3} v f_{\mathrm{e}}(\vec{x},\vec{v},\tau),\qquad 
n_{\mathrm{i}}(\vec{x},\tau) = n_{0} \int d^{3} v f_{\mathrm{i}}(\vec{x},\vec{v},\tau),
\label{BZ4}\\
&& \vec{v}_{\mathrm{e}}(\vec{x},\tau) = n_{0} \int d^3 v\, \vec{v}\,f_{\mathrm{e}}(\vec{x},\vec{v},\tau),\qquad 
\vec{v}_{\mathrm{i}}(\vec{x},\tau) = n_{0} \int d^{3} v\, \vec{v}\, f_{\mathrm{i}}(\vec{x},\vec{v},\tau),
\label{BZ5}
\end{eqnarray}
the evolution of the zeroth-order moment of Eq. (\ref{VL1}) implies the evolution equation 
of the charge concentrations:
\begin{eqnarray}
\partial_{\tau} n_{\mathrm{i}} + \theta_{\mathrm{i}} n_{\mathrm{i}} + \vec{v}_{\mathrm{i}}\cdot \vec{\nabla} n_{\mathrm{i}} =0,
\qquad \partial_{\tau} n_{\mathrm{e}}+ \theta_{\mathrm{e}} n_{\mathrm{e}} + \vec{v}_{\mathrm{e}}\cdot \vec{\nabla} n_{\mathrm{e}} =0,
\label{MX6}
\end{eqnarray}
where $\theta_{\mathrm{i}} = \vec{\nabla}\cdot \vec{v}_{\mathrm{i}}$ and  
$\theta_{\mathrm{e}} = \vec{\nabla}\cdot \vec{v}_{\mathrm{e}}$ are the 
three-divergences of the comoving three-velocities. 

Even if  the equations for the velocities follow in a similar manner from Eqs. (\ref{VL1}) and (\ref{BZ4})--(\ref{BZ5}), the same results of the Vlasov-Landau approach can be derived by perturbing (to first order) the covariant momentum conservation:
\begin{eqnarray}
&& \nabla_{\mu} T^{\mu\nu}_{(\mathrm{i})} = F^{\nu\alpha} j_{\alpha}^{(\mathrm{i})} + (\rho_{\gamma} + p_{\gamma}) \, \Gamma_{\gamma\mathrm{i}} [u_{(\gamma)}^{\nu} - u_{(\mathrm{i})}^{\nu}] + \rho_{\mathrm{e}} \Gamma_{\mathrm{ei}} [u_{(\mathrm{e})}^{\nu} - u_{(\mathrm{i})}^{\nu}],
\label{STR1}\\
&& \nabla_{\mu} T^{\mu\nu}_{(\mathrm{e})}  = F^{\nu\alpha} j_{\alpha}^{(\mathrm{e})} + (\rho_{\gamma} + p_{\gamma}) \, \Gamma_{\gamma\mathrm{e}} [u_{(\gamma)}^{\nu} - u_{(\mathrm{e})}^{\nu}] + \rho_{\mathrm{e}} \Gamma_{\mathrm{ei}} [u_{(\mathrm{i})}^{\nu} - u_{(\mathrm{e})}^{\nu} ],
\label{STR2}\\
&& \nabla_{\mu} T^{\mu\nu}_{(\gamma)}  = \Gamma_{\gamma \mathrm{i}} (\rho_{\gamma} + p_{\gamma}) [u_{(\mathrm{i})}^{\nu} - u_{(\gamma)}^{\nu}]
+  \Gamma_{\gamma \mathrm{e}} (\rho_{\gamma} + p_{\gamma}) [u_{(\mathrm{e})}^{\nu} - u_{(\gamma)}^{\nu}],
\label{STR3}
\end{eqnarray}
where the expressions of $T^{\mu\nu}_{(\mathrm{i})}$, $T^{\mu\nu}_{(\mathrm{e})}$ and $T^{\mu\nu}_{(\gamma)}$ have been already
introduced in Eqs. (\ref{COV5}) and (\ref{COV6}). Since the first-order scalar fluctuations of $\nabla_{\mu} T^{\mu\nu}$ for a generic energy-momentum tensor is given by\footnote{The terms with overlines in Eq. (\ref{STR3a}) denote the background values of the corresponding quantity.}:
\begin{equation}
\delta_{\mathrm{s}} \nabla_{\mu} T^{\mu\nu}= \partial_{\mu} \delta_{\mathrm{s}} T^{\mu\nu}+ 
\delta_{\mathrm{s}} \Gamma^{\mu}_{\mu\alpha} \overline{T}^{\alpha\nu} + \overline{\Gamma}^{\mu}_{\mu\alpha} \delta_{\mathrm{s}} T^{\alpha\nu} 
+ \delta_{\mathrm{s}} \Gamma^{\nu}_{\alpha\beta} \overline{T}^{\alpha\beta}+ 
\overline{\Gamma}^{\nu}_{\alpha\beta} \delta_{\mathrm{s}} T^{\alpha\beta},
\label{STR3a}
\end{equation} 
the evolution equations of the velocities and of the density contrasts 
can be obtained from the scalar fluctuations of the Christoffel connections expressed in the longitudinal gauge of Eq. (\ref{STR4}) (see e.g. \cite{primer}). The notation $\delta_{\mathrm{s}}$ has been already introduced after Eq. (\ref{FLUC1}).
In appendix \ref{APPC} the main equations of the present section will be studied 
in the synchronous coordinate system already defined in Eq. (\ref{SYN2}). 

If the free tensor index $\nu$ of  Eqs. (\ref{STR1})--(\ref{STR3}) is space-like,  
we obtain the evolution equations of the peculiar velocities for electrons, 
ions and photons:
\begin{eqnarray}
&& \vec{v}_{\mathrm{e}}^{\,\prime} + {\mathcal H}\,\vec{v}_{\mathrm{e}} = - \frac{e}{m_{\mathrm{e}} \, a} [ \vec{E} + \vec{v}_{\mathrm{e}} \times \vec{B}] - \vec{\nabla} \phi 
+ 
\frac{4}{3} \frac{\rho_{\gamma}}{\rho_{\mathrm{e}}} a 
\Gamma_{\gamma \, \mathrm{e}} (\vec{v}_{\gamma} - \vec{v}_{\mathrm{e}}) + a \Gamma_{\mathrm{e\,i}} ( \vec{v}_{\mathrm{i}} - \vec{v}_{\mathrm{e}}),
\label{STR5}\\
&&  \vec{v}_{\mathrm{i}}^{\, \prime} + {\mathcal H}\,\vec{v}_{\mathrm{i}} =  \frac{e}{m_{\mathrm{i}} \, a} [ \vec{E} + \vec{v}_{\mathrm{i}} \times \vec{B}] - \vec{\nabla} \phi 
+ 
\frac{4}{3} \frac{\rho_{\gamma}}{\rho_{\mathrm{i}}} a 
\Gamma_{\gamma \, \mathrm{i}} (\vec{v}_{\gamma}-\vec{v}_{\mathrm{i}} ) + a \Gamma_{\mathrm{e\,i}} \frac{\rho_{\mathrm{e}}}{\rho_{\mathrm{i}}}( \vec{v}_{\mathrm{e}} - \vec{v}_{\mathrm{i}}),
\label{STR6}\\
&& \vec{v}_{\gamma}^{\,\prime} = - \frac{1}{4} \vec{\nabla} \delta_{\gamma} - \vec{\nabla} \phi 
+ a \Gamma_{\gamma\mathrm{i}} (\vec{v}_{\mathrm{i}} - \vec{v}_{\gamma}) + 
a \Gamma_{\gamma\mathrm{e}}  ( \vec{v}_{\mathrm{e}} - \vec{v}_{\gamma}),
\label{STR7}
\end{eqnarray}
where $\Gamma_{\gamma\mathrm{e}} $, $\Gamma_{\gamma\mathrm{i}}$ and $\Gamma_{\mathrm{e\,i}}$ are, respectively, 
the electron-photon, the ion-photon and the electron-ion interaction rates \cite{spitzer,krall}.
The peculiar velocities of the various species follow the notations already introduced in Eqs.(\ref{S4}) and (\ref{S4c}). In Eq. (\ref{STR7}) and in what follows the density contrasts and the divergence of the three-velocities will be denoted, respectively, by   by $\delta_{X} = \delta_{\mathrm{s}}\rho_{X}/\rho_{X}$ and  $\theta_{X} = \vec{\nabla}\cdot \vec{v}_{X}$ where $X$ is one of the different species of the plasma.  
Obviously the density contrasts and the peculiar velocities change from one gauge to the other and, in particular, from the longitudinal to the synchronous gauges 
and vice versa (see, in particular, Eqs. (\ref{APB2}) and (\ref{APB3})). 

When the free tensor index appearing in Eqs. (\ref{STR1}), (\ref{STR2}) and (\ref{STR3}) 
is taken to be time-like (i.e. $\nu=0$) the evolution equations of the 
density contrasts of the strongly interacting species turn out to be:
\begin{eqnarray}
&& \delta_{\mathrm{e}}' = - \theta_{\mathrm{e}} + 3 \psi' 
- \frac{e}{m_{\mathrm{e}} a}\vec{E} \cdot \vec{v}_{\mathrm{e}},\qquad \vec{\nabla} \cdot \vec{v}_{\mathrm{e}}= \theta_{\mathrm{e}},
\label{STR11}\\
&& \delta_{\mathrm{i}}' = - \theta_{\mathrm{i}} + 3 \psi' 
+ \frac{e}{m_{\mathrm{i}} a}\vec{E} \cdot \vec{v}_{\mathrm{i}},\qquad \vec{\nabla} \cdot \vec{v}_{\mathrm{i}} = \theta_{\mathrm{i}},
\label{STR12}\\
&& \delta_{\gamma}' = 4\psi' - \frac{4}{3} \theta_{\gamma}, \qquad \vec{\nabla} \cdot \vec{v}_{\gamma}= \theta_{\gamma},
\label{STR13}
\end{eqnarray}
with the same notations already employed in Eq. (\ref{MX6}). 
All in all  Eqs. (\ref{STR5})--(\ref{STR7}) and Eqs. (\ref{STR11})--(\ref{STR13})
describe a three-fluid system formed by photons, electrons and ions. 

\subsubsection{The photon-lepton-baryon fluid}
While the Thomson rate increases with the temperature, the 
Coulomb rate decreases. More precisely, for physical temperatures larger than the eV, we have that 
$\Gamma_{\mathrm{ei}}/H \propto 10^{11} (\widetilde{T}/\mathrm{eV})^{-1/2}$
while $\Gamma_{\gamma\mathrm{e}}/H \propto 10^{4} (\widetilde{T}/\mathrm{eV})$. The meeting point of the two rates occurs close to the MeV so that, in the regime described by Fig. \ref{figurescale}, photons are strongly coupled to the baryon-lepton (or simply baryon) fluid. The governing equations of the baryon fluid follow from the equations for the electrons and the ions. In particular, by summing 
Eq. (\ref{STR5}) (multiplied by $m_{\mathrm{e}}$) and Eq. (\ref{STR6}) (multiplied by $m_{\mathrm{i}}$) 
the baryon and the photon velocities will obey, respectively, the following pair of equations:
\begin{eqnarray}
&& \vec{v}_{\mathrm{b}}^{\,\prime} + {\mathcal H}  \vec{v}_{\mathrm{b}} = \frac{\vec{J}\times 
\vec{B}}{a^4 \rho_{\mathrm{b}}(1 + m_{\mathrm{e}}/m_{\mathrm{i}})} - \vec{\nabla}\phi +
 \frac{\epsilon^{\prime}}{R_{\mathrm{b}}} (\vec{v}_{\gamma} - \vec{v}_{\mathrm{b}}),
\label{STR16}\\
&& \vec{v_{\gamma}}^{\prime} = -\frac{1}{4} \vec{\nabla} \delta_{\gamma} - \vec{\nabla}\phi +\epsilon^{\prime} ( \vec{v}_{\mathrm{b}} - \vec{v}_{\gamma}), 
\label{STR17}
\end{eqnarray}
$R_{\mathrm{b}}$ and $\epsilon^{\prime}$ denote, respectively, the baryon to photon ratio of  Eq. (\ref{Rbdef}) and the differential optical depth 
introduced in  Eq.(\ref{diffop}). The two new one-fluid variables of 
Eqs. (\ref{STR16})--(\ref{STR17}) are the centre of mass 
velocity of the electron-ion system and the total comoving current:
\begin{equation}
\vec{v}_{\mathrm{b}} = \frac{m_{\mathrm{e}} \vec{v}_{\mathrm{e}} + m_{\mathrm{i}} \vec{v}_{\mathrm{i}}}{m_{\mathrm{e}} + m _{\mathrm{i}}},\qquad 
\vec{J}= e \, n_{0} ( \vec{v}_{\mathrm{i}} - \vec{v}_{\mathrm{e}}),
\label{STR17a}
\end{equation}
where $\vec{J}$ (already introduced in Eqs. (\ref{COV8})--(\ref{S4})) has been 
explicitly written in the globally neutral case. The evolution equation of the baryonic density contrast $\delta_{\mathrm{b}}$ follows instead from the sum of  Eqs. (\ref{STR11}) and (\ref{STR12}) :
\begin{equation}
\delta_{\mathrm{b}}' = - \theta_{\mathrm{b}} + 3 \psi' + \frac{\vec{J} \cdot \vec{E}}{a^4 \rho_{\mathrm{b}}}, \qquad 
\delta_{\mathrm{b}} = \frac{m_{\mathrm{e}}}{m_{\mathrm{i}} + m_{\mathrm{e}}} \delta_{\mathrm{e}} + \frac{m_{\mathrm{i}}}{m_{\mathrm{i}} + m_{\mathrm{e}}} \delta_{\mathrm{i}},
\label{STR18}
\end{equation}
where, as before, $\theta_{\mathrm{b}} = \vec{\nabla}\cdot\vec{v}_{\mathrm{b}}$ and 
 $\rho_{\mathrm{b}}= \widetilde{n}_{0} (m_{\mathrm{i}} + m_{\mathrm{e}} )$ denotes the physical 
 (not comoving) baryonic matter density in the globally neutral case. It is relevant to remark, as already mentioned in section \ref{sec2} that the system is not Weyl invariant. The most convenient
solution to exploit the remaining symmetries of the equations is therefore to use as much as possible comoving currents and comoving electromagnetic fields (as in Eqs. (\ref{STR16}), (\ref{STR17}) and (\ref{STR18})) together with the physical energy densities and pressures which appear directly in the background evolution equations (\ref{COV11}) and (\ref{COV12}).

Since the photon and the baryon velocities are quickly  synchronized because of the 
hierarchy between the photon-electron rate and the Hubble rate, prior to decoupling the system of Eqs. (\ref{STR16}), (\ref{STR17}) and (\ref{STR18}) can be further combined to eliminate 
the momentum exchange between photons and baryons:
\begin{eqnarray}
&& \delta_{\gamma}' = 4\psi' - \frac{4}{3} \theta_{\gamma\mathrm{b}},\qquad \delta_{\mathrm{b}}' = 3 \psi' - \theta_{\gamma\mathrm{b}} 
+ \frac{\vec{J} \cdot \vec{E}}{a^4 \rho_{\mathrm{b}}},
\label{STR21}\\
&& \theta_{\gamma\mathrm{b}}^{\,\prime} + \frac{{\mathcal H} R_{\mathrm{b}}}{R_{\mathrm{b}} +1} \theta_{\gamma\mathrm{b}}  - 
\frac{\eta\,  \nabla^2\theta_{\gamma\mathrm{b}} }{\rho_{\gamma} ( 1 + R_{\mathrm{b}})}= \frac{3 \vec{\nabla}\cdot(\vec{J} \times \vec{B}) }{4 a^4 \rho_{\gamma} ( R_{\mathrm{b}} + 1)}- \frac{\nabla^2 \delta_{\gamma}}{4 ( 1 + R_{\mathrm{b}})} -\nabla^2\phi ,
\label{STR22}
\end{eqnarray}
where $\theta_{\gamma\mathrm{b}}= \vec{\nabla}\cdot\vec{v}_{\gamma\mathrm{b}}$ is the three-divergence of the corresponding peculiar velocity. The shear viscosity term 
$\eta = (4/15) \rho_{\gamma} \lambda_{\gamma\mathrm{e}}$
(see Eq. (\ref{STR22})), depends upon the photon mean free path $\lambda_{\gamma\mathrm{e}}$ which is inversely proportional to the differential optical depth $\epsilon^{\prime}$.

While in the limit $\vec{J} \to 0$ Eqs. (\ref{STR21}) and (\ref{STR22}) reproduce 
the conventional system of equations, in the present situation 
the electric fields and the total currents are essential.
The simplest way of addressing this problem is to resort to 
the magnetohydrodynamical reduction \cite{krall,goed,guide2}, 
as originally suggested in \cite{MOD1,MOD3} and subsequently discussed 
by different authors (see e.g. \cite{MOD4,MOD4a}). 
The cumbersome expression obtained by taking the difference of  Eq. (\ref{STR5}) 
(multiplied by $e\,n_{\mathrm{i}}$) and of Eq. (\ref{STR6}) 
(multiplied by $e\,n_{\mathrm{e}}$) can be simplified in the 
limit $(m_{\mathrm{e}}/m_{\mathrm{i}})\ll 1$. Furthermore, since the plasma is globally 
neutral, the final result of this procedure translates into a differential form of the Ohm law reading \cite{MHD1}:
\begin{eqnarray}
\partial_{\tau} \vec{J} + \biggl({\mathcal H} + a\Gamma_{\mathrm{ie}} + 
\frac{4 \rho_{\gamma}\Gamma_{\mathrm{e}\gamma}}{ 3 n_{0} \,m_{\mathrm{e}}}\biggr) \vec{J} &=& \frac{\overline{\omega}_{\mathrm{pe}}^2 }{4\pi} \biggl(\vec{E} + \vec{v}_{\mathrm{b}} \times \vec{B} + \frac{\vec{\nabla} p_{\mathrm{e}}}{e\,n_{0}}- \frac{\vec{J}\times 
\vec{B}}{e n_{0}}\biggr) 
\nonumber\\
&+& \frac{4 e \rho_{\gamma} \Gamma_{\mathrm{e}\gamma}}{3 m_{\mathrm{e}}} 
(\vec{v}_{\mathrm{b}} - \vec{v}_{\gamma}).
\label{STR23}
\end{eqnarray}
The terms $\partial_{\tau}\vec{J}$ and ${\mathcal H} \vec{J}$ are 
comparable in magnitude; moreover they are both smaller than $\Gamma_{\mathrm{ie}}$ and 
$\Gamma_{\mathrm{e}\gamma}$, i.e. 
${\mathcal H} \vec{J} \simeq \partial_{\tau} \vec{J} < (4/3) (\rho_{\gamma}/m_{\mathrm{e}}) \Gamma_{\mathrm{e}\gamma} <a\Gamma_{\mathrm{ie}}$. Since Eq. (\ref{STR23}) is dominated 
by the Coulomb rate $\Gamma_{\mathrm{ie}}$,  at the right-hand side the term 
containing $(\vec{v}_{\mathrm{b}} - \vec{v}_{\gamma})$ is estimated by subtracting 
Eqs. (\ref{STR16}) and (\ref{STR17}); this term
 is driven exponentially to zero at a rate controlled by $a \Gamma_{\gamma\mathrm{e}} (1 + R_{\mathrm{b}}^{-1})$. The Ohm's law (in its asymptotic 
 form) becomes then\footnote{In Eq. (\ref{APP4}) the term containing the gradient of the electron pressure is the curved-space counterpart of the  thermoelectric term \cite{spitzer} while the term proportional to the vector product of the current and of the magnetic field is the curved-space counterpart of the Hall term.}:
\begin{equation}
\vec{J} = \sigma \biggl(\vec{E} + \vec{v}_{\gamma\mathrm{b}} \times \vec{B} + \frac{\vec{\nabla} p_{\mathrm{e}}}{e\,n_{0}}- \frac{\vec{J}\times \vec{B}}{e\,n_{0}}\biggr),
\label{APP4}
\end{equation}
where $\sigma$ denotes the comoving conductivity whose explicit form turns out to be:
\begin{equation}
\sigma = \frac{\overline{\omega}_{\mathrm{pe}}^2 }{4\pi\{a \Gamma_{\mathrm{ie}} + (4/3)[\rho_{\gamma}/(n_{0} m_{\mathrm{e}})] \Gamma_{\mathrm{e}\gamma}\}}
\to  \frac{9}{8 \pi \sqrt{3}} \frac{T}{e^2} \sqrt{\frac{T}{m_{\mathrm{e}} a}} [\ln{\Lambda_{\mathrm{C}}(T)}]^{-1},
\label{COLL4}
\end{equation}
where, as in Eq. (\ref{hier1}), $\Lambda_{\mathrm{C}} = 1.102 \times (h_{0}^2 \Omega_{\mathrm{b}0}/0.02273)^{-1/2}$ is the argument of the Coulomb logarithm; the second limit in Eq. (\ref{COLL4}) follows when the Coulomb rate dominates.

The hierarchy of scales leading to the simplification of Eq. (\ref{APP4}) 
also implies that the displacement current of Eq. (\ref{S4}) is negligible 
in comparison with the Ohmic current\footnote{ For the same reason 
one can easily show  from  Eq. (\ref{STR23}) that 
$\partial_{\tau}\vec{J} \ll \overline{\omega}_{\mathrm{p\,e,i}}^2 \vec{E}$) 
and that $\partial_{\tau}^2\vec{J} \ll \overline{\omega}_{\mathrm{pe}}^2 \vec{J}$.}. Since $4\pi \vec{J} \gg \partial_{\tau}\vec{E}$  the general form of Eqs. (\ref{S1}) and (\ref{S4}) gets much simpler since  
the electric fields, the magnetic fields and the Ohmic current are all solenoidal:
\begin{equation}
 \vec{\nabla} \cdot \vec{E} =0, \qquad \vec{\nabla} \cdot \vec{B} =0, \qquad  \vec{\nabla} \cdot \vec{J} =0.
\label{COLL5}
\end{equation}
The term $\vec{J}\cdot\vec{E}$ appearing in Eq. (\ref{STR21}) can be approximated, in the baryon rest frame, as:
\begin{equation}
\frac{\vec{J}\cdot\vec{E}}{a^4 \rho_{\mathrm{b}}} = \frac{(\vec{\nabla}\times\vec{B})^2}{4 \pi \sigma a^4 \rho_{\mathrm{b}}}, \qquad \vec{E} \simeq \frac{\vec{\nabla}\times \vec{B}}{4 \pi \sigma}.
\end{equation}
Moreover, since Eq. (\ref{COLL4}) implies that $\sigma/\overline{\omega}_{\mathrm{pe}} \gg 1$ 
(and also that $\sigma/T \gg 1$), the large-scale electric fields are highly suppressed 
by powers of $\sigma^{-1}$ in the baryon rest frame \cite{MHD1} and can be estimated, from Eq. (\ref{APP4}). The corresponding evolution equation for the magnetic field becomes
then: 
\begin{equation}
\partial_{\tau} \vec{B} = \vec{\nabla} \times (\vec{v}_{\gamma\mathrm{b}} \times \vec{B})+ \frac{1}{4\pi \sigma} \nabla^2 \vec{B} 
+ \vec{\nabla} \times \biggl[ \frac{\vec{\nabla} p_{e}}{e n_{0}}\biggr] - \vec{\nabla}\times\biggl[ \frac{(\vec{\nabla} \times \vec{B}) \times \vec{B}}{4 \pi e n_{0}}\biggr],
\label{COLL8}
\end{equation}
where, to avoid ambiguities, the primes (denoting a derivation with respect to $\tau$) 
have been replaced by $\partial_{\tau}$. 
If the thermoelectric and Hall terms are neglected in Eq. (\ref{COLL8}) we obtain the standard form of the 
magnetic diffusivity equation 
where the bulk velocity of the plasma coincides with the baryon-photon velocity.  
By making explicit Eq. (\ref{APP4}) in terms of the electric field and 
by taking the first conformal time derivative of the obtained expression we have an equation 
formally similar to Eq. (\ref{COLL8}):
\begin{equation}
\partial_{\tau} \vec{E}= -  \partial_{\tau}(\vec{v}_{\gamma\mathrm{b}}\times \vec{B}) + \frac{\partial_{\tau}\vec{J}}{\sigma} - \partial_{\tau}\biggl[ \frac{\vec{\nabla} p_{e}}{e n_{0}}\biggr] -\partial_{\tau}\biggl[ \frac{(\vec{\nabla} \times \vec{B}) \times \vec{B}}{4 \pi e n_{0}}\biggr], 
\label{COLL9}
\end{equation}
where, to lowest order, $\partial_{\tau}\vec{J} = - \vec{\nabla}\times(\vec{\nabla}\times \vec{E})/(4 \pi)$.
Neglecting now the thermoelectric and the Hall terms in Eqs. (\ref{COLL8}) and (\ref{COLL9})
we can obtain the following simplified description:
\begin{equation}
 \partial_{\tau} \vec{B} = \vec{\nabla} \times (\vec{v}_{\gamma\mathrm{b}} \times \vec{B})+ \frac{1}{4\pi \sigma} \nabla^2 \vec{B}, \qquad \partial_{\tau}\vec{E} = -\partial_{\tau} (\vec{v}_{\mathrm{b}} \times \vec{B}) + \frac{1}{4\pi \sigma} \nabla^2 \vec{E}.
\label{APP8}
\end{equation}

Before recombination denoting with $\tau_{\mathrm{c}}$ the inverse of the differential 
optical depth (i.e. $\tau_{\mathrm{c}} = 1/\epsilon^{\prime}$), 
an explicit evolution equations for the velocity differences between 
baryons and photons can be derived by including the effects 
of the magnetic random fields.  In the tight-coupling approximation, 
 we have that the baryon and photon velocities 
coincide, at least approximately, i.e. $\theta_{\gamma\mathrm{b}} \simeq \theta_{\gamma} \simeq \theta_{\mathrm{b}}$. When the tight-coupling approximation breaks down it is possible to derive an evolution equation  
for $(\theta_{\mathrm{b}} - \theta_{\gamma})$ valid for $|\tau_\mathrm{c}/\tau| \ll 1$ and $k\tau_c\ll1$. 
The idea of treating the problem in this way is due to Peebles and Yu \cite{pee3} and has been subsequently 
rediscovered, for numerical purposes, in Ref. \cite{bertschingerma,MOD2a,MOD2c,MOD2d}.

\subsubsection{Weakly interacting constituents}
While the evolution of the CDM fluctuations only consists of the following two equations:
\begin{equation}
\delta_{\mathrm{c}}' = 3 \psi' - \theta_{\mathrm{c}},
\qquad \theta_{\mathrm{c}}^{\,\prime} + {\mathcal H} \theta_{\mathrm{c}} = - \nabla^2 \phi,
\label{NU1}
\end{equation}
the neutrinos obey the collisionless Boltzmann equation whose lowest 
multipoles obey the following triplet of equations in Fourier space
\begin{eqnarray}
&& \delta_{\nu}' = - \frac{4}{3} \theta_{\nu} + 4 \psi',
\label{NU2}\\
&& \theta_{\nu}' = - k^2 \sigma_{\nu} + \frac{k^2}{4} \delta_{\nu} 
+ k^2\phi,
\label{NU3}\\
&& \sigma_{\nu}^{\prime} = \frac{4}{15} \theta_{\nu} - \frac{3}{10} k {\mathcal F}_{\nu3},
\label{NU4}
\end{eqnarray}
where ${\mathcal F}_{\nu3}$ is the octupole of the full neutrino phase space distribution. The 
collisionless Boltzmann equation supplemented by the contribution of the geometry is given by
\begin{equation}
\partial_{\tau} {\mathcal  F}_{\nu} + i k\mu {\mathcal  F}_{\nu} = 4 (\psi' - i k \mu \phi).
\label{NU5} 
\end{equation}
The notations used in Eqs. (\ref{NU2})--(\ref{NU4}) and (\ref{NU5}) coincide with the ones 
of Refs. \cite{MOD1,MOD3}; these notations are essentially the ones of Ref. \cite{bertschingerma}
(the signature of the metric (\ref{OM1}) is however different and the longitudinal potentials of Eq. (\ref{STR4}) are defined in a slightly different manner). 
If we expand the angular dependence of the reduced phase-space distribution  in series of Legendre polynomials
\begin{equation}
{\mathcal  F}_{\nu}( \vec{k}, \hat{n}, \tau) = \sum_{\ell} (-i)^{\ell} ( 2 \ell + 1) {\mathcal   F}_{\nu\ell}(\vec{k},\tau) P_{\ell}(\mu),
\label{NU6}
\end{equation}
Eqs. (\ref{NU2}), (\ref{NU3}) and (\ref{NU4}) correspond, up to trivial numerical factors, to the evolution of the monopole, dipole and quadrupole of ${\mathcal F}_{\nu}$. For higher multipoles (i.e. $\ell \geq 3$) 
\begin{equation}
{\mathcal  F}_{\nu\ell}' = \frac{k}{2\ell +1} [ \ell {\mathcal  F}_{\nu,(\ell-1)}  - (\ell+1) {\mathcal  F}_{\nu (\ell+1)}].
\label{NU7}
\end{equation} 
Note that in Eqs. (\ref{NU3}) and (\ref{NU4}), $\sigma_{\nu}$ is related 
to the neutrino anisotropic stress as $\partial_{i}\partial_{j} \Pi_{\mathrm{f}}^{ij}= (p_{\nu} + \rho_{\nu})\nabla^2 \sigma_{\nu}$ where $\Pi_{\mathrm{f}}^{ij}$ is the anisotropic stress of the fluid. These two 
quantities coincide within the $\Lambda$CDM paradigm where the only source 
of anisotropic stress of the fluid is represented by the massless neutrinos. 
While in the concordance scenario the dark energy component does not fluctuate, it is 
often interesting to investigate the stability of the magnetized initial conditions in the presence 
of a fluctuating dark energy component which, among other things, may affect 
the integrated Sachs-Wolfe effect. Two supplementary evolution equations 
must be considered in this case and they correspond to the evolution of the dark energy density contrast and 
of the peculiar velocity\footnote{A possible approach is to 
select a particular frame, the so-called dark energy rest frame \cite{dore}, where the sound speed is 
independently assigned: in this case there will be two supplementary parameters in the game, i.e. 
the barotropic index and the sound speed of dark energy.}.

\subsubsection{The fluctuations of the geometry}
The weakly interacting species are affected by
the magnetic random fields through the evolution of the 
fluctuations of the geometry obeying the inhomogeneous Einstein 
equations. By perturbing both 
sides of Eq. (\ref{COV1}) with respect to the scalar fluctuations of the geometry 
(\ref{OM1}) we have\footnote{For the interested reader we mention that the explicit form of the scalar fluctuations of the Ricci and Einstein tensors can be found, for 
instance, in appendices C and D of Ref. \cite{primer}.} 
\begin{equation}
\delta_{\mathrm{s}} R_{\mu}^{\nu} - \frac{1}{2} \delta_{\mu}^{\nu} \, \delta_{\mathrm{s}} R = \ell_{P}^2 \,\delta_{\mathrm{s}} T_{\mu}^{\nu},
\label{PertS1}
\end{equation}
where $\delta_{\mathrm{s}}$ denotes the scalar fluctuation of the coresponding tensor, exactly as in
 Eq. (\ref{FLUC1}).
The $(00)$ and  $(0i)$ components of Eq. (\ref{PertS1}) 
relate the scalar fluctuations of the geometry to their first derivatives. 
In particular, denoting by $\delta_{\mathrm{s}}\rho_{\mathrm{t}}$ the total fluctuation of the energy density in the gauge (\ref{STR4}),
\begin{equation}
\delta_{\mathrm{s}}\rho_{\mathrm{t}} = \delta_{\mathrm{s}} \rho_{\mathrm{c}} + \delta_{\mathrm{s}}\rho_{\nu} + \delta_{\mathrm{s}}\rho_{\gamma} + \delta_{\mathrm{s}}\rho_{\mathrm{b}},
 \label{GEOM2}
 \end{equation}
the Hamiltonian constraint 
\begin{equation}
 \nabla^2 \psi - 3 {\mathcal H} ( {\mathcal H} \phi + \psi') = \frac{\ell_{P}^2 a^2}{2} \biggl[ 
 \delta_{\mathrm{s}} \rho_{\mathrm{t}} + \delta_{\mathrm{s}} \rho_{\mathrm{B}} + 
 \delta_{\mathrm{s}} \rho_{{\mathrm{E}}}\biggr],
\label{HAMS}
\end{equation}
follows from the $(00)$ component of Eq. (\ref{PertS1}). The fluctuations of the electromagnetic energy density and of the corresponding pressure are preceded by $\delta_{\mathrm{s}}$ since, unlike the corresponding anisotropic stresses, they only affect the evolution of the scalar modes:
\begin{equation}
\delta_{\mathrm{s}} \rho_{\mathrm{B}} = \frac{B^2}{8\pi a^4}, \qquad 
\delta_{\mathrm{s}} \rho_{\mathrm{E}} = \frac{E^2}{8\pi a^4},\qquad  
\delta_{\mathrm{s}} p_{\mathrm{B}} = \frac{\delta_{\mathrm{s}} \rho_{\mathrm{B}}}{3}, \qquad  
\delta_{\mathrm{s}} p_{\mathrm{E}} = \frac{\delta_{\mathrm{s}} \rho_{\mathrm{E}}}{3}.
\label{GEOM1}
\end{equation}
The magnetic and electric energy densities are customarily referred to the energy density of the photon background \cite{MOD1,MOD3} by introducing the following dimensionless variables: 
\begin{equation}
\Omega_{\mathrm{E}}(\vec{x},\tau) = \frac{ \delta_{s}\rho_{\mathrm{E}}(\vec{x},\tau)}{\rho_{\gamma}(\tau)},\qquad 
 \Omega_{\mathrm{B}}(\vec{x},\tau) = \frac{ \delta_{s}\rho_{\mathrm{B}}(\vec{x},\tau)}{\rho_{\gamma}(\tau)}.
\label{GEOM1a}
\end{equation}
Since the total three-velocity field is {\em not} solenoidal in the case of the scalar modes, from the $(0i)$ component of Eq. (\ref{PertS1}) we can write the three-divergence of the momentum constraint
\begin{equation}
\nabla^2 ( {\mathcal H} \phi + \psi') + \frac{\ell_{P}^2 a^2}{2}\biggl[ (p_{\mathrm{t}} + \rho_{\mathrm{t}}) \theta_{\mathrm{t}} + \frac{\vec{\nabla}\cdot(\vec{E} \times \vec{B})}{4 \pi a^4}\biggr]=0,
\label{MOMS}
\end{equation}
where  $ \theta_{\mathrm{t}}$ denotes the total velocity field:
 \begin{eqnarray}
 (p_{\mathrm{t}} + p_{\mathrm{t}}) \theta_{\mathrm{t}} &=& \sum_{a} (p_{\mathrm{a}} + \rho_{\mathrm{a}})\theta_{\mathrm{a}} \equiv \frac{4}{3}\rho_{\nu} \theta_{\nu} + \frac{4}{3} \rho_{\gamma} \theta_{\gamma} + 
 \rho_{\mathrm{c}} \theta_{\mathrm{c}} + \rho_{\mathrm{b}} \theta_{\mathrm{b}}
 \nonumber\\
 &=&  \rho_{\mathrm{c}} \theta_{\mathrm{c}} +\frac{4}{3}\rho_{\nu} \theta_{\nu} + 
 \frac{4}{3}\rho_{\gamma} \theta_{\gamma\mathrm{b}} ( 1 + R_{\mathrm{b}}).
  \label{GEOM3}
 \end{eqnarray}
The expression appearing in the second line of Eq. (\ref{GEOM3}) holds in the tight 
coupling approximation while the two preceding expressions are general. 
When the free indices are both space-like, the trace-full part of Eq. (\ref{PertS1}) is dynamical and it is given by:
\begin{equation}
\psi'' + {\mathcal H} (\phi' + 2 \psi') + ({\mathcal H}^2 + 2 {\mathcal H}') \phi  + \frac{1}{3} \nabla^2 (\phi - \psi) = \frac{\ell_{P}^2 a^2}{2} \biggl[ 
\delta_{\mathrm{s}} p_{\mathrm{t}} + \delta_{\mathrm{s}} p_{\mathrm{B}}
+ \delta_{\mathrm{s}} p_{\mathrm{E}}\biggr],
\label{GEOM5}
\end{equation}
where $\delta_{\mathrm{s}} p_{\mathrm{t}}$ denotes the fluctuation of the total pressure 
which can always be expressed as the sum of an adiabatic contribution (containing the total 
sound speed $c_{\mathrm{st}}$) supplemented by a non-adiabatic (or entropic) fluctuation:
\begin{equation}
\delta_{\mathrm{s}} p_{\mathrm{t}} = c_{\mathrm{st}}^2 \delta \rho_{\mathrm{t}} + \delta p_{\mathrm{nad}},\qquad c_{\mathrm{st}}^2 = \frac{p_{\mathrm{t}}^{\prime}}{\rho_{\mathrm{t}}^{\prime}}.
\label{GEOM6}
\end{equation}
Equation (\ref{GEOM6}) stipulates that the fluctuations of the total pressure can either arise as fluctuations 
of the total energy density or as fluctuations of the chemical content of the plasma: 
$\delta p_{\mathrm{nad}}$ denotes the non-adiabatic fluctuation of the total pressure \cite{nad1,nad2,nad3}
and it can only appear if the plasma contains, at least, two different species. 
The last equation to be discussed is the traceless part of the  $(ij)$ component of Eq. (\ref{PertS1}):
\begin{equation}
\partial^{i} \partial^{j}(\phi - \psi) - \frac{\delta^{ij}}{3} \nabla^2(\phi- \psi)= \ell_{P}^2 a^2 \Pi^{ij}_{\mathrm{tot}},\qquad \Pi^{ij}_{\mathrm{tot}}=  \Pi^{ij}_{\mathrm{f}} + \Pi_{\mathrm{E}}^{ij}  + \Pi_{\mathrm{B}}^{ij},
\label{GEOM8}
\end{equation}
where $\Pi^{ij}_{\mathrm{tot}}$ denotes the total anisotropic stress of the system written as the 
sum of a fluid component and of the electromagnetic components:
\begin{equation}
\Pi_{\mathrm{E}\,i}^{j} = \frac{1}{4\pi a^4} \biggl[ E_{i} E^{j} - \frac{\delta_{i}^{j}}{3} E^2 \biggr],\qquad 
 \Pi_{\mathrm{B}\,i}^{j} = \frac{1}{4\pi a^4} \biggl[
 B_{i} B^{j} - \frac{\delta_{i}^{j}}{3} B^2 \biggr].
\label{GEOM11}
\end{equation}
A transparent relation between the longitudinal 
fluctuations and the sources of anisotropic stress can be obtained by applying two spatial gradients to both sides of Eq. (\ref{GEOM8}):
\begin{eqnarray}
\nabla^2 (\phi - \psi) = \frac{3}{2} \ell_{P}^2 a^2 \Pi_{\mathrm{tot}}, \qquad \partial_{i} \partial_{j} \Pi^{ij}_{\mathrm{tot}} = \nabla^2 \Pi_{\mathrm{tot}},\qquad \Pi_{\mathrm{tot}}= \Pi_{\mathrm{E}}+ \Pi_{\mathrm{B }} + \Pi_{\mathrm{f}}.
\label{GEOM11a}
\end{eqnarray}
It is practical to express the electric and magnetic anisotropic stresses in dimensionless terms \cite{MOD1,MOD3}.
Recalling that in the concordance paradigm the dominant source of anisotropic stress 
is provided by massless neutrinos the anisotropic stresses of Eq. (\ref{GEOM11a}) 
can also be expressed as:
\begin{eqnarray}
&&\partial_{i} \partial_{j} \Pi^{ij}_{\mathrm{f}} = \nabla^2 \Pi_{\mathrm{f}}, \qquad \Pi_{\mathrm{f}} \equiv \Pi_{\nu}=  (\rho_{\nu} + p_{\nu}) \sigma_{\nu},
\label{GEOM10}\\
&&\partial_{i} \partial_{j} \Pi^{ij}_{\mathrm{E}}= \nabla^2 \Pi_{\mathrm{E}}= (p_{\gamma} + \rho_{\gamma}) \nabla^2 \sigma_{\mathrm{E}},\qquad 
\partial_{i} \partial_{j} \Pi^{ij}_{\mathrm{B}}= \nabla^2 \Pi_{\mathrm{B}}=(p_{\gamma} + \rho_{\gamma}) \nabla^2 \sigma_{\mathrm{B}}.
\label{GEOM12}
\end{eqnarray}
Since $\sigma_{\mathrm{E}}$ and $\sigma_{\mathrm{B}}$ are both dimensionless they are 
the analog of the dimensionless ratios defined in Eq. (\ref{GEOM1a})  for the electric and 
magnetic energy densities. The two sets of variables are related by the vector identities discussed in Eqs. (\ref{VECID1}) and (\ref{VECID4}). The relation $\Pi_{\mathrm{B}}=(p_{\gamma} + \rho_{\gamma}) \sigma_{\mathrm{B}}$ is also practical since $\sigma_{\mathrm{B}}$ is the magnetic analog of $\sigma_{\nu}$ introduced in the neutrino case \cite{MOD1,MOD3} (see also \cite{bertschingerma} for analog notations in the 
case of neutrinos\footnote{There is a notational ambiguity in Ref. \cite{chone34am}. When reviewing the theoretical results 
the authors use a definition for $\sigma_{\mathrm{B}}$ which is dimensionfull while the original definition employed in Ref. \cite{MOD1,MOD3} (and consistent with the notations of Ref. \cite{bertschingerma}) was dimensionless.}).  

According to the vector identities discussed in appendix \ref{APPA} (see, in particular, Eqs. (\ref{VECID4}) and (\ref{VECID7})) the magnetohydrodynamical Lorentz force appearing in Eqs. (\ref{STR22}) can be solely expressed in terms of $\sigma_{\mathrm{B}}$ and $\Omega_{\mathrm{B}}$. Indeed, from Eq. (\ref{STR22}) 
we have that 
\begin{equation}
\frac{3 \vec{\nabla}\cdot(\vec{J} \times \vec{B}) }{4 a^4 \rho_{\gamma} (R_{\mathrm{b}} + 1)} = \frac{1}{4 (R_{\mathrm{b}}+1)}\biggl[ 4 \nabla^2 \sigma_{\mathrm{B}} - \nabla^2 \Omega_{\mathrm{B}}\biggr], \qquad \vec{J} = \frac{\vec{\nabla} \times \vec{B}}{4 \pi},
\label{GEOM13a}
\end{equation}
where the right hand side follows from the left hand side thanks to Eq. (\ref{VECID4}). This observation reduces the number 
of correlators by a factor of $2$ \cite{MOD1,MOD3}. Equation (\ref{GEOM13a}) is however not universal but only holds in the absence of Ohmic electric fields which can be however included with their relative power spectra. It is finally relevant to mention that there exist magnetic field configurations which are force-free. A typical 
example are the eigenvecors of the curl, i.e. $\vec{\nabla}\times \vec{B} =\alpha \vec{B}$ originally 
studied, in the case of constant $\alpha$ within the so-called Chandrasekhar-Kendall representation \cite{chan1}.
More realistic configurations where the gyrotropy does decrease at large distance scales can be found \cite{chan2}.
These situations are even simpler to analyze since, as discussed in \cite{MOD1}, the vanishing of the total 
current implies, according to Eqs. (\ref{GEOM13a}) and (\ref{VECID4}) 
that $\Omega_{\mathrm{B}} = 4 \sigma_{\mathrm{B}}$.

\subsubsection{Gauge-invariant quasi-normal modes}
In the absence of magnetic and electric fields the Cauchy data of the temperature and polarization anisotropies can be determined with standard methods \cite{bertschingerma,MOD2a,MOD2c,MOD2d}) 
by expanding the lowest multipoles of the Einstein-Boltzmann hierarchy in powers of $| k\tau | <1 $. While the conventional semianalytic approaches \cite{semian0,semian1,semian2,semian3} can be generalized to include the presence of magnetic random fields, a more transparent strategy is to solve directly the (gauge-invariant) evolution of the quasi-normal modes of the system and then deduce the metric fluctuations 
in the wanted coordinate system. 

To derive the evolution equations of the quasi-normal modes we first take the difference between Eq. (\ref{GEOM5}) and the Hamiltonian constraint of Eq. (\ref{HAMS}) (multiplied by $c_{\mathrm{st}}^2$).
After some algebra the following first-order equation can be readily obtained:
\begin{equation}
 {\mathcal R}' = \Sigma_{{\mathcal R}} - \frac{2 a^2 \nabla^2 \psi}{\ell_{P}^2 {\mathcal H} z_{t}^2}, \qquad z_{t}= \frac{a^2 \sqrt{\rho_{\mathrm{t}} + p_{\mathrm{t}}}}{{\mathcal H}\, c_{\mathrm{st}}},
\label{NMODE2}
\end{equation}
where ${\mathcal R}$ is the curvature perturbation on comoving orthogonal hypersurfaces defined in Eq. (\ref{NMR4}); $\Sigma_{{\mathcal R}}$ contains the contribution of the electromagnetic fields, of the non-adiabatic pressure fluctuations and of the total anisotropic stress:
\begin{equation}
\Sigma_{{\mathcal R}}= - \frac{{\mathcal H}}{p_{t} + \rho_{t}} \delta p_{\mathrm{nad}} + \frac{{\mathcal H}}{p_{t} + \rho_{t}} \biggl[ \biggl( c_{\mathrm{st}}^2 - \frac{1}{3}\biggr)(\delta_{\mathrm{s}}\rho_{E} + \delta_{\mathrm{s}}\rho_{B})  + \Pi_{\mathrm{tot}} \biggr],
\label{NMODE4}
\end{equation}
where, as in Eqs. (\ref{GEOM8}) and (\ref{GEOM11a}), $\Pi_{\mathrm{tot}}$ contains both 
the electromagnetic and the fluid contributions.
By eliminating the longitudinal fluctuations of the metric (i.e. $\phi$ and $\psi$) through 
 the Eq. (\ref{NMR4}), the conformal time derivative of Eq. (\ref{NMODE2}) leads to 
 a decoupled equation for ${\mathcal R}$:
\begin{equation}
{\mathcal R}^{\prime\prime} + 2 \frac{z_{t}'}{z_{t}} {\mathcal R}^{\prime} - c_{\mathrm{st}}^2 \nabla^2 {\mathcal R} = \Sigma_{{\mathcal R}}' + 2 \frac{z_{t}'}{z_{t}} \Sigma_{{\mathcal R}} + \frac{3 a^4}{z_{t}^2} \Pi_{\mathrm{tot}}.
\label{NMODE5}
\end{equation}
Eq. (\ref{NMODE5}) follows by using 
the background equations of Eqs. (\ref{COV11})--(\ref{COV12}) together with Eqs. (\ref{NMR4}) and (\ref{GEOM11a}). The result of Eq. (\ref{NMODE5}) has been originally derived in the appendix of Ref. \cite{NoM1} and subsequently applied to various problems \cite{NoM2} involving the inflationary initial 
conditions of the magnetized curvature perturbations. When $\delta p_{\mathrm{nad}} \to 0$, $\Pi_{\mathrm{tot}} \to 0$ and in the absence of electromagnetic contributions, Eq. (\ref{NMODE5}) coincides with the evolution equation of the normal mode of a relativistic irrotational fluid derived by Lukash \cite{lukash,strokov,luk2}.

Equation (\ref{NMODE5}) must be supplemented by the evolution equation for the anisotropic stress. In the concordance scenario the only source of anisotropic stress comes from the neutrino sector; we can then take a conformal time derivative of both sides of Eq. (\ref{NU4}) and obtain:
\begin{equation}
\sigma_{\nu}^{\prime\prime} = \frac{k^2}{15} \delta_{\nu} + \frac{4}{15} k^2 \phi - \frac{11}{21}  k^2 \sigma_{\nu},
\label{NMODE6}
\end{equation}
where the neutrino hierarchy has been truncated, for illustration, to the octupole (notice, however, that ${\mathcal F}_{\nu\, 3}' \neq 0$). 
By taking a further conformal time derivative of Eq. (\ref{NMODE6}), $\delta_{\nu}^{\prime}$ can be eliminated by means of Eq. (\ref{NU2}); the term $\phi^{\prime}$ shall be traded for the explicit definition of the total anisotropic stress (\ref{GEOM12}). Finally the remaining parts of the equation (proportional to $\psi^{\prime}$) will be easily 
eliminated by using the second relation of Eq. (\ref{NMR4}) (holding in the longitudinal gauge) 
together with Eq. (\ref{NMODE2}). The final result for the evolution 
equation of $\sigma_{\nu}$ will then be \cite{NoM2}
\begin{equation}
\sigma_{\nu}''' + \frac{8}{5} {\mathcal H}^2 R_{\nu} \Omega_{R} \sigma_{\nu}' - \frac{6}{7} \nabla^2 \sigma_{\nu}' - 
\frac{8 {\mathcal H}}{5 \overline{M}_{\mathrm{P}}^2} \Pi_{\mathrm{tot}}
= \frac{4 z_{t}^2 }{15 \overline{M}_{\mathrm{P}}^2}  \biggl[ \biggl(\frac{{\mathcal H}}{a^2}\biggr)'  ({\mathcal R}' - \Sigma_{{\mathcal R}}) +  c_{\mathrm{st}}^2 \biggl(\frac{{\mathcal H}}{a^2}\biggr)\nabla^2 {\mathcal R} \biggr],
\label{NMODE7}
\end{equation}
where, by definition, $\Pi_{\mathrm{tot}} = [(p_{\nu} + \rho_{\nu}) \sigma_{\nu} + (p_{\gamma} + \rho_{\gamma})\sigma_{\mathrm{B}}]$. Note that Eq. (\ref{NMODE7}) has been derived under the hypothesis that $R_{\nu} \neq 0$ since we effectively divided both
sides of the equation by $R_{\nu}$.

The form of the evolution equations of the quasi-normal modes depend on the field content of the background. While Eq. (\ref{NMODE5}) has been obtained in the case of a perfect and irrotational fluid,
if the field content changes the quasi-normal modes may obey a slightly 
different equation.  For instance when the gauge kinetic term is coupled, in the action, to a single scalar field as $\lambda(\varphi) F_{\alpha\beta} F^{\alpha\beta}$ the evolution of the magnetized curvature perturbations can be explicitly written in terms of the auxiliary variable $\overline{\Delta}_{{\mathcal R}}$:
\begin{equation}
\overline{\Delta}_{{\mathcal R}} = \Delta_{{\mathcal R}} - \frac{{\mathcal H} a^2}{\varphi^{\prime\,2}} P, \qquad P = \frac{\vec{\nabla}\cdot( \vec{E}\times \vec{B})}{4 \pi a^4},
\label{fourth}
\end{equation}
where $\Delta_{{\mathcal R}}$ is the Laplacian of the curvature perturbations on comoving orthogonal hypersurfaces (i.e. $\Delta_{{\mathcal R}} = \nabla^2 {\mathcal R}$) 
and $P$ is the three-divergence of the Poynting vector. The equation obeyed by $\overline{\Delta}_{{\mathcal R}}$ is given by: 
\begin{equation}
\overline{\Delta}_{{\mathcal R}}^{\prime\prime} + 2 \frac{z^{\prime}_{\varphi}}{z_{\varphi}} \overline{\Delta}_{{\mathcal R}}^{\prime} - \nabla^2 \overline{\Delta}_{{\mathcal R}}= {\mathcal S}, \qquad z_{\varphi}= \frac{a \varphi^{\prime}}{{\mathcal H}}.
\label{fourtha} 
\end{equation}
The source term ${\mathcal S}$ of Eq. (\ref{fourtha}) is a functional of $P$ and of the fluctuations of the electromagnetic energy density:
\begin{equation}
{\mathcal S} = \frac{a^2}{2 \overline{M}_{P}^2}\biggl[ P^{\prime} - \biggl( 2 \frac{{\mathcal H}^{\prime}}{{\mathcal H}} + 2 \frac{a^2}{\varphi^{\prime}} V_{,\, \varphi}\biggr) P+ 
\nabla^2 (\delta_{\mathrm{s}}\rho_{B} + \delta_{\mathrm{s}}\rho_{E}) \biggr] + \frac{ 2 a^2 {\mathcal H} {\mathcal F} }{\varphi^{\prime\, 2}}\nabla^2(\delta_{\mathrm{s}}\rho_{B} - \delta_{\mathrm{s}}\rho_{E} ),
\label{fourthb}
\end{equation}
where $V_{,\, \varphi} \equiv \partial V/\partial\varphi$ and $V(\varphi)$ is the scalar potential. When two scalar fields are present, Eq. (\ref{fourtha}) will be replaced by the evolution of two 
coupled equations \cite{NoM1} which can be more easily deduced in the uniform curvature gauge \cite{hw1,hw2}. Equation (\ref{fourthb}) can be used in a variety 
of situations and, for instance, during inflation\footnote{Equation (\ref{fourthb}) has been recently used with its tensor 
analog to show that the tensor to scalar ratio $r_{T}$ cannot be too small 
if the adiabatic contribution is to dominate against the magnetic contribution during inflation. More specifically  $10^{-3} < r_{T} < 0.1$ 
gauge fields are directly coupled to the inflaton in a single-field scenario \cite{NoM4}.}. 
When the electromagnetic contribution vanishes, Eq. (\ref{fourtha}) reduces to the results of \cite{KS,chibisov}
quoted in Eq. (\ref{NM3}). The continuity of the magnetized perturbations across the inflationary transition has been analyzed in terms of these variables and the related temperature and 
polarization anisotropies have been computed in \cite{NoM1,NoM2}. 

\subsubsection{Complementary gauge-invariant descriptions} 
Since it is well known that the equations obeyed by the two Bardeen potentials\footnote{The gauge-invariant generalizations of the longitudinal fluctuations of the metric are given by $\Phi$ and $\Psi$
(i.e. the so-called Bardeen potentials) that are defined as 
$\Phi = \phi + ( F - G^{\prime})^{\prime} + {\mathcal H}( F- G^{\prime})$ and as 
$\Psi = \psi - {\mathcal H} ( F - G^{\prime})$. Recall, in this respect, the explicit form of the 
gauge transformations of Eq. (\ref{FE}) and the notion of longitudinal gauge (i.e. Eq. (\ref{STR4})).} \cite{bard1}  coincide with the equations 
written in the longitudinal gauge where  $\Psi \equiv \psi$ and $\Phi \equiv \phi$, every 
equation written in the longitudinal gauge can be swiftly translated in gauge-invariant 
terms by using the explicit form of the Bardeen potentials. So, for instance the Hamiltonian constraint of Eq. (\ref{HAMS}) and the definition of ${\mathcal R}$ of Eq. (\ref{NMR4}) can be directly expressed in terms of $\Phi$ and $\Psi$ as:
\begin{equation}
\nabla^2 \Psi - 3 {\mathcal H} ({\mathcal H}\Phi + \Psi') = \frac{\ell_{P}^2 a^2}{2} \biggl(\delta^{\mathrm{(gi)}}_{\mathrm{s}} \rho_{\mathrm{t}} + \delta_{\mathrm{s}} \rho_{\mathrm{B}} +\delta_{\mathrm{s}}\rho_{\mathrm{E}}\biggr),\qquad {\mathcal R} = - \Psi - \frac{{\mathcal H}  ( {\mathcal H} \Phi + \Psi')}{{\mathcal H}^2 - {\mathcal H}Õ},
\label{NMODE9}
\end{equation}
where $\delta^{\mathrm{(gi)}}_{\mathrm{s}} \rho_{\mathrm{t}} = \delta_{\mathrm{s}} \rho_{\mathrm{t}}
+ \rho_{\mathrm{t}}^{\prime} (F - G^{\prime})$ denotes the gauge-invariant fluctuation 
of the total energy density. The remaining equations of the longitudinal 
description can all be written in terms of the Bardeen potentials but this extension does not change the properties of the governing equations: if $\phi$ and $\psi$ 
diverge, the same will be true for $\Phi$ and $\Psi$. 

A complementary gauge-invariant description of the magnetized curvature perturbations can be obtained in terms of the density contrast on uniform curvature 
hypersurfaces: 
\begin{equation}
\zeta = -   \psi - {\mathcal H} \frac{\delta_{\mathrm{s}}\rho_{\mathrm{t}} + \delta_{\mathrm{s}}\rho_{\mathrm{B}}+ \delta_{\mathrm{s}}\rho_{\mathrm{E}}}{\rho_{\mathrm{\mathrm{t}}}'}\equiv - \Psi - {\mathcal H} \frac{\delta_{\mathrm{s}}^{(\mathrm{gi})}\rho_{\mathrm{t}} + \delta_{\mathrm{s}}\rho_{\mathrm{B}}+ \delta_{\mathrm{s}}\rho_{\mathrm{E}}}{\rho_{\mathrm{\mathrm{t}}}'},
\label{NMODE11}
\end{equation}
where the second equality in Eq. (\ref{NMODE11}) follows from the correspondence between Bardeen potentials and longitudinal 
fluctuations. Alternatively $\zeta$ describes the curvature fluctuations in the hypersurfaces 
where the total energy density is uniform. In the conventional case many 
authors do not make any difference between $\zeta$ and ${\mathcal R}$. Indeed from the definition of ${\mathcal R}$ and $\zeta$ (i.e. Eqs. (\ref{NMR4}) and 
(\ref{NMODE11})) the Hamiltonian constraint of Eq. (\ref{HAMS}) implies:
\begin{equation}
 \zeta - {\mathcal R} = \frac{2 \nabla^2 \Psi}{ 3 \ell_{P}^2 a^2 ( p_{t} + \rho_{t})} =  \frac{\Sigma_{\mathcal R} - {\mathcal R}'}{3 {\mathcal H} c_{\mathrm{st}}^2},
\label{NMODE10}
\end{equation} 
where the first equation follows from the Hamiltonian constraint while the second relation is obtained 
by using Eq. (\ref{NMODE2}). Thanks to Eq. (\ref{NMODE10}) the second-order equation obeyed by $\zeta$ is far more involved than Eq. (\ref{NMR4})  even if the two equations coincide in the $k\tau\ll 1$ limit. 
The best strategy is probably to compute ${\mathcal R}$ from Eq. (\ref{NMODE5}) and then calculate ${\mathcal R}'$; with these two ingredients  $\zeta$ can be immediately derived from Eq. (\ref{NMODE10}).  The first-order equation obeyed by $\zeta$ has been discussed at length in Ref. \cite{MOD3} (see in particular second paper) 
and it is given by 
\begin{equation}
\zeta^{\prime} = \Sigma_{\mathcal R} - \frac{{\mathcal H}}{p_{\mathrm{t}} + \rho_{\mathrm{t}}} - \frac{\theta_{\mathrm{t}}}{3},
\label{NMODE16}
\end{equation}
where $\theta_{\mathrm{t}}$ denotes the three-divergence of the total velocity field (see Eqs. (\ref{MOMS}) and (\ref{GEOM3})) and the other quantities have been already defined\footnote{The decoupled equation for $\zeta$ (analog to Eq. (\ref{NMODE5})) is formally non-local since it contains the inverse of the function $f(k,\tau) =1 + k^2/[ 3({\mathcal H}^2 - {\mathcal H}^{\prime})]$. To lowest order in  $k \tau < 1$ we have that $f(k,\tau) \to 1$: in this limit $\zeta$ and ${\mathcal R}$ evolve at the same rate, as implied by Eq.  (\ref{NMODE10}). See, in this respect, Ref. \cite{suy}.}.

\subsection{Adiabatic and non-adiabatic initial conditions} 
In the concordance paradigm, when the dark energy does not fluctuate, there are, overall five different sets of Cauchy data: one adiabatic \cite{bertschingerma} and four non-adiabatic \cite{nad1,nad2,nad3} initial conditions.  It can be actually shown on a general ground  that the field equations for cosmological perturbations in the Newtonian gauge always have an adiabatic solution, for which ${\mathcal R}$ is nonzero and constant in the limit of large wavelength \cite{weinberg}. The four non-adiabatic modes are the CDM radiation mode, the baryon-entropy mode, the neutrino entropy mode and the neutrino isocurvature 
velocity mode.  The solutions of the Einstein-Boltzmann hierarchy must be regular. If they are 
divergent they may be unphysical unless they can be smoothly described in at least 
one gauge. The explicit solutions for the adiabatic and non-adiabatic modes can be obtained from Eqs. (\ref{NMODE5}) and (\ref{NMODE7}) by setting to zero the electromagnetic contributions.  

If electromagnetic contribution vanishes in  Eq. (\ref{NMODE4}) the only source of Eq. (\ref{NMODE5}) comes from the entropic fluctuations of plasma and the general form of $\delta p_{\mathrm{nad}}$ is \cite{nad3}: 
\begin{eqnarray}
\delta p_{\mathrm{nad}}(\vec{x},\tau) &=& 
 \frac{1}{6 {\mathcal H} \rho_{\mathrm{t}}'} \sum_{\mathrm{m\,n}} \rho_{\mathrm{m}}' \rho_{\mathrm{n}}' 
(c_{\mathrm{sm}}^2 - c_{\mathrm{sn}}^2) {\mathcal S}_{\mathrm{m n}},\qquad {\mathcal S}_{\mathrm{m n}} = - (\zeta_{\mathrm{m}} - \zeta_{\mathrm{n}}),
\nonumber\\
 \zeta_{\mathrm{m}}  &=&  - \Psi + \frac{\delta^{(\mathrm{gi})}_{\mathrm{m}}}{3(w_{\mathrm{m}} + 1)},\qquad \zeta_{\mathrm{n}} = - \Psi + \frac{\delta^{(\mathrm{gi})}_{\mathrm{n}}}{3(w_{\mathrm{n}} + 1)},
\label{GEOM7}
\end{eqnarray}
where the indices m and n are not tensor indices but denote 
two generic species of the plasma; $c_{\mathrm{sm}}^2$ and $c_{\mathrm{sn}}^2$ 
are their associated  sound speeds while $w_{\mathrm{m}}$ and $w_{\mathrm{n}}$ are the corresponding 
barotropic indices. In Eq. (\ref{GEOM7}) $\zeta_{\mathrm{m}}$ and $\zeta_{\mathrm{n}}$ 
the density contrasts (for each independent fluid) on the hypersurfaces 
where the curvature is uniform and their explicit form is:
\begin{equation}
\zeta_{\nu} = - \Psi + \frac{\delta^{\mathrm{(gi)}}_{\nu}}{4}, \qquad \zeta_{\gamma} = - \Psi + \frac{\delta^{\mathrm{(gi)}}_{\gamma}}{4}, \qquad 
\zeta_{\mathrm{c}} = - \Psi + \frac{\delta^{\mathrm{(gi)}}_{\mathrm{c}}}{3},\qquad \zeta_{\rm b} = - \Psi + \frac{\delta^{\mathrm{(gi)}}_{\mathrm{b}}}{3}.
\label{GEOMzeta}
\end{equation}
Since the adiabatic solution must have, 
by definition, vanishing $\delta p_{\mathrm{nad}}$ we must require: 
\begin{equation}
\zeta_{\nu} (k,\tau) = \zeta_{\gamma}(k, \tau) = \zeta_{\mathrm{c}}(k,\tau) = \zeta_{\mathrm{b}}(k,\tau) = {\mathcal R}_{*}(k) + {\mathcal O}(k^2 \tau^2),
\label{AD2}
\end{equation}
for $\tau < \tau_{eq}$ and $ k\tau \ll 1$. 
Equations (\ref{STR18})--(\ref{STR21}) and (\ref{NU1})--(\ref{NU2}) imply that 
the variables of Eq. (\ref{GEOMzeta}) obey 
\begin{equation}
\zeta_{\gamma} '= - \frac{\theta_{\gamma{\mathrm{b}}}}{3},\qquad \zeta_{\nu}' = - \frac{\theta_{\nu}}{3}, \qquad 
\zeta_{\mathrm{c}}' = - \frac{\theta_{\mathrm{c}}}{3},\qquad \zeta_{{\mathrm{b}}}' = - \frac{\theta_{\gamma{\mathrm{b}}}}{3}. 
\label{GEOMzeta2}
\end{equation}
When the entropic contribution vanishes  Eq. (\ref{AD2}) together with Eq. (\ref{GEOMzeta2}) 
imply that the three-divergences of the peculiar velocities are all coincident and of the order 
of $k^2\tau$:
\begin{equation}
\theta_{\gamma{\mathrm{b}}}(k,\tau) = \theta_{\nu}(k,\tau)= \theta_{\mathrm{c}}(k,\tau) = \theta_{\mathrm{c}}(k,\tau) = {\mathcal O}(k^2 \tau),
\label{GEOMzeta3}
\end{equation}
for $k \tau < 1 $ and $\tau< \tau_{\mathrm{eq}}$. The initial conditions of Eq. (\ref{AD2}) during the radiation-dominated phase can be bootstrapped from the inflationary solution and, in the sudden reheating approximation they read:
\begin{equation}
{\mathcal R}(k,\tau) = {\mathcal R}_{*}(k) + \biggl(\frac{a_{r}}{a} \biggr) \frac{(1 + \eta + \epsilon)( 3 + \eta + \epsilon)}{(3 - \epsilon)} {\mathcal R}_{*}(k) 
\biggl(\frac{a_{*} \, H_{*}}{a_{r} \, H_{r}}\biggr)^3,
\label{AD5}
\end{equation}
where the star denotes the time at which the given wavelength crossed the Hubble radius during inflation\footnote{ Note that $\epsilon= - \dot{H}/H^2$ has been already introduced after Eq. (\ref{int1}); in Eq. (\ref{AD5}) $\eta = \ddot{\varphi}/(H\dot{\varphi})$ is the other standard slow-roll parameter. }.

If the large-scale magnetic fields are excited during an inflationary stage of expansion the continuity 
of the curvature perturbations across the inflationary transition may correct 
the constant adiabatic solution. In particular the analysis of Refs. \cite{NoM1,NoM2} (and, later, of Ref. \cite{dur}) showed that at the onset of the radiation dominated phase we rather have 
\begin{equation}
{\mathcal R}(k,\tau) ={\mathcal R}_{*}(k) + c_{B}(\epsilon) R_{\gamma} \overline{\Omega}_{\mathrm{B}}(k) +  d_{B}(\epsilon) R_{\gamma} \overline{\sigma}_{\mathrm{B}}(k) + {\mathcal O}(k^2 \tau^2),
\label{AD5a}
\end{equation}
where $c_{B}(\epsilon)$ and $d_{B}(\epsilon)$ are of the same order and 
at most ${\mathcal O}(1/\epsilon)$ where $\epsilon$ is the slow-roll parameter;
note also that $\overline{\Omega}_{\mathrm{B}}$ and $\overline{\sigma}_{\mathrm{B}}$ 
are conceptually the same quantities defined in Eqs. (\ref{GEOM1a}) and (\ref{GEOM11}) but they are quantitively different. The same kind of result holds also in the tensor case \cite{NoM4}. 
While it would superficially seem that the inflationary contributions simply renormalize 
the standard adiabatic mode, the new terms in Eq. (\ref{AD5a}) should be used 
to set joined bounds on the magnetic field intensity, on the total number of efolds \cite{NoM1} (see also \cite{suy}) or even on the tensor to scalar ratio when, for instance, the inflaton is 
directly coupled to the gauge fields. Note finally that in the models of inflationary magnetogenesis 
the inflaton might not be directly coupled to the gauge fields but to one (or more) 
spectator fields. In these cases Eq. (\ref{AD5a}) will have to be supplemented by the 
corresponding entropic contributions \cite{NoM1}.

\subsubsection{Regular solutions and the magnetized adiabatic mode}
In the conventional case the initial conditions of the Einstein-Boltzmann hierarchy are set during 
radiation and not directly at the onset of inflation. The rationale for this sensible physical choice 
stems from the observation that in spite of the early features 
of the inflationary phase the adiabatic solution is pretty general. 
Something similar also happens in the case when magnetic random fields are present. 
Therefore, during the radiation epoch $c_{\mathrm{st}} = 1/\sqrt{3}$, and the canonical form of  Eqs. (\ref{NMODE5})--(\ref{NMODE7}) can be directly written in terms  $y = k\tau$:
\begin{eqnarray}
&& \frac{d^2 {\mathcal R}}{d y^2} + \frac{2}{y} \frac{d {\mathcal R}}{d y} + \frac{{\mathcal R}}{3} = \frac{1}{y} \biggl( \frac{d \sigma_{t}}{d y} + \frac{2}{y} \sigma_{t} \biggr),
\label{AD6}\\
&& \frac{d^3 \sigma_{t}}{d y^3} + \biggl(\frac{6}{7} + \frac{8 R_{\nu}}{5 y^2} \biggr)\frac{d \sigma_{t}}{d y} - 16 \frac{R_{\nu}}{y^3}\sigma_{t}
 + \frac{16 R_{\nu}}{5 y^2 } \biggl( 3 \frac{d {\mathcal R}}{d y} +  \frac{y {\mathcal R}}{3}  \biggr) =0.
\label{AD7}
\end{eqnarray}
Since ${\mathcal H} = 1/\tau$ during the radiation epoch, we have that $y = k/{\mathcal H} = k/(a H)$ and $y$ is 
in fact the ratio between the particle horizon and the physical wavelength of the fluctuation. Equations (\ref{AD6}) and (\ref{AD7}) show that the system depends on a single scaling variable. 
Deep in the radiation-dominated epoch and for typical wavelengths larger than the Hubble radius 
(see Fig. \ref{figurescale}) the total anisotropic stress should obey the conditions 
\begin{equation}
\sigma_{t}(k,\tau) \ll 1 ,\qquad \sigma_{t}'(k,\tau) \ll 1, \qquad  \sigma_{t}''(k,\tau) \ll 1,
\label{AD3}
\end{equation}
where in the magnetized  $\Lambda$CDM scenario we have $\sigma_{t} =  R_{\nu} \sigma_{\nu} + R_{\gamma}\sigma_{B}$. The solutions of Eqs.  (\ref{AD6}) and (\ref{AD7}), subjected to the initial conditions of Eq. (\ref{AD3}),  are regular in the limit $y < 1$ provided:
\begin{equation}
\sigma_{t}(y) = {\mathcal A} y^{\gamma}  + {\mathcal O}(y^{2 + \gamma}), \qquad 
{\mathcal R}(y) = {\mathcal R}_{*}(k) + {\mathcal B} y^{\delta} + {\mathcal O}(y^{2 + \delta}),
\label{AD8}
\end{equation}
where $\gamma >0$ and $\delta> 0$. While the conditions of Eq. (\ref{AD8}) hold
 in the case of  adiabatic (or quasi-adiabatic) initial conditions 
 they may accidentally work also for various entropic solutions provided the 
appropriate form of $\delta p_{\mathrm{nad}}$ is considered.
Sticking, for the moment, to the case of the adiabatic initial conditions we have that Eqs. (\ref{AD6}) and (\ref{AD7}) imply that the parameters of Eq. (\ref{AD8}) must obey 
$ 12 {\mathcal A} = (18 {\mathcal B} + {\mathcal R}_{*})$ with $\delta = \gamma = 2$.
Once ${\mathcal R}$ is known Eqs. (\ref{NMODE2}), (\ref{NMR4}) and (\ref{GEOM11a}) written in a radiation-dominated epoch give the explicit expressions 
of the Bardeen potentials $\Phi$ and $\Psi$:
\begin{equation}
\frac{d {\mathcal R}}{d y} = \frac{\sigma_{t}}{y}  + \frac{y \Psi}{6}, \qquad {\mathcal R} = - \Psi - \frac{\Phi}{2}, \qquad (\Psi - \Phi) = \frac{6 \sigma_{\mathrm{t}}}{y^2},
\label{AD10}
\end{equation}
Since $\Psi$ and $\Phi$ are both constant to leading order, 
Eq. (\ref{AD8}) also implies ${\mathcal A} = (\Psi - \Phi)/6 $.
The full gauge-invariant solution for the magnetized adiabatic mode is:
\begin{eqnarray}
{\mathcal R}(y) = {\mathcal R}_{*} + {\mathcal B} y^2 + {\mathcal O}(y^4),
\qquad
\sigma_{\nu} ( y) = - \frac{R_{\gamma}}{R_{\nu}} \sigma_{B} +  \frac{18 {\mathcal B} + {\mathcal R}_{*}}{12} y^2 + {\mathcal O}(y^4).
\label{AD11}
\end{eqnarray}
In the limit $\sigma_{B}\to 0$, the solution of Eq. (\ref{AD11}) reproduces the standard adiabatic mode. From Eq. (\ref{AD11}) we can also derive the solution for $\zeta$; in fact Eq. (\ref{NMODE10}) implies
\begin{equation}
\zeta(y) = {\mathcal R}(y) + \sigma_{t}(y) - y \frac{d {\mathcal R}}{d y}  \equiv {\mathcal R}_{*}+ \frac{1}{2}\biggl( {\mathcal B} + \frac{{\mathcal R}_{*}}{6}\biggr) y^2 + {\mathcal O}(y^4),
\label{AD12}
\end{equation}
where the second equality follows after inserting the solution of Eq. (\ref{AD11}) into the first relation of Eq. (\ref{AD12}).By finally taking the difference between $\zeta$ and ${\mathcal R}$ 
we correctly obtain a result ${\mathcal O}(y^2)$, as independently required by the Hamiltonian 
constraint and by Eq. (\ref{NMODE10}):
\begin{equation}
\zeta(y) - {\mathcal R}(y) = \frac{1}{2} \biggl( \frac{{\mathcal R}_{*}}{6} - {\mathcal B}\biggr) y^2 \equiv - \frac{\Psi_{*}}{6} y^2 +  {\mathcal O}(y^4).
\label{AD13}
\end{equation}
Thanks to the regular form of the solution (\ref{AD11}) all the remaining variables 
can be simply determined. The solution for the magnetized adiabatic mode can then be written in terms of the Bardeen potentials and in terms of the other gauge-invariant plasma fluctuations. The result can be expressed as:
\begin{eqnarray} 
\Phi(k,\tau) &=& -\frac{10 \,{\mathcal R}_{*}(k)}{4 R_{\nu} + 15} -  \frac{8 R_{\gamma}}{4 R_{\nu} +15} \biggl\{  \biggl[\sigma_{\mathrm{B}}(k) + \sigma_{E}(k)\biggr] 
\nonumber\\
&-& \frac{R_{\nu}}{4} \biggl[\Omega_{\mathrm{B}}(k) + \Omega_{\mathrm{E}}(k)\biggr]\biggr\} + {\mathcal O}(k^2 \tau^2),
\nonumber\\
\Psi(k,\tau) &=& \biggl( 1 + \frac{2}{5} R_{\nu}\biggr) \Phi(k,\tau)  + \frac{4 R_{\gamma}}{5} \biggl\{  \biggl[\sigma_{\mathrm{B}}(k)+  \sigma_{\mathrm{E}}(k)\biggr] 
\nonumber\\
&-& \frac{R_{\nu}}{4} \biggl[\Omega_{\mathrm{B}}(k) + \Omega_{\mathrm{E}}(k)\biggr]\biggr\} + {\mathcal O}(k^2 \tau^2),
\nonumber\\
\delta^{(\mathrm{gi})}_{\gamma}(k,\tau) &=& -2 \Phi(k,\tau) - R_{\gamma} \biggl[\Omega_{\mathrm{B}}(k)  + \Omega_{\mathrm{E}}(k)\biggr] + {\mathcal O}(k^2 \tau^2),
\nonumber\\
\delta^{(\mathrm{gi})}_{\nu}(k,\tau) &=& -2 \Phi(k,\tau) - R_{\gamma} \biggl[\Omega_{\mathrm{B}}(k) + \Omega_{\mathrm{E}}(k)\biggr] + {\mathcal O}(k^2 \tau^2),
\nonumber\\
\delta^{(\mathrm{gi})}_{\mathrm{c}}(k,\tau) &=& - \frac{3}{2} \Phi(k,\tau) - \frac{3}{4}R_{\gamma} \biggl[\Omega_{\mathrm{B}}(k) + \Omega_{\mathrm{E}}(k)\biggr]+ {\mathcal O}(k^2 \tau^2),
\nonumber\\
 \delta^{(\mathrm{gi})}_{\mathrm{b}}(k,\tau) &=& - \frac{3}{2} \Phi(k,\tau) - \frac{3}{4}R_{\gamma} \biggl[\Omega_{\mathrm{B}}(k) + \Omega_{\mathrm{E}}(k)\biggr]+ {\mathcal O}(k^2 \tau^2),
\nonumber\\
\sigma^{(\mathrm{gi})}_{\nu}(k,\tau) &=& - \frac{R_{\gamma}}{R_{\nu}} \biggl[ \sigma_{\mathrm{B}}(k) +\sigma_{\mathrm{E}}(k)\biggr]+ \frac{k^2 \tau^2}{6 R_{\nu}} [ \Psi(k,\tau) - \Phi(k,\tau)] +  {\mathcal O}(k^4 \tau^4),
\nonumber\\
\theta^{(\mathrm{gi})}_{\gamma\mathrm{b}}(k,\tau) &=& \frac{k^2 \tau}{2} \biggl[ \Phi(k,\tau) + \frac{R_{\nu}}{2}  \Omega_{\mathrm{B}}(k)  - 
\frac{R_{\gamma}}{2} \Omega_{\mathrm{E}}(k) - 2 \sigma_{\mathrm{B}}(k) \biggr] + {\mathcal O}(k^3 \tau^2),
\nonumber\\
\theta^{(\mathrm{gi})}_{\nu}(k,\tau) &=& \frac{k^2 \tau}{2}\biggl[ \Phi(k,\tau) - \frac{R_{\gamma} \Omega_{\mathrm{B}}(k)+\Omega_{\mathrm{E}}(k)}{2} + 2 \frac{R_{\gamma}}{R_{\nu}}( \sigma_{\mathrm{B}}(k) +\sigma_{\mathrm{E}}(k)) \biggr] + {\mathcal O}(k^3 \tau^2),
\nonumber\\
\theta^{(\mathrm{gi})}_{\mathrm{c}}(k,\tau) &=& \frac{k^2 \tau}{2} \Phi(k,\tau)+ {\mathcal O}(k^3 \tau^2).
\label{L4}
\end{eqnarray}
The solution of Eq. (\ref{L4}) is gauge-invariant and it can be expressed in any specific coordinate system. From Eq. (\ref{L4}) we can also obtain the magnetized 
adiabatic mode in the synchronous gauge by expressing the Bardeen potentials and the other gauge-invariant fluctuations in the synchronous coordinate system.

The magnetized adiabatic mode has been originally discussed in Refs. \cite{MOD1,MOD3} (see also \cite{MOD4,MOD4a}). In all these analyses the magnetic field has been taken to be fully inhomogeneous in order not to break explicitly the isotropy\footnote{See however \cite{MOD5,MOD5b} for a complementary perspective where a uniform component is discussed.}.  A specific strategy for the semi-analytical and for the numerical study of the magnetized temperature and polarization anisotropies has been presented in \cite{MOD6} and subsequently analyzed with dedicated and independent numerical 
approaches \cite{MOD7,MOD8,MOD7a,MOD8a,MOD9,MOD10,MOD11}. 
The explicit form of the magnetized adiabatic mode in the synchronous gauge can be 
explicitly found in \cite{MOD1,MOD3,NoM2,MOD7}.

While the magnetized adiabatic mode of Eq. (\ref{L4}) shares some of the properties of bona 
fide adiabatic modes but not all of them and this is the reason why in \cite{MOD1,MOD3} 
the terminology quasi-adiabatic mode has been suggested. Even if this terminology 
did not encounter much success in the papers dealing with data analysis, it is 
however relevant to remark that a genuine adiabatic mode must satisfy Eqs. (\ref{AD2})
and (\ref{GEOMzeta3}). Equation (\ref{L4}) does not satisfy Eq. (\ref{GEOMzeta3}) 
but does satisfy Eq. (\ref{AD2}). In fact, inserting the solution (\ref{L4}) into
 Eq. (\ref{GEOMzeta}) the following equation can be readily obtained:
 \begin{equation}
\zeta_{\nu} (k,\tau) = \zeta_{\gamma}(k, \tau) = \zeta_{\mathrm{c}}(k,\tau) = \zeta_{\mathrm{b}}(k,\tau) = {\mathcal R}_{*}(k) - \frac{R_{\gamma}}{4} [ \Omega_{\mathrm{B}}(k) + 
\Omega_{\mathrm{E}}(k)] +
 {\mathcal O}(k^2 \tau^2).
\label{QA1}
\end{equation}
The condition (\ref{QA1}), as its adiabatic counterpart of Eq. (\ref{AD2}), always 
implies that $\delta p_{\mathrm{nad}}\to 0$. If we however consider Eq. (\ref{GEOMzeta3})
we see clearly that it is not satisfied in the same way. More precisely, from Eq. (\ref{L4}) 
we have that the peculiar velocities are all ${\mathcal O}(k^2 \tau)$ to leading order but 
their values are different. Let us finally compute the total $\zeta$ introduced in Eq. 
(\ref{NMODE11}); using the definitions of Eq. (\ref{GEOMzeta}) into Eq. (\ref{NMODE11}) 
we have that 
\begin{equation}
\zeta= \frac{\rho_{\mathrm{c}} \zeta_{\mathrm{c}} + \rho_{\mathrm{b}} \zeta_{\mathrm{b}} 
+ 4 \rho_{\gamma} \zeta_{\gamma}/3 +  4\rho_{\nu} \zeta_{\nu}/3} {\rho_{\mathrm{t}} + p_{\mathrm{t}}}
+ \frac{\delta_{\mathrm{s}} \rho_{\mathrm{B}} + \delta_{\mathrm{s}} \rho_{\mathrm{E}}}{3 (\rho_{\mathrm{t}} + p_{\mathrm{t}})},
\label{QA2}
\end{equation}
where $p_{\mathrm{t}}$ and $\rho_{\mathrm{t}}$ have been 
explicitly given in Eq. (\ref{F10a}).
If we now insert Eq. (\ref{QA1}) into Eq. (\ref{QA2}) we obtain that $\zeta(k,\tau) = {\mathcal R}_{*}(k) + {\mathcal O}(k^2\tau^2)$ as implied by the standard adiabatic mode. 
It is therefore true that also the magnetized adiabatic mode does not lead to entropic fluctuations 
but the properties of this mode are not exactly the ones of the conventional adiabatic solution.

\subsubsection{Divergent solutions}
In the absence of the neutrino anisotropic stress the system of Eqs. (\ref{AD6}) and (\ref{AD7}) leads to a divergent solution. In this case the equation to integrate is simply given by:
\begin{equation}
\frac{d^{2} {\mathcal R}}{d y^2} + \frac{2}{y} \frac{d R}{d y} + c_{\mathrm{st}}^2 {\mathcal R} = \frac{ 2 R_{\gamma} \sigma_{B}}{y^2}.
\label{ANN17}
\end{equation}
By now introducing the auxiliary variable $q(y) = y {\mathcal R}(y)$, Eq. (\ref{ANN17}) 
has the following solution:
\begin{eqnarray}
{\mathcal R}(y) &=& {\mathcal R}_{*} \frac{\sin{(c_{\mathrm{st}} y)}}{c_{\mathrm{st}} y} + 2 R_{\gamma} \sigma_{B} \biggl[ \biggl( \mathrm{Ci}(c_{\mathrm{st}} y) -   \mathrm{Ci}(c_{\mathrm{st}} y_{i}) \biggr) \frac{\sin{(c_{\mathrm{st}} y)}}{c_{\mathrm{st}} y} 
\nonumber\\
&+& \biggl( \mathrm{Si}(c_{\mathrm{st}} y_{i}) -   \mathrm{Si}(c_{\mathrm{st}} y) \biggr) \frac{\cos{(c_{\mathrm{st}} y)}}{c_{\mathrm{st}} y} \biggr],
\label{ANN20}
\end{eqnarray}
where $y = k \tau$ and $y_{i} = k \tau_{i}$; the two arbitrary constants of the homogeneous equation have been fixed by requiring 
that ${\mathcal R}(y_{i}) \to {\mathcal R}_{*}$ for $y_{i} \ll 1$. In Eq. (\ref{ANN20}) $c_{\mathrm{st}} = 1/\sqrt{3}$ denotes the total sound speed of the plasma (which is constant in the radiation-dominated epoch); the functions $\mathrm{Ci}(z) $ and $\mathrm{Si}(z)$ are 
the standard cosine integral and sine integral functions \cite{abr1,abr2}.
The expression at the right hand side of Eq. (\ref{ANN20}) can be expanded in powers of $y$ and $y_{i}$ with the result that:
\begin{equation}
{\mathcal R}(y) = {\mathcal R}_{*} + 2 R_{\gamma} \sigma_{B} [\ln{(y/y_{i})} -1] + 2 R_{\gamma} \sigma_{B} y_{i}/y + {\mathcal O}(y^2) + {\mathcal O}(y^2 y_{i}) + {\mathcal O}(y_{i}^2) 
\label{ANN21}
\end{equation}
where we recall that $y \geq y_{i}$. From Eq. (\ref{ANN20})  the Bardeen potentials are:
\begin{eqnarray}
\Psi(k,\tau) &=& \frac{6 R_{\gamma} \sigma_{\mathrm{B}}}{k^2 \tau^2} \biggl[ 1 - 2 \biggl( \frac{\tau_{i}}{\tau}\biggr) - c_{\mathrm{st}}^2 k^2 \tau_{i} \tau + 
{\mathcal O}(k^4\tau^4)\biggr],
\nonumber\\
\Phi(k,\tau) &=&  \frac{6 R_{\gamma} \sigma_{\mathrm{B}}}{k^2 \tau^2}  \biggl[ - 2 \biggl( \frac{\tau_{i}}{\tau}\biggr) - c_{\mathrm{st}}^2 k^2 \tau_{i} \tau + 
{\mathcal O}(k^4\tau^4)\biggr],
\label{ANN22}
\end{eqnarray}
and they both diverge for $k \tau <1$; furthermore,  to all orders in $k\tau$, 
$(\Psi- \Phi) = 6 R_{\gamma} \sigma_{\mathrm{B}}/(k^2 \tau^2)$. 
While the magnetized adiabatic mode 
is a regular mode in any gauge, it has been known for long time that large-scale magnetic fields also admit divergent modes. This possibility has been pointed out long time ago (see, in particular, Eqs. (8.22)-(8.23) of  Ref. \cite{magnetized} and discussion therein). The divergence of the Bardeen potentials 
implies similar singularities in the plasma fluctuations. It is desirable 
to find a description where no singularities appear in the same way as, in the conventional situation, the divergent modes are cured by going 
to the synchronous gauge\footnote{This happens in the case of the divergent 
isocurvature modes arising in the neutrino sector \cite{DIV1}.}. 
However the mode of Eq. (\ref{ANN21}) does also diverge in the synchronous gauge. 

Taken at face value, the previous results suggest that it is not possible to use the divergent modes as initial conditions 
of the Einstein-Boltzmann hierarchy after neutrino decoupling. In the case of the 
divergent modes the initial conditions should be given at early time 
but different models of the origin of large scale magnetic fields suggest different 
normalization times. Fortunately, even if ${\mathcal R}$ grows 
logarithmically before neutrino decoupling,  after neutrino decoupling, 
the anisotropic stress of the fluid turns the divergent mode into a regular one. 

This approximate picture can be justified in terms of explicit analytic solutions and numerical integrations (see e.g. Fig. 1 in \cite{NoM2}).  The numerical results can be analytically understood by solving Eqs. (\ref{AD6}) and (\ref{AD7}) in general terms, i.e. without imposing the condition (\ref{AD3}). For this purpose Eqs. (\ref{AD6}) and (\ref{AD7}) can be written in terms of the new variable $w = (x -x_{i})$ where 
$x= \ln{y}$ and $x_{i} = \ln{y_{i}}$ corresponds to the moment at which initial conditions are set. Since $w$ evolves from $0$ we can Laplace transform Eqs. (\ref{AD6}) and (\ref{AD7}), solve the obtained algebraic conditions and then anti-transform the general solution that depends on the anisotropic stress and its derivatives; this analysis 
has been discussed in the second paper of Ref. \cite{NoM2}. Since the divergent modes turn anyway into regular (or compensated) modes it makes more sense to constrain directly the magnetized adiabatic mode (and its non-adiabatic generalization). This is, after all, the logic followed in the conventional case when the magnetic fields are absent. 

The magnetized adiabatic mode introduced in \cite{MOD1,MOD3} has been referred to as compensated mode in some observational analyses \cite{chone34am}. 
The divergent solutions discussed here and firstly analyzed in \cite{magnetized} have been dubbed passive modes in \cite{DIV2}. Incidentally, in Ref. \cite{DIV2} the authors included the 
masses of the neutrinos in the Boltzmann hierarchy as previously suggested \cite{DIV3} 
with rather different quantitive results.  We shall not denote the magnetized adiabatic mode 
as compensated since this terminology is ambiguous and may conflict with a different 
sort of compensation previously pointed out and discussed hereunder (see Eq. (\ref{expl2}) and discussion 
thereafter).

\subsubsection{Post-equality evolution}
While the initial conditions of the Einstein-Boltzmann hierarchy are assigned prior to decoupling 
and in the regime of Fig. \ref{figurescale}, the subsequent evolution must  be followed by means of numerical integration. For instance in Refs. \cite{MOD1,MOD3,MOD9}  some specific integration methods have been explored and the use of Boltzmann codes for the computation of the temperature and polarization anisotropies is common (see e.g. \cite{MOD7,MOD8,MOD7a,MOD8a}). Since numerical results without 
a specific analytic understanding are often difficult to decipher,  it is useful to mention some 
simple analytical approximations that can be used, for instance, in the estimate 
of the Sachs-Wolfe \cite{MOD1,MOD3} and integrated Sachs-Wolfe \cite{MOD11} effects. 
These methods typically apply for large-scales  (i.e. $\ell < \sqrt{z_{\mathrm{rec}}}$) but can be extended to smaller scales \cite{MOD3,MOD9} to understand with semianalytic methods the distortions of the acoustic peaks caused by magnetic random fields. The approaches of \cite{MOD3,MOD9} generalize to the magnetized case the classic analytic strategies employing approximate Gaussian forms of the 
visibility function \cite{naselskyb,weinbergb,primer} (see also \cite{semian0,semian1,semian2,semian3}). The obtained results have been compared with the ones Boltzmann solvers and they can be usefully employed for 
a qualitative understanding of the typical distortions induced by the magnetic random fields 
on the angular power spectra \cite{MOD3,MOD8a}.

Even to analyze the simplest large-scale effects (like the Sachs-Wolfe or the integrated Sachs-Wolfe effects) we need to integrate Eq. (\ref{NMODE5}) across matter-radiation equality since 
the photon decoupling occurs during the matter-dominated stage of expasnion.  The modes interested by the present calculation are the ones that will reenter the Hubble radius after equality. This means that the spatial gradients in Eq. (\ref{NMODE5}) can be neglected so that 
the approximate form of the equation becomes 
\begin{equation}
\partial_{\tau} \biggl[ z_{\mathrm{t}}^2 \biggl({\mathcal R}^{\prime} - \Sigma_{{\mathcal R}}\biggr) \biggr] = 3 a^4 \Pi_{\mathrm{tot}}.
\label{post1}
\end{equation}
Equation (\ref{post1}) has been explicitly solved in a number of different ways (see e.g. \cite{MOD3,MOD8a}). If we set initial conditions in the radiation epoch by considering, for instance,
the magnetized adiabatic mode\footnote{This is not a 
restrictive approximation since, as we saw, different initial conditions  must relax to regular 
modes and, among them, a particular role is played by the magnetized adiabatic mode. } of Eq. (\ref{L4}) we have that $\Pi_{\mathrm{tot}} = {\mathcal O}(k^2 \tau^2)$ (and hence negligible around equality). Equation (\ref{post1}) can then be rephrased as: 
\begin{equation}
\frac{\partial {\mathcal R}}{\partial \ln{\alpha}} = - \frac{c_{\mathrm{st}}^2\rho_{\mathrm{M}}}{\rho_{\mathrm{t}} + p_{\mathrm{t}}} {\mathcal S}_{*}
+ \biggl(c_{\mathrm{st}}^2 - \frac{1}{3}\biggr) \frac{\delta_{\mathrm{s}}\rho_{\mathrm{B}}}{p_{\mathrm{t}} + \rho_{\mathrm{t}}},
\label{post2}
\end{equation}
where $\alpha= a/a_{\mathrm{eq}}$ is the scale factor normalized at equality and $\rho_{\mathrm{M}}$ is the 
total energy density of non-relativistic matter. In Eq. (\ref{post2}) we also added for immediate comparison a matter-radiation isocurvature mode. 
Equation (\ref{post2}) is easily solvable  by recalling the standard results for the total barotropic index 
$w_{\mathrm{t}}$, for the total sound speed $c_{\mathrm{st}}$ and for the critical fractions 
of matter (i.e. $\Omega_{\mathrm{M}}$) and radiation (i.e. $\Omega_{\mathrm{R}}$):
\begin{equation}
w_{\mathrm{t}} = \frac{1}{3 (\alpha + 1)}, \qquad c_{\mathrm{st}}^2 = \frac{4}{3 (3\alpha + 4)}, \qquad 
\Omega_{\mathrm{M}} = \frac{1}{\alpha+1}, \qquad \Omega_{\mathrm{R}} = \frac{\alpha}{\alpha+1}.
\label{post2a}
\end{equation}
Using repeatedly the results of Eq. (\ref{post2a}) in Eq. (\ref{post2}) we obtain after simple algebra:
\begin{equation}
{\mathcal R}(k,\alpha)= {\mathcal R}_{*}(k) - \frac{\alpha [ 4 {\mathcal S}_{*}(k) + 3 R_{\gamma} \Omega_{\mathrm{B}}(k)]}{4 (3 \alpha + 4)},
\label{zetaalpha}
\end{equation}
where ${\mathcal R}_{*}(k)$ denotes the constant adiabatic mode;
 well before and well after equality, Eq. (\ref{zetaalpha}) implies, respectively, 
\begin{eqnarray}
&& \lim_{\alpha\gg 1} {\mathcal R}(k,\alpha) = {\mathcal R}_{*}(k) - \frac{{\mathcal S}_{*}(k)}{3} - \frac{R_{\gamma} \Omega_{\mathrm{B}}(k)}{4},
\label{alphagg}\\
&& \lim_{\alpha\ll 1}  {\mathcal R}(k,\alpha) = {\mathcal R}_{*}(k) - \frac{{\mathcal S}_{*}(k)}{4}\alpha - \frac{3R_{\gamma} \Omega_{\mathrm{B}}(k)}{16}\alpha.
\label{alphall}
\end{eqnarray}
From Eqs. (\ref{alphagg}) and (\ref{alphall}) and from the definition of curvature perturbations we can determine the Bardeen potential 
\begin{eqnarray}
\Psi(k,\alpha) &=& - \frac{{\mathcal R}_{*}(k)}{15\alpha^3} \{ 16 [ \sqrt{\alpha + 1} -1] + \alpha [ \alpha ( 9\alpha + 2) -8]\} 
\nonumber\\
&+& \frac{4 {\mathcal S}_{*}(k) + 3 R_{\gamma} \Omega_{\mathrm{B}}(k)}{20 \alpha^3} \{ 16 [ 1 - \sqrt{\alpha + 1}] + \alpha [ 8 + \alpha (\alpha -2)]\}.
\label{expl1}
\end{eqnarray}
As in the case of Eqs. (\ref{alphagg}) and (\ref{alphall}) Eq. (\ref{expl1}) can be evaluated well after and well before equality:
\begin{eqnarray}
&& \lim_{\alpha \gg 1} \Psi(k,\alpha)= - \frac{3}{5} {\mathcal R}_{*}(k) + 
\frac{ 4 {\mathcal S}_{*}(k) + 3 R_{\gamma} \Omega_{\mathrm{B}}(k)}{20},
\nonumber\\
&& \lim_{\alpha \ll 1}\Psi(k,\alpha) = - \frac{2}{3}  {\mathcal R}_{*}(k) 
+ \frac{\alpha}{32} \biggl[ 
\frac{4}{3} {\mathcal R}_{*}(k) + 4 {\mathcal S}_{*}(k) + 3 R_{\gamma}\Omega_{\mathrm{B}}(k)\biggr].
\label{expl2}
\end{eqnarray}
These results have been used for various semi-analytical determinations of the temperature and polarization anisotropies and demonstrate that the terminology ``compensated mode'' for what the magnetized adiabatic mode may be confusing in some cases. Non-adiabatic modes in the presence of magnetic fields may indeed lead to 
different compensating effects specifically analyzed in \cite{nad3}. The idea in short is that at the level of the initial data the 
non-adiabatic modes may erase either partially or totally the contribution of the magnetic random fields. For instance in Eq. (\ref{zetaalpha}) we could imagine that the non-adiabatic and magnetic fluctuations could compensate in such a way that the net effect will be either erased of substantially reduced (i.e. $|4 {\mathcal S}_{*}(k) + 3 R_{\gamma} \Omega_{\mathrm{B}}(k)|\ll 1 $). If this happens it can be shown numerically \cite{nad3} that the potential distortions in the temperature and polarization anisotropies may be strongly reduced. 

\subsection{Magnetized vector modes}
Since the baryon velocity is not solenoidal the system of the scalar modes of the geometry and of the 
plasma is compressible because the fluctuations may compress the plasma when 
$\vec{\nabla}\cdot \vec{v}_{\mathrm{b}} \neq 0$. The vector modes of the geometry are instead affected 
by the solenoidal  component of the total velocity field which will be denoted hereunder by $\vec{{\mathcal V}}$ (or, in gauge-invariant terms by $\vec{V}$). By definition we will have the that 
the total velocity, the velocities of the various species and the fluctuations of the geometry 
will all be solenoidal. In the vector case the fluctuations of the plasma and of the geometry 
cannot compress the plasma and are hence called incompressible \cite{krall,goed}. 
The compressible and the incompressible closures are physically rather different.

\subsubsection{Evolution equations and momentum constraint}
The vector fluctuations of the geometry are parametrized in terms of two divergence-less 
vectors\footnote{We remind that $\delta_{\mathrm{v}} g_{0i} = - a^2 Q_{i}$ and 
$\delta_{\mathrm{v}} g_{ij} = a^2 (\partial_{i} W_{j} + \partial_{j} W_{i})$ with 
$\partial_{i}W^{i}= \partial_{i} Q^{i} =0$.} which have been already introduced
in Eqs. (\ref{FLUC3})  and (\ref{QW}).  Since the vector fluctuations cannot compress the plasma the Hamiltonian constraint disappears; furthermore the momentum constraint can be easily solved.
Finally the vector components of the electromagnetic energy-momentum tensor is technically less cumbersome than in the scalar case (see, for instance, Eqs. (\ref{PItot})--(\ref{PIT}) and discussion therein).  In analogy with Eq. (\ref{PertS1}) we can consider the vector fluctuations 
of Eq. (\ref{COV1}) implying 
\begin{equation}
\delta_{\mathrm{v}} R_{\mu}^{\nu} = \ell_{P}^2 \delta_{\mathrm{v}} T_{\mu}^{\nu}, \qquad 
\delta_{\mathrm{v}} \biggl(\nabla_{\mu} T^{\mu}_{\nu}\biggr) =0,
\label{PertV1}
\end{equation}
where  $\delta_{\mathrm{v}}$ denotes a vector fluctuation of the corresponding tensor and has been already introduced after Eq. (\ref{FLUC1}). The only relevant equations for the vector problem are, respectively, the  $(0i)$ and $(i\neq j)$ components of the perturbed Einstein equations (\ref{PertV1}):
\begin{equation} 
\delta_{\mathrm{v}} R_{0}^{i}  = \ell_{P}^2 \delta_{\mathrm{v}} T_{0}^{i},\qquad \delta_{\mathrm{v}} R_{i}^{j}  = \ell_{P}^2 \delta_{\mathrm{v}} T_{i}^{j},
\label{VEC0}
\end{equation}
where $\delta_{\mathrm{v}} R_{i}^{j}$ and $\delta_{\mathrm{v}} R_{0}^{i}$ 
are the components of the 
Ricci tensor perturbed to first-order in the vector fluctuations defined in Eqs. Eqs. (\ref{FLUC3})  and (\ref{QW}). The explicit form of $\delta_{\mathrm{v}} R_{i}^{j}$ and $\delta_{\mathrm{v}} R_{0}^{i}$ can be found, for instance, in appendix D of Ref. \cite{primer}. 

It is more practical to write the momentum constraint of (\ref{VEC0}) by taking directly the curl of the obtained expression and the 
result is\footnote{In what follows we shall often use the notation $\vec{\omega}_{X}$ 
for the vorticity of a given vector; this notation cannot be confused with the one
used for the angular frequencies thanks to the presence of the vector symbol.}
\begin{eqnarray}
 \nabla^2 \vec{\omega}_{Z} &=& 2 a^2 \ell_{P}^2 (\rho_{\mathrm{t}} + p_{\mathrm{t}}) \vec{\omega}_{V} + \frac{\ell_{P}^2}{2 \pi a^2} \vec{\nabla}\times\biggl(\vec{E} \times \vec{B}\biggr),
\label{VEC1}\\
\vec{\omega}_{Z} &=& \vec{\nabla}\times \vec{Z}, \qquad \vec{\omega}_{V} = \vec{\nabla} \times \vec{V},
\label{VEC1a}
\end{eqnarray}
The total vorticity $\vec{\omega}_{V} = \vec{\nabla} \times \vec{V}$ of the plasma obeys: 
\begin{equation}
(p_{\mathrm{t}} + \rho_{\mathrm{t}}) \vec{\omega}_{V} = \frac{4}{3} \rho_{\gamma} \vec{\omega}_{\gamma} + \frac{4}{3} \rho_{\nu} \vec{\omega}_{\nu}+ \rho_{\mathrm{e}} \vec{\omega}_{\mathrm{e}} +\rho_{\mathrm{i}} \vec{\omega}_{\mathrm{i}}  + \rho_{\mathrm{c}} \vec{\omega}_{\mathrm{c}}, 
\label{VEC4}
\end{equation}
where the vorticities of the different species have been introduced. Note that the two vectors $\vec{Z}$ and $\vec{V}$ (and their corresponding vorticities $\vec{\omega}_{Z}$ 
and $\vec{\omega}_{V}$)
are gauge-invariant and they are given by:
\begin{equation}
\vec{Z} = \partial_{\tau} \vec{W} + \vec{Q}, \qquad \vec{V} = \vec{{\mathcal V}}+ \vec{Q}.
\label{VEC3}
\end{equation}
Indeed, recalling Eq. (\ref{QW}), we have that $\vec{Z}$ is automatically gauge-invariant.  Since total vector
fluctuation of the fluid and the related gauge shift are given, respectively, by 
\begin{equation}
\delta_{\mathrm{v}} T_{0}^{i} = ( p_{\mathrm{t}} + \rho_{\mathrm{t}}) {\mathcal V}^{i}, \qquad {\mathcal V}^{i} \to \widetilde{{\mathcal V}}^{i} = {\mathcal V}^{i} + \partial_{\tau} \zeta^{i},
\label{VEC3a}
\end{equation}
it follows from Eq. (\ref{QW}) that also $\vec{V}$ and $\vec{\omega}_{V}$ are explicitly gauge-invariant. 
The $(i\neq j)$ components of the perturbed Einstein equations (\ref{PertV1}) can be finally written as:
\begin{equation}
\biggl(\partial_{i}Z_{j} + \partial_{j} Z_{i}\biggr)^{\prime} + 2 {\mathcal H} \biggl(\partial_{i}Z_{j} + \partial_{j} Z_{i}\biggr) = - 2\ell_{P}^2 a^2 \biggl[\Pi_{ij}^{(vec,\mathrm{B})} + \Pi_{ij}^{(vec,\nu)}\biggr].
\label{VEC2}
\end{equation}
Since the vector fluctuations cannot compress the plasma, the diagonal part of the perturbed energy-momentum tensor only affects the scalar modes; in Eq. (\ref{VEC2}) $\Pi_{ij}^{(vec,\mathrm{B})}$ and $\Pi_{ij}^{(vec,\nu)}$ denote, respectively, the vector component of the magnetic anisotropic stress (see also Eqs. (\ref{PItot})--(\ref{PIV}) and discussion therein) and the anisotropic stress of the fluid which is only due to neutrinos, as it happens in the concordance paradigm. The most notable 
difference between the compressible and the incompressible case is the possible 
presence of turbulence. Even if there are situations where the turbulence may be 
compressible, isotropic turbulence is often considered to be incompressible 
when the characteristic velocities are much smaller than the sound 
speed of the medium. 

The hypothesis of primeval turbulence is inextricably bound to the vector modes of the geometry and
has been a recurrent theme since the first speculations 
on the origin of the light nuclear elements. The implications of turbulence for galaxy formation 
have been pointed out in the fifties by Von Weizs\"aker and Gamow \cite{VW}. 
They have been scrutinized in the sixties and early seventies by various authors \cite{turb1} 
(see also \cite{peebles,barrow} and discussions therein).
The first  connection between vector modes of the plasma and 
large-scale magnetism dates back to the seminal 
contributions of Harrison \cite{harrison}. While the idea of Harrison 
was to use turbulence to generate large-scale magnetic fields also the opposite process, leading to Alfv\'en waves was implicitly discussed. We remind that in weakly coupled plasmas the incompressible closure is associated with the presence of Alfv\'en waves \cite{krall,biskamp,goed}. 
The potential presence of Alfv\'en waves prior to decoupling has been 
convincingly scrutinized in recent times \cite{VV2b} already with the WMAP 5-years 
data release \cite{chone30}.  Since Alfv\'en waves are typically treated 
in the presence of a uniform magnetic fields \cite{VV2b} (see also \cite{VV2a}) 
the vector fluctuations break explicitly the isotropy of the background by inducing off-diagonal 
correlations in multipole space. For this reasons the bounds on the 
primordial vector modes are closely related, in this case, to the studies
aiming at testing the statistical isotropy of the microwave background \cite{VV4,VV5}.

\subsubsection{Initial conditions of the vector problem}
In the absence of magnetic random fields and fluid anisotropic stress the vector modes of the geometry are always suppressed as  $a^{-2}$, as long as the background geometry expands. Indeed, in this situation,  Eqs. (\ref{VEC1}) and (\ref{VEC2}) imply
\begin{equation}
\partial_{\tau} \vec{Z}+ 2 {\mathcal H} \vec{Z} =0, \qquad \vec{V} = \frac{\nabla^2 \vec{Z}}{2 a^2 \ell_{P}^2 (\rho_{\mathrm{t}} + p_{\mathrm{t}})}.
\label{VEC2a}
\end{equation}
The solution of Eq. (\ref{VEC2a}) depends on the evolution of the background. 
In the case of a post-inflationary evolution dominated by a perfect fluid with constant 
barotropic index $w$, in Fourier space Eq. (\ref{VEC2a}) implies:
\begin{equation}
\vec{Z}(k,\tau) = \vec{Z}(k,\tau_{e}) \biggl(\frac{a}{a_{i}}\biggr)^{-2}, \qquad 
\vec{V}(k,\tau) = - \biggl(\frac{k}{a_{e} H_{e}}\biggr)^2  \frac{  \vec{Z}(k,\tau_{e})}{6 (1+ w)} \biggl(\frac{a}{a_{e}}\biggr)^{3w -1},
\label{VEC2b}
\end{equation}
where Eqs. (\ref{COV11}) and (\ref{COV12}) have been used together with
 the identity ${\mathcal H} = a H$; in Eq. (\ref{VEC2}) $\tau_{r}$ denotes some initial times coinciding, for instance, with the end of inflation. 
 In the concordance paradigm the evolution of the background is 
 dominated by radiation almost immediately after inflation and it is therefore clear 
 that, in this case, primeval turbulence (or simply a growing mode of the velocity) 
 are categorically excluded. While this is the reason why vector modes never appear in the concordance paradigm, two exceptions can be envisaged. If the 
 evolution of the post-inflationary background deviates from $w=1/3$, then 
 the vector modes of the geometry will always decay but the total velocity and the total vorticity 
 may potentially increase as long as $w> 1/3$. In particular when $w\to 1$  the total velocity increases 
 as $a^2$. This possibility has been suggested long ago by Barrow as a potential 
 source of primeval turbulence \cite{barrow}. A similar possibility arises 
in various classes of bouncing scenarios  where, for some time, the universe contracts  and vector modes can grow especially in the presence of dynamical extra-dimensions \cite{mgvect1,mgvect2}. 

The same analysis leading to Eq. (\ref{VEC2b}) can be generalized to the case 
of inflation where the evolution of $\vec{Z}(k,\tau)$ is the same but the evolution 
of the velocity is slightly different:
\begin{equation}
\vec{Z}(k,\tau) = \vec{Z}(k,\tau_{i}) \biggl(\frac{a}{a_{i}}\biggr)^{-2}, \qquad 
\vec{V}(k,\tau) = - \frac{1}{4 \epsilon} \biggl(\frac{k}{a_{i} H_{i}}\biggr)^2  \vec{Z}(k,\tau_{i}) \biggl(\frac{a}{a_{i}}\biggr)^{-4}.
\label{VEC2c}
\end{equation}
We finally mention that the contribution of the Poynting flux in Eq. (\ref{VEC1}) 
does not alter significantly the conclusions of Eqs. (\ref{VEC2b}) and (\ref{VEC2c}).
All in all we can say that the vector modes of the geometry are always 
strongly suppressed both during and after inflation. 

Let us now examine the situation where magnetic random fields appear in 
Eqs. (\ref{VEC1}) and (\ref{VEC2}) either alone or in the presence of the anisotropic stress of 
the fluid.  From Eqs. (\ref{VEC1}) and (\ref{VEC2}) we can argue that, as in the 
scalar case, two broad categories of initial conditions are possible. In the first case the fluid anisotropic stress vanishes in Eq. (\ref{VEC2}).
In contrast with the scalar case, however, $\vec{Z}$ does not diverge. In fact 
Eq. (\ref{VEC2}) leads, after trivial algebra, to the equation $\partial_{\tau}(a^2 Z^{i}) = \ell_{P}^2 K^{i}$ 
where $K^{i}$ is the source vector which depends on the comoving magnetic fields and it is constant in time, at least 
in the simplest situation.  This means that $Z^{i}(k,\tau)$ always decays; for instance it decays as $1/a(\tau)$ during radiation\footnote{More generally
$Z^{i}(k,\tau)$ decays as $\tau^{\beta}$ where $\beta = 3 (w-1)/(3w +1)$ and, as above, $w$ is the barotropic index of the dominant component of the plasma. This expression is a simple consequence of Eqs. (\ref{COV11})--(\ref{COV12})}. 

In the second class of initial conditions the anisotropic stress of the neutrinos and of the magnetic fields are simultaneously present.
A regular solution of Eq. (\ref{VEC2}) then follows by requiring that the total anisotropic stress vanishes when conditions of Fig. \ref{figurescale}
are met, namely 
\begin{equation}
\Pi_{ij}^{(vec,\mathrm{B})} + \Pi_{ij}^{(vec,\nu)} = {\mathcal O}(k^2 \tau^2), \qquad \tau \ll \tau_{eq}, \qquad k\tau \ll 1.
\label{VEC3b}
\end{equation}
Note that Eq. (\ref{VEC3b}) is in fact the vector analog of the (regular) magnetized adiabatic mode introduced in Eq. (\ref{AD8}), (\ref{AD11}) and 
illustrated in Eq. (\ref{L4}). The regular initial  condition of Eq. (\ref{VEC3b}) is sometimes referred to as compensated vector 
mode but in fact it is probably the only relevant vector initial condition of the temperature and polarization 
anisotropies. Besides Eq. (\ref{VEC3b}) we then need to discuss the neutrino hierarchy; in the vector 
case  and for massless neutrinos the collisionless part of the Boltzmann equation can be written as: 
\begin{equation}
\partial_{\tau} {\mathcal  F}^{(vec)}_{\nu} +\hat{n}^{i} \partial_{i} {\mathcal  F}^{(vec)}_{\nu} = - \hat{n}^{i}\hat{n}^{j} \partial_{i} Z_{j},
\label{VEC3c} 
\end{equation}
which is the vector analog of Eq. (\ref{NU5}). Note that the collisionless Boltzmann equation in the vector case has been 
already written explicitly in Eqs. (\ref{BRDEC7})--(\ref{BRDEC10}) when discussing the brightness perturbations. Note that 
the first expression in Eq. (\ref{BRDEC10}) is actually written in the gauge $Q_{i}=0$ and coincides, up to a sign, with the right hand side of Eq. (\ref{VEC3c})
(see also \cite{mg2010}).  The effects of the magnetic fields on the vector modes of the geometry have been numerically analyzed in \cite{VV2}
(see also \cite{VV1} for earlier semi-analytical estimates).  In \cite{VV2} the effect of the neutrinos has been accurately included.

\subsubsection{Strongly interacting species}
The evolution of the vorticities of the electrons, of the ions and of the photons can be written as:
\begin{eqnarray}
\partial_{\tau}\vec{\omega}_{\mathrm{e}}+ {\mathcal H}\,\vec{\omega}_{\mathrm{e}} &=& \frac{e n_{\mathrm{e}}}{\rho_{\mathrm{e}} \, a^{4}} \biggl[ \partial_{\tau} \vec{B} 
+ (\vec{v}_{\mathrm{e}}\cdot\vec{\nabla}) \vec{B} + \theta_{\mathrm{e}} \vec{B} - (\vec{B} \cdot\vec{\nabla})\vec{v}_{\mathrm{e}}\biggr]  
\nonumber\\
&+& 
\frac{4}{3} \frac{\rho_{\gamma}}{\rho_{\mathrm{e}}} a 
\Gamma_{\gamma \, \mathrm{e}} (\vec{\omega}_{\gamma} - \vec{\omega}_{\mathrm{e}}) + a \Gamma_{\mathrm{e\,i}} ( \vec{\omega}_{\mathrm{i}} - \vec{\omega}_{\mathrm{e}}),
\label{VEC6}\\
\partial_{\tau} \vec{\omega}_{\mathrm{i}} + {\mathcal H}\,\vec{\omega}_{\mathrm{i}} &=&  - \frac{e n_{\mathrm{i}}}{\rho_{\mathrm{i}} \, a^{4}} 
\biggl[ \partial_{\tau} \vec{B} + (\vec{v}_{\mathrm{i}}\cdot\vec{\nabla}) \vec{B} + \theta_{\mathrm{i}} \vec{B} - (\vec{B} \cdot\vec{\nabla})\vec{v}_{\mathrm{i}}\biggr]
\nonumber\\
&+& \frac{4}{3} \frac{\rho_{\gamma}}{\rho_{\mathrm{i}}} a 
\Gamma_{\gamma \, \mathrm{i}} (\vec{\omega}_{\gamma}-\vec{\omega}_{\mathrm{i}} ) 
+ a \Gamma_{\mathrm{e\,i}} \frac{\rho_{\mathrm{e}}}{\rho_{\mathrm{i}}}( \vec{\omega}_{\mathrm{e}} - \vec{\omega}_{\mathrm{i}}),
\label{VEC7}\\
\partial_{\tau}\vec{\omega}_{\gamma} &=&  a \Gamma_{\gamma\mathrm{i}} (\vec{\omega}_{\mathrm{i}} - \vec{\omega}_{\gamma}) + 
a \Gamma_{\gamma\mathrm{e}}  ( \vec{\omega}_{\mathrm{e}} - \vec{\omega}_{\gamma}),
\label{VEC8}
\end{eqnarray}
Note that we used  $\vec{\nabla}\cdot\vec{E}=0$ in Eqs. (\ref{VEC6}) and (\ref{VEC7}) since the plasma 
is globally neutral.
At the right hand sides of Eqs. (\ref{VEC6}) and (\ref{VEC7}) various higher-order 
terms have been kept to illustrate the possible coupling of the vorticity with the compressible part of the comoving velocity.  These terms, however, all contain spatial gradients and will 
therefore be negligible over large length-scales, as argued long ago by Harrison \cite{harrison}.

Equations (\ref{VEC6}), (\ref{VEC7}) and (\ref{VEC8}) have three different scales of vorticity exchange: the photon-ion, the photon-electron and the electron ion rates whose relative magnitude determines the terms that are potentially subleading in the different dynamical regimes. By taking the ratios of the two rates appearing at the right hand side of Eqs. (\ref{VEC6}) and 
 (\ref{VEC7}) the following two dimensionless ratios can be constructed
 \begin{equation}
\frac{3 \rho_{\mathrm{e}} \, \Gamma_{\mathrm{e\, i}}}{4 \, \rho_{\gamma} \Gamma_{\gamma\mathrm{e}}} = \biggl(\frac{T}{T_{\mathrm{e}\gamma}}\biggr)^{-5/2}, \qquad 
\frac{3 \rho_{\mathrm{e}} \, \Gamma_{\mathrm{e\, i}}}{4 \, \rho_{\gamma} \Gamma_{\gamma\mathrm{i}}} =\biggl(\frac{T}{T_{\mathrm{i}\gamma}}\biggr)^{-5/2}.
\label{VEC10}
\end{equation}
The effective temperatures 
$T_{\mathrm{e}\gamma}$ and $T_{\mathrm{i}\gamma}$ appearing in Eq. (\ref{VEC10}) 
are defined as:
\begin{equation}
T_{\mathrm{e}\gamma} = m_{\mathrm{e}} \, {\mathcal N}^{2/5} \, \eta_{\mathrm{b}0}^{2/5}, \qquad 
T_{\mathrm{i}\gamma} = m_{\mathrm{e}}^{-1/5} m_{\mathrm{p}}^{4/5}  {\mathcal N}^{2/5} \, \eta_{\mathrm{b}0}^{2/5},
\qquad {\mathcal N} = \frac{270 \overline{\zeta}(3)}{32 \, \pi^5} \ln{\Lambda_{\mathrm{C}}},
\label{VEC11}
\end{equation}
where where $\overline{\zeta}(3)$  has been already introduced after Eq. (\ref{gplasma}) and the ion mass has been estimated through the proton mass. In more explicit terms $T_{\mathrm{e}\gamma}$ and $T_{\mathrm{i}\gamma}$ can be estimated as:
\begin{equation}
T_{\mathrm{e}\gamma} = 88.6\, \biggl(\frac{h_{0}^2 \Omega_{\mathrm{b}0}}{0.02258}\biggr)^{2/5} \, \mathrm{eV},\qquad 
T_{\mathrm{i}\gamma} = 36.08\, \biggl(\frac{h_{0}^2 \Omega_{\mathrm{b}0}}{0.02258}\biggr)^{2/5} \, \mathrm{keV}.
\label{VEC12}
\end{equation}
In the regime $T> T_{\mathrm{i}\gamma}$ the Coulomb rate can be neglected in comparison 
with the Thomson rates and the vorticities of photons, electrons and ions approximately 
coincide. For $T_{\mathrm{e}\gamma}< T < T_{\mathrm{i}\gamma}$  the evolution equations of the vorticities of the ions and of the photons are, up to spatial gradients, 
\begin{equation}
 \partial_{\tau} \vec{\omega}_{\mathrm{i}} + {\mathcal H} \vec{\omega}_{\mathrm{i}} = 
- \frac{e n_{\mathrm{i}}}{\rho_{\mathrm{i}} a^4} \partial_{\tau} \vec{B},\qquad 
\partial_{\tau} \vec{\omega}_{\gamma} =a \Gamma_{\gamma\mathrm{e}} ( \vec{\omega}_{\mathrm{e}}-\vec{\omega}_{\gamma}),
\label{VEC17}
\end{equation}
while the evolution of the magnetic field is 
\begin{eqnarray}
\partial_{\tau} \vec{B} = \vec{\nabla} \times (\vec{v}_{\mathrm{e}} \times \vec{B}) + \frac{\nabla^2 \vec{B}}{4\pi \sigma} - \frac{4}{3} \frac{\rho_{\gamma}}{\rho_{\mathrm{b}}} \, a^2 \, \frac{m_{\mathrm{i}}}{e} \Gamma_{\mathrm{e}\gamma} (\vec{\omega}_{\gamma} - \vec{\omega}_{\mathrm{e}}).
\label{VEC17a}
\end{eqnarray}
By eliminating the electron-photon rate between Eqs. (\ref{VEC17}) and (\ref{VEC17a}) and by neglecting the 
spatial gradients in the second relation of Eq. (\ref{VEC17a}), the following pair of approximate conservation laws can be obtained:
\begin{equation}
\partial_{\tau} \biggl( a \vec{\omega}_{\mathrm{i}} + \frac{e}{m_{\mathrm{i}}} \vec{B}\biggr) =0,\qquad 
\partial_{\tau} \biggl( \frac{e}{m_{\mathrm{i}}} \vec{B} - \frac{a}{R_{\mathrm{b}}} \vec{\omega}_{\gamma}\biggr) =0,
\label{VEC19}
\end{equation}
where, as usual,  $R_{\mathrm{b}}$ is the baryon to photon ratio.
By further combining the relations of Eq.  (\ref{VEC19}) the vorticity of the photons can be directly related 
to the vorticity of the ions since $\partial_{\tau}[ R_{\mathrm{b}} \vec{\omega}_{\mathrm{i}} + \vec{\omega}_{\gamma}] =0$. By assuming that 
at a given time $\tau_{\mathrm{r}}$ the primordial value of the vorticity in the 
electron photon system  is $\vec{\omega}_{\mathrm{r}}$ and that $\vec{B}(\tau_{\mathrm{r}})=0$
we shall have that 
\begin{equation}
a_{\mathrm{r}} \vec{\omega}_{\mathrm{i}}(\tau_{\mathrm{r}}) + 
\frac{4}{3} \frac{\rho_{\gamma}(\tau_{\mathrm{r}})}{\rho_{\mathrm{b}}(\tau_{\mathrm{r}})} 
a_{\mathrm{r}}  \vec{\omega}_{\gamma}(\tau_{\mathrm{r}}) = \vec{\omega}_{\mathrm{r}}.
\label{VEC20}
\end{equation}
Thus the solution of Eq. (\ref{VEC17}) with the initial condition (\ref{VEC20}) can be written as:
\begin{equation}
\vec{\omega}_{\mathrm{i}}(\vec{x},\tau) = - \frac{e}{m_{\mathrm{i}}} \frac{\vec{B}(\vec{x},\tau)}{a(\tau)} + \frac{a_{\mathrm{r}}}{a(\tau)} \vec{\omega}_{\mathrm{r}},\qquad  \vec{\omega}_{\gamma}(\vec{x},\tau) = \frac{R_{\mathrm{b}}(\tau)}{a(\tau)} [ \vec{\omega}_{\mathrm{r}} - a(\tau) 
\vec{\omega}_{\mathrm{i}}(\vec{x},\tau)].
\label{VEC22}
\end{equation}
The approximate conservation laws of Eqs. (\ref{VEC19}) can also be phrased in terms of the physical vorticities $\vec{\Omega}_{X}(\vec{x},\tau) = a(\tau) \vec{\omega}_{X}(\vec{x},\tau)$ where $X$ denotes a generic subscript. Note that while $\vec{\omega}_{X}$ is related to $\vec{B}$, the physical vorticity $\vec{\Omega}_{X}$ is directly proportional to $\vec{{\mathcal B}}$. For instance, in the treatment of  \cite{harrison} the use of the physical vorticity and of the physical magnetic field is preferred. 

For typical temperatures $T< T_{\mathrm{e}\gamma}$ the electrons and the ions are more strongly coupled than the electrons and the photons. This means that the effective evolution can be described in terms of the one-fluid magnetohydrodynamical (MHD in what folllows) equations  where, on top of the total current $\vec{J}$  the center of mass vorticity of the electron-ion system is introduced 
\begin{equation}
\vec{\omega}_{\mathrm{b}} = \frac{m_{\mathrm{i}} \vec{\omega}_{\mathrm{i}} + m_{\mathrm{e}} \vec{\omega}_{\mathrm{e}}}{m_{\mathrm{e}} + m_{\mathrm{i}}}.
\label{VEC23}
\end{equation}
Equation (\ref{VEC6}) (multiplied by $m_{\mathrm{e}}$) and Eq. (\ref{VEC7}) (multiplied 
by $m_{\mathrm{i}}$) can therefore be summed up so that the effective set of evolution equations becomes, in this regime, 
\begin{eqnarray}
&& \partial_{\tau} \vec{\omega}_{\mathrm{b}} + {\mathcal H} \vec{\omega}_{\mathrm{b}} = 
\frac{\vec{\nabla}\times(\vec{J} \times \vec{B})}{a^4 \, \rho_{\mathrm{b}}} + \frac{\epsilon'}{R_{\mathrm{b}}} 
(\vec{\omega}_{\gamma} - \vec{\omega}_{\mathrm{b}}),
\label{VEC26}\\
&& \partial_{\tau} \vec{B} = \vec{\nabla}\times(\vec{v}_{\mathrm{b}}\times \vec{B}) + \frac{\nabla^2 \vec{B}}{4 \pi \sigma} 
+ \frac{m_{\mathrm{i}} a}{e \, R_{\mathrm{b}}} \epsilon' (\vec{\omega}_{\mathrm{b}} - \vec{\omega}_{\gamma}),
\label{VEC27}\\
&& \partial_{\tau} \vec{\omega}_{\gamma} = \epsilon' (\vec{\omega}_{\mathrm{b}} - \vec{\omega}_{\gamma}).
\label{VEC28}
\end{eqnarray}
In the tight coupling limit Eqs. (\ref{VEC26}), (\ref{VEC27}) and (\ref{VEC28}) 
imply that $\vec{\omega}_{\mathrm{b}\gamma} \simeq \vec{\omega}_{\mathrm{b}} \simeq \vec{\omega}_{\gamma}$ 
while $\vec{\omega}_{\mathrm{b}\gamma}$ obeys 
\begin{equation}
\partial_{\tau} \vec{\omega}_{\mathrm{b}\gamma} + \frac{{\mathcal H} R_{\mathrm{b}}}{R_{\mathrm{b}} + 1} 
\vec{\omega}_{\mathrm{b}\gamma} = R_{\mathrm{b}}\frac{\vec{\nabla}\times (\vec{J} \times \vec{B})}{\rho_{\mathrm{b}} \, a^4
 (R_{\mathrm{b}} + 1)}.
 \label{VEC29}
 \end{equation}
In analogy with what has been done above, a conservation laws can be derived by combining Eqs. (\ref{VEC26}) and (\ref{VEC27}) 
\begin{equation}
\partial_{\tau} \biggl( \vec{B} + \frac{m_{\mathrm{i}}}{e}\, a \,\vec{\omega}_{\mathrm{b}}\biggr)  = 
\vec{\nabla} \times (\vec{v}_{\mathrm{b}} \times \vec{B}) + \frac{\nabla^2 \vec{B}}{4\pi \sigma} + 
\frac{m_{\mathrm{i}}}{e} \frac{\vec{\nabla} \times (\vec{J} \times \vec{B})}{a^3 \rho_{\mathrm{b}}}.
\label{VEC30}
\end{equation}
From Eqs. (\ref{VEC27}) and (\ref{VEC28}) and by neglecting the spatial 
gradients it also follows 
\begin{equation}
\partial_{\tau} \biggl( \vec{B} - \frac{a}{R_{\mathrm{b}}} \frac{m_{\mathrm{i}}}{e} \vec{\omega}_{\gamma} \biggr) = 0.
\label{VEC31}
\end{equation}
Equations (\ref{VEC30}) and (\ref{VEC31})
are separately valid, but, taken together and in the limit of tight baryon-photon coupling, 
they imply that the magnetic filed must be zero when the tight-coupling is exact (i.e. $\vec{\omega}_{\gamma} =\vec{\omega}_{\mathrm{b}}$).

In  the different physical regimes discussed above the key point is to find a suitable source of large-scale vorticity which could be converted, in some way into a large-scale magnetic field \cite{harrison}.  The conversion can not only occur prior to matter-radiation equality but also after \cite{vort1} in the regime where, as explained,  the baryon-photon coupling becomes weak. Indeed, Eqs. (\ref{VEC19}) and (\ref{VEC30}) have the same 
dynamical content when the spatial gradients are neglected and the only difference involves the coupling to the photons. A source of large-scale vorticity may reside in the spatial gradients of the geometry and of the electromagnetic sources \cite{vort2} but the total vorticity (estimated to lowest order in the spatial gradients) turns out to be negligible for cosmological standards. 

\subsubsection{Turbulence?}

Since one of the motivations of the analysis o vector modes prior to decoupling has been the possible presence of turbulence in the early Universe \cite{harrison} it may be relevant to understand, at least in the framework of the concordance paradigm, what are the typical values of the Reynolds numbers prior to decoupling.  Let us start by reminding 
that in a magnetized plasma the kinetic and magnetic Reynolds numbers are defined as \cite{biskamp,goed}
\begin{equation}
R_{\mathrm{kin}} = \frac{v_{\mathrm{rms}}\, L_{v}\, }{\nu_{\mathrm{th}}}, \qquad R_{\mathrm{magn}} = \frac{v_{\mathrm{rms}}\, L_{B}\, }{\nu_{\mathrm{magn}}},
\label{R0} 
\end{equation}
where $v_{\mathrm{rms}}$ estimates the bulk velocity of the plasma while 
$\nu_{\mathrm{th}}$ and $\nu_{\mathrm{magn}}$ are the coefficients of thermal and magnetic diffusivity; $L_{v}$ and $L_{B}$ are, respectively, the correlation scales of the velocity field and of the magnetic field.  What matters in various situations are not the absolute values of the Reynolds numbers but rather their ratio $Pr_{\mathrm{magn}} = R_{\mathrm{magn}}/R_{\mathrm{kin}}$ which is called Prandtl number. 

In the first obvious situation we have that $R_{\mathrm{kin}} \gg 1$ and 
$R_{\mathrm{magn}} \gg 1$: in this case the Universe 
is both magnetically and kinetically turbulent. Prior to electron-positron annihilation (i.e. $T \geq \mathrm{MeV}$) the coefficient of thermal diffusivity can be estimated as 
 $\nu_{\mathrm{th}} \sim (\alpha_{\mathrm{em}}^2 T)^{-1}$ from the two-body scattering of relativistic species with significant momentum transfer. The conductivity 
of the plasma is different from the ones considered before and it is given by $\sigma \sim T/\alpha_{\mathrm{em}}$; the magnetic diffusivity becomes then $\nu_{\mathrm{magn}} = \alpha_{\mathrm{em}} (4\pi T)^{-1}$. Assuming, for sake of simplicity, thermal and kinetic equilibrium of all relativistic species (which is not exactly the case for $T\sim \mathrm{MeV}$) the kinetic Reynolds number turns out to be $R_{\mathrm{kin}} \simeq {\mathcal O}(10^{16})$, 
the magnetic Reynolds number is $R_{\mathrm{magn}} \simeq 4\pi/\alpha_{\mathrm{em}}^3 R_{\mathrm{kin}} \sim {\mathcal O}(10^{24})$ and $Pr_{\mathrm{magn}} \sim 10^{7}$. The latter estimates have been obtained by assuming, in Eq. (\ref{R0}), 
 $L_{v} \simeq L_{B} \sim H^{-1}$ (where $H^{-1}$ is the Hubble radius at the corresponding 
 epoch). In the symmetric phase of the standard electroweak theory where all the species (including the Higgs boson and the top quark) are in thermal and kinetic equilibrium, 
 $\nu_{\mathrm{th}}$ and $\nu_{\mathrm{magn}}$ can be estimated in analog terms and $R_{\mathrm{kin}} \sim {\mathcal O}(10^{11})$ and $R_{\mathrm{magn}} \sim {\mathcal O}(10^{17})$.
 
 A direct calculation of the magnetic and kinetic Reynolds numbers shows that prior to decoupling we are 
in the situation where $R_{\mathrm{kin}} = {\mathcal O}(10^{-4})$  and $R_{\mathrm{magn}} = {\mathcal O}(10^{15})$ for $z = {\mathcal O}(1500)$ implying $Pr_{\mathrm{magn}} = {\mathcal O}(10^{19})$ \cite{mgrey}. Prior to decoupling the plasma is not kinetically turbulent but the largeness of the magnetic Reynolds number guarantees the conservation of the magnetic flux and of the helicity. Indeed, when $R_{\mathrm{magn}} \gg 1$ the two Alfv\'en theorems hold true and they imply the conservation of the total magnetic flux and of the total helicity:
\begin{eqnarray}
&& \frac{d}{d\tau} \int_{\Sigma} \vec{B} \cdot d\vec{\Sigma}=- \nu_{\mathrm{magn}} \int_{\Sigma} \vec{\nabla} \times(\vec{\nabla}
\times\vec{B})\cdot d\vec{\Sigma},
\nonumber\\
&& \frac{d}{d \tau}\int_{V} d^3 x \vec{A}~\cdot \vec{B} = - 2 \nu_{\mathrm{magn}} \int_{V} d^3 x
{}~\vec{B}(\cdot\vec{\nabla} \times\vec{B}),
\label{FC}
\end{eqnarray}
where $V$ and $\Sigma$ are a fiducial volume and a fiducial surface moving with the conducting fluid; $\vec{B}$ and $\vec{A}$ 
denote the comoving magnetic field and the comoving vector potential. In the 
ideal hydromagnetic limit (i.e. $\sigma \to \infty$, $\nu_{\mathrm{magn}} \to 0$ and $R_{\mathrm{magn}} \to \infty$) the flux is exactly conserved and the number of links and twists in the magnetic flux lines is also preserved by the time evolution.  

 We have therefore to admit, at least in the context 
of the $\Lambda$CDM paradigm, that what dominates is not the incompressible flow but rather the compressible velocity field.  Indeed in the absence of the magnetic fields 
we have that the monopole and the dipole of the scalar hierarchy in the tight-coupling approximation have both amplitudes ${\mathcal O}({\mathcal R}_{*})$ and oscillate, respectively, as $\cos{k r_{s}(\tau)}$ and as $\sin{k r_{s}(\tau)}$ \cite{hus} 
(see also \cite{semian0,semian2,semian3}) where $r_{s}(\tau)$ is the sound horizon of Eq. (\ref{soundhor}). This means that the physically meaningful initial conditions for the vector modes in the magnetized case are not the ones of Eq. (\ref{VEC3b}) but rather the ones where the magnetically induced vector modes are strongly suppressed.

\subsection{Magnetized tensor modes}
The scalar modes of the geometry are subjected to the Hamiltonian and to the momentum 
constraint and their evolution may compress the plasma. The vector modes only 
experience the momentum constraint and are therefore incompressible.
The tensor modes are subjected neither to the Hamiltonian nor to the momentum 
constraint and their evolution equation is: 
\begin{equation}
h_{ij}^{\prime\prime} + 2 {\mathcal H} h_{ij}^{\prime} - \nabla^2 h_{ij} = - 2 \ell_{P}^2 a^2 \Pi_{ij}^{(tens)},
\label{TENS1}
\end{equation}
where the total anisotropic stress $\Pi_{ij}^{(tens)}$ is given as the sum of the magnetic and of the fluid components:
\begin{equation}
 \Pi_{ij}^{(tens)} = \Pi_{ij}^{(tens,\nu)} + \Pi_{ij}^{(tens, \mathrm{B})},\qquad \partial_{i} \Pi^{i j}_{(tens)} = \Pi_{i}^{(tens)\,i}=0.
\label{TENS1a}
\end{equation}
For the explicit form of $ \Pi_{ij}^{(tens, \mathrm{B})}$ see Eqs. (\ref{PItot})--(\ref{PIT}) and discussion therein. The power spectrum of the total anisotropic stress can always be introduced from the two-point function with the 
same notations used in the case of the tensor modes:
\begin{equation}
\langle \Pi_{ij}^{(t)}(\vec{k},\tau)\, \Pi_{mn}^{(t)}(\vec{p},\tau) \rangle = \frac{2 \pi^2}{k^3} {\mathcal S}_{i j m n}(\hat{k}) P_{\Pi}(k,\tau) \delta^{(3)}(\vec{k}+ \vec{p}),
\label{TENS2}
\end{equation}
where ${\mathcal S}_{i j m n}(\hat{k})$ has been already defined in Eqs. (\ref{IRFTEN4})--(\ref{IRFTEN5}).
Equation (\ref{TENS1}) can be also written, in explicit terms, by considering, separately, the polarisations of the 
gravitational wave. In particular, by defining three mutually orthogonal unit vectors 
$\hat{a}$, $\hat{b}$ and $\hat{k}$, the polarisations of the gravitational wave can be written as 
\begin{equation}
\hat{e}^{\oplus}_{ij}= (\hat{a}_{i} \hat{a}_{j} - \hat{b}_{i} \hat{b}_{j}), \qquad \hat{e}^{\otimes}_{ij}= (\hat{a}_{i} \hat{b}_{j} + \hat{a}_{i} \hat{b}_{j}).
\label{TENS3}
\end{equation}
In this case Eq. (\ref{TENS1}) becomes
\begin{equation}
h_{X}^{\prime\prime} + 2 {\mathcal H} h_{X}^{\prime} - \nabla^2 h_{\oplus} = - 2 \ell_{P}^2 a^2 \Pi_{X}^{(tens)},
\label{TENS4}
\end{equation}
where $X$ coincides either with $\otimes$ or with $\oplus$ and where
 $\Pi_{ij}^{(tens)}$ is decomposed as $\Pi_{ij}^{(tens)}= \Pi^{(tens)}_{\oplus} \hat{e}_{ij}^{\oplus} + \Pi^{(tens)}_{\otimes} \hat{e}_{ij}^{\otimes}$. The evolution equation of the perturbed  phase-space distribution, in the tensor case\footnote{Equation (\ref{TENS6}) follows from 
the collisionless Boltzmann equation for the perturbed neutrino phase space distribution. Note that between ${\mathcal F}_{\nu}$ (or ${\mathcal F}_{\gamma}$)
and the brightness perturbations $\Delta_{\mathrm{I}}$ discussed in section \ref{sec2} there is a numerical factor (i.e. ${\mathcal F}_{\gamma} = 4 \Delta_{\mathrm{I}}$).
This difference entails a different numerical factor in from of the metric fluctuation appearing in the corresponding equations (see also \cite{primer} for further details).} and for massless neutrinos, is
\begin{equation}
\partial_{\tau} {\mathcal F}_{\nu}^{(tens)} + i k \mu {\mathcal F}_{\nu}^{(tens)} + 2 \hat{n}^{i} \hat{n}^{j} \partial_{\tau} h_{ij}=0.
\label{TENS6}
\end{equation}
 Inserting Eq. (\ref{TENS3}) into Eq. (\ref{TENS6}) we obtain, as usual,
\begin{equation}
\partial_{\tau} {\mathcal F}_{\nu}^{(tens)}+ i k \mu {\mathcal F}_{\nu}^{(tens)} + 2 [ (\hat{n}\cdot\hat{a})^2 - (\hat{n}\cdot\hat{b})^2] \partial_{\tau} h_{\oplus} + 
4 (\hat{n}\cdot\hat{a}) (\hat{n}\cdot\hat{b}) \partial_{\tau} h_{\otimes} =0,
\label{TENS7}
\end{equation}
which can also be written in more explicit terms by recalling that $(\hat{n}\cdot\hat{a})^2 - (\hat{n}\cdot\hat{b})^2= (1 - \mu^2)\cos{2 \varphi}$ and that
$ 2  (\hat{n}\cdot\hat{a}) (\hat{n}\cdot\hat{b}) = (1 - \mu^2) \sin{2 \varphi}$. The various moments of the distribution can then be 
obtained by separating in ${\mathcal F}_{\nu}^{(tens)}$ the contribution of the two polarization in full analogy 
with the discussion of the tensor brightness perturbations. 

\subsubsection{Initial conditions for the tensor problem}

A naive argument would suggest that if the magnetic contribution is absent the problem 
should be trivial since the only source of anisotropic stress is provided by the massless neutrinos which could be safely and completely neglected.  This case, typical of the concordance paradigm, for long time was considered of pure academic interest until Weinberg \cite{tens1} suggested that the free-streaming of the neutrinos 
could provide a significant source of damping of the tensor modes with an effect of the order of $10\%$.  This effect  has been subsequently analyzed with numerical  \cite{tens2,tens3} and analytical \cite{tens4,tens5} methods. If  the anisotropic stress of the neutrinos and of the magnetic fields are simultaneously present, a regular solution of Eq. (\ref{TENS1}) then follows by requiring that the total anisotropic stress vanishes when conditions of 
Fig. \ref{figurescale} are met, namely 
\begin{equation}
\Pi_{ij}^{(tens,\mathrm{B})} + \Pi_{ij}^{(tens,\nu)} = {\mathcal O}(k^2 \tau^2), \qquad \tau \ll \tau_{eq}, \qquad k\tau \ll 1,
\label{TENS8}
\end{equation}
in full analogy with the vector case (see Eq. (\ref{VEC3b})). As already mentioned Eqs. (\ref{VEC3b}) and (\ref{TENS8})
are nothing but the analog of the (regular) magnetized adiabatic mode \cite{MOD1,MOD3} introduced in Eq. (\ref{AD8}), 
(\ref{AD11}) and illustrated in Eq. (\ref{L4}). The  initial  condition of Eq. (\ref{TENS8}) is sometimes referred to as compensated tensor 
mode which has been specifically analyzed in \cite{VV2} but in fact it is probably the only relevant tensor initial condition of the temperature and polarization anisotropies.

In the last class of initial conditions the only source of anisotropic stress is represented by the magnetic fields. This kind of initial 
condition is realized when the magnetic fields are (for some reason) dominant and anyway prior to neutrino decoupling approximately occurring for temperatures of the order of the MeV. Equation (\ref{TENS1}) can be rephrased in terms of the tensor normal mode $\mu_{ij}$:
\begin{equation}
\mu_{ij}^{\prime\prime} - \nabla^2 \mu_{ij} - \frac{a^{\prime\prime}}{a} \mu_{ij} = - 2 \ell_{P}^2 a^3(\tau) \Pi^{(tens)}_{ij}, \qquad \mu_{ij} = a h_{ij}.
\label{TENS9}
\end{equation}
The solution of Eq. (\ref{TENS1}) and (\ref{TENS9}) obviously depends on the functional form of the scale factor. For instance during a radiation-dominated evolution we have that: 
\begin{eqnarray}
h_{ij}(\vec{x},\tau) &=& \overline{h}_{ij}(\vec{x},\tau) - \frac{ 2 \ell_{P}^2}{a(\tau)} \int d^{3} x^{\prime} \, \int_{\tau_{*}}^{\tau} d\xi \,{\mathcal G}(\vec{x}, \vec{x}^{\,\prime}; \tau, \xi) \, a^3(\xi) \, \Pi^{(tens)}_{ij}(\vec{x}^{\prime}, \xi),
\nonumber\\
{\mathcal G}(\vec{x}, \vec{x}^{\,\prime}; \tau, \xi) &=& \frac{1}{(2\pi)^3} \int\frac{ d^{3} k}{k} \,e^{- i \vec{k}\cdot(\vec{x} - \vec{x}^{\,\prime})} \, \sin{[ k (\xi - \tau)]},
\label{TENS10}
\end{eqnarray}
where $\overline{h}_{ij}(\vec{x},\tau)$ denotes the solution of the homogeneous equation. By going to Fourier space we then have
\begin{equation}
h_{ij}(\vec{k},\tau) = \overline{h}_{ij}(k,\tau_{*}) \frac{\sin{k\tau}}{k\tau} - \frac{2 \ell_{P}^2}{a(\tau) k} \int_{\tau_{*}}^{\tau} d\xi \, a^3(\xi) \, \Pi_{ij}^{(tens)}(\vec{k},\xi) \sin{k[(\xi- \tau)]}, 
\label{TENS11}
\end{equation}
which can be easily solved if $\Pi_{ij}^{(tens)}(\vec{k},\xi) \equiv \Pi_{ij}^{(tens,\mathrm{B})}(\vec{k},\xi)$ and, moreover, $\Pi_{ij}^{(tens,\mathrm{B})}(\vec{k},\xi) = \overline{\Pi}_{ij}(k)/a^4(\xi)$. In the latter case the full solution can be expressed in terms of sine integrals 
and cosine integrals. The result mirrors exactly what has been 
already discussed in the case of the divergent quasi-normal mode of the scalar problem 
in Eq. (\ref{ANN20}) and in the related discussion. In the case of a radiation-dominated 
evolution the solution will then be logarithmically divergent at early times 
as it happens in the case of Eq. (\ref{ANN21}).

\subsubsection{Gravitational waves at intermediate frequencies}
The initial conditions in the absence of neutrino anisotropic stress have been explored from the first time in Refs. \cite{knot1} (see also \cite{knot2}). The idea was rather simple:
owing to the two Alfv\'en theorems  
(see Eq. (\ref{FC})) helical configurations of the magnetic field possibly present in the early 
may survive over cosmological scales given the largeness of the magnetic Reynolds number.
If this happens, the helical configurations that maximize 
the gyrotropy \cite{chan1,chan2} may affect the cosmic graviton spectrum and lead to a polarized background of relic gravitational waves at intermediate frequencies \cite{knot1,knot2} 
(see also \cite{knot3} for the same kind of idea). This idea has been 
specifically explored at the electroweak scale \cite{knot2}. It has actually been argued 
that inside the electroweak particle horizon, hypermagnetic knots (HK in what follows)
can be pictured as a collection of flux tubes (closed because of transversality) but characterized by a non-vanishing  gyrotropy (i.e. $\vec{B} \cdot \vec{\nabla} \times \vec{B}$ where $\vec{B}$ will denote, for the moment, the comoving hypermagnetic field). 
The dynamical production of HK and Chern-Simons waves suggested in the past a viable mechanism for the generation of the baryon asymmetry of the Universe \cite{knot2} (see also \cite{knot5}). Earlier ideas along this direction (even suggesting a connection with gravitational waves) can be found in Ref. \cite{knot6}. 

Using Eq. (\ref{TENS11}) the energy density of the gravitational waves induced by hypermagnetic knots can be explicitly 
computed. Recently (see last paper of Ref. \cite{knot2}), using two complementary approaches, a physical template family for the emission of the gravitational 
radiation produced by the HK has been constructed and the energy density of the relic gravitons can be parametrized as:
\begin{equation}
\Omega_{gw}(\nu,\tau_0)=  \frac{\Omega_{B}^2}{(1 + z_{\mathrm{eq}}) (1 + z_{\Lambda})^3}  \biggl(\frac{\overline{\nu}}{\overline{\nu}_{ew}}\biggr)^{\alpha} e^{ - 2 (\overline{\nu}/\overline{\nu}_{\sigma})^2},\qquad \overline{\nu}\geq \overline{\nu}_{ew},
\label{TENS12}
\end{equation}
where $\Omega_{gw}$ is the energy density of the gravitational waves in critical units; $z_{\mathrm{eq}}$ is the redshift to equality and $z_{\Lambda}$ is the redshift to $\Lambda$-dominance. In Eq. (\ref{TENS12}) $\Omega_{B}$ is a dimensionless amplitude computable from the HK configuration.  
The spectral energy density ranges between $\overline{\nu}_{ew} = {\mathcal O}(20)\, \mu\mathrm{Hz}$ and $\overline{\nu}_{\sigma} =  {\mathcal O}(50)\, \mathrm{kHz}$ which 
are the frequencies corresponding, respectively, to the 
electroweak Hubble radius and to the dissipation scale 
inside the electroweak horizon\footnote{In the following lines the 
frequencies mentioned in the discussion are not frequencies of the photons but rather 
of the gravitons.}. While between $\nu_{ew}$ and $\nu_{\sigma}$ the inflationary contribution implies a spectral energy density $h_{0}^2 \Omega_{gw}^{(inf)} = {\mathcal O}(10^{-17})$,
the signal due to hypermagnetic knots can be as large as  $h_{0}^2 \Omega_{gw}^{(knots)} = {\mathcal O}(10^{-8})$ without conflicting with current bounds applicable to stochastic backgrounds of gravitational radiation like the big-bang nucleosynthesis 
bound \cite{knot7} and the pulsar timing bound \cite{knot8}.

The intermediate frequency range of the spectrum of relic gravitational radiation goes from few $\mu$Hz to $10$ kHz. This intermediate range encompasses the operating windows of space-borne interferometers (hopefully available twenty years from now) 
and of terrestrial detectors (already available but still insensitive to stochastic backgrounds of relic gravitons of cosmological origin). This statement can be understood by comparing the quoted 
sensitivities of the Ligo/Virgo experiments with the constraints imposed by the big-bang 
nucleosynthesis bound. Moreover, between few $\mu$Hz and $10$ kHz, the conventional  inflationary models lead to relic gravitons whose spectral energy density can be (at most) of the order of $10^{-17}$.

The lack of observation of gravitational waves between few $\mu$Hz and 
$10$ kHz will potentially exclude the presence of hypermagnetic knots configurations at the electroweak scale. Conversely the 
observation of a signal in the range that encompasses the operating windows of space-borne and terrestrial wide-band detectors will not necessarily confirm the nature of the source. Further scrutiny will be needed but the signal of the hypermagnetic knots can be 
disambiguated since the stochastic background of gravitational 
waves produced by the hypermagnetic knots is polarized. Last but not least a gravitational signal coming from maximally gyrotropic configurations of the hypercharge may offer an indirect test of the equations of anomalous magnetohydrodynamics whose spectrum includes hypermagnetic knots and Chern-Simons waves as low-frequency excitations \cite{knot9}.

\renewcommand{\theequation}{4.\arabic{equation}}
\setcounter{equation}{0}
\section{Magnetized angular power spectra}
\label{sec4}
In the minimal version of the magnetized $\Lambda$CDM 
 the dominant source of large-scale inhomogeneities are the curvature fluctuations of Eq. (\ref{SPS}) (also discussed in Eqs. (\ref{IRFSC1})--(\ref{IRFSC3})):
\begin{equation}
\langle {\mathcal R}(\vec{k},\tau) \,{\mathcal R}(\vec{p},\tau) \rangle = \frac{2 \pi^2}{k^3} P_{{\mathcal R}}(k,\tau)
\delta^{(3)}(\vec{k} + \vec{p}), \qquad P_{{\mathcal R}}(k,\tau) = {\mathcal A}_{{\mathcal R}}   \biggl(\frac{k}{k_{\mathrm{p}}}\biggr)^{n_{\mathrm{s}}-1},
\label{PS4a}
\end{equation} 
where ${\mathcal A}_{{\mathcal R}} = {\mathcal O}(2.4)\times 10^{-9}$ is the spectral amplitude at the pivot scale 
$k_{\mathrm{p}} = 0.002 \,\, \mathrm{Mpc}^{-1}$. According to Eq. (\ref{PS4a}) the scale-invariant limit occurs for $n_{\mathrm{s}} \to 1$. Had we chosen to include the factor $1/k^3$ directly in the definition of the power spectrum, the scale-invariant limit would have been shifted by three units (i.e. $\overline{n}_{\mathrm{s}} \to - 3$). The second concurrent source of large-scale inhomogeneities will be the magnetic random fields whose two-point function, in Fourier space, reads:
\begin{equation}
\langle B_{i}(\vec{k},\tau) \,B_{j}(\vec{p},\tau) \rangle = \frac{2 \pi^2}{k^3} P_{B}(k,\tau) \, p_{ij}(\hat{k}) \delta^{(3)}(\vec{k} + \vec{p}), \qquad P_{\mathrm{B}}(k,\tau) = A_{\mathrm{B}} \biggl(\frac{k}{k_{\mathrm{L}}}\biggr)^{n_{\mathrm{B}} -1}, 
\label{PSM1}
\end{equation}
where $k_{\mathrm{L}}=\mathrm{Mpc}^{-1}$ is commonly referred to as the magnetic pivot scale. In the parametrizations of Eqs. (\ref{PS4a}) and (\ref{PSM1}) the power spectra $P_{{\mathcal R}}(k,\tau)$ and $P_{\mathrm{B}}(k,\tau)$ have the same dimensions of the corresponding correlation functions in real space. In particular $A_{\mathrm{B}}$ has the dimensions of an energy density as the two-point function of Eq. (\ref{OM2}). 

Equation (\ref{PSM1}) introduces only two supplementary parameters (i.e. $n_{\mathrm{B}}$ and $A_{\mathrm{B}}$) in comparison  with the six parameters of the concordance paradigm. 
We note that in the current literature the power spectra of magnetic fields are often assigned in such a way that their scale-invariant limit would correspond to $\overline{n}_{\mathrm{B}} \to - 3$ (and not to $n_{\mathrm{B}} \to 1$). As explained after Eq. (\ref{PS4a}) 
 there is nothing deep with this choice: it amounts to including 
 or excluding a factor  $1/k^3$ from the definition of the  power spectrum. We shall not follow this convention which seems deliberately confusing. The scale-invariant limit of magnetic random
 fields will then correspond here to $n_{\mathrm{B}}\to 1$ 
 exactly as in the case of the curvature inhomogeneities (i.e.
 $n_{\mathrm{s}} \to 1$). 

\subsection{Blue and red magnetic power spectra}
According to the standard terminology the curvature inhomogeneities are characterized red spectral indices (i.e. $n_{\mathrm{s}} < 1$): this means that their corresponding two-point function slightly increases for large length-scales, as it can be argued from the 
preceding sections (see, in particular, Tab. \ref{TABLEAA} and discussion therein).  Since the scale-invariant limit of the magnetic power spectrum coincides with $n_{\mathrm{B}} \to 1$, when $n_{\mathrm{B}} > 1$ we shall be 
talking about {\em blue magnetic spectra} while when $n_{\mathrm{B}} < 1$ we have the case of {\em red spectra}. In some papers (see e.g. \cite{chone34am}) 
the spectra $\overline{n}_{B} <0 $ are referred to as red spectra; this terminology is peculiar since when the scale-invariant limit 
is $\overline{n}_{B} \to - 3$ the spectra $-3 < \overline{n}_{B} <0$ are charecterized by a two-point function which decreases 
for large length-scales. Thus the spectra $ -3 < \overline{n}_{\mathrm{B}} <0$ are in fact blue and not red.

It is customary to assign the magnetic energy density 
over a typical comoving scale of the order of $k_{\mathrm{L}}^{-1}$. This procedure is described in detail in the appendix 
\ref{APPA} but, for the present purposes, what matters is that $A_{\mathrm{B}}$ can be traded for the regularized 
magnetic energy density defined in Eqs. (\ref{REG1}), (\ref{REG2}) and (\ref{REG3}) 
\begin{eqnarray} 
 A_{\mathrm{B}} &=& \frac{(2\pi)^{n_{\mathrm{B}} -1}}{\Gamma[(n_{\mathrm{B}} -1)/2]}\,  
 B_{\mathrm{L}}^2, \qquad \mathrm{for} \qquad n_{\mathrm{B}}>1,
 \label{PSM2}\\
 A_{\mathrm{B}} &=& \biggl(\frac{ 1 -n_{\mathrm{B}}}{2}\biggr) \biggl(\frac{k_{0}}{k_{\mathrm{L}}}\biggr)^{(1 - n_{\mathrm{B}})} \,  B_{\mathrm{L}}^2,\qquad \mathrm{for}\qquad 
 n_{\mathrm{B}} < 1.
 \label{PSM3}
 \end{eqnarray}
As $k_{\mathrm{L}}$ is related to the ultraviolet cut-off  (necessary in the case of blue spectra), in Eq.  (\ref{PSM3}) $k_{0}$ is related to the infrared cut-off which is typically chosen between $k_{\mathrm{p}}$ and the Hubble scale, i.e. $H_{0} < k_{0}< k_{\mathrm{p}}$. In the case of white spectra (i.e. $n_{\mathrm{B}} =1$) the two-point function is logarithmically divergent in real space and this is fully analog to what happens in Eq. (\ref{IRFSC3}) when $n_{\mathrm{s}} =1$, i.e. the Harrison-Zeldovich (scale-invariant) spectrum.  If the amplitude of the two-point function is assigned in terms of $B_{\mathrm{L}}$ the differences in the definition of the power spectrum are immaterial\footnote{There could be some who would 
like to change the present conventions and write $B_{i}(\vec{x},\tau) = \int d^{3} k e^{- i \vec{k}\cdot \vec{x}} B_{i}(\vec{k},\tau)$ (in contrast with our conventions expressed in Eq. (\ref{IRFVEC3})). There could also be some others who would like to assign the power spectrum as $\langle B_{i}(\vec{k}, \tau) 
\langle B_{i}(\vec{p}, \tau)\rangle = (2\pi)^3 \delta^{(3)}(\vec{k} + \vec{p}) p_{ij}(\hat{k}) \overline{P}_{\mathrm{B}}(k,\tau)$ with 
$P_{\mathrm{B}}(k,\tau) = \overline{A}_{\mathrm{B}} k^{\overline{n}_{B}}$ (in contrast with our conventions 
expressed in Eq. (\ref{PSM1}) and in appendix \ref{APPA}). In spite of its peculiar units,  $\overline{A}_{\mathrm{B}}$ 
can always be traded for $B_{\mathrm{L}}^2$ which is always the same in spite of the conventions. Note however, as already mentioned, that $n_{\mathrm{B}} = (\overline{n}_{\mathrm{B}} +4) $. }.

The magnetic random fields affect, in different ways, all the measured temperature and polarization anisotropies and, in particular, the temperature autocorrelations (for short $TT$ correlations), the polarization autocorrelations (in the jargon $EE$ correlations) and their mutual cross-correlation (i.e. the $TE$ power spectrum). In the conventional $\Lambda$CDM paradigm the  tensor modes (possibly inducing a $B$-mode power spectra) are by definition disregarded. In the magnetized $\Lambda$CDM model there is indeed a further possible source of $B$-mode polarisation which is represented by the Faraday effect and which will be examined in the following section.  
\begin{table}[!ht]
\begin{center}
\begin{tabular}{||l|c|c|c|c|c||}
\hline
\hline
\rule{0pt}{4ex} References & Pivotal data  & $B_{\mathrm{L}}$ &$n_{\mathrm{B}}$& Rationale \\
\hline
\cite{MOD1} & WMAP 1-year      &$ B_{\mathrm{L}}< 10^{-7.8} $ G      & $n_{\mathrm{B}} > 1.1$ &SW  \\
\cite{MOD3} & WMAP 3-years               &$ B_{\mathrm{L}} < 8$ nG                & $n_{\mathrm{B}} >1$     &  TE peak \\ 
\cite{MOD3} & WMAP 3-years                &$ B_{\mathrm{L}} < 0.5$ nG          & $n_{\mathrm{B}} <1$     &   TE peak \\
\cite{MOD4} & WMAP 3-years &$ B_{\mathrm{L}} < 7.7$ nG         & $ 1 < n_{\mathrm{B}} < 1.5 $& TT/TE \\
\cite{MOD6}   & WMAP 3-years               &$B_{\mathrm{L}} < 2$ nG             & $n_{\mathrm{B}} > 1.1$ & TT/TE \\
\cite{MOD6}   &WMAP 3-years              &$B_{\mathrm{L}} < 0.1$ nG  & $n_{\mathrm{B}} < 1$ & TT \\
\cite{MOD7}   & WMAP 3-years               & $B_{\mathrm{L}} < 5$ nG & $n_{\mathrm{B}} >1.1$ & TT/TE \\
\cite{MOD8}   & WMAP 3-years               & $B_{\mathrm{L}} < 5$ nG & $n_{\mathrm{B}} >1.1$ & TT/TE \\                   
\cite{MOD7a} & WMAP 3-years               &$B_{\mathrm{L}} < 5 $ nG & $n_{\mathrm{B}} > 1.1$ & TT/TE \\
\cite{MOD8b} & WMAP 5-years                & $B_{\mathrm{L}} < 4.5 $ nG & $ 1.1 \leq n_{\mathrm{B}} < 2.5$  & TT/TE\\
\cite{MOD9}   & WMAP 5-years                & $B_{\mathrm{L}} < 3 $ nG & $ 1.1 \leq n_{\mathrm{B}} < 2.5$  & TT/TE\\
\cite{MOD10} & WMAP 5-years                & $B_{\mathrm{L}} < 9 $ nG & $n_{\mathrm{B}} \to 1$ & TT/TE \\
\cite{MOD11}  & WMAP 5-years                & $B_{\mathrm{L}} < 5 $ nG & $1.1 \leq n_{\mathrm{B}} < 2.5$ & TT/TE \\
\cite{NoM2} & WMAP 9-years         & $B_{\mathrm{L}} < 5$ nG & $1.1 \leq n_{\mathrm{B}} < 2.5$ & TT, TE\\
\hline
\end{tabular}
\caption{Bounds on the intensity of magnetic random fields from the temperature and polarization anisotropies 
for of initial conditions of the Einstein-Boltzmann hierarchy.}
\label{TABLE4}
\end{center}
\end{table}
Through the years specific bounds on $B_{\mathrm{L}}$ and $n_{\mathrm{B}}$ have been derived from the analysis of the temperature and of the polarisation anisotropies.
In Tab. \ref{TABLE4} we report the bounds on magnetic random fields from various different papers, not always homogeneous. Note, for instance, that some references did include the estimates 
of the $EE$ correlations (not measured by the WMAP collaboration) from various polarization experiments 
like for instance the Quad experiment \cite{chone40,chone41,quad1,quad2}. As already mentioned we remind 
that through the years few bounds on magnetic random fields have been derived from the (unobserved) 
non-Gaussianities \cite{MOD2}. More recently these bounds have been refined 
\cite{MOD2cc} but it is not always clear which are the initial conditions of the Einstein-Boltzmann hierarchy 
assumed in the analysis. 

The bounds of the Planck collaboration \cite{chone34am} 
are more consistent than the scattered results 
of Tab. \ref{TABLE4}. By this we mean that the $TT$, $TE$ and $EE$ correlations, at least in the 
last data release, all come from the same experiment. 
They can be summarized  by saying that from the $TT$ correlations 
alone and from the $TE$, $EE$  angular power spectra the data suggest $B_{\mathrm{L}} < {\mathcal O}(4.4)\, \mathrm{nG})$. Slightly more constraining values seem to be obtained in the case of nearly scale-invariant spectra in the range $B_{\mathrm{L}} < {\mathcal O}(2.1)\, \mathrm{nG}$. The range of spectral indices analyzed in \cite{chone34am}
 is $1.1 < n_{\mathrm{B}} < 6$. For $n_{\mathrm{B}} \to 6$ the bounds lead to $B_{\mathrm{L}} < 0.011$ nG.
In Ref. \cite{chone34am} the scale invariant limit is reached 
 for $\overline{n}_{\mathrm{B}} \to - 3$; in terms of the present conventions 
 $n_{\mathrm{B}} = 4 + \overline{n}_{\mathrm{B}}$. 
What is interesting to remark is that 
the inclusion of the magnetic fields in the initial data of the Einstein-Boltzmann hierarchy immediately brings the bounds to be of the order of the nG. There is a simple physical rationale for this occurrence which can be appreciated by looking at the approximate solutions of the Einstein-Boltzmann hierarchy\footnote{The authors of Ref. \cite{AJ} claimed somehow stronger bounds 
by looking at small-scale effects. It is however unclear which kind of initial conditions they are adopting for the Einstein-Boltzmann 
hierarchy. It seems that the authors assume exactly the results of the present discussion by neglecting the 
large-scale effects of stochastic fields. In this sense the approach of Ref. \cite{AJ} is orthogonal to the one of this paper.}.

\subsection{Sachs-Wolfe and integrated Sachs-Wolfe effects} 
The visibility function ${\mathcal K}(\tau)$ can be 
expressed, in general terms, as  
\begin{equation}
{\mathcal K}(\tau) = \epsilon' e^{- \epsilon(\tau,\tau_{0})},\qquad \epsilon(\tau,\tau_{0})  = \int_{\tau}^{\tau_{0}}\,\, x_{\mathrm{e}}(\tau')\, \,
\sigma_{\mathrm{e}\gamma} \,\,\tilde{n}_{\mathrm{e}}(\tau') a(\tau')\, d\tau',
\label{visdef}
\end{equation}
where $\epsilon(\tau,\tau_{0})$ denotes the optical depth as opposed to the differential optical depth $\epsilon^{\prime}$.
Barring for the reionization peak (which is relevant for much lower redshifts)
the visibility function vanishes  for  $\tau \gg \tau_{\mathrm{rec}}$
and has a maximum around recombination. The finite thickness effects of the last scattering surface 
are customarily taken into account by approximating ${\mathcal K}(\tau)$ with a Gaussian profile \cite{semian0,semian1,semian2,semian3} 
centered at $\tau_{\mathrm{rec}}$, i.e. 
\begin{eqnarray}
{\mathcal K}(\tau) &=& {\mathcal N}(\sigma_{\mathrm{rec}}) e^{- \frac{(\tau - \tau_{\mathrm{rec}})^2}{2 \sigma_{\mathrm{rec}}^2}},\qquad \int_{0}^{\tau_{0}} {\mathcal K}(\tau) d\tau = 1, 
\label{rot5}\\
{\mathcal N}(\sigma_{\mathrm{rec}})&=& \sqrt{\frac{2}{\pi}} \frac{1}{\sigma_{\mathrm{rec}}} 
\biggl[ \mathrm{erf}\biggl( \frac{\tau_{0} - \tau_{\mathrm{rec}}}{\sqrt{2} \sigma_{\mathrm{rec}}} \biggr)+   \mathrm{erf}\biggl( \frac{ \tau_{\mathrm{rec}}}{\sqrt{2} \sigma_{\mathrm{rec}}} \biggr)\biggr]^{-1}, 
\label{rot6}\\
\mathrm{erf}(z) &=& \frac{2}{\sqrt{\pi}} \int_{0}^{z} e^{- t^2} dt.
\end{eqnarray}
The overall normalization ${\mathcal N}(\sigma_{\mathrm{rec}})$ is 
fixed by normalizing to $1$ the integral of ${\mathcal K}(\tau)$ since
 the visibility function gives the probability that a photon last scatters between $\tau$ and $\tau + d\tau$. In the limits $\tau_{0} \gg \tau_{\mathrm{rec}}$ and $ \tau_{0} \gg \sigma_{\mathrm{rec}}$, Eq. (\ref{rot6}) simplifies since the error 
 functions go  to a constant and ${\mathcal N}(\sigma_{\mathrm{rec}}) \to\sigma_{\mathrm{rec}}^{-1}\, \sqrt{2/\pi}$. In the latter limit, the thickness of the last scattering surface, i.e. $\sigma_{\mathrm{rec}}$, is of the order of $\tau_{\mathrm{rec}}$. 
 
While the parametrizations of Eqs. (\ref{rot5}) and (\ref{rot6}) can be 
used in various situations to derive semi-analytic expressions of the temperature 
and polarization anisotropies \cite{semian0,semian1,semian2,semian3} (see also \cite{MOD1,MOD3}), for typical multipoles $\ell \leq \sqrt{z_{\mathrm{rec}}}$ the finite width of the visibility function is immaterial: for sufficiently small $\ell$ 
everything goes as if the opacity suddenly drops at recombination and this
implies that ${\mathcal K}(\tau)$  presents a sharp (i.e. infinitely thin) peak at the recombination time.  Thus, since the visibility is proportional to a Dirac delta 
function and $e^{- \epsilon(\tau,\tau_{0})}$ is proportional to an Heaviside theta function, the line of sight solution of the evolution equation for $\Delta_{\mathrm{I}}$ leads to a clear separation between Sachs-Wolfe (SW) and integrated Sachs-Wolfe (ISW) contributions:
\begin{eqnarray}
 \Delta_{\mathrm{I}}(k,\mu,\tau_{0}) &=& \Delta_{\mathrm{I}}^{(\mathrm{SW})}(k,\mu,\tau_{0})  + \Delta_{\mathrm{I}}^{(\mathrm{ISW})}(k,\mu,\tau_{0}), 
\label{SW1}\\
\Delta_{\mathrm{I}}^{(\mathrm{SW})}(k,\mu,\tau_{0}) &=& \biggl[ \frac{\delta_{\gamma}}{4} + \phi \biggr]_{\tau_{\mathrm{rec}}} e^{- i \mu x(\tau_{\mathrm{rec}})},
\label{SW2}\\
\Delta_{\mathrm{I}}^{(\mathrm{ISW})}(k,\mu,\tau_{0}) &=& \int_{\tau_{\mathrm{rec}}}^{\tau_{0}} 
(\phi' +\psi') e^{- i \mu x(\tau)} \, d\tau,
\label{SW3}
\end{eqnarray}
where we recall that, by definition, $\delta_{\gamma}(k,\tau) = 4\Delta_{\mathrm{I}0}(k,\tau)$.  If the visibility is infinitely thin and the phase appearing in Eq. (\ref{SW3}) is $\tau$-independent (i.e. 
$i\mu x(\tau) \simeq i k \mu (\tau_{0} - \tau_{\mathrm{rec}})$) and it coincides 
with the phase of the SW term. Moreover, according to  Eqs. (\ref{zetaalpha}) and (\ref{alphagg}), the large-scale curvature perturbations become 
\begin{equation}
{\mathcal R}(k,\tau_{rec}) = {\mathcal R}_{*} - \frac{{\mathcal S}_{*}}{3} - \frac{R_{\gamma}}{4} \Omega_{B}, \qquad \mathrm{for} \qquad \frac{a_{\mathrm{rec}}}{a_{\mathrm{eq}}} \gg 1.
\label{SW4}
\end{equation}
The same limit leading to Eq. (\ref{SW4}) also implies
 that $\delta_{\gamma}^{\prime} \simeq 4 \psi^{\prime}$ and 
the expression for the Sachs-Wolfe contribution becomes:
\begin{equation}
 \biggl[ \frac{\delta_{\gamma}(k,\tau)}{4} + \psi(k,\tau) \biggr]_{\tau_{\mathrm{rec}}} = 
 2 \psi(k,\tau_{\mathrm{rec}})  - \frac{3}{2} \psi(k,\tau_{*}) \equiv  2 \psi(k,\tau_{\mathrm{rec}}) + {\mathcal R}_{*}(k)  - \frac{R_{\gamma}}{4} \Omega_{\mathrm{B}}(k),
 \label{SW5}
 \end{equation}
 where $\tau_{*}$ denotes here the moment at which the large-scale 
curvature perturbations are normalized prior to equality. 
In Eq. (\ref{SW5}) we used that $\delta_{\gamma}^{\prime} \simeq 4 \psi^{\prime}$ as it follows from Eqs. (\ref{STR13}) and (\ref{STR21}). By direct integration we also have
\begin{equation}
\delta_{\gamma}(k,\tau_{\mathrm{rec}}) = 4 \psi(k,\tau_{\mathrm{rec}}) + \delta_{\gamma}(k, \tau_{*}) - 4 \psi(k,\tau_{*}) + {\mathcal O}(k^2 \tau^2),
\label{SW5a}
\end{equation}
To determine the contribution $\delta_{\gamma}(k, \tau_{*})$ we recall from Eq. (\ref{GEOMzeta}) that in the absence of Ohmic electric fields (irrelevant 
for the SW and ISW effects) the density contrast on uniform curvature hypersurfaces can be expressed, in the longitudinal gauge, as $\zeta = - \psi - {\mathcal H}(\delta \rho_{t} + \delta \rho_{B})/\rho_{t}^{\prime}$; this implies that around $\tau_{*}$, 
\begin{equation}
\delta_{\gamma}(k, \tau_{*})  = 4[\zeta_{*}(k) + \psi(k,\tau_{*})]  - R_{\gamma} \Omega_{B}(k).
\label{SW6}
\end{equation}
Putting all together we have that the Sachs-Wolfe 
contribution can be written as:
\begin{equation}
\Delta_{\mathrm{I}}^{(\mathrm{SW})}(k,\mu,\tau_{0}) =\biggl[- \frac{{\mathcal R}_{*}}{5} + \frac{2}{5} {\mathcal S}_{*} - \frac{R_{\gamma}}{20} \Omega_{B}\biggr] e^{- i k\mu \tau_{0}}.
\label{SW7}
\end{equation}
From the (comoving) magnetic field intensity $B_{\mathrm{L}}$ we can easily 
compute the associated energy density referred to the photon background  
(i.e. $\overline{\Omega}_{\mathrm{B L}} = B_{\mathrm{L}}^2/(8\pi \overline{\rho}_{\gamma})$). Dividing $\overline{\Omega}_{\mathrm{B L}}$ by ${\mathcal A}_{{\mathcal R}}$ we obtain:
\begin{equation} 
\frac{\overline{\Omega}_{\mathrm{B L}}}{{\mathcal A}_{\mathcal R}}=  39.56\, \biggl(\frac{B_{\mathrm{L}}}{\mathrm{nG}}\biggr)^{2} \, \biggl(\frac{T_{\gamma0}}{2.725\,\mathrm{K}}\biggr)^{-4} \, \biggl(\frac{{\mathcal A}_{{\mathcal R}}}{2.41\times 10^{-9}}\biggr)^{-1}.
\label{SW8}
\end{equation}
 Equation (\ref{SW8}) demonstrates qualitatively why  the bounds on the magnetic fields from the temperature autocorrelations (see Tab. \ref{TABLE4}) are in the nG range.  To improve Eq. (\ref{SW7}) we can note that $\alpha_{\mathrm{rec}} = a_{\mathrm{rec}}/a_{\mathrm{eq}} = {\mathcal O}(3)$; to be more precise we can then write the full functions describing the large-scale evolution of the SW contribution \cite{MOD1,MOD3}
\begin{eqnarray}
\Delta^{(\mathrm{SW})}_{\mathrm{I}}(k,\mu,\tau_{0}) &=& \biggl[- \frac{{\mathcal R}_{*}(k)}{5} 
{\mathcal S}{\mathcal W}_{{\mathcal R}}(\alpha_{\mathrm{rec}}) + \frac{R_{\gamma} \Omega_{\mathrm{B}}(k)}{20} \,{\mathcal S}{\mathcal W}_{\mathrm{B}}(\alpha_{\mathrm{rec}})\biggr] e^{- i \mu y_{\mathrm{rec}}},
\label{SW9}\\
{\mathcal S}{\mathcal W}_{{\mathcal R}}(\alpha) &=& 1 + \frac{4}{3 \alpha} - \frac{16}{3 \alpha^2}
+ \frac{16( \sqrt{\alpha +1} -1)}{3 \alpha^3},
\nonumber\\
{\mathcal S}{\mathcal W}_{\mathrm{B}}(\alpha) &=& 1 - \frac{12}{\alpha} + \frac{48}{\alpha^2} + 
\frac{32(  1 - \sqrt{\alpha +1})}{\alpha^3},
\label{SW10}
\end{eqnarray}
where, for simplicity, the non-adiabatic contribution has been neglected. 
The SW contribution typically 
peaks for comoving wavenumbers $k \simeq 0.0002 \, \mathrm{Mpc}^{-1}$ while the ISW effect contributes between $k_{\mathrm{min}} =0.001\, \mathrm{Mpc}^{-1}$ and $k_{\mathrm{max}}= 0.01\, \mathrm{Mpc}^{-1}$. 
Even if  both contributions are reasonably separated in scales, the SW and ISW effects may partially compensate in the presence of a 
fluctuating dark energy background which is however not the case of the 
concordance paradigm \cite{dore,MOD11}. Further improvements 
of Eqs. (\ref{SW8}) and (\ref{SW10}) have been studied through 
the years but will not be discussed here. We rather stress that 
the direct bounds on the magnetic random fields obtainable from the SW effect and from the $TT$ correlations (see Tab. \ref{TABLE4}) are qualitatively 
consistent with more accurate determinations of the temperature and polarization anisotropies.

\subsection{Temperature and polarization observables}
Some qualitative features of the temperature and polarization anisotropies will now be illustrated. These themes have been discussed 
through the years in various analyses \cite{MOD1,MOD2,
MOD3,MOD4,MOD5b,MOD6,MOD7,MOD8,MOD7a,
MOD8a,MOD8b,MOD9,MOD10,MOD11}.  For this purpose
there are two complementary strategies. The first possibility 
is to include all the initial conditions without bothering about 
their physical significance. This is the approach often taken in 
data analyses where all sorts of initial data are compared.
The second possibility is to select the initial conditions 
which are simultaneously minimal and physically 
justified on the basis of the underlying concordance paradigm. 
According to this second viewpoint no further parameters must be added besides the ones of Eq. (\ref{PS4a}): this excludes, for instance, the divergent modes where the evolution of the anisotropic stress must be specified, at least, from the end on inflation down to neutrino decoupling.

Since in the concordance paradigm the scalar modes are the dominant 
source of inhomogeneity, we can also expect that the regular 
magnetized mode of Eq. (\ref{L4}) will be the more relevant 
over large-scales. The vectors are likely not to play a role unless the rotational velocity field dominates against its irrotational counterpart. As argued in section \ref{sec3} this is very different from what happens in the case 
of the $\Lambda$CDM scenario and would probably require, prior to decoupling, a turbulent flow which seems excluded by the smallness 
of the kinetic Reynolds number prior to matter-radaition equality. 
For the same reasons we are led to exclude, in the first approximation, the tensor modes which may however lead to a further source of $B$-mode polarization, as we shall see in section \ref{sec5}. 
Furthermore, if the magnetized tensor modes are regular before equality, they are also subdominant.

With these specifications we shall now focus on the scalar brightness perturbations and remind that $\Delta_{\mathrm{I}}(\hat{n},\tau)$ can be easily expanded in ordinary spherical harmonics so that the angular power spectrum of the temperature autocorrelations is defined as:
\begin{equation}
 C_{\ell}^{(\mathrm{TT})} = \frac{1}{2\ell + 1} \sum_{m} \langle 
 a^{(\mathrm{T})*}_{\ell\,m} a^{(\mathrm{T})}_{\ell\,m}\rangle,\qquad 
 a^{(\mathrm{T})}_{\ell m} 
=  \int d \hat{n}\, Y_{\ell m}^{*}(\hat{n}) \,\Delta_{\mathrm{I}}(\hat{n}, \tau),
\label{E1a}
\end{equation}
where $Y_{\ell m}(\hat{n})$ are the (scalar) spherical harmonics. 
Conversely $\Delta_{\mathrm{Q}}(\hat{n},\tau)$ and $\Delta_{\mathrm{U}}(\hat{n},\tau)$ are naturally expanded in terms of spin-$2$ spherical harmonics \cite{sud,zalda}. The orthogonal combinations
\begin{equation}
\Delta_{\pm}(\hat{n},\tau) = \Delta_{\mathrm{Q}}(\hat{n},\tau) \pm i \Delta_{\mathrm{U}}(\hat{n},\tau),
\label{int1a}
\end{equation}
transform, respectively, as fluctuations of spin-weight $\pm 2$ \cite{sud,zalda,GS}. Owing to this observation,  $\Delta_{\pm}(\hat{n},\tau)$ can be expanded in terms of spin-$\pm2$ spherical harmonics 
$_{\pm 2}Y_{\ell\,m}(\hat{n})$, i.e. 
\begin{equation}
\Delta_{\pm}(\hat{n},\tau) = \sum_{\ell \, m} a_{\pm2,\,\ell\, m} \, _{\pm 2}Y_{\ell\, m}(\hat{n}).
\label{int2a}
\end{equation}
The $E$- and $B$-modes are, up to a sign, the real and the imaginary 
parts of $a_{\pm 2,\ell\,m}$, i.e. 
\begin{equation}
a^{(\mathrm{E})}_{\ell\, m} = - \frac{1}{2}(a_{2,\,\ell m} + a_{-2,\,\ell m}), \qquad  
a^{(\mathrm{B})}_{\ell\, m} =  \frac{i}{2} (a_{2,\,\ell m} - a_{-2,\,\ell m}).
\label{int3a}
\end{equation}
In real space (as opposed to Fourier space), the fluctuations constructed from 
$a^{(\mathrm{E})}_{\ell\,m}$ and $a^{(\mathrm{B})}_{\ell\,m}$ have the 
property of being invariant under rotations on a plane orthogonal 
to $\hat{n}$.  They can therefore be expanded in terms of (ordinary) spherical harmonics:
\begin{equation}
\Delta_{\mathrm{E}}(\hat{n},\tau) = \sum_{\ell\, m} N_{\ell}^{-1} \,  a^{(\mathrm{E})}_{\ell\, m}  \, Y_{\ell\, m}(\hat{n}),\qquad 
\Delta_{\mathrm{B}}(\hat{n},\tau) = \sum_{\ell\, m} N_{\ell}^{-1} \,  a^{(\mathrm{B})}_{\ell\, m}  \, Y_{\ell\, m}(\hat{n}),
\label{int4a}
\end{equation}
where $N_{\ell} = \sqrt{(\ell - 2)!/(\ell +2)!}$.  
The $EE$ and $BB$ angular power spectra are then defined as: 
\begin{equation}
C_{\ell}^{(EE)} = \frac{1}{2\ell + 1} \sum_{m = -\ell}^{\ell} 
\langle a^{(\mathrm{E})*}_{\ell m}\,a^{(\mathrm{E})}_{\ell m}\rangle,\qquad 
C_{\ell}^{(BB)} = \frac{1}{2\ell + 1} \sum_{m=-\ell}^{\ell} 
\langle a^{(\mathrm{B})*}_{\ell m}\,a^{(\mathrm{B})}_{\ell m}\rangle.
\label{int5aa}
\end{equation}
In the minimal version of the $\Lambda$CDM paradigm the adiabatic fluctuations of the scalar curvature lead to a polarization which is characterized exactly by the condition $a_{2,\,\ell m} = a_{-2,\,\ell m}$, i.e. $a_{\ell m}^{(\mathrm{B})} =0$. 
It is however true that a $B$-mode polarization is induced through the lensing of the primary anisotropies; this secondary $B$-mode polarization has been already detected by the South Pole Telescope \cite{SPTpol}. In the presence 
of magnetic random fields the further sources of $B$-mode polarization 
can be envisaged (see section \ref{sec5}) but, for the moment, we shall 
just focus on the basic observables detected so far. 
By focussing the attention on the primary anisotropies we can therefore only define a further power spectrum given by the cross-correlation of the temperature and of the $E$-mode polarization:
\begin{equation}
C_{\ell}^{(TE)} = \frac{1}{2\ell + 1} \sum_{m = -\ell}^{\ell} 
\langle a^{(\mathrm{T})*}_{\ell m}\,a^{(\mathrm{E})}_{\ell m}\rangle.
\label{int6a}
\end{equation}
In the standard terminology Eqs. (\ref{E1a}) and (\ref{int5aa}) give, respectively, the $TT$ and $EE$ correlations while  Eq. (\ref{int6a}) gives 
instead the $TE$ power spectrum. The normalized temperature and polarization autocorrelations (i.e., respectively, $TT$ and $EE$ angular 
power spectra) and their mutual cross-correlations (i.e. the $TE$ angular power spectra) can be written, with shorthand notation, as: 
\begin{equation}
 G^{(\mathrm{TT})}_{\ell} =  \frac{\ell (\ell  +1)}{2 \pi} C_{\ell}^{(\mathrm{TT})}, \qquad 
 G^{(\mathrm{EE})}_{\ell} =  \frac{\ell (\ell  +1)}{2 \pi} C_{\ell}^{(\mathrm{EE})},\qquad  G^{(\mathrm{TE})}_{\ell} =  \frac{\ell (\ell  +1)}{2 \pi} C_{\ell}^{(\mathrm{TE})},
\label{int7a}
\end{equation}
and are measured in $(\mu \mathrm{K})^2$. It is now useful to discuss 
the qualitative features of the temperature and polarization anisotropies 
when the magnetic random fields are included according to the 
minimal logic spelled out above in this section. The initial conditions 
of the Einstein-Boltzmann hierarchy are therefore 
fixed in terms of the magnetized adiabatic mode of Eq. (\ref{L4}).

\subsection{Magnetized temperature autocorrelations}

For magnetic fields with $B_{\mathrm{L}} = {\mathcal O}(\mathrm{few})$ 
nG and spectral indices $n_{\mathrm{B}} = {\mathcal O}(1)$ the first Doppler peak of the temperature autocorrelations increases sharply. Already for $0.1 \mathrm{nG}< B_{\mathrm{L}} < 2 \,\mathrm{nG}$ 
the third peak increases while the second peak becomes less 
pronounced. As soon as $B_{\mathrm{L}} \geq 2$ nG the second peak 
practically disappears and it is replaced by a sort of hump. 
\begin{figure}[!ht]
\centering
\includegraphics[height=6.1cm]{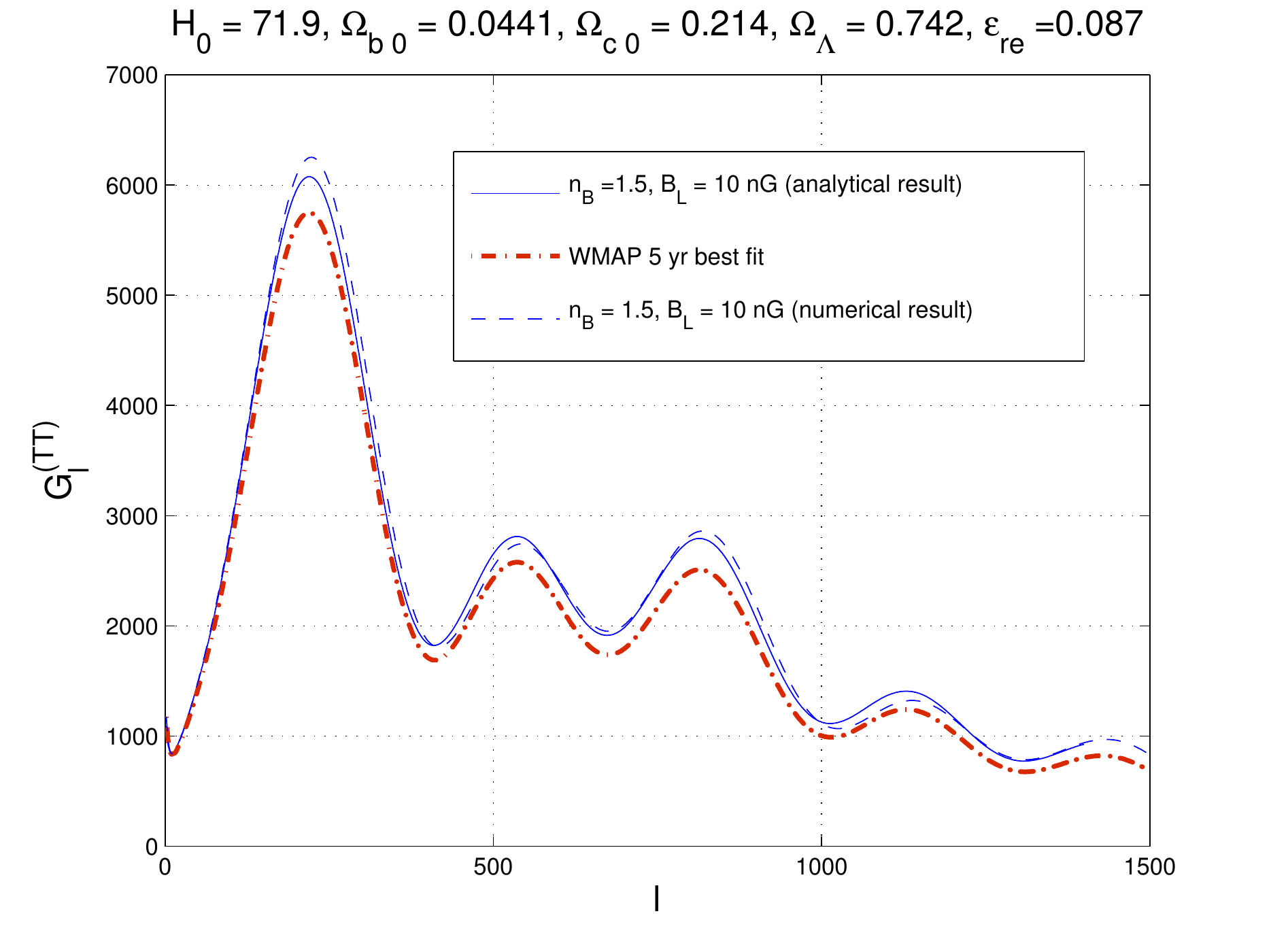}
\includegraphics[height=6.1cm]{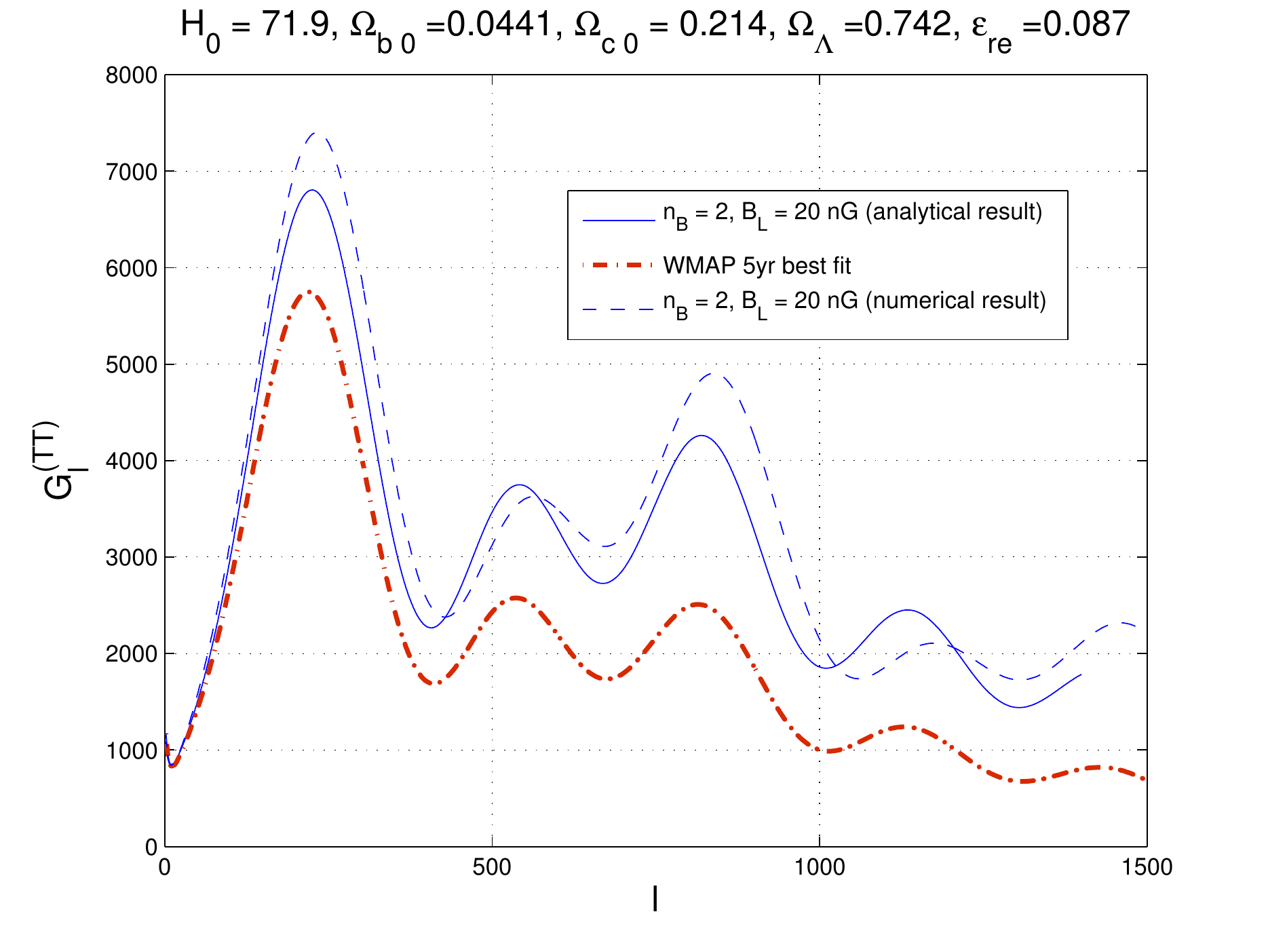}
\caption[a]{The temperature autocorrelations are illustrated in the case of rather 
large values of the magnetic field (i.e. $10$ and $20$ nG) with the purpose 
of emphasizing the typical patterns of distortions 
induced by the magnetized adiabatic mode of Eq. (\ref{L4}). For comparison 
the WMAP-5y best fit has been also included. These plots as the following ones are adapted 
from Ref. \cite{MOD8a}.}
\label{figure3}      
\end{figure}
In Fig. \ref{figure3} we illustrate the $TT$ autocorrelations for a deliberately extreme set of parameters: this choice will make more apparent the effect of the increase of the magnetic field intensity. 
The inclusion of a magnetized background has a threefold 
effect on the temperature autocorrelations: the height of the 
first acoustic peak increases, the second peak is distorted and it eventually 
turns into a hump for sufficiently large values of $B_{\mathrm{L}}$ (or of $n_{\mathrm{B}}$). The third 
peak is, at the same time, distorted and raised. 
In Fig. \ref{figure3} as we move from the plot at the left to the plot at the right 
the spectral index increases. The increase of the spectral slope entails 
also an increase of the distortions, as we can see from Fig. \ref{figure3}.

As suggested in \cite{MOD8a,MOD8b} the impact of the magnetic field parameters follows from the relative ratios of the first three acoustic peaks which  defined as:
\begin{equation}
\overline{H}_{1} = \frac{{\mathcal G}^{(\mathrm{TT})}_{\ell_{1}}}{{\mathcal G}^{(\mathrm{TT})}_{\ell =10}}
\qquad \overline{H}_{2} = \frac{{\mathcal G}^{(\mathrm{TT})}_{\ell_{2}}}{{\mathcal G}^{(\mathrm{TT})}_{\ell_{1}}}, \qquad 
\overline{H}_{3} = \frac{{\mathcal G}^{(\mathrm{TT})}_{\ell_{3}}}{{\mathcal G}^{(\mathrm{TT})}_{\ell_{2}}},
\label{int8a}
\end{equation}
where $\ell_{1} ={\mathcal O}(220)$,  $\ell_{2} = {\mathcal O}(535)$ and  $\ell_{3} = {\mathcal O}(816)$ 
are, respectively, the locations of the first three acoustic peaks. When the magnetic field intensity increases we have that, typically, $\overline{H}_{1}$ and $\overline{H}_{3}$ increase more than $\overline{H}_{2}$ so that 
the resulting shape of the $TT$ correlations changes qualitatively. 
\begin{figure}[!ht]
\centering
\includegraphics[height=6.1cm]{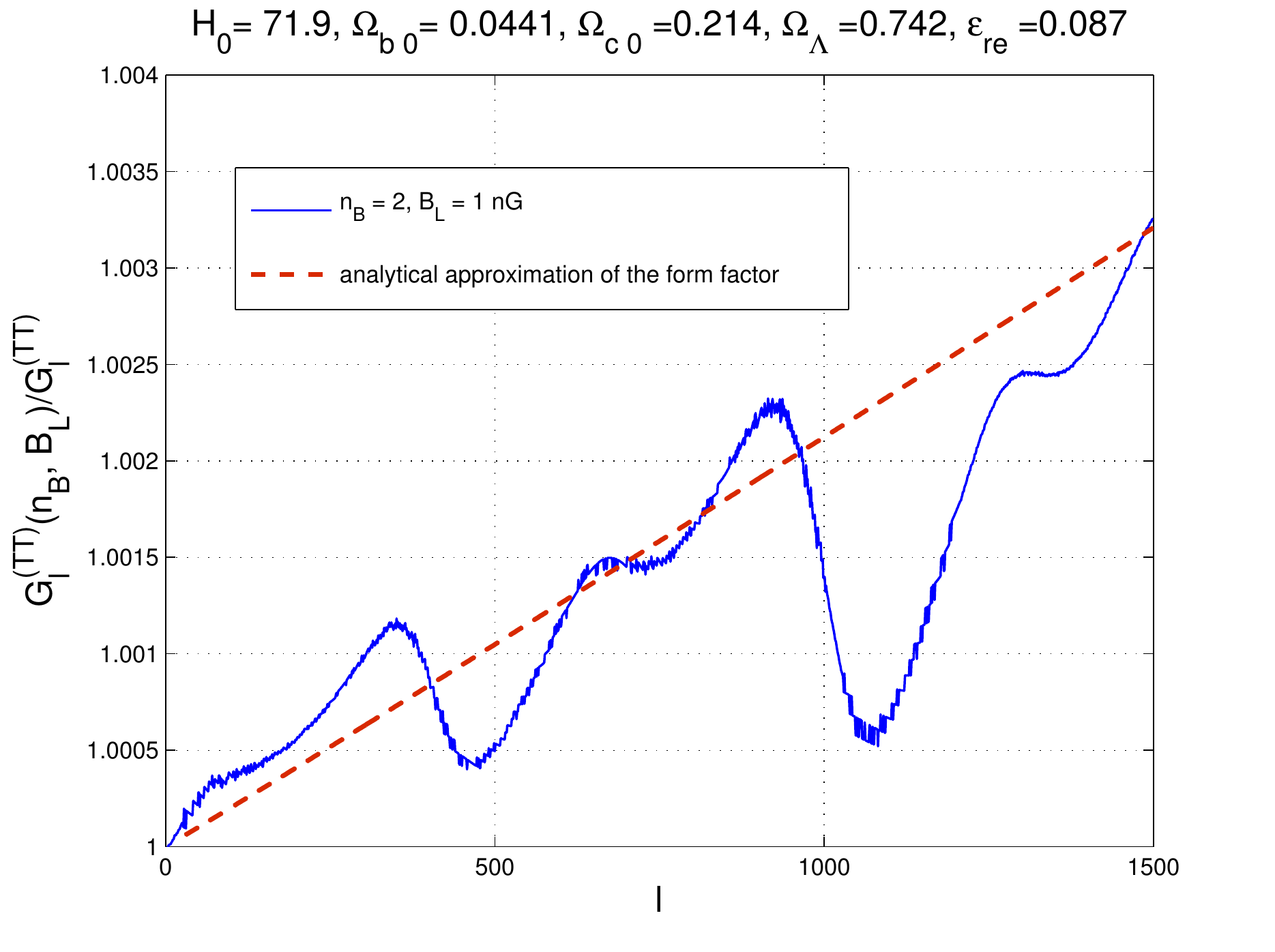}
\includegraphics[height=6.1cm]{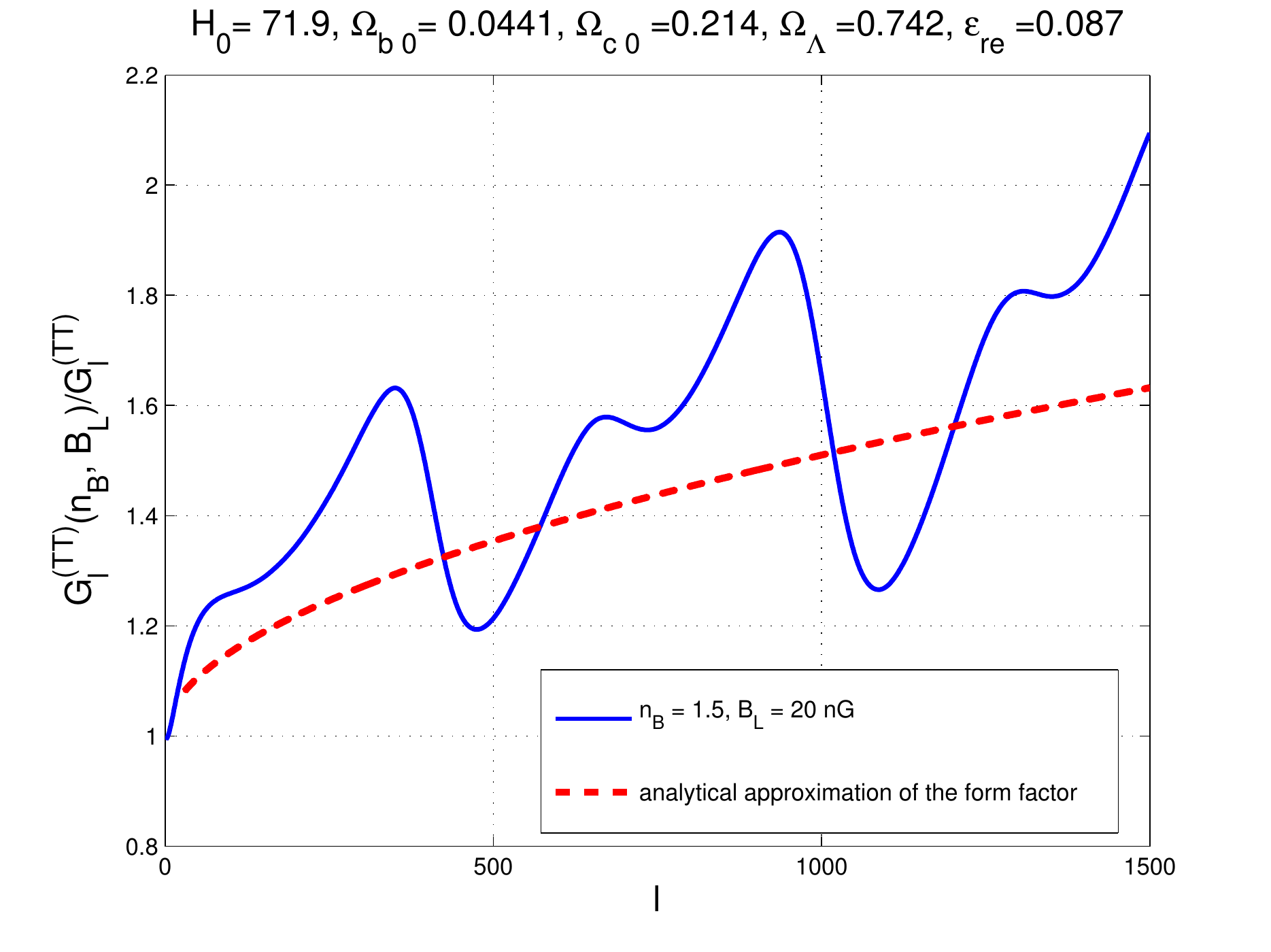}
\caption[a]{The scaling properties of the magnetized  $TT$ correlations.}
\label{figure4}      
\end{figure}
To gauge the effects of the magnetic random fields on the $TT$ correlations
a practical strategy is to fix the $\Lambda$CDM parameters to their 
best-fit values and then construct the ratio ${\mathcal G}^{(\mathrm{TT})}(n_{\mathrm{B}}, B_{\mathrm{L}})/{\mathcal G}^{(\mathrm{TT})}$
where ${\mathcal G}^{(\mathrm{TT})}(n_{\mathrm{B}}, B_{\mathrm{L}})$ is the angular power spectrum computed for a particular set of parameters of the magnetized background and ${\mathcal G}^{(\mathrm{TT})}$ is the 
angular power spectrum for the best fit of the $\Lambda$CDM parameters. 
In Fig. \ref{figure4} this ratio is illustrated for different values of the intensity and of the spectral index. What is apparent is that a magnetic field $B_{\mathrm{L}} = {\mathcal O}(1)$ nG induces a modification 
${\mathcal O}(10^{-3})$ on the shape of the angular power spectrum (see 
\cite{MOD8a} for further details). 

The dashed lines in Fig. \ref{figure4} and the full lines in Fig. \ref{figure3} illustrate the results of the analytic approximation to the temperature autocorrelations. The idea, in short, is to include consistently 
the contributions of the magnetic fields in the evolution of the monopole and of the dipole of the brightness 
perturbations and to evaluate the angular power spectra by using a semi-analytic form of the visibility function (like the one outlined in Eq. (\ref{rot5})). Assuming tight coupling between photons, electrons and baryons, 
the evolution of the monopole and of the dipole of the brightness perturbations determines the source term in the temperature 
and polarization anisotropies. The monopole and the dipole obey, in Fourier space, the following pair of equations:
\begin{eqnarray}
&& (\psi - \Delta_{\mathrm{I}0})' = k \Delta_{\mathrm{I}1},
\label{int9a}\\
&&[(R_{\mathrm{b}} + 1) \Delta_{\mathrm{I}1}]'  + 2 \frac{k^2}{k_{\mathrm{D}}^2} (R_{\mathrm{b}} +1)
 \Delta_{\mathrm{I}1}= 
\frac{k}{3} \Delta_{\mathrm{I}0} + \frac{k(R_{\mathrm{b}} +1)}{3}\phi + \frac{k (\Omega_{\mathrm{B}} - 4 \sigma_{\mathrm{B}})}{12},
\label{int10a}
\end{eqnarray}
where $k_{\mathrm{D}}$ is the wave-number corresponding to diffusive 
damping.To lowest-order in the photon-baryon coupling the diffusive 
damping is $k_{\mathrm{D}}^{-2} = \eta/[\rho_{\gamma} ( 1 + R_{\mathrm{b}})]$ where $\eta$ has 
been defined right after Eq. (\ref{STR22}); more precisely we have \cite{pee3,naselskyb,primer}:
\begin{equation}
\frac{1}{k^2_{\mathrm{D}}} = \frac{2}{5} \int_{0}^{\tau} c_{\mathrm{sb}}(\tau') \frac{ a_{0} d\tau'}{a(\tau')\,\,x_{\mathrm{e}} n_{\mathrm{e}} \sigma_{\mathrm{e}\gamma}}.
\label{int11a}
\end{equation}
The estimates based on shear 
viscosity can be improved by going to higher order in the tight-coupling expansion 
and by further refining the estimates depending upon the explicit values 
of the $\Lambda$CDM parameters. To second order in the tight-coupling expansion the inclusion of the polarization allows one to estimate \cite{TC2,TC3}:
\begin{equation}
\frac{1}{k_{\mathrm{D}}^2 }= \int_{0}^{\tau} \frac{d\tau'}{6 (R_{\mathrm{b}} + 1) \epsilon'} \biggl[\frac{16}{15} + \frac{R_{\mathrm{b}}^2}{R_{\mathrm{b}} + 1}\biggr].
\label{int12a}
\end{equation}
The  factor $16/15$ arises since the polarization fluctuations are taken consistently 
into account in the derivation. This difference is physically relevant. Grossly speaking we can indeed say that
more polarization implies more anisotropy (and vice versa); more polarization implies a faster damping by diffusion. 
Note that $k_{\mathrm{D}}$ provides an effective ultra-violet cut-off for the magnetic energy spectra and will be used later on.

The evolution of the monopole and of the dipole can be determined from the 
WKB solution of Eqs. (\ref{int9a}) and (\ref{int10a}), i.e. 
\begin{eqnarray}
&& \Delta_{\mathrm{I}0}(k,\tau) + \phi(k,\tau) = {\mathcal L}(k,\tau) + \sqrt{c_{\mathrm{sb}}} {\mathcal M}(k,\tau) 
\cos{[k r_{\mathrm{s}}(\tau)]} \, e^{- \frac{k^2}{k_{\mathrm{D}}^2}},
\label{int13a}\\
&& \Delta_{\mathrm{I}1}(k,\tau) = c_{\mathrm{sb}}^{3/2} {\mathcal M}(k,\tau) 
\sin{[k\, r_{\mathrm{s}}(\tau)]} e^{- \frac{k^2}{k_{\mathrm{D}}^2}},
\label{int14a}
\end{eqnarray}
where ${\mathcal L}(k,\tau)$ and ${\mathcal M}(k,\tau)$ are fixed once 
the initial conditions of the Einstein-Boltzmann hierarchy are specified. In what 
follows, as already mentioned, the initial conditions shall correspond 
to the magnetized adiabatic mode. More details on the semi-analytic 
approaches for the analysis of the temperature and polarization 
anisotropies can be found in a series of papers \cite{MOD3,MOD8a}.

The coefficients $a_{\ell m}^{(\mathrm{T})}$ and $a_{\ell m}^{(\mathrm{E})}$ are determined in terms of the monopole, dipole and quadrupole 
of the intensity and in terms of the monopole and quadrupole of the polarization. More specifically the coefficient $a_{\ell m}^{(\mathrm{T})}$ is
\begin{equation}
a^{(\mathrm{T})}_{\ell m} = \frac{\sqrt{4 \pi}}{(2\pi)^{3/2}} \, (-i)^{\ell} \, \sqrt{2 \ell + 1} 
\int d^3 k e^{- \frac{k^2}{k_{\mathrm{t}}^2}} \biggl[ 
(\Delta_{\mathrm{I}0} + \phi) j_{\ell}(x) + 3 \Delta_{\mathrm{I}1} \biggl(\frac{d j_{\ell}}{dx}\biggr)\biggr],
\label{int15a}
\end{equation}
where $ x = k(\tau_{0} - \tau_{*})$ and  where $j_{\ell}(x)$ are the spherical Bessel functions \cite{abr1,abr2} of argument $x$. In Eq. (\ref{int15a})  $k_{\mathrm{t}} = \sqrt{3}/\sigma_{*}$ arises from the integration over $\tau$ of the Gaussian visibility function. The  coefficient  $a_{\ell m}^{(\mathrm{E})}$ turns out to be:
\begin{equation}
a_{\ell m}^{(\mathrm{E})}= \frac{3}{4} \frac{(-i)^{\ell}}{(2\pi)^{3/2}} 
\sqrt{\frac{(\ell -2)!}{(\ell + 2)!}} \sqrt{4\pi} \sqrt{2\ell + 1} 
\int d^{3} k\,\, x^2 \,\, [ ( 1 + \partial_{x}^2)^2] j_{\ell}(x) \int_{0}^{\tau_{0}}
{\mathcal K}(\tau) S_{\mathrm{P}}(k,\tau) d\tau.
\label{int16a}
\end{equation}
Using Eqs. (\ref{int15a}) and (\ref{int16a}) in Eqs. (\ref{E1a}), (\ref{int5aa}) and (\ref{int6a}) we can obtain 
the semianalytic forms of the $TT$, $TE$ and $EE$ angular power spectra. This is the way 
the dashed lines in Fig. \ref{figure4} have been computed. As we shall see 
the same approximations lead to some interesting results for the $TE$ and $EE$ correlations. 

\subsection{Magnetized polarisation correlations and cross-correlations}

The $TE$ cross-correlations together with the $TT$ spectra 
are among the most useful indicators of the adiabatic nature of the CMB initial 
conditions. The polarization observations are therefore a rather sensitive tool which can be used for the scrutiny of a magnetized component. 
\begin{figure}[!ht]
\centering
\includegraphics[height=6.1cm]{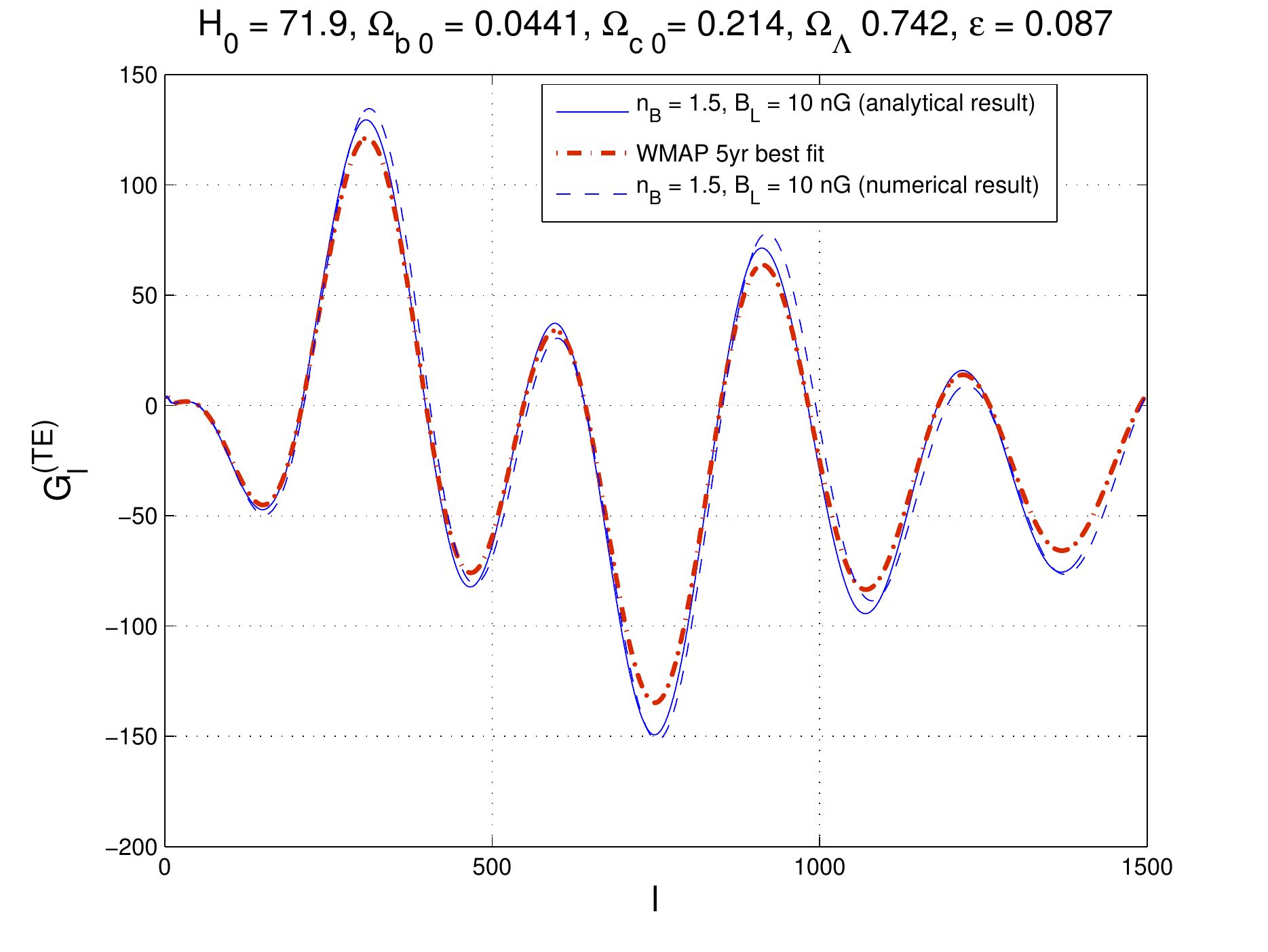}
\includegraphics[height=6.1cm]{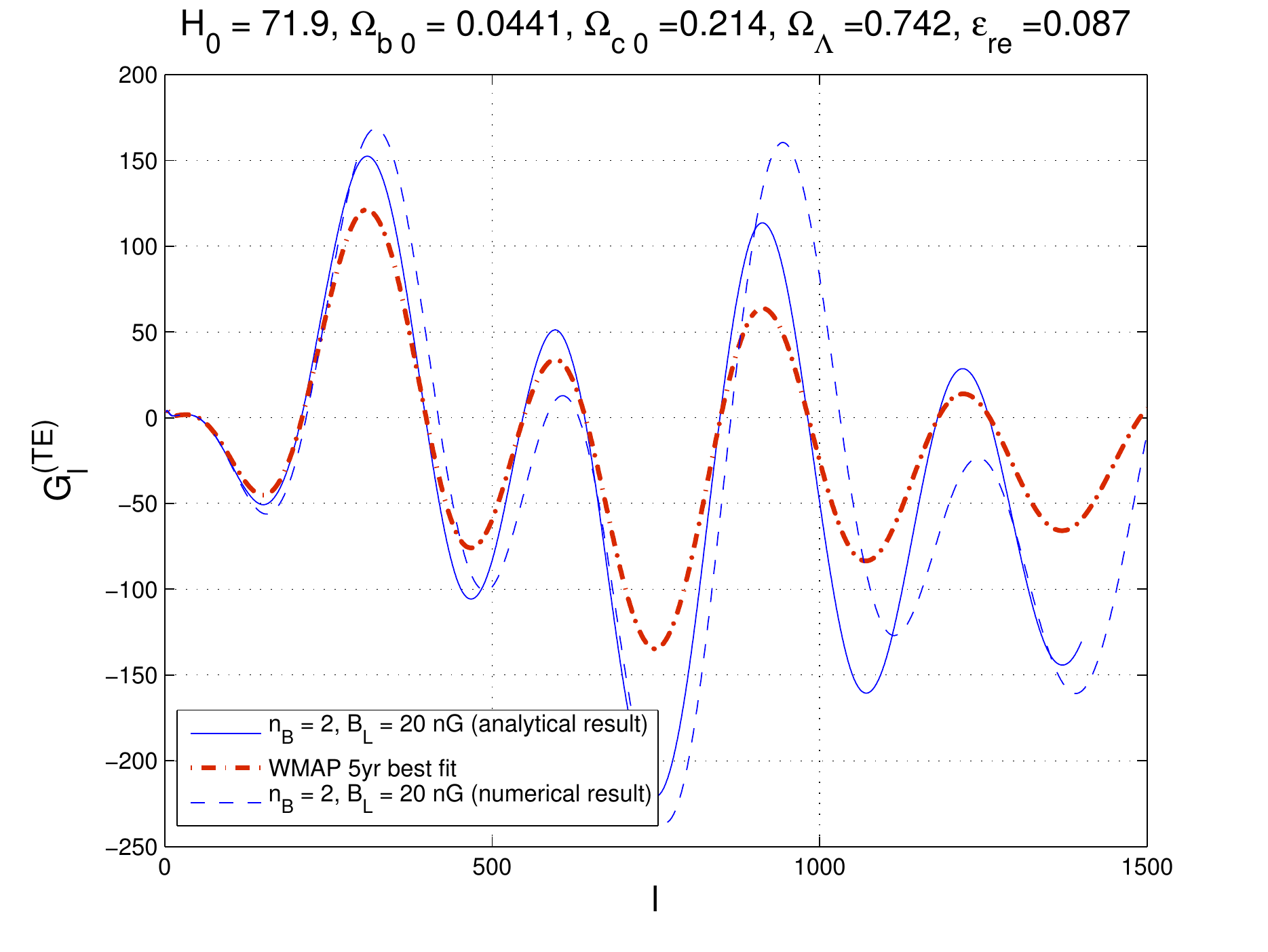}
\caption[a]{The $TE$ correlations are illustrated for the same set of extreme 
magnetic field parameters already employed in Fig. \ref{figure4}. In both plots 
with the full lines we reported the analytical results obtained 
on the basis of the approximation scheme mentioned in Eqs. (\ref{int13a}) and (\ref{int14a}).}
\label{figure5}      
\end{figure}
In Fig. \ref{figure5} the $TE$ correlations are illustrated for the case of blue magnetic spectral indices
$1 < n_{\mathrm{B}} < 5/2$. On a purely qualitative ground when the magnetic field is of the order of $0.1$ nG the magnetized
TE correlations  cannot be distinguished from the angular power spectra computed in the absence of magnetic fields. This observation shows once more the typical sensitivities of this type of approach. 

\begin{figure}[!ht]
\centering
\includegraphics[height=6.1cm]{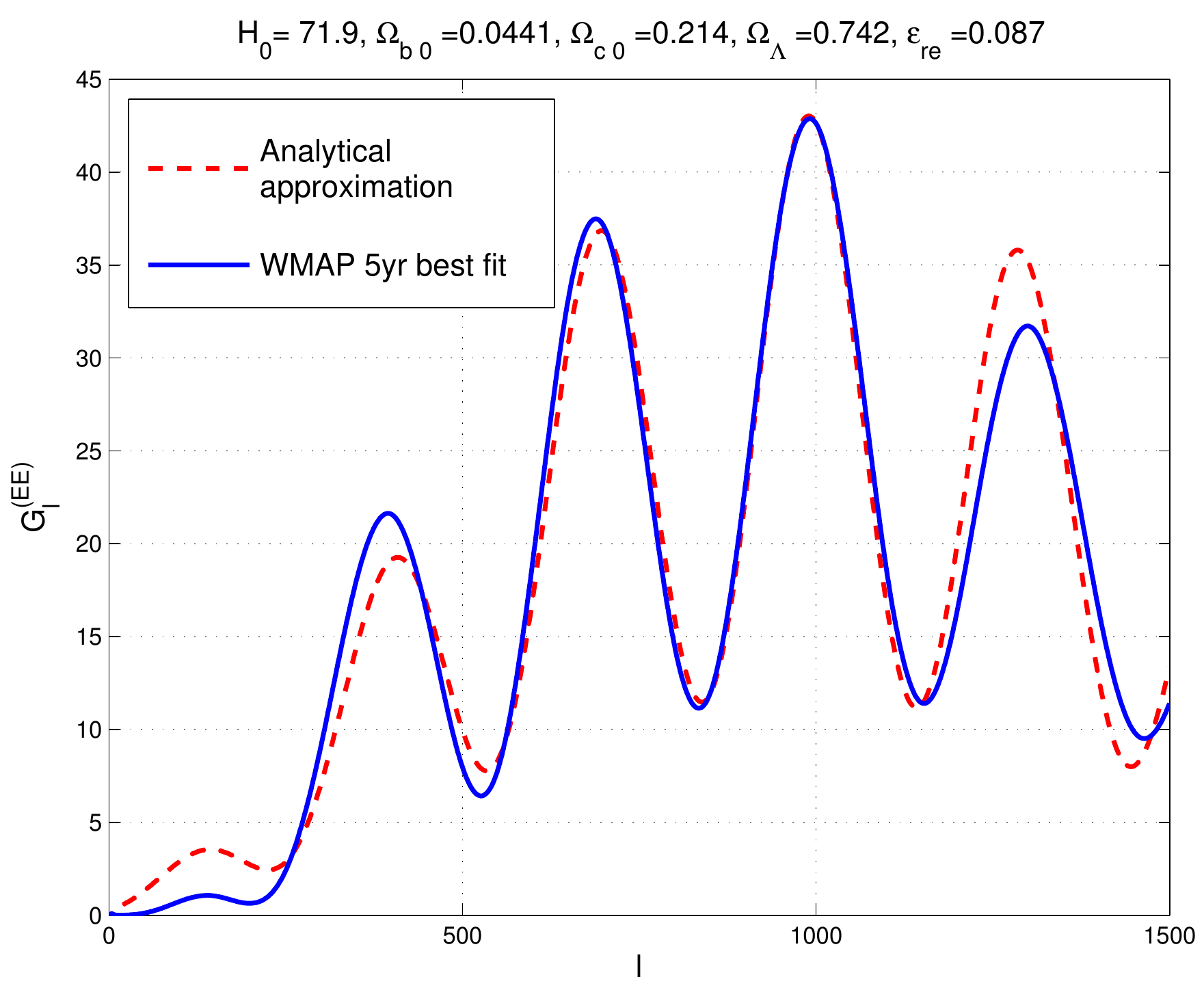}
\includegraphics[height=6.1cm]{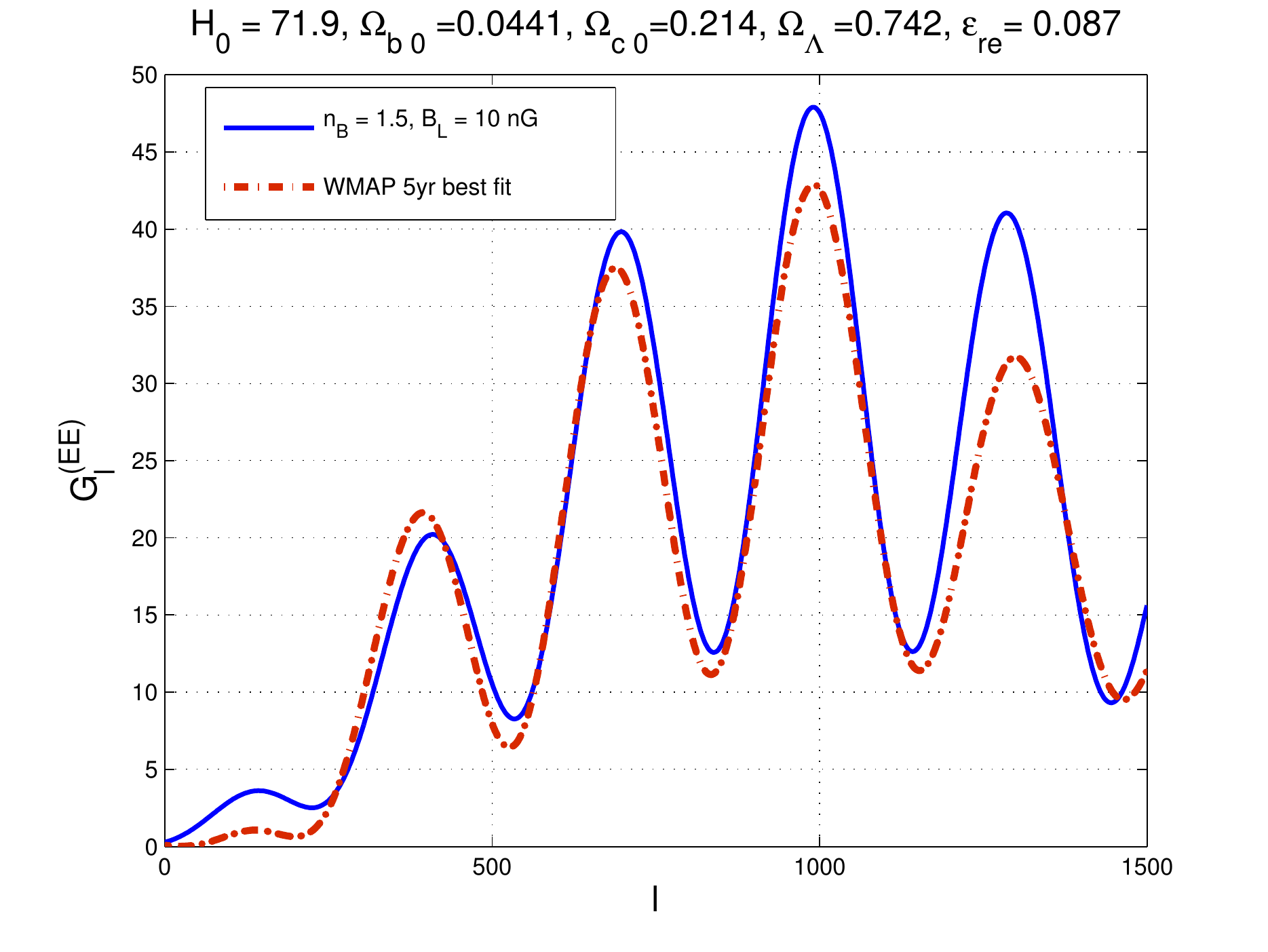}
\caption[a]{The $EE$ correlations are illustrated for the same sets 
of parameters of Figs. \ref{figure4} and \ref{figure5}. Note that in the left plot, with the dashed line, we 
illustrate the semi-analytic approximation in the absence of magnetic fields.}
\label{figure6}      
\end{figure}

Unlike the case of the temperature autocorrelations (where the position of the Doppler peak cannot be moved by a stochastic magnetic field)  there is an observable shift of the second and third (correlation) peaks of the $TE$ spectra. This distortion also entails a shift 
of the position  of the corresponding peaks. A similar effect is observed in the magnetized $EE$ correlations which are 
reported in Fig. \ref{figure6} (see in particular the plot at the  right). 

Figures \ref{figure5} and \ref{figure6} show, a posteriori, that the magnetic fields also affect the polarization observables even without a Faraday rotation term (which will be specifically discussed in section \ref{sec5}). 
The physical reason of the obtained result can be easily 
understood: to zeroth-order in the tight-coupling expansion, 
the magnetic field affects the dipole of the brightness perturbation 
for the intensity. Always to zeroth order, this contribution is reflected 
in a further source term for the monopole. But both the $TE$ and $EE$ 
power spectra arise to first-order in the tight-coupling expansion 
and are proportional to the first-order dipole through a term which is, up to a numerical factor, $k/\epsilon'$. This shows why we also get an effect on the polarization observables even if the Faraday rotation term is absent. 
The results illustrated in this section show that it is possible to obtain accurate 
estimates of the temperature autocorrelations and of the polarization correlations also in the presence of a magnetized background.

\renewcommand{\theequation}{5.\arabic{equation}}
\setcounter{equation}{0}
\section{Faraday rotation}
\label{sec5}
In a magnetized plasma the linear polarization of the microwave background 
rotates thanks to the Faraday effect and this suggestion already appeared in a 
relevant remark of Ref. \cite{rees} where the author argued that repeated 
Thomson scatterings of the primeval radiation during the early phases of a 
slightly anisotropic universe would modify the black-body spectrum and produce 
linear polarization. The CMB polarization could then be modified by the depolarizing effect of a uniform magnetic field \cite{rees} possibly accommodated in an axisymmetric background geometry. After the first observations of the temperature anisotropies, the Faraday rotation of the microwave background has been analyzed in different frameworks \cite{far1,far2,far3}. By developing the original perspective of Ref. \cite{rees} the initial analyses have been conducted within the uniform field approximation \cite{far1,far2,far3} (see also \cite{far4}). In the case of magnetic random fields the Faraday rotation is however characterized by a specific power spectrum \cite{far5} (see also 
\cite{far6,far7,far8}) and the uniform field approximation is not always suitable. 
The Faraday effect may also mix with the birefringence induced by pseudo-scalar particles \cite{far8}  (see also \cite{far8a,far8b}). In this case the frequency dependence of the signal differs from the situation of the Faraday effect alone. From the viewpoint of the temperature and polarization angular power spectra the Faraday effect rotates the $E$-mode polarization into a $B$-mode that must be computed from the appropriate magnetized initial conditions \cite{far9} already introduced in the previous section. In this case the total $B$-mode polarization will be a convolution of the Faraday rotation spectrum and of the $E$-mode autocorrelation determined, in its turn, by the magnetized initial conditions explored 
in sections \ref{sec3} and \ref{sec4}. In a more realistic perspective the 
Faraday rotation occurs while the polarization is formed without an explicit scale separation between the two moments; the theoretical analysis of this description is still not completely clear.

\subsection{Pivotal frequencies} 

Since the Faraday effect leads to a frequency-dependent $B$-mode power spectrum it is sometimes 
relevant to define a pivot frequency at which the signal is computed. 
This possible need of a pivotal frequency scale is rarely mentioned 
in the current literature and it is therefore useful to clarify this aspect. 
The energy density of the CMB spectrum in critical units can be written as 
\begin{equation}
 \frac{1}{\rho_{\mathrm{c}}} \frac{d\rho_{\gamma}}{d\ln{\overline{\omega}}}= 
\frac{15}{\pi^4} h_{0}^2 \Omega_{\gamma 0}\, g(x), \qquad 
h_{0}^2 \Omega_{\gamma 0} = 2.47 \times 10^{-5},
\label{CMBmax1}
\end{equation}
where $g(x)=  x^{4}/(e^{x} -1)$; note that $x = \overline{\omega}/T_{\gamma 0}$ and $\overline{\omega} = 2 \pi \overline{\nu}$.
The maximum of the spectral energy density (\ref{CMBmax1}) follows from the extremum of $g(x)$ and the result is 
$x_{\mathrm{max}}^{(\rho)} = 3.920$. In analog terms the maximum of the brightness of the microwave background spectrum follows from the extremum of $f(x) = x^{3}/(e^{x} -1)$ corresponding  to $x_{\mathrm{max}}^{(b)} = 2.821$. Consequently, two motivated pivotal frequencies for the normalization of the $B$-mode autocorrelation induced by the Faraday effect are, respectively, the maximum of the brightness and  of the energy density\footnote{This pivot  
frequency has no relation with the pivot wavenumber $k_{\mathrm{p}}$ used to assign 
the spectrum of large-scale curvature inhomogeneities. However the logic for introducing these two concepts are similar: 
since the $B$-mode induced by the Faraday effect depends on the frequency it is necessary to fix a conventional frequency to compare different signals and different bounds.}
\begin{equation}
\overline{\nu}^{(b)}_{\mathrm{max}} = 160.3 \,\mathrm{GHz}, \qquad \overline{\nu}^{(\rho)}_{\mathrm{max}} = 222.6\, \mathrm{GHz},
\label{maxfreq}
\end{equation}
where $\overline{\nu}^{(b)}$ and $\overline{\nu}_{\mathrm{max}}$ denote respectively the maximum of 
the brightness and of the spectral energy density. When needed, we shall preferentially use $\overline{\nu}^{(\rho)}_{\mathrm{max}}$ as pivotal frequency for the Faraday rotation signal. 

The rationale for the introduction of the pivot frequency
also depends on the widely different observational channels exploited by the various experiments.
For instance the WMAP experiment observed the microwave sky in five frequency channels centered at 
$23$, $33$, $41$, $61$ and $94$ in units of GHz. The Planck experiment observed the microwave sky in  nine frequency channels: three frequency channels (i.e. $30,\,44,\,70$ GHz) belonged to the low frequency instrument (LFI);  six 
channels (i.e. $100,\,143,\,217,\,353,\,545,\,857$ GHz) belonged to the high 
frequency instrument (HFI). Since not all the channels see the same $B$-mode angular power spectrum it is desirable that observational papers would clearly specify the pivot frequency when comparing their results to potential theoretical signals coming from the Faraday effect. This is rarely done since, understandably, each collaboration just considers its own operating frequency (for instance $70$ GHz in the case of Ref. \cite{chone34am} or $148$ GHz in the case of Ref. \cite{chone44}).
We point out that this practice can be improved by referring the angular power spectra either 
to the maximum of the spectral energy density or to the maximum of the brightness, 
as suggested in Eq. (\ref{maxfreq}). 

\subsection{Dispersion relations}
\label{sec5sub1}
When the frequency of the (polarized) CMB
photons exceeds the plasma frequency the baryon fluid is no longer a 
sound dynamical approximation and a two-fluid description 
is mandatory. This is the essence of Faraday rotation and of 
the other dispersive phenomena that can be analyzed 
by studying the propagation of high-frequency waves 
in magnetized plasmas.

The relevant dispersion relations are derived by linearizing Eqs. (\ref{S1})--(\ref{S4}) 
in the presence of a background magnetic field $\vec{B}(\vec{x})$. By separating 
the background fields from the propagating electromagnetic waves we have 
\begin{eqnarray}
&& n_{\rm e,\,i}(\vec{x},\tau) = n_{0} + \delta n_{\rm e,\,i}(\vec{x},\tau),
\qquad \vec{B}(\vec{x},\tau) = \vec{B}(\vec{x}) + \vec{b}(\vec{x},\tau),
\nonumber\\
&& \vec{v}_{\rm e,\,i}(\vec{x},\tau) = 
\delta \vec{v}_{\rm e,\,i} (\vec{x},\tau),\qquad
\vec{E}(\vec{x},\tau)=  \vec{e}(\vec{x},\tau),
\label{fluct}
\end{eqnarray}
where $\vec{e}(\vec{x},\tau)$ and $\vec{b}(\vec{x},\tau)$ are, respectively,
the electric and magnetic fields of the wave while 
 $\vec{B}(\vec{x})$ is the large-scale magnetic random 
 field. The evolution of the concentrations 
and of the velocities for electrons and ions 
\begin{eqnarray}
&& \delta n_{\rm e}^{\,\prime} + n_{0} \vec{\nabla} \cdot \delta \vec{v}_{\rm e} =0,
\qquad  \delta\vec{v}_{\rm e}^{\,\prime} + {\cal H} \delta \vec{v}_{\rm e} = 
- \frac{e }{m_{\rm e} a} \biggl[\vec{e} 
+ \delta\vec{v_{\rm e}} \times \vec{B}\biggr],
\label{deltanep}\\
&& \delta n_{\rm i}^{\,\prime} + n_{0} \vec{\nabla} \cdot \delta \vec{v}_{\rm i} =0,\qquad \delta\vec{v}_{\rm i}^{\,\prime} + {\cal H} \delta \vec{v}_{\rm i} = \frac{e}{m_{\rm i}a} \biggl[ \vec{e} 
+ \delta\vec{v_{\rm i}} \times \vec{B}\biggr],
\label{deltavep}
\end{eqnarray}
can be used to solve for the propagation of the electromagnetic waves in the plasma 
by using the following pair of equations: 
\begin{equation}
\vec{\nabla} \times \vec{e} = -  \vec{b}^{\,\,\prime},\qquad \vec{\nabla}\times  \vec{b} =  
\vec{e}^{\,\,\prime}+ 4\pi\,e\,n_{0} ( \delta \vec{v}_{\rm i} - \delta \vec{v}_{\rm e}),
\label{deltadc}
\end{equation}
by recalling that the corresponding electric and magnetic fields are both divergenceless 
(i.e. $\vec{\nabla} \cdot \vec{e} = 0$ and $\vec{\nabla} \cdot \vec{b} = 0$).

Waves that are locally parallel (i.e. $\parallel$) or orthogonal (i.e. $\perp$) to the magnetic field $\vec{B}$  obey different dispersion relations. As already mentioned in connection with the plasma hierarchies, the trajectory of the charged species can be determined as a perturbation around the centre of the particle orbit \cite{guide1,guide2} by using an expansion of the magnetic random field in spatial gradients and by keeping the leading order result. In the present context the magnetic field is uniform over the typical scale of the particle orbit, which also means that the field experienced by the electron in traversing a Larmor orbit is almost constant. By taking the  Fourier and the Laplace transforms the evolution equations 
(\ref{deltadc}) can be recast in the following form
\begin{equation}
k^2 \vec{e}_{\vec{k},\overline{\omega}} - (\vec{k}\cdot\vec{e}_{\vec{k},\overline{\omega}}) \vec{k} =\epsilon(\overline{\omega},\alpha) \overline{\omega}^2 \vec{e}_{\vec{k},\overline{\omega}},\qquad \alpha = i {\mathcal H}/\overline{\omega} = H/\omega\ll 1,
\label{relation}
\end{equation}
where $\alpha$ accounts for the curved-space corrections.
The dielectric tensor $\epsilon(\overline{\omega},\alpha)$ 
can be written in a generalized matrix notation as:
\begin{equation}
\epsilon(\overline{\omega},\alpha) 
= \left(\matrix{\epsilon_{\perp 1}(\overline{\omega},\alpha)
& i\epsilon_{\perp 2}(\overline{\omega},\alpha) & 0&\cr
-i \epsilon_{\perp 2}(\overline{\omega},\alpha) & \epsilon_{\perp 1}(\overline{\omega},\alpha) &0&\cr
0&0&\epsilon_{\parallel}(\overline{\omega},\alpha) }\right),
\label{epstens}
\end{equation}
where the various entries of the matrix (\ref{epstens}) are:
\begin{eqnarray}
&& \epsilon_{\parallel}(\overline{\omega},\alpha) = 1 - \frac{ \overline{\omega}_{\rm p\,i}^2}{\overline{\omega}^2 (1 +\alpha)} - \frac{ \overline{\omega}_{\rm p\,e}^2}{\overline{\omega}^2 (1 +\alpha)},
\label{epspar}\\
&& \epsilon_{\perp 1}(\overline{\omega},\alpha) = 1 - \frac{\overline{\omega}^2_{\rm p\, i} (\alpha + 1) }{\overline{\omega}^2 (\alpha+ 1)^2 - \overline{\omega}_{\rm B\, i}^2} -
\frac{\overline{\omega}^2_{\rm p\, e} (\alpha + 1) }{\overline{\omega}^2 (\alpha+ 1)^2 - \overline{\omega}_{\rm B\, e}^2},
\label{epsperp1}\\
&& \epsilon_{\perp 2}(\omega,\alpha) = 
\frac{\overline{\omega}_{\rm B\, e}}{\overline{\omega}} \frac{\overline{\omega}^2_{\rm p\, e } }{\overline{\omega}^2 (\alpha+ 1)^2 - \overline{\omega}^2_{\rm B\, e}} 
-  \frac{\overline{\omega}_{\rm B\, i}}{\overline{\omega}} \frac{\overline{\omega}^2_{\rm p\, i } }{\overline{\omega}^2 (\alpha+ 1)^2 - \overline{\omega}^2_{\rm B\, i}}.
\label{epsperp2}
\end{eqnarray}
Introducing now $\vec{k}_{\parallel} =k \cos{\vartheta}$ and $\vec{k}_{\perp} = k \sin{\vartheta}$,
the relevant dispersion relations are readily deduced from Eq. (\ref{relation}):
\begin{equation}
2 \epsilon_{\parallel} \cos^2{\theta}[ (n^2 - \epsilon_{-}) (n^2 - \epsilon_{+})]
= \sin^{2}{\theta} (\epsilon_{\parallel} - n^2) [ n^2 (\epsilon_{+} + \epsilon_{-})
- 2 \epsilon_{+} \epsilon_{-}].
\label{dispersion}
\end{equation}
In Eq. (\ref{dispersion}), as usual, the refractive 
index\footnote{The refractive index must not be confused with the unit vector $\hat{n}$. While another potential ambiguity concerns the comoving concentrations (e. g. $n_{0}$) the different variables should be rather clear from the context because of the presence 
of specific subscripts.} is defined as $\overline{\omega}/k= 1/n$.
Equation (\ref{dispersion}) is the Appleton-Hartree dispersion relation \cite{stix,krall} and 
$\epsilon_{\pm}(\omega,\alpha) = \epsilon_{\perp 1}(\omega,\alpha) \pm \epsilon_{\perp 2}(\omega,\alpha)$ are the 
dielectric tensors in the circular basis. 

If the propagation occurs along the magnetic field direction [i.e. $\theta =0$ in Eq. (\ref{dispersion})]  the waves with positive helicity (i.e. $\hat{e}_{+}$) and negative helicity (i.e. $\hat{e}_{-}$) experience two different phase velocities 
$v_{\pm}(\omega,\alpha) = 1/n_{\pm}(\overline{\omega},\alpha)$
with $n_{\pm}(\overline{\omega},\alpha) = \sqrt{\epsilon_{\pm}(\overline{\omega},\alpha)}$.
Conversely when the propagation is orthogonal to the magnetic field direction there is an {\em ordinary}  and an {\em extraordinary} wave whose
dispersion relations are given, respectively, by
\begin{eqnarray}
&& k_{\mathrm{O}} = \overline{\omega}\sqrt{1 - \frac{\overline{\omega}_{\mathrm{pe}}^2}{\overline{\omega}^2 (1 + \alpha)}},
\label{OW}\\
&& k_{\mathrm{E}} = \overline{\omega} \sqrt{\frac{\overline{\omega}
[\overline{\omega}^2 (\alpha + 1)^2  - \overline{\omega}_{\mathrm{Be}}^2] - 2 \overline{\omega}_{\mathrm{pe}}^2 \overline{\omega}^2 (\alpha + 1) + \overline{\omega}_{\mathrm{pe}}^4}{\overline{\omega}^2 [ \overline{\omega}^2 (\alpha+1)^2 - \overline{\omega}_{\mathrm{Be}}^2 - \overline{\omega}_{\mathrm{pe}}^2 (\alpha + 1)]}}.
\label{EW}
\end{eqnarray}
Equations (\ref{OW}) and (\ref{EW}) imply, in the physical range of frequencies, that $\overline{\omega}^2 \to k^2$. The typical CMB angular frequency of Eq. (\ref{maxfreq}) is actually ${\mathcal O}(200)$ GHz  and hence much larger than the plasma and Larmor frequencies (see Eqs. (\ref{FR6})--(\ref{FR7})) which are instead in the MHz or even kHz ranges.
Furthermore, in the physical 
range of parameters  the plasma and Larmor frequencies of the electrons are always much larger than the corresponding frequencies of the ions. Thus the relevant physical regime of the dispersion relations can be summarized by the following hierarchies between the various frequencies:
\begin{equation}
\frac{\overline{\omega}_{\mathrm{pe}}}{\overline{\omega}_{\mathrm{Be}}}\gg 1, \qquad \frac{\overline{\omega}_{\mathrm{pi}}}{\overline{\omega}_{\mathrm{Bi}}} \gg 1, 
\qquad \frac{\overline{\omega}_{\mathrm{pe}}}{\overline{\omega}_{\mathrm{pi}}} \gg 1, \qquad \frac{\overline{\omega}}{\overline{\omega}_{\mathrm{pe}}} \gg 1,
\label{freq8}
\end{equation}
where, as already mentioned, $\overline{\omega}$ denotes the comoving angular frequency of the microwave background photons estimated, for instance, from the maximum of the spectral energy density (see Eq. (\ref{maxfreq})). Under the conditions expressed by Eq. (\ref{freq8}) the expressions of $\epsilon_{\pm}(\overline{\omega},\alpha)$ greatly simplifies and the result is
\begin{equation}
\epsilon_{\pm}(\overline{\omega},\alpha) = 1 - \frac{\overline{\omega}_{\mathrm{pe}}^2}{\overline{\omega}[ \overline{\omega}(\alpha + 1) \pm \overline{\omega}_{\mathrm{Be}}]}, 
\label{freq9}
\end{equation}
implying that the dispersion relations of the ordinary and extraordinary wave are $\overline{\omega}^2 \to k^2$, as it follows also in different plasmas \cite{krall}.

\subsection{Microwave background polarization and Faraday screening}
\label{sec2sub2}
For a monochromatic wave polarized along $\hat{e}_{1}$ at $\tau =0$
(e.g.  $\vec{e}(z, \tau) = E_{0} \hat{e}_{1} e^{- i (\overline{\omega}\tau - k z)}$) 
the positive and negative helicities are defined as 
$\hat{e}_{\pm} = (\hat{e}_{1} \pm i \hat{e}_{2})/\sqrt{2}$. 
Consequently the linear polarization is effectively composed by 
two circularly polarized waves, one with positive helicity 
(propagating with wavenumber $k_{+} = \sqrt{\epsilon_{+}(\overline{\omega},\alpha)} 
\,\,\overline{\omega}$) the other with negative helicity 
(propagating with wavenumber $k_{-} = \sqrt{\epsilon_{-}(\overline{\omega},\alpha)} \,\, \overline{\omega}$). 
We have therefore that after a conformal time $\tau$ the electric field of the wave will be
\begin{eqnarray}
\vec{e}(z,\tau) &=& \frac{E_{0}}{\sqrt{2}} \biggl[ \hat{e}_{+} e^{- i (\overline{\omega}\tau - k_{+} z)} + \hat{e}_{-} e^{- i (\overline{\omega}\tau - k_{+} z)}\biggr] 
\nonumber\\
&=&
\frac{E_{0}}{2} \biggl[ \hat{e}_{1} \biggl( e^{i k_{+} z} + e^{i k_{-} z}\biggr) + i \hat{e}_{2}  \biggl( e^{i k_{+} z} - e^{i k_{-} z}\biggr)\biggl]e^{- i \overline{\omega}\tau},
\label{FF2}
\end{eqnarray}
where the second equality follows from the definitions of $\hat{e}_{\pm}$. 
We can now compute easily the four Stokes parameters of the wave which are, by definition,  
\begin{eqnarray}
&& I = |\vec{e}\cdot\hat{e}_{1}|^2 + |\vec{e}\cdot\hat{e}_{2}|^2,
\qquad  V = 2 \,\mathrm{Im}[(\vec{e}\cdot\hat{e}_{1})^{*} (\vec{e} \cdot\hat{e}_{2})],
\label{IandV}\\
&& Q =  |\vec{e}\cdot\hat{e}_{1}|^2 - |\vec{e}\cdot\hat{e}_{2}|^2,
\qquad  U=  2 \,\mathrm{Re}[(\vec{e}\cdot\hat{e}_{1})^{*} (\vec{e} \cdot\hat{e}_{2})].
\label{QandU}
\end{eqnarray}
When the polarization plane of the incoming wave is 
rotated, two out of four Stokes parameter will be rotated; more specifically 
while $I$ and $V$ are left invariant, $Q$ and $U$ are rotated.
Inserting Eq. (\ref{FF2}) into Eqs. (\ref{IandV}) and (\ref{QandU}) 
 the only Faraday rotated Stokes parameters are\footnote{In the case the initial wave is polarized along $\hat{e}_{1}$ 
we have that $U^{(\mathrm{in})}=0$ and $Q^{(\mathrm{in})}=E_{0}^2$. Note that $\Phi$ has been used in section \ref{sec3} to denote one of the two Bardeen potentials. Since 
the Faraday rate and the Bardeen potential never appear in the same discussion, no confusion is possible.}:
\begin{eqnarray}
&&Q^{(\mathrm{F})} = Q^{(\mathrm{in})} \cos{( 2 \Delta \Phi)} +  U^{(\mathrm{in})} 
\sin{( 2 \Delta\Phi)},
\nonumber\\
&& U^{(\mathrm{F})} = - Q^{(\mathrm{in})} \sin{( 2 \Delta \Phi)} +  U^{(\mathrm{in})} 
\cos{( 2 \Delta \Phi)}.
\label{rot3}
\end{eqnarray}
where, by definition,
\begin{equation}
\Delta \Phi = \frac{\overline{\omega}}{2}\biggl[ \sqrt{\epsilon_{+}(\overline{\omega}, \alpha)} - 
\sqrt{\epsilon_{-}(\overline{\omega}, \alpha)}\biggr] \Delta z.
\label{rot2}
\end{equation}
The rate of rotation per unit time is called Faraday rotation rate  and it is given by:
\begin{eqnarray}
{\mathcal F}(\hat{n}) &=& \frac{d\Phi}{d\tau} = \frac{\overline{\omega}}{2}\biggl[\sqrt{\epsilon_{+}(\overline{\omega},\alpha)} -\sqrt{\epsilon_{-}(\overline{\omega},\alpha)}\biggr]
\nonumber\\
&=&\frac{\overline{\omega}_{\mathrm{Be}}}{2} \biggl(\frac{\overline{\omega}_{\mathrm{pe}}}{\overline{\omega}}\biggl)^2 \equiv  \frac{e^3}{2\pi m_{\mathrm{e}}^2} a \tilde{n}_{\mathrm{e}} x_{\mathrm{e}} \biggl(\frac{\vec{B}\cdot \hat{n}}{\overline{\nu}^2}\biggr),
\nonumber\\
\mathrm{F}(\hat{n}) &=& \frac{{\mathcal F}(\hat{n})}{\epsilon^{\prime}} = \frac{ 3}{16 \pi^2 e} \frac{\hat{n} \cdot \vec{B}}{\overline{\nu}^2}. 
\label{varphi}
\end{eqnarray}
The first line of Eq. (\ref{varphi}) is  the definition of Faraday rotation rate; the second line of Eq. (\ref{varphi}) follows by recalling the hierarchies of  Eq. (\ref{freq8}) i.e.  $|\overline{\omega}/\overline{\omega}_{\mathrm{pe}}|\gg 1$ and for $ |\overline{\omega}/\overline{\omega}_{\mathrm{Be}}|\gg 1$.  Finally, the third line of Eq. (\ref{varphi}) follows by taking into account the definition of the differential optical depth, (i.e. $\epsilon' = a \tilde{n}_{\mathrm{e}} x_{\mathrm{e}}\sigma_{\mathrm{e}\gamma}$) and by making the cross section of Eq. (\ref{diffop}) explicit.

In the conventional lore the polarization is only generated very near the surface 
of last scattering as the photons begin to decouple from the electrons and generate a quadrupole 
moment through free-streaming \cite{far1,far2,far3}. According to this perspective 
the polarization of the microwave background is first generated and then it is rotated by a Faraday screen.
From Eqs. (\ref{rot3}) and (\ref{varphi}) we can take the total time derivative supposing that the 
initial polarization is independent on time. From the Faraday rotation rate we then get: 
\begin{equation}
\Delta_{\mathrm{Q}}' +  n^{i} \partial_{i} \Delta_{\mathrm{Q}} = 2 \epsilon' F(\hat{n}) \Delta_{\mathrm{U}},\qquad \Delta_{\mathrm{U}}' + n^{i} \partial_{i} \Delta_{\mathrm{U}} = -2 \epsilon' F(\hat{n}) \Delta_{\mathrm{Q}}.
 \label{mix2}
\end{equation}
Even if the  polarization and its rotation are two concurrent phenomena, the approximation
of the Faraday screening is considered satisfactory as long as the Faraday rotation rate 
is sufficiently small. Equations  (\ref{mix2}) describe the Faraday rotation mixing
which can be computed within two complementary strategies. 
In the first approach the $E$-mode polarization is computed as if the magnetic fields 
were absent from the initial conditions of the Einstein-Boltzmann hierarchy (see e.g. \cite{far1,far2,far3,far4, far7,far10,far10a}). In a more realistic strategy the $E$-mode polarization is computed by taking into account the appropriate magnetized initial conditions \cite{far9}. While within the first approach the calculation just assumes the conventional adiabatic mode of the concordance paradigm, in the second approach the magnetic random fields not only affect the 
Faraday rotation rate but also the $E$-mode power spectrum.
The total angle of rotation at some reference time $\tau= \tau_{*}$ is defined in terms of the visibility 
function of Eq. (\ref{visdef}):
\begin{equation}
\Phi(\hat{n},\tau_{*}) = 
{\mathcal A}(\overline{\nu}) \int_{0}^{\tau_{*}} {\mathcal K}(\tau) \,[\hat{n} \cdot 
\vec{B}(\vec{x},\tau)]\,\, d\tau,\qquad {\mathcal K}(\tau) = \epsilon'(\tau) e^{- \epsilon(\tau,\tau_{*})}.
\label{rot1a}
\end{equation}
where ${\mathcal A}(\overline{\nu}) = 3 /(16 \pi^2 e \overline{\nu}^{2})$ depends upon the comoving frequency $\overline{\nu}$.  Given an explicit form of the visibility function (see e.g. Eqs. (\ref{rot6})) the total rotation rate can be approximately evaluated \cite{far6,far9}. In the sudden approximation the visibility function is sharply peaked around the decoupling time with practically zero width so that, in this limit, $\Phi(\hat{n},\tau_{*})\to F(\hat{n},\tau_{*})$. 

\subsection{Induced $B$-mode polarization}
The $B$-mode polarization induced by the Faraday rate is more easily 
estimated in real space; the angular power spectrum is obtained 
after a simple but lengthy algebra. The starting point 
of the derivation are the two orthogonal combinations of the brightness perturbations 
already introduced in Eq. (\ref{int2a}), namely
 $\Delta_{\pm}(\hat{n},\tau) = \Delta_{\mathrm{Q}}(\hat{n},\tau) \pm i \Delta_{\mathrm{U}}(\hat{n},\tau)$; in terms of these 
quantities the $E$-mode and the $B$-mode polarizations are given by: 
\begin{eqnarray}
\Delta_{\mathrm{E}} (\hat{n},\tau) &=& - \frac{1}{2}\biggl\{( 1 -\mu^2) \partial_{\mu}^2 (\Delta_{+} + \Delta_{-}) - 4 \mu  \partial_{\mu}(\Delta_{+} + \Delta_{-}) - 2  (\Delta_{+} + \Delta_{-}) 
\nonumber\\
&-& 
\frac{\partial_{\varphi}^2 (\Delta_{+} + \Delta_{-})}{1 - \mu^2 } 
+
2 i\biggl[ \partial_{\varphi}  \partial_{\mu}(\Delta_{+} - \Delta_{-}) - \frac{\mu}{1 - \mu^2} \partial_{\varphi}  (\Delta_{+} - \Delta_{-})\biggr] \biggr\},
\label{BF2}\\
 \Delta_{\mathrm{B}} (\hat{n},\tau) &=& \frac{i}{2} \biggl\{( 1 -\mu^2) \partial_{\mu}^2(\Delta_{+} - \Delta_{-}) - 4 \mu  \partial_{\mu}(\Delta_{+} - \Delta_{-}) - 2  (\Delta_{+} - \Delta_{-})  
\nonumber\\
&-& 
\frac{\partial_{\varphi}^2  (\Delta_{+} - \Delta_{-})}{1 - \mu^2 }
+ 
2 i\biggl[ \partial_{\varphi}  \partial_{\mu}(\Delta_{+} + \Delta_{-}) - \frac{\mu}{1 - \mu^2} \partial_{\varphi}  (\Delta_{+} + \Delta_{-})\biggr] \biggr\},
\label{BF3}
\end{eqnarray}
where, as usual,  $\partial_{\mu}$ denotes a derivation 
with respect to $\mu = \cos{\vartheta}$ while $\partial_{\varphi}$ denotes a derivation 
with respect to $\varphi$. According to Eqs. (\ref{BF2}) and (\ref{BF3}) the initial B-mode polarization is absent 
(i.e. $\Delta_{\mathrm{B}}( \hat{n},\tau) =0$) provided 
$\Delta_{+}(\hat{n},\tau)= \Delta_{-}(\hat{n},\tau)$ and when the brightness perturbation does not depend on $\varphi$ 
[i.e. $\partial_{\varphi}( \Delta_{+} + \Delta_{-}) =0$]. Once $\Delta_{\mathrm{E}}(\hat{n},\tau) $ and $\Delta_{\mathrm{B}}(\hat{n},\tau) $ are determined, it is sufficient to recall that:
\begin{equation}
a^{(\mathrm{E})}_{\ell\, m} =   N_{\ell} \int d\hat{n} \,\Delta_{\mathrm{E}}(\hat{n},\tau) \, Y^{*}_{\ell\, m}(\hat{n}),\qquad 
a^{(\mathrm{B})}_{\ell\, m} =   N_{\ell} \int d\hat{n} \,\Delta_{\mathrm{B}}(\hat{n},\tau) \, Y^{*}_{\ell\, m}(\hat{n}),
\label{aEaB}
\end{equation}
where $N_{\ell} = \sqrt{(\ell - 2)!/(\ell +2)!}$. In terms of Eq. (\ref{aEaB})
 the $EE$ and $BB$ angular power spectra can be easily computed from their definition:
 \begin{equation}
C_{\ell}^{(\mathrm{EE})} = \frac{1}{2\ell + 1} \sum_{m = -\ell}^{\ell} 
\langle a^{(\mathrm{E})*}_{\ell m}\,a^{(\mathrm{E})}_{\ell m}\rangle,\qquad 
C_{\ell}^{(\mathrm{BB})} = \frac{1}{2\ell + 1} \sum_{m=-\ell}^{\ell} 
\langle a^{(\mathrm{B})*}_{\ell m}\,a^{(\mathrm{B})}_{\ell m}\rangle,
\label{int5}
\end{equation}
where, as usual,  $\langle ...\rangle$ denotes the ensemble average. If the Faraday rate operates after the linear polarization 
is effectively produced, to lowest order in the 
rate the $E$-mode and the $B$-mode in real space are, respectively,
\begin{eqnarray}
\Delta_{\mathrm{E}}(\hat{n},\tau) &=& - \partial_{\mu}^2[(1 - \mu^2) \Delta_{\mathrm{P}}(\hat{n},\tau)], 
\label{BF7}\\
\Delta_{\mathrm{B}}(\hat{n},\tau) &=& 2 \, \partial_{\mu}^2[ (1 - \mu^2)\,\Phi(\vec{x},\hat{n},\tau)\,
 \Delta_{\mathrm{P}}(\hat{n},\tau)].
\label{BF8}
\end{eqnarray}
The result of Eq. (\ref{BF8}) can then be expressed $\ell$-space (see second paper of Ref. \cite{far9}); the result obtained in this 
way coincide other derivations \cite{far5,far6,far7,far8}. Denoting with $C_{\ell}^{(\mathrm{FF})}$ the angular power spectrum of Faraday rotation \cite{far5,far6,far7,far8,far9}
the final result for the $B$-mode polarization is
\begin{equation}
{\mathcal C}_{\ell}^{(\mathrm{BB})} = \sum_{\ell_{1},\,\,\ell_{2}}  {\mathcal Z}(\ell, \ell_1, \ell_{2}) C_{\ell_{2}}^{(\mathrm{EE})} C_{\ell_{1}}^{(\mathrm{F})}
\label{F14a}
\end{equation}
where ${\mathcal Z}(\ell, \ell_1, \ell_{2})$ is a function of the multipole moments containing a Clebsch-Gordon coefficient (see Eqs. (\ref{F13aa}) and (\ref{CBBF})); in Eq. (\ref{F14a}) $C_{\ell_{2}}^{(\mathrm{EE})}$ denotes, as usual, the angular power spectrum 
of the (magnetized) $E$-mode autocorrelation. While the explicit expression for ${\mathcal Z}(\ell, \ell_{1},\ell_{2})$ is reported in Eq. 
(\ref{CBBF}),  the sum of Eq. (\ref{F14a}) must be conducted in compliance with the constraints stemming from the triangle inequality $|\ell_{1} - \ell_{2}| \leq \ell \leq \ell_{1}+ \ell_{2}$. 

Recalling the results (\ref{F1})--(\ref{F10}) and using the shorthand notation $F(\hat{n},\tau) = {\mathcal A}(\overline{\nu}) \vec{B}\cdot \hat{n}$, the 
two-point function of the Faraday rate is given by:
\begin{equation}
\langle F(\hat{n}_{1},\tau) F(\hat{n}_{2},\tau)\rangle = \frac{1}{4\pi}\sum_{\ell} (2 \ell + 1)C_{\ell}^{(\mathrm{FF})} P_{\ell}(\hat{n}_{1} \cdot
\hat{n}_{2}),
\label{APSfar}
\end{equation}
where the angular power spectrum of Faraday rotation is
\begin{equation}
C_{\ell}^{(\mathrm{FF})} = 4\pi {\mathcal A}^2(\overline{\nu}) \ell (\ell +1) \int \frac{d k}{k} P_{B}(k,\tau) \frac{j_{\ell}^2(k\tau_{0})}{k^2 \tau_{0}^2}.
\label{F13a}
\end{equation}
Note that $\tau_{0}$ has been already introduced when discussing the line of sight solutions of the brightness perturbations (see e.g. Eq. (\ref{SW3})) and it comes by approximating 
$i\mu x(\tau) \simeq i k \mu (\tau_{0} - \tau_{\mathrm{rec}}) \simeq  i k \mu \tau_{0}$. When the magnetic 
power spectrum is a power-law, Eq. (\ref{F13a}) can be analytically integrated and the final result is \cite{far9}
 \begin{eqnarray}
C_{\ell}^{(\mathrm{FF})} &=& \overline{C}^{(\mathrm{FF})} \,\, \ell(\ell + 1)\, \, {\mathcal I}_{\ell}(n_{\mathrm{B}},k_{d}),
\label{CCF1}\\
\overline{C}^{(\mathrm{FF})}  &=& 1.015\times 10^{-5}\,\,\bigg(\frac{B_{\mathrm{L}}}{\mathrm{nG}}\biggr)^2 \biggl(\frac{\overline{\nu}}{\overline{\nu}^{(\rho)}_{\mathrm{max}}}\biggr)^{-4}\,
\biggl(\frac{k_{0}}{k_{\mathrm{L}}}\biggr)^{n_{\mathrm{B}} -1} \, \frac{(2\pi)^{n_{\mathrm{B}}-1}}{\Gamma[(n_{\mathrm{B}}-1)/2]},
\label{CCF2}\\
{\mathcal I}_{\ell}(n_{\mathrm{B}},k_{d}) &=& 
\int_{0}^{\infty} z^{n_{\mathrm{B}} -4} j_{\ell}^2(z) e^{ - 2 (z/x_{\mathrm{d}})^2}\,\, d z,
\label{CCF3}
\end{eqnarray}
where $x_{\mathrm{d}}= k_{\mathrm{d}} \tau_{0}$
and $k_{\mathrm{d}}^{-2} =( k_{\mathrm{D}}^{-2} + k_{\sigma}^{-2} + k_{\mathrm{t}}^{-2})$;
note that $k_{\mathrm{D}}$ and $k_{\sigma}$ parametrize, respectively, 
the effects of the thermal and magnetic diffusivities while $k_{\mathrm{t}}$ is of the order 
of $\tau_{\mathrm{rec}}^{-1}$ and parametrizes the finite thickness effects of the last scattering 
surface. The results of the integral of Eq. (\ref{CCF3}) are 
expressible  generalized hypergeometric functions since  $j_{\ell}(z)$ are the usual spherical Bessel functions \cite{abr1}. 

It is possible to estimate analytically Eq. (\ref{F14a}) at small angular scales (i.e.  $\ell_{1}\gg 1$, $\ell_{2} \gg 1$ and $\ell \gg 1$) where the Clebsch-Gordon coefficient appearing inside
 ${\mathcal Z}(\ell, \ell_1, \ell_{2})$ can be evaluated in analogy with the semiclassical limit in non relativistic quantum mechanics. This approach to the asymptotics of the Clebsch-Gordon coefficients was originally studied in Ref. \cite{far11} by exploiting the connection of the Clebsch-Gordon coefficients with the Wigner $3j$ and $6j$ symbols (see also \cite{far12}).
This analytical technique  has been exploited in \cite{far9a} (see also the last paper of Ref. \cite{far9}) 
for the explicit estimates of ${\mathcal Z}(\ell, \ell_1, \ell_{2})$.
\begin{figure}[!ht]
\centering
\includegraphics[height=6.7cm]{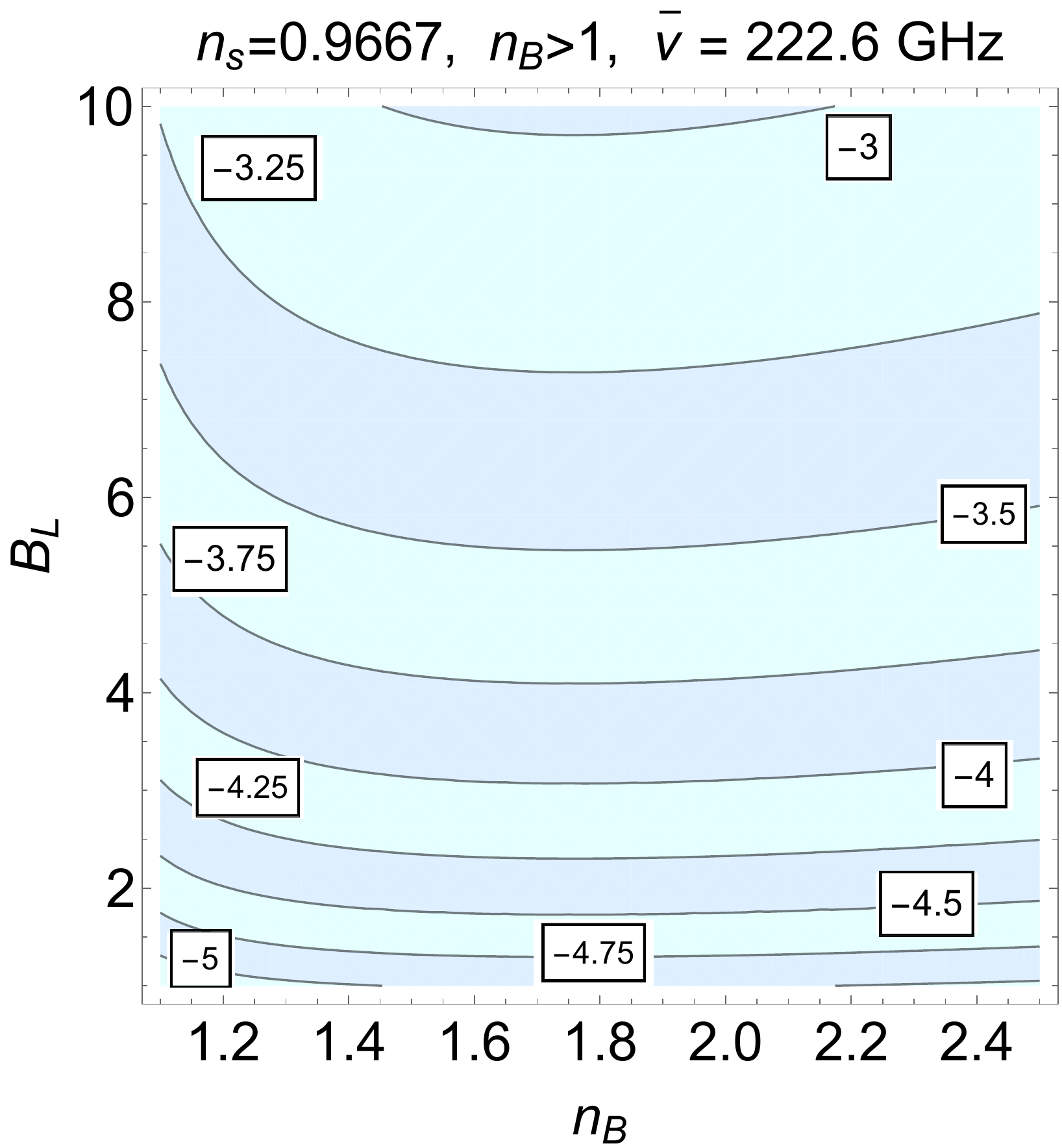}
\includegraphics[height=6.7cm]{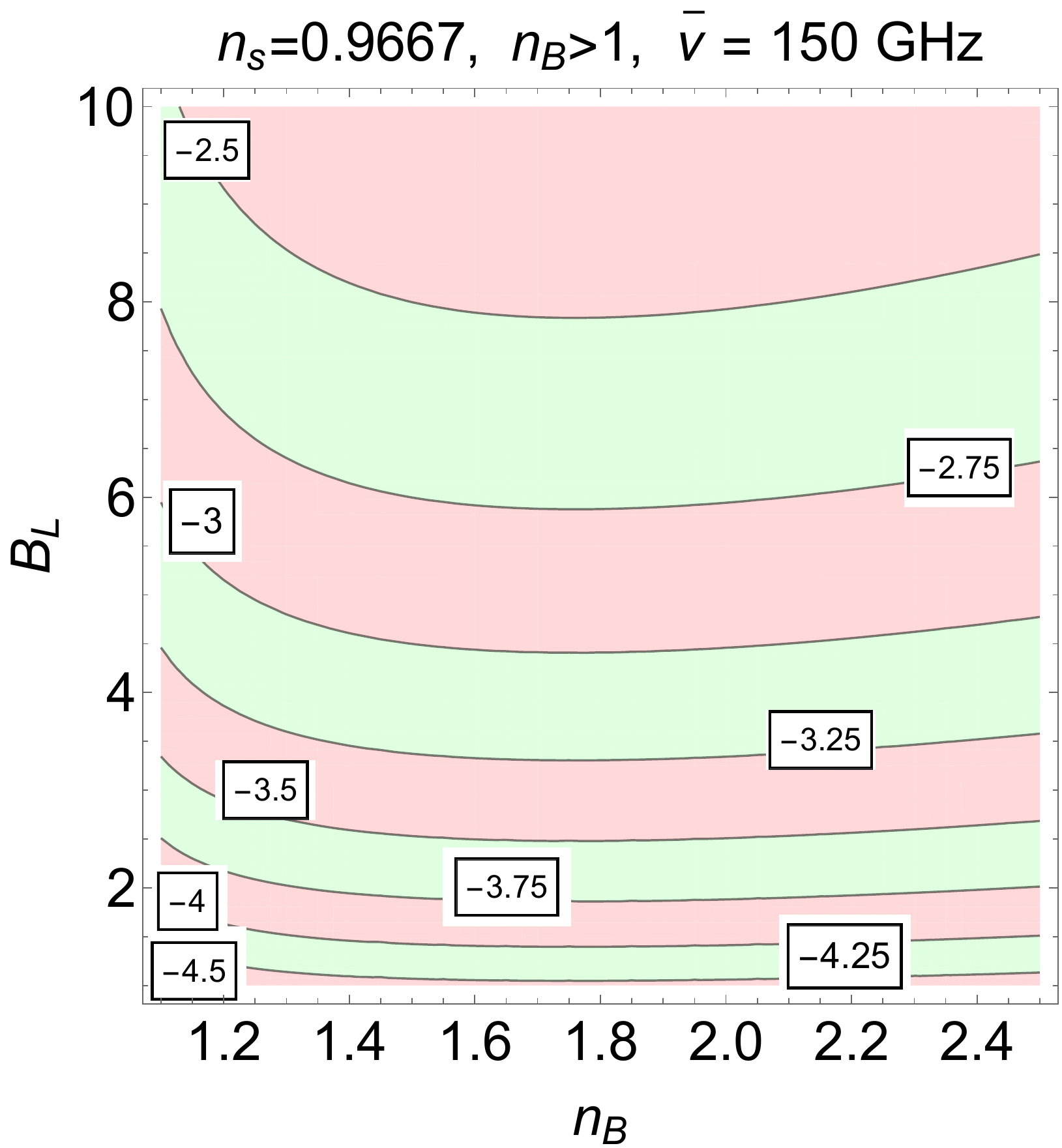}
\caption[a]{The angular power spectrum of the $B$-mode autocorrelation induced by the Faraday effect 
is illustrated for two different frequencies (i.e. $222.6$ GHz and $150$ GHz). The two plots refer to the case of {\em blue} spectra of the magnetic random field. The different labels in both plots denote the common logarithm of $\ell(\ell +1) C_{\ell}^{(\mathrm{BB})}/(2\pi)$ in units of $(\mu \mathrm{K})^2$ and for a $\ell = 1000$ which 
roughly corresponds to the maximum of the $E$-mode autocorrelation entering directly the expression of Eq. (\ref{F14a}). Note that, on the vertical axes, $B_{L}$ is measured in nG.}
\label{figureBM1}      
\end{figure}

\subsection{Orders of magnitudes of the $B$-mode autocorrelations}

\begin{figure}[!ht]
\centering
\includegraphics[height=6.7cm]{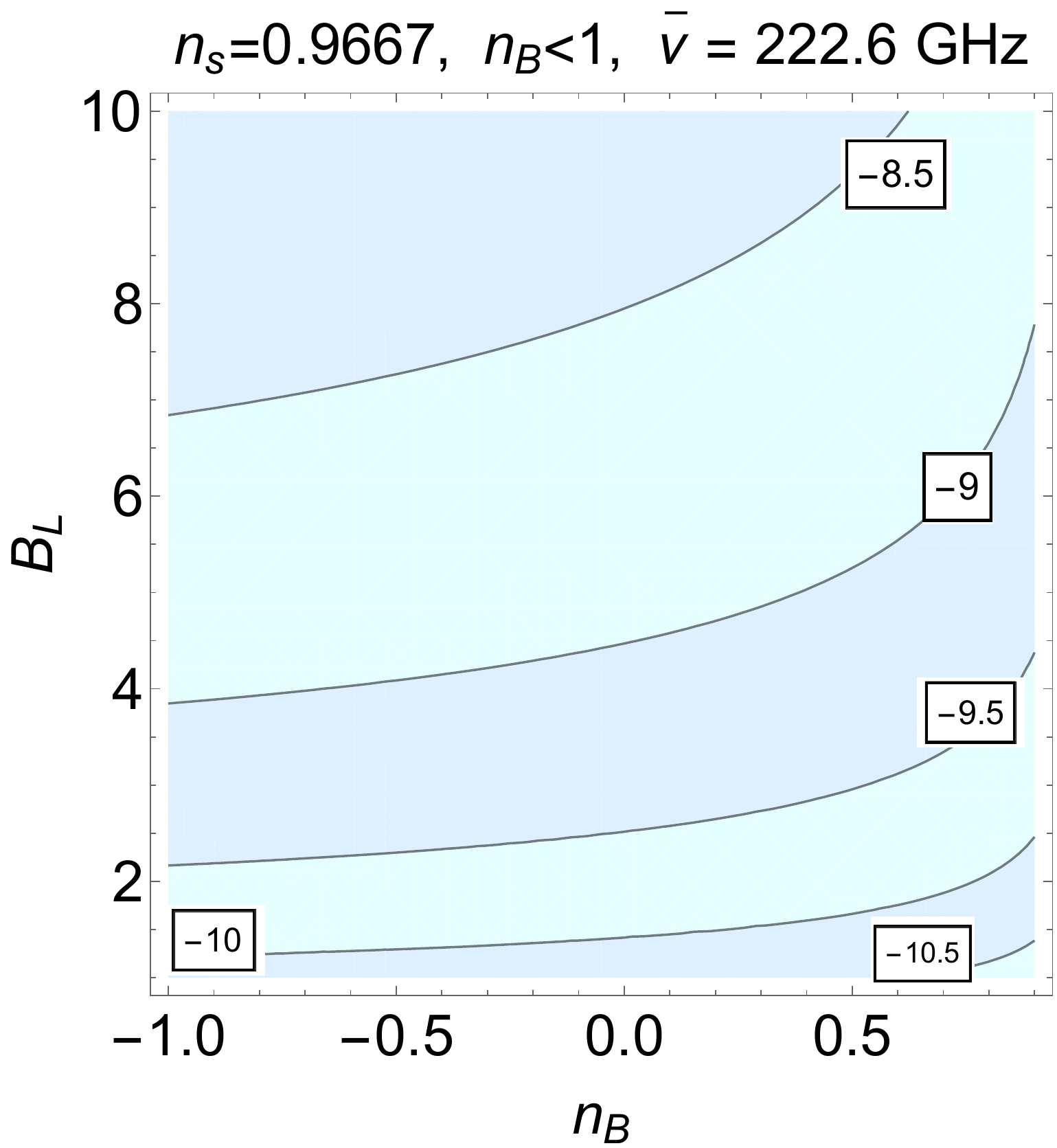}
\includegraphics[height=6.7cm]{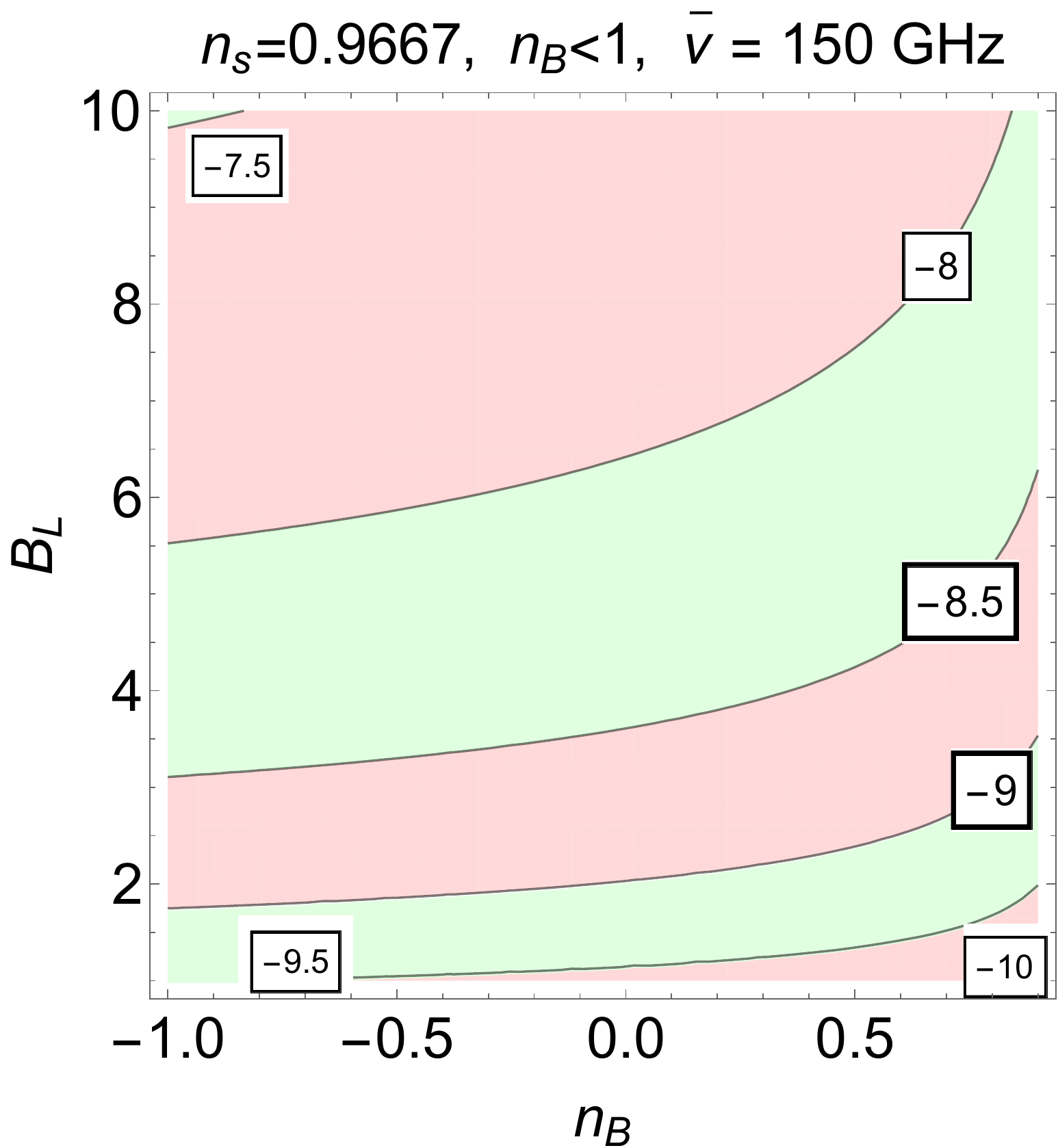}
\caption[a]{The angular power spectrum of the $B$-mode autocorrelation induced by the Faraday effect 
is illustrated for two different frequencies (i.e. $222.6$ GHz and $150$ GHz) and in the case of {\em red} spectra of the magnetic random field. As in the case of Fig. \ref{figureBM2}, the different labels in both plots denote the common logarithm of $\ell(\ell +1) C_{\ell}^{(\mathrm{BB})}/(2\pi)$ in units of $(\mu \mathrm{K})^2$ and for a $\ell = 1000$. As in Fig. \ref{figureBM1}, $B_{L}$ is measured in nG.}
\label{figureBM2}      
\end{figure}

To estimate the $B$-mode autocorrelation we can use 
Eq. (\ref{F14a}) and then evaluate the obtained result for $\ell = {\mathcal O}(1000)$, roughly corresponding to the maximum of the $E$-mode autocorrelation. By following this strategy, the common logarithm of the $B$-mode autocorrelation 
induced by the Faraday effect is illustrated in Fig. \ref{figureBM1}  (in the case of blue spectral indices, i.e.  $n_{\mathrm{B}} > 1$) and in Fig. \ref{figureBM2} (for red spectral indices, i.e.  $n_{\mathrm{B}} <1$). In both figures the left plots correspond to a frequency channel coinciding with the maximum of the energy density of the microwave background radiation discussed in Eqs. (\ref{CMBmax1}) and (\ref{maxfreq}). Conversely, in the right plots of Figs. \ref{figureBM1} and \ref{figureBM2},
the frequency is instead $150$ GHz coinciding, incidentally, with the operating window of the Bicep2 experiment \cite{chone43}. The various labels in the plots report the common logarithm of the $BB$ power spectrum computed from Eq. (\ref{F14a}) and evaluated for $\ell = 1000$. 

\begin{figure}[!ht]
\centering
\includegraphics[height=6.5cm]{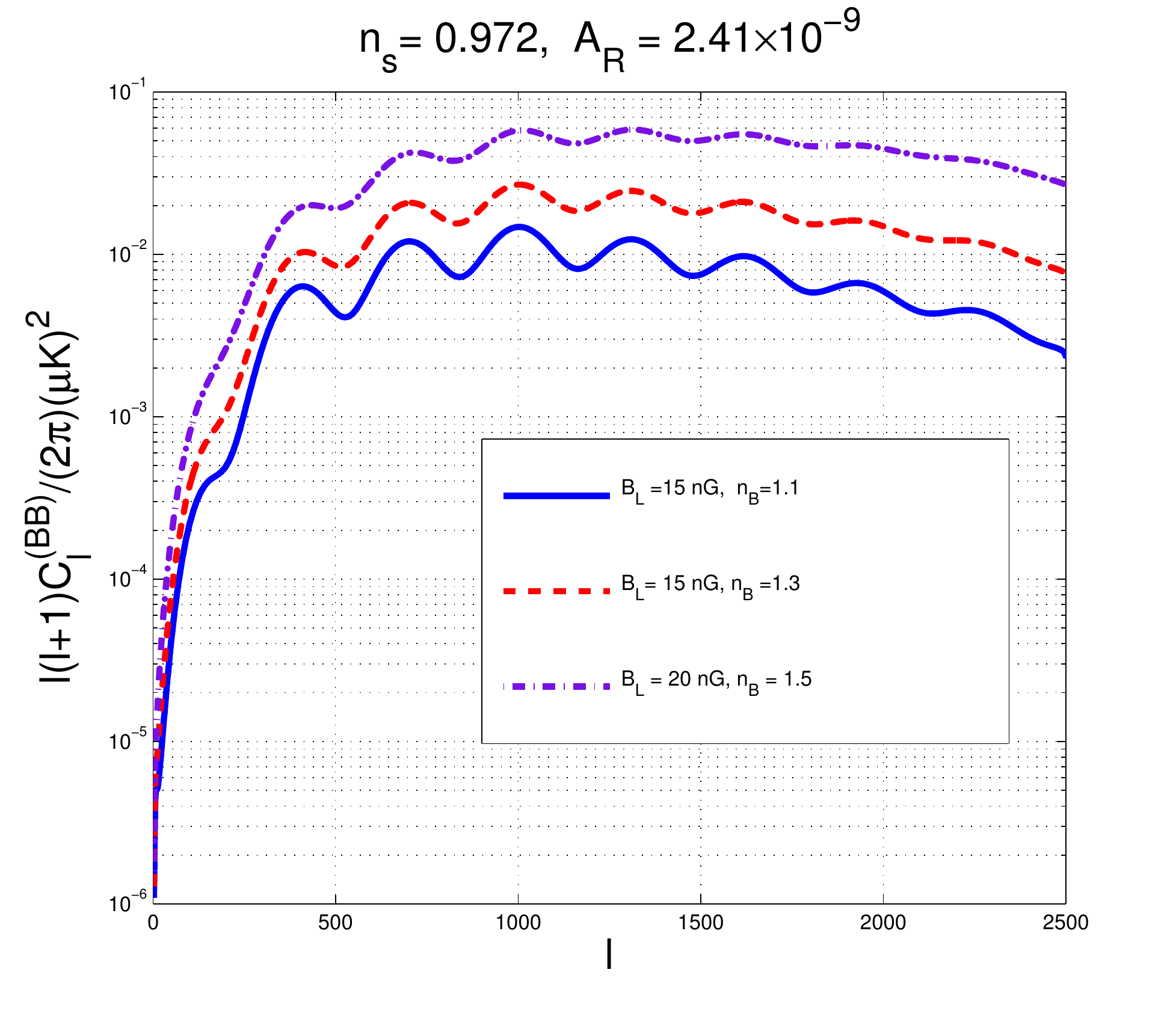}
\includegraphics[height=6.5cm]{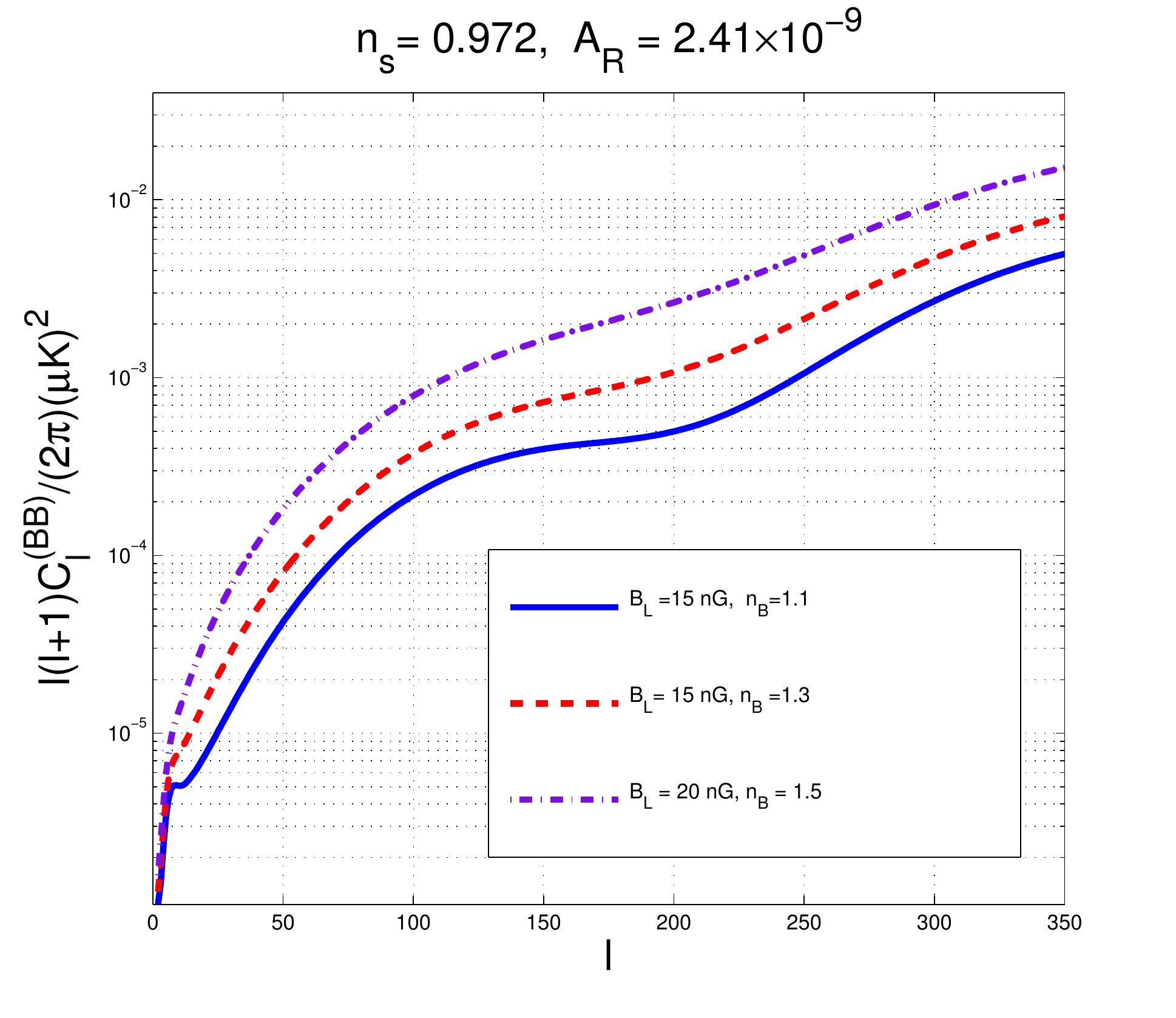}
\caption[a]{The $B$ mode polarization induced by the Faraday effect  in the magnetized  
$\Lambda$CDM scenario with no tensors and with different choices of magnetic field parameters. The plot
on the right illustrates the large angular scales. The axes are semilogarithmic and the frequency is $150$ GHz. The plots are adapted from Ref. \cite{far9a}.}
\label{figure8}      
\end{figure}

\begin{figure}[!ht]
\centering
\includegraphics[height=6.5cm]{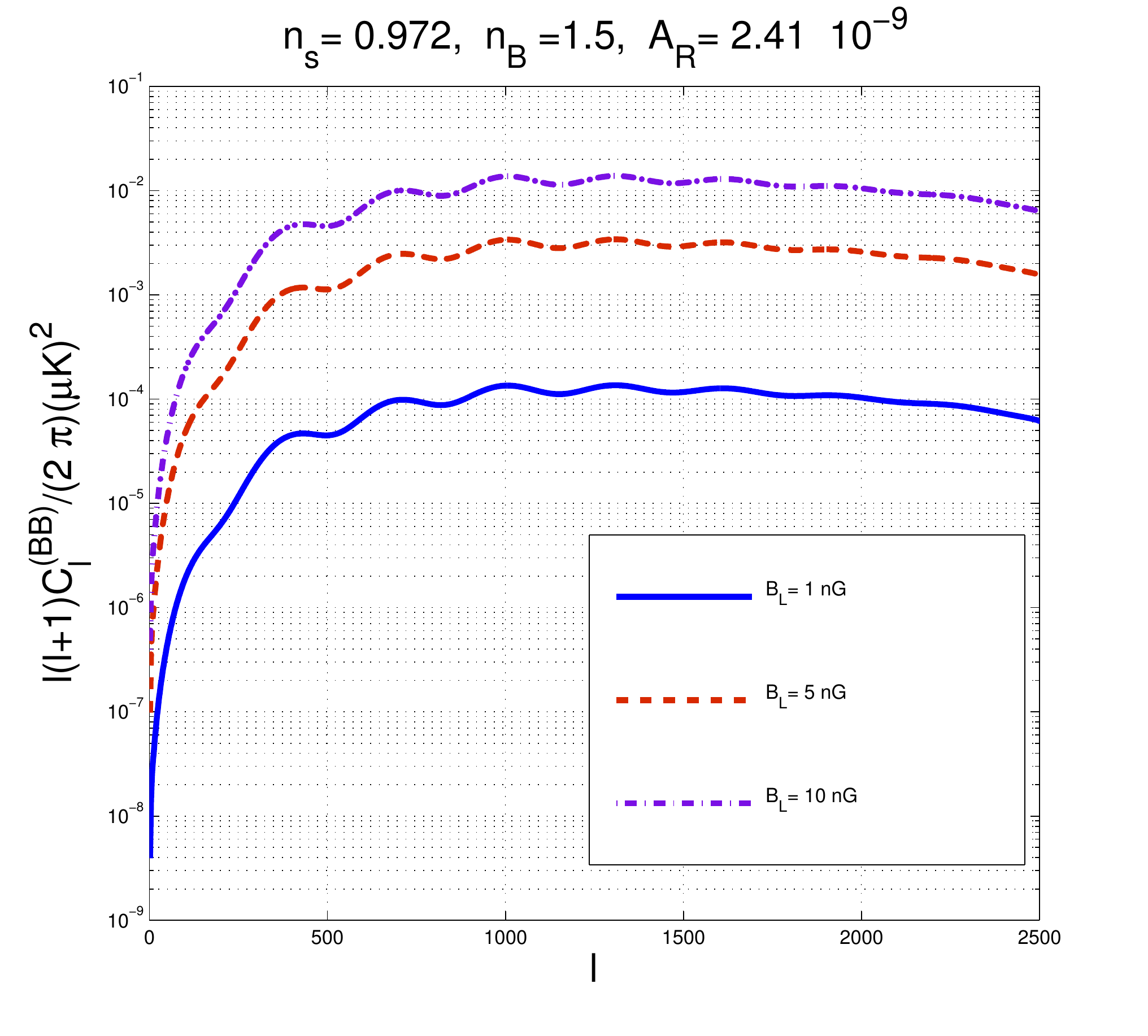}
\includegraphics[height=6.5cm]{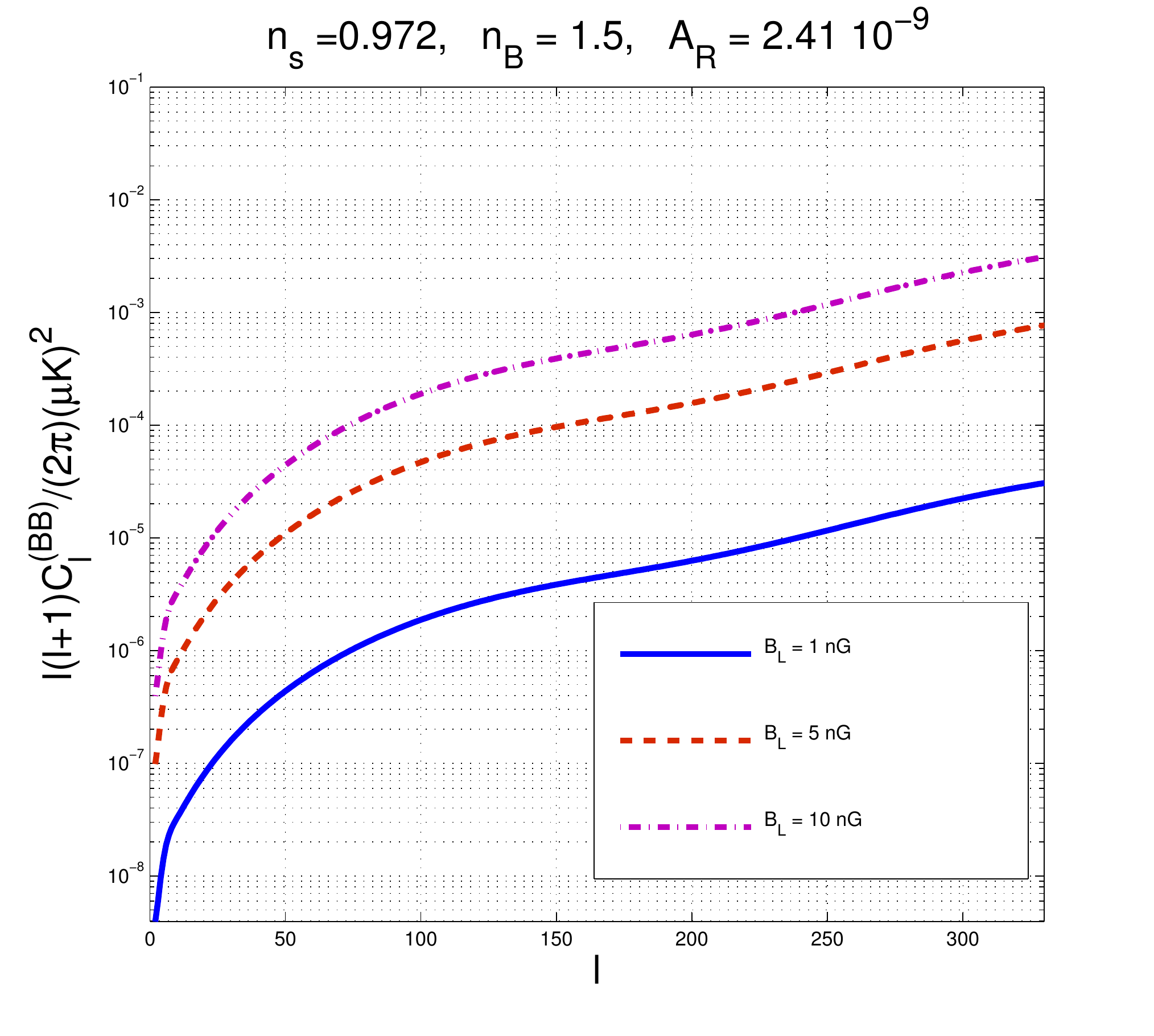}
\caption[a]{The $B$ mode polarization induced by the Faraday effect  in the magnetized  
$\Lambda$CDM. In both plots the magnetic spectral index is $n_{\mathrm{B}} = 1.5$. The plot
on the right describes large angular scales. The axes are semilogarithmic in both plots.}
\label{figure9}      
\end{figure}

\begin{figure}[!ht]
\centering
\includegraphics[height=6.5cm]{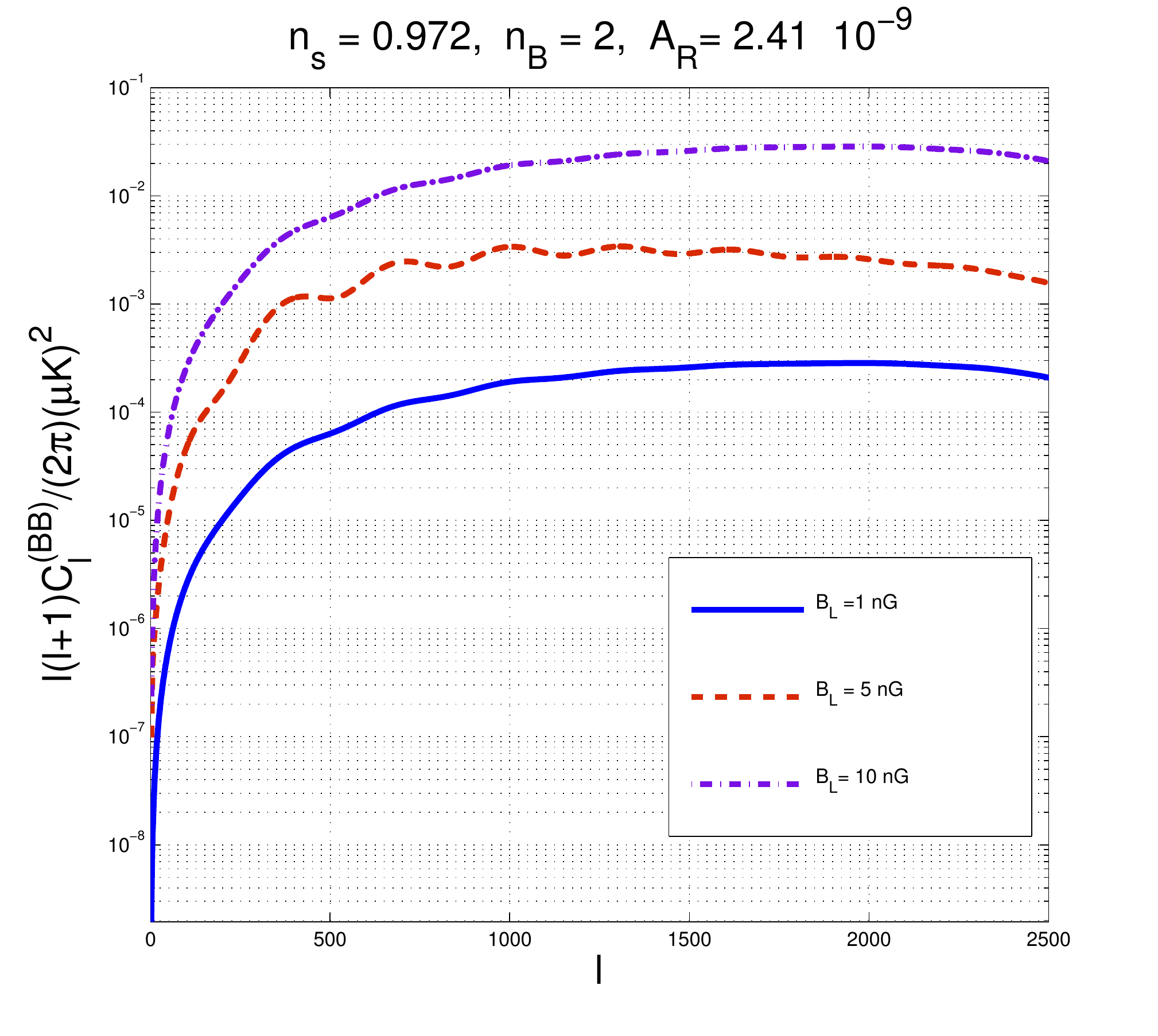}
\includegraphics[height=6.5cm]{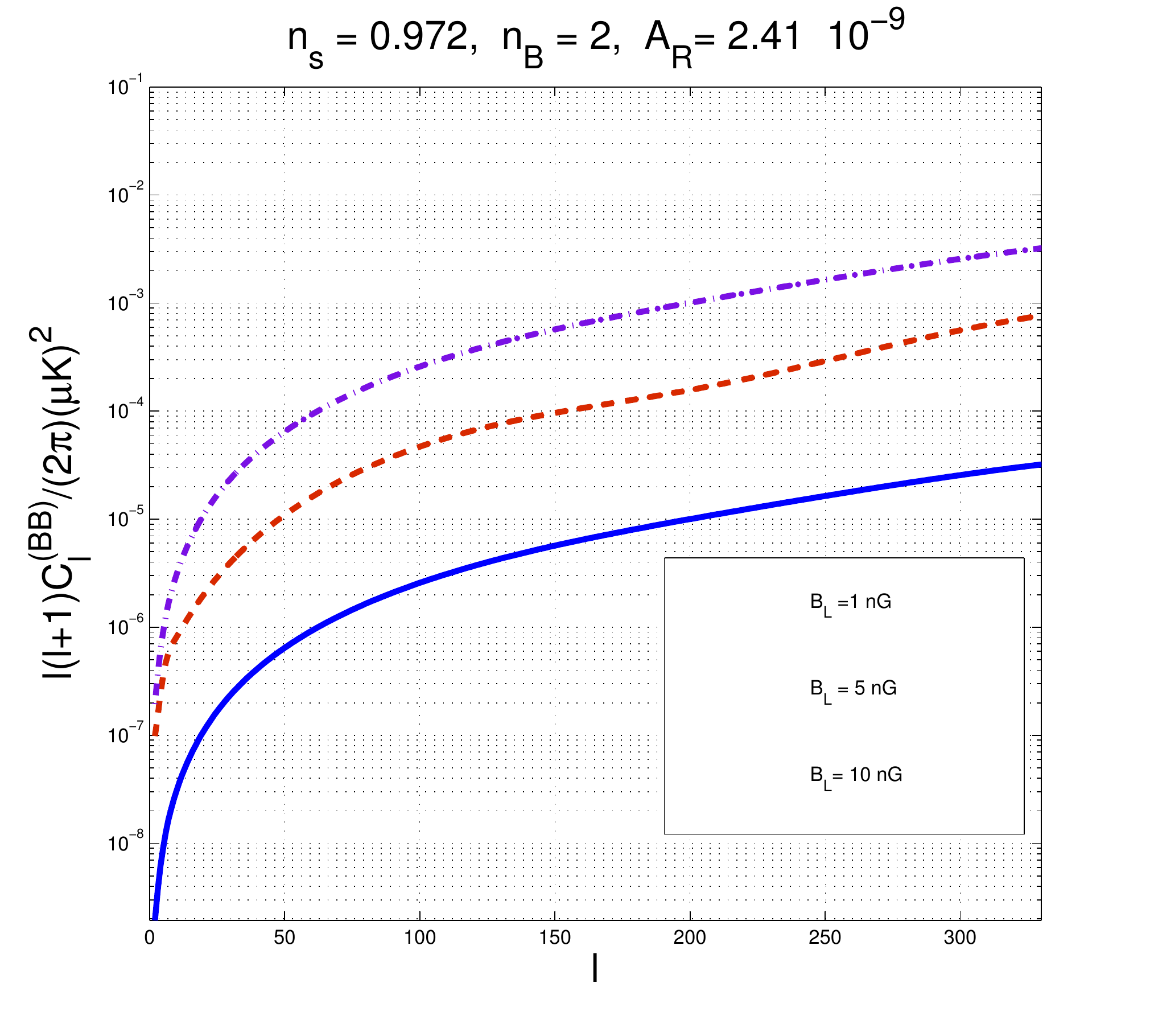}
\caption[a]{The $B$ mode polarization induced by the Faraday effect  in the magnetized  
$\Lambda$CDM. In both plots the magnetic spectral index is $n_{\mathrm{B}} = 2$. The plot on the right describes large angular scales. The axes are semilogarithmic in both plots.}
\label{figure10}      
\end{figure}

The  orders of magnitude of the $BB$ correlations of Fig. \ref{figureBM1} can be qualitatively understood in rather simple terms. 
Assuming that $\overline{{\mathcal G}}^{(\mathrm{EE})}_{\ell} = \ell(\ell + 1) C_{\ell}^{(EE)}/(2 \pi)$ is well estimated by the measured angular power spectrum, we can (approximately) calculate the $E$-mode autocorrelation as:
\begin{equation}
\overline{{\mathcal G}}^{(\mathrm{EE})}_{\ell} \simeq 50\,\mu\mathrm{K}^2, \qquad \mathrm{for} \,\, \ell \simeq \ell_{\mathrm{max}} = 1000.
\label{estBB1}
\end{equation}
The actual value of the maximum of $\overline{{\mathcal G}}^{(\mathrm{EE})}_{\ell}$ is slightly overestimated but Eq. (\ref{estBB1}) is purposely generous with the aim of establishing an order of magnitude estimate valid for $\ell = {\mathcal O}(\ell_{\mathrm{max}})$. From the analysis of Ref. \cite{far9a} and from the 
explicit analytic expression of Eqs. (\ref{F14a}) and (\ref{F13a}), the order of magnitude of the $B$-mode autocorrelation is then given by: 
\begin{equation}
{\mathcal G}_{\ell}^{(\mathrm{BB})}  \simeq 4.9 \times 10^{-4} \times \biggl(\frac{B_{\mathrm{L}}}{\mathrm{nG}}\biggr)^2 \biggl(\frac{150\, \mathrm{GHz}}{\overline{\nu}}\biggr)^{4} \,\mu\mathrm{K}^2\,\,\,n_{\mathrm{B}} >1.
\label{estBB2}
\end{equation}
For smaller multipoles we have that ${\mathcal G}^{(\mathrm{EE})}_{\ell}< 5 \,\mu \mathrm{K}^2 $. 
This means that Eq. (\ref{estBB2}) should be corrected by a factor ${\mathcal O}(10^{-4})$; in other words 
for $\ell < 100$ we will have that ${\mathcal G}_{\ell}^{(\mathrm{BB})}  < 10^{-5} (B_{\mathrm{L}}/\mathrm{nG})^2 
[\overline{\nu}/( 150 \mathrm{GHz})]^{-4}\,\, \mu\mathrm{K}^2$. This point can be appreciated by looking at  Figs. \ref{figure8}, \ref{figure9} and \ref{figure10} where the explicit angular power spectra have been illustrated by adapting the results of Ref. \cite{far9a}.

The benchmark frequency of Eq. (\ref{estBB2}) is close to the maximum of the 
brightness of the CMB spectrum and it coincides, for immediate convenience, 
with the Bicep2 \cite{chone43} operating frequency. 
As we saw in section \ref{sec4} the value of $1$ nG for the magnetic field intensity is 
barely compatible with the distortions produced by a magnetic random field on the 
measured temperature and polarization anisotropies: 
this is the reason why we used the nG strength as a reference value in Eq. (\ref{estBB2}).
When the magnetic power spectrum is red (i.e. $n_{\mathrm{B}} <1$) the values of the $B$-mode autocorrelation are ${\mathcal O}(10^{-4})$ smaller than in the case of blue spectra (see Fig. \ref{figureBM2}); they can be 
qualitatively captured by the analog of Eq. (\ref{estBB2}):
\begin{equation}
{\mathcal G}_{\ell}^{(\mathrm{BB})}  \simeq 5 \times 10^{-8} \times \biggl(\frac{B_{\mathrm{L}}}{\mathrm{nG}}\biggr)^2 \biggl(\frac{150\, \mathrm{GHz}}{\overline{\nu}}\biggr)^{4} \,\mu\mathrm{K}^2\,\,\,n_{\mathrm{B}} <1.
\label{estBB2a}
\end{equation}
Even Eqs. (\ref{estBB2}) and (\ref{estBB2a}) 
can only be correct within an order of magnitude, they are 
useful together with Figs. \ref{figureBM1} and \ref{figureBM2}
for swift qualitative estimates. For instance the Bicep2 measurement claimed a $B$-mode detection \cite{chone43}
which quickly turned out to be a foreground effect.  Even if confirmed,
the Bicep2 measurement (i.e. ${\mathcal G}_{\ell}^{(\mathrm{BB})} \simeq (5.07 \pm 1.13)\times 10^{-2}\, \,\mu\mathrm{K}^2$ for $\ell \simeq 248$) could not have been the effect of Faraday rotation \cite{far9a}. This 
conclusion follows from a qualitative analysis of Eq. (\ref{estBB2}): the purported Bicep2 observation could have been only reproduced for $B_{\mathrm{L}} = {\mathcal O}(10)$ nG which is forbidden by the analysis of the magnetized temperature and polarization anisotropies.  This conclusion fits well with the results of more sophisticated discussions like the ones illustrated in Figs. \ref{figure8},  \ref{figure9} and  \ref{figure10}.

In Figs. \ref{figure8},  \ref{figure9} and  \ref{figure10} with the full, dashed and dot-dashed lines we report the results for the $BB$ spectrum induced by the Faraday effect and numerically computed on the basis of Eq. (\ref{F14a}) after having included the magnetic fields in the Einstein-Boltzmann hierarchy. 
Both plots of Fig. \ref{figure8} share the same parameters but the plot on the right is focussed on the large  angular scales while the plot on the left illustrates the small angular scales. Semilogarithmic scales are used in both plots.
We can already see that the angular power spectra can only be ${\mathcal O}(10^{-2})\, \mu\mathrm{K}^2$ 
when $B_{\mathrm{L}} = {\mathcal O}(15)$ nG. 
This trend is confirmed by Figs. \ref{figure9} and \ref{figure10}.
 In Fig. \ref{figure9} the magnetic spectral index has been fixed at $n_{\mathrm{B}}= 1.5$ while in Fig. \ref{figure10} the spectral index has been fixed to $n_{\mathrm{B}}= 2$. The full, dashed and dot-dashed curves in the various plots of Figs. \ref{figure9} and \ref{figure10} denote, respectively magnetic field intensities of  $1$, $5$ and $10$ nG.

We finally remind that the tensor modes of the geometry could also 
produce a $B$-mode polarization. The effect of the tensor modes 
of inflationary origin is customarily parametrized in term of the tensor 
to scalar ratio $r_{\mathrm{T}}$ introduced in Eq. (\ref{int1}) (see also the last 
line of Tab. \ref{TABLEAA} and the discussion therein). 
For a tensor to scalar ratio $r_{\mathrm{T}} = {\mathcal O}(0.1)$ (which 
corresponds to the most recent limits) the tensor modes of inflationary 
origin are always larger than the one of the Faraday effect 
(at least for $B_{\mathrm{L}} < \mathrm{nG}$). One could however argue that even in the 
absence of tensor modes of inflationary origin the magnetic random fields may induce 
tensor modes and hence a $B$-mode. Using the same accuracy of Eq. (\ref{estBB2}) 
we can estimate that, in this case, the $BB$ power spectrum will be ${\mathcal G}_{\ell}^{(\mathrm{BB})} \simeq 10^{-4} (B_{\mathrm{L}}/\mathrm{nG})^4$. Since, however, $B_{\mathrm{L}}$ is smaller than the 
nG the Faraday rotation signal always dominates being proportional to 
$(B_{\mathrm{L}}/\mathrm{nG})^2$ rather than to $(B_{\mathrm{L}}/\mathrm{nG})^4$.

\subsection{Faraday scaling}
The frequency scaling  induced by the Faraday effect is the most powerful tool  to disambiguate 
the possible origin of the $B$-mode polarization. The $BB$ angular power spectrum induced 
by the tensor modes is frequency independent. Conversely, given the signal at a certain pivot frequency $\nu_{p}$ 
the B mode polarization induced by Faraday rotation at a different observational frequency 
$\nu$ can be obtained in terms of  this simple scaling law:
 \begin{equation}
{\mathcal G}^{(\mathrm{BB})}_{\ell}(\overline{\nu}) = \biggl(\frac{\overline{\nu}_{p}}{\overline{\nu}}\biggr)^4 {\mathcal G}^{(\mathrm{BB})}_{\ell}(\overline{\nu}_{p}).
\label{eq1BB}
\end{equation}
A direct application of Eq. (\ref{eq1BB}) involves the possibility of distinguishing a potential 
$B$-mode signal from known unwanted foregrounds. For instance both the synchrotron 
and the free-free emissions lead to a frequency dependence of the $B$-mode angular 
power spectrum which is however very different from the one of the Faraday 
rotated $E$-mode polarization.

 Equation (\ref{eq1BB}) could be used to infer the origin of the $B$-mode signal. 
 To give an example of this second possibility let us suppose
that an experiment measured ${\mathcal G}^{(\mathrm{BB})}_{\ell} ={\mathcal O}(10^{-2})\,\,\mu\mathrm{K}^2$ at a pivot frequency $\nu_{p}=150$ GHz. Since direct upper limits on the 
$B$-mode autocorrelation  have been presented, over different observational frequencies
we could simply ask if the various upper limits are simultaneously compatible with this purported signal
and with the scaling law provided by Eq. (\ref{eq1BB}). 
The answer to this question is negative \cite{far14} and it has been used to exclude 
Faraday rotation as a possible origin of the purported Bicep2 signal \cite{chone43} 
which was instead due to a foreground contamination.

The pivot frequencies of the microwave background polarization experiments can be conventionally divided into two ranges conventionally denoted hereunder by $\nu_{low}$ and $\nu_{high}$:
\begin{equation}
26 \,\, \mathrm{GHz} \leq \nu_{low} \leq 36\,\, \mathrm{GHz}, \qquad
100 \,\, \mathrm{GHz} \leq \nu_{high} \leq 150\,\, \mathrm{GHz}.
\label{nulowhigh}
\end{equation}
The Dasi (Degree Angular Scale Interferometer) \cite{dasi1,dasi2} and the Cbi (Cosmic Background Imager) \cite{cbi} experiments were both working in a range coinciding exactly with $\nu_{low}$. 
Four other experiments have been conducted around $\nu_{high}$ and they are:
{\it i)} Boomerang (Balloon Observations of Millimetric Extragalactic Radiation and Geophysics) working at $145$ GHz with 
four pairs of polarization sensitive bolometers \cite{boom}; {\it ii)}
Maxipol (Millimiter Anisotropy experiment Imaging array) working at $140$ GHz with 12 polarimeters
 \cite{maxipol}; {\it iii)} Quad\footnote{An acronym or a contraction between the Quest (Q and U extragalactic sub-mm telescope) and the Dasi experiments.}
working with 31 pairs of polarization sensitive bolometers: 12 at $100$ GHz and $19$ at $150$ GHz \cite{quad1,quad2}; 
{\it iv)} Bicep2 \cite{BICEP2,bicep2new} and its precursor Bicep1 \cite{bicep1}  working, respectively, at $150$ GHz and $100$ GHz.

\begin{table}[!ht]
\begin{center}
\begin{tabular}{||l|c|c|c|c|c||}
\hline
\hline
\rule{0pt}{4ex} References & Data  & $B_{\mathrm{L}}$ &$n_{\mathrm{B}}$& Frequency \\
\hline
\cite{chone34am} & Planck    &$ B_{\mathrm{L}}< 1.38 \, \mu \mathrm{G}$      & $ \mathrm{undetermined} $ & $70$ GHz \\
\cite{chone44} & Polarbear   &$ B_{\mathrm{L}} <  93 $ nG                & $n_{\mathrm{B}} \to 1$     &  $148$ GHz \\ 
\cite{bicep2new} & Bicep2 &$ B_{\mathrm{L}} < 30$ nG         & $ n_{\mathrm{B}} \to 1 $& $150$ GHz \\
\cite{far9,far14} & Cbi               & $B_{\mathrm{L}} < 15 $ nG & $ 1.1 \leq n_{\mathrm{B}} < 2.5$  & $30$ GHz\\
 \cite{far9,far14} & Capmap         & $B_{\mathrm{L}} < 10 $ nG & $ 1.1 \leq n_{\mathrm{B}} < 2.5$  &$35$--$46$ GHz\\
\hline
\end{tabular}
\caption{The bounds inferred from direct limits on the $B$-mode polarization at low frequencies are compared with 
the most recent (and less restrictive) limits over larger frequencies.}
\label{TABLEBB}
\end{center}
\end{table}
There are finally four polarization sensitive experiments working in mixed or intermediate frequency ranges. They include:
{\it a)} the WMAP experiment \cite{chone31} (see also \cite{WMAP9}) spanning five frequencies from $23$ 
to $94$ GHz; {\it b)} the Capmap experiment (Cosmic Anisotropy Polarization Mapper) \cite{capmap}, with $12$ receivers operating between $84$ and $100$
GHz and four receivers operating between $35$ and $46$ GHz; {\it c)} the Quiet (Q/U imager experiment) \cite{chone41a,quiet} operating at $43$ GHz (during the first season of the experiment) and at $95$ GHz (during the second season of the experiment). Finally we have the Planck experiment \cite{chone33a,chone34a}: the three low frequency channels (i.e. $30,\,44,\,70$ GHz) belonged to the low frequency instrument (LFI);  six channels (i.e. $100,\,143,\,217,\,353,\,545,\,857$ GHz) belonged to the high 
frequency instrument (HFI). 

The $BB$ angular power spectrum induced by the Faraday effect gets larger (and even much 
larger) when the frequency decreases. It is therefore expected that limits obtained 
for frequencies between $70$ GHz and $150$ GHz may be less constraining than the 
limits obtained over much smaller frequencies. 
The observational frequencies of Planck \cite{chone34am}, Polarbear \cite{chone44} and Bicep2 \cite{bicep2new} 
give bounds on the Faraday rotated $E$-mode polarization which are  sometimes less constraining than
 some of the previous polarization experiments operating over much smaller frequencies. This theme 
might suggest some useful reflections which are summarized in Tab. \ref{TABLEBB}. It is finally relevant to mention that 
the next generation of radio-telescopes (like the daring 
project\footnote{The collecting 
area of SKA, as the name suggest, will be of $10^{6}\, {\rm m}^2$. The specifications for the SKA require an angular resolution of $0.1$ arcsec at $1.4$ GHz, a frequency capability of $0.1$--$25$ GHz, and a field of view of at 
least $1\,{\rm deg}^2$ at $1.4$ GHz \cite{SKA}.  The number of independent beams is expected to be larger than $4$ and the number of instantaneous 
pencil beams will be roughly 100 with a maximum 
primary beam separation of about $100$ $\mathrm{deg}$ at low frequencies 
(becoming $1$ $\mathrm{deg}$ at high frequencies, i.e. of the order of $1$ GHz).These specifications will allow full sky surveys of Faraday Rotation.}  of the Square Kilometre Array \cite{SKA}) may get their frequency capability up to $25$ GHz.
The overlap between radio-astronomy and microwave background will then be observationally accessible in a frequency range where the signal due to a Faraday-induced $B$-mode  is maximal.

\subsection{Stochastic Faraday mixing as a Markov Process}
In the standard lore the polarization of the microwave background is
 first generated and then it is rotated by a Faraday screen.
The two steps of the process, however, cannot be neatly separated.
The idea explored in Ref. \cite{far13} (see also \cite{far9a}) is therefore to 
describe the Faraday rate as a random, stationary and approximately Markovian process.
In this approach the Faraday rate $X_{F}(\vec{x},\tau)$ is not a deterministic variable but rather a stochastic process which is stationary insofar as the autocorrelation function $\Gamma(\tau_{1},\tau_{2}) = \langle X_{F}(\tau_{1}) X_{F}(\tau_{2}) \rangle$ only depends on time differences i.e. $\Gamma(\tau_{1},\tau_{2}) = \Gamma(|\tau_{1}- \tau_{2}|)$. 
The evolution of the brightness perturbations 
\begin{equation}
\Delta_{\pm}' + (\epsilon' + n^{i} \, \partial_{i} ) \Delta_{\pm} = {\mathcal M}(\vec{x},\tau) \mp 2 i \, X_{F}(\vec{x},\tau) \Delta_{\pm},
\label{re3}
\end{equation}
becomes then a stochastic differential equation. The simplest approximation is to consider $X_{F}$ as a random variable characterized by a given probability distribution; this case has been already analyzed in the framework of the synchrotron emission \cite{SYNC1,SYNC1a,SYNC2} and will not be specifically analyzed here. 

As an example of stationary process not delta-correlated consider the case where $\Gamma(\tau_{1} - \tau_{2}) = \langle X_{F}(\tau_{1}) \, X_{F}(\tau_{2}) \rangle$  can take only two values $\overline{x}_{F}^2$ and $ - \overline{x}_{F}^2$ and let us suppose that $X_{F}(\tau)$ has switched an even number of times in the interval between $\tau_{1}$ and $\tau_{2}$ so that $\Gamma(\tau_{1} - \tau_{2}) = \overline{x}_{F}^2$ whereas the correlation 
function gives $- \overline{x}_{F}^2$ if there have been an odd number of switches. If $p(n, \Delta\tau)$ is the probability of $n$ switches in the interval $\Delta\tau = \tau_{1} - \tau_{2}$, it follows that 
\begin{equation}
\Gamma(\Delta\tau) = \overline{x}_{F}^2 \sum_{n = 0,\, 2,\, 4\,...}^{\infty} p(n,\Delta\tau) -  \overline{x}_{F}^2  \sum_{n = 1,\, 3,\, 5\,...}^{\infty} p(n,\Delta\tau)
= \overline{x}_{F}^2 \sum_{n =0}^{\infty} (-1)^{n} \, p(n,\Delta\tau).
\label{dich}
\end{equation}
As the switches are random with average rate $r$, $p(n,\Delta\tau) $ is nothing but a Poisson distribution with mean number of switches $\overline{n} = r \, \Delta\tau $, i.e. 
$p_{n} = \overline{n}^{n} e^{- \overline{n}}/n!$. This means that $\Gamma(\Delta\tau) = \overline{x}_{F}^2 \exp{[ - 2 r\, \Delta\tau]}$. This is an example of dichotomic Markov process \cite{stoch1,stoch1a} applied to the case of stochastic Faraday rate. 

Using the technique of the cumulant expansion \cite{stoch1,stoch1a} and in the absence of a primordial 
tensor contribution the angular power spectra of the $E$-mode and of the $B$-mode polarizations can be derived and they are:
\begin{equation}
C_{\ell}^{(EE)}(\omega_{F}) = e^{- \omega_{F} }\, \cosh{\omega_{F}} \,\overline{C}_{\ell}^{(EE)}, \qquad
C_{\ell}^{(BB)}(\omega_{F}) = e^{- \omega_{F} }\, \sinh{\omega_{F}} \,\overline{C}_{\ell}^{(EE)};
\label{cor6}
\end{equation}
where 
\begin{equation}
 \omega_{F} = 4 \int_{\tau_{*}}^{\tau} \, d\tau_{1} \,  \int_{\tau_{*}}^{\tau} \, d\tau_{2} \langle X_{F}(\tau_{1}) \, X_{F}(\tau_{2}) \rangle.
 \label{cor1}
 \end{equation}
In Eq. (\ref{cor1}) $\tau_{*}$ denotes the photon decoupling. Even if $X_{F} \leq 1$, $\omega_{F}$ is 
not bound to be smaller than $1$. However if $|\omega_{F} |<1$, from Eq. (\ref{cor1}) $C_{\ell}^{(EE)} \simeq 
\overline{C}_{\ell}^{(EE)}$ while $C_{\ell}^{(BB)} \simeq \omega_{F} \overline{C}_{\ell}^{(EE)}$.

If the $B$-mode polarization 
induced by the tensor modes of the geometry is instead present \cite{apprais}, the stochastic Faraday mixing also affects 
the tensor modes of the geometry and the analog of the result mentioned above is given by:
\begin{eqnarray}
C_{\ell}^{(EE)} &=& e^{-\omega_{F}} \cosh{\omega_{F}} \, \biggl( \overline{C}_{\ell}^{(EE)} + {\mathcal C}_{\ell}^{(EE)} \biggr) + e^{- \omega_{F}}\, \sinh{\omega_{F}}\,
 {\mathcal C}_{\ell}^{(BB)},
 \nonumber\\
 C_{\ell}^{(BB)} &=& e^{-\omega_{F}} \sinh{\omega_{F}} \biggl( \overline{C}_{\ell}^{(EE)} + {\mathcal C}_{\ell}^{(EE)} \biggr) + e^{-\omega_{F}} \, \cosh{\omega_{F}}
\, {\mathcal C}_{\ell}^{(BB)};
\label{INT3}
\end{eqnarray}
where $\overline{C}_{\ell}^{(EE)}$ denotes the $E$ mode power spectrum coming from the scalar modes of the geometry while
${\mathcal C}_{\ell}^{(BB)}$ and ${\mathcal C}_{\ell}^{(EE)}$ (both in calligraphic style) denote, respectively, the polarization observables induced by the tensor modes of the geometry. Both the $E$ mode and the $B$ mode polarization are frequency dependent since $\omega_{F}$ is proportional to the square of the rate and, ultimately, to the fourth power of the comoving wavelength. The stochastic approach to the Faraday rate represents an ideal framework for deriving a set of scaling laws only involving the measured polarization power spectra \cite{far13,far14}.  Note that Eq. (\ref{INT3}) not only describes the rotation of an initial $E$-mode polarization but also the inverse effect, i.e. the rotation of an initial $B$-mode polarization 
of tensor origin. 

\renewcommand{\theequation}{6.\arabic{equation}}
\setcounter{equation}{0}
\section{Circular polarizations?}
\label{sec6}
Large-scale magnetic fields prior to decoupling may also circular polarizations.  
Direct limits on the $V$-mode power spectrum
between $10^{-7}\,\,\mu\mathrm{K}^2 $ and $10^{-4}\,\, \mu\mathrm{K}^2$ could 
directly rule out (or rule in) pre-decoupling magnetic fields in the range of $0.1$--$1$ nG for 
typical frequencies between $10$ GHz and $30$ GHz. 
 
\subsection{The $V$-mode polarization of the microwave background}

The intensity (i.e. $I= |\vec{E}\cdot\hat{e}_{1}|^2 +  |\vec{E}\cdot\hat{e}_{2}|^2$) 
and the circular polarizations (i.e. $V = 2 \,\mathrm{Im}[ (\vec{E}\cdot\hat{e}_{1})^{*} (\vec{E}\cdot\hat{e}_{2})]$),
are both invariant for a rotation of $\hat{e}_{1}$ and $\hat{e}_{2}$ 
in the plane orthogonal to the direction of propagation of the radiation 
(see also Eqs. (\ref{IandV}) and (\ref{QandU})). 
The $V$-mode polarization is then described by two supplementary power spectra: the $V$-mode autocorrelation 
(i.e. the $VV$ spectrum) and the cross-correlation with the temperature (i.e. the $VT$ spectrum). 
The $VV$ and the $VT$ power spectra are the analog of the $EE$ and $TE$ power spectra arising in the case of the 
linear polarization. The $TT$,  $VT$ and the $VV$ power spectra shall be preferentially considered hereunder
since they are anyway larger than the $VE$ and $VB$ correlations.

In the absence of pre-decoupling magnetic field the primeval circular polarization is 
decoupled from the temperature fluctuations and from the linear polarization. 
A computable  amount of circular polarization is then
generated when the electron-photon scattering takes place in a magnetized environment, 
as previously discussed in section \ref{sec2}. In this case the circular polarization 
directly affects both the temperature anisotropies as well as the $E$-mode and $B$-mode polarizations.  
Thus a primordial $V$-mode polarization (possibly present prior to decoupling) 
can be constrained by using the magnetized plasma as a polarimeter. 
Conversely if the circular polarization vanishes initially the magnetic field acts effectively 
as a polarizer.

When the curvature perturbations are the dominant source of large-scale inhomogeneity,
the evolution equations for the brightness perturbations can be derived 
from Eqs. (\ref{BRI}), (\ref{BRQ}), (\ref{BRU}) and (\ref{BRV}). 
In this case, the evolution of relevant the brightness perturbations is given by:
\begin{eqnarray}
\Delta_{\mathrm{I}}' + ( i k \mu + \epsilon') \Delta_{\mathrm{I}} &=& \psi' - i k\mu \phi + \epsilon'\biggl[ \Delta_{\mathrm{I}0} + 
\mu v_{\mathrm{b}} - \frac{P_{2}(\mu)}{2} S_{\mathrm{P}}\biggr] 
\nonumber\\
&-& \frac{3}{2} i\, \epsilon' \, f_{\mathrm{e}}(\overline{\omega}) \,( 1 + \mu^2) \Delta_{\mathrm{V}1} 
\label{deltaI}\\
 \Delta_{\mathrm{P}}' + ( i k \mu + \epsilon') \Delta_{\mathrm{P}} &=& \frac{3}{4} ( 1 - \mu^2) \epsilon' S_{\mathrm{P}}  - \frac{3}{2} i \epsilon' f_{\mathrm{e}}(\overline{\omega}) (\mu^2 -1) \Delta_{\mathrm{V}1},
\label{deltaP}\\
  \Delta_{\mathrm{V}}' + ( i k \mu + \epsilon') \Delta_{\mathrm{V}}  &=& \epsilon' \mu \biggl\{ f_{\mathrm{e}}(\overline{\omega}) [ 2 \Delta_{\mathrm{I}0} - S_{\mathrm{P}}] - \frac{3}{2} i \Delta_{\mathrm{V}1}\biggr\},
\label{deltaV}
\end{eqnarray}
where $f_{\mathrm{e}}(\overline{\omega})$ denotes, 
the ratio between the Larmor frequency of the electrons and the angular frequency of the 
observational channel (see also Eq. (\ref{FEa})). The limit $f_{\mathrm{e}}(\overline{\omega}) \to 0$ 
corresponds to the standard situation where the plasma is not magnetized.
While the tensor and the vector modes may also affect the $V$-mode polarization,
their role is less relevant at least in the light of the concordance paradigm.

The analysis of the $V$-mode polarization calls for direct
measurements of the circular polarizations of the CMB \cite{C1,C2}. 
While diverse circularly polarized foregrounds may exist \cite{C2c} (see also, for instance, \cite{C2a}),
they are qualitatively different from the ones customarily considered in the case of linear 
polarizations.  The bounds on circular polarizations coming from direct searches have a rather long history 
which can be traced back to the seminal contribution of Ref. \cite{lubin1} (see also \cite{lubin2,lubin3}) and of Ref. \cite{part1} (see also \cite{part2,part3}). The measurements of  \cite{part1,part2,part3}
 were conducted for a typical wavelength of $6$ cm (corresponding to $\nu= 4.9$ GHz) and 
 used the Very Large Array radio-telescope in Socorro (New Mexico). 
 Conversely the limits of \cite{lubin1,lubin2,lubin3} used a $\nu=33$ GHz 
 radiometer (corresponding to a wavelength of $9$ mm)  which used a Faraday rotator to switch 
 between orthogonal and linear polarization states. The bounds of Ref. \cite{lubin1,lubin2,lubin3} 
 and \cite{part1,part2,part3} have been used in Refs. \cite{C1,C2}. Recently further measurements 
 appeared in the literature. Improved upper limits on the 
 circular polarization on large angular scale have been presented 
 in \cite{C2c,C2b}. Direct constraints at intermediate 
 angular scales appeared in the literature \cite{C3} thanks to the Spider 
 collaboration. To be relevant for the ideas conveyed in this section the 
 present sensitivities should be improved, in the future, by at least $6$ or even $7$ orders 
 of magnitude. 

Besides the presence of a magnetic field only few other sources of circular polarization 
have been discussed in the literature. They 
include photon-photon interactions \cite{C4} and pseudo-scalar particles \cite{far8,C5}.
The $V$-mode of the CMB has been also suggested as a probe 
for the first stars \cite{C6}. The circular polarization is finally invoked as the result 
of the Faraday conversion of linearly polarized radiation. 
Faraday conversion (typical of relativistic jets) should not be confused with Faraday rotation. 
In the presence of relativistic electrons linearly polarized radiation can be Faraday converted into 
circularly polarized radiation.  The Faraday {\em conversion} and Faraday {\em rotation} have a different dependence 
upon the magnetic field intensity and upon the frequency \cite{SYNC1,SYNC1a,C10}. 
For the latter mechanism to operate, relativistic electrons must be present in the 
system and this can happen only as a secondary effect when CMB photons pass through 
magnetized clusters; this is however not the idea pursued here since the pre-decoupling 
plasma is cold and the charge carriers are non-relativistic. 

\subsection{The magnetized plasma as a polarizer}

\begin{figure}[!ht]
\centering
\includegraphics[height=6cm]{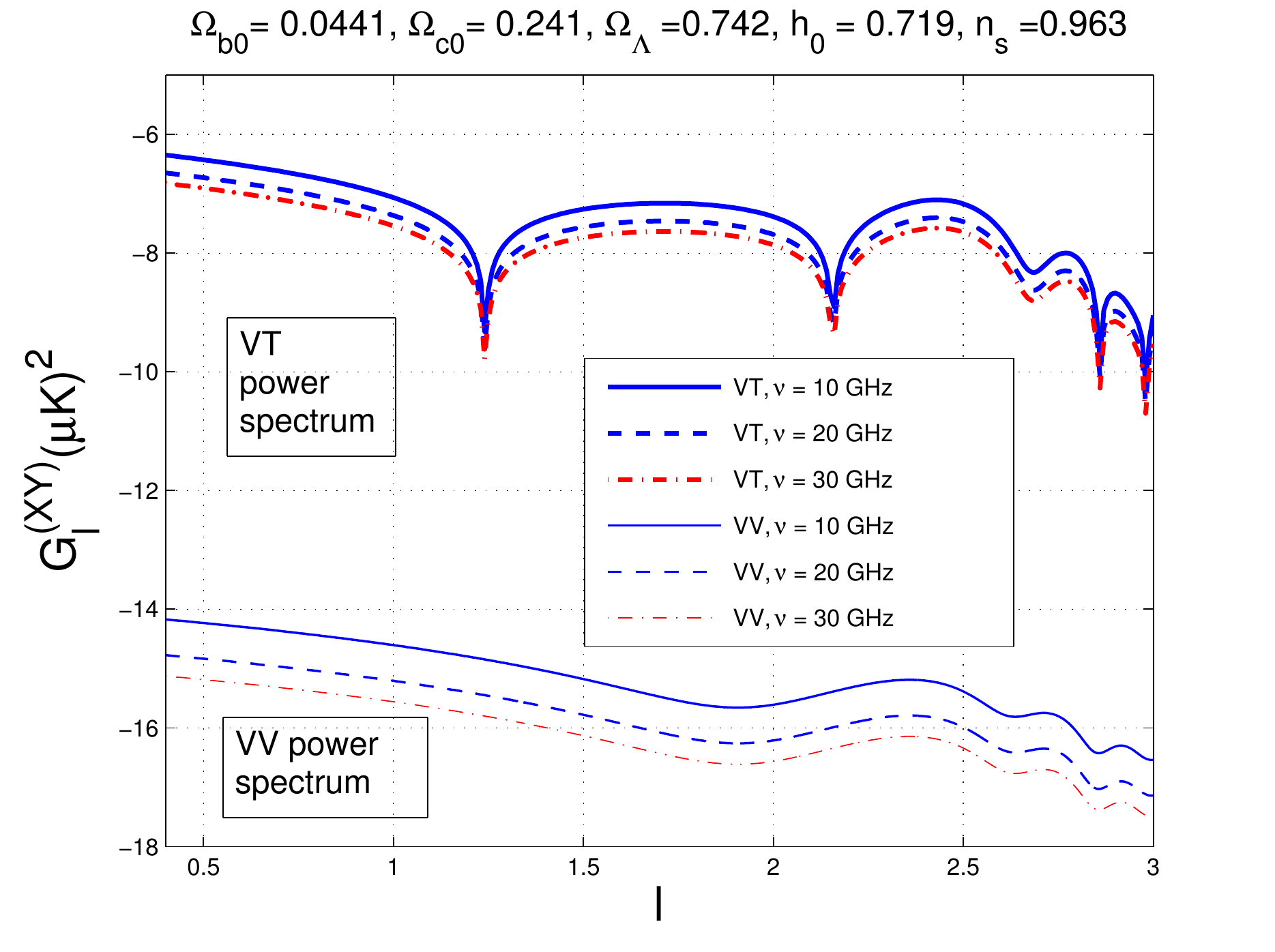}
\includegraphics[height=6cm]{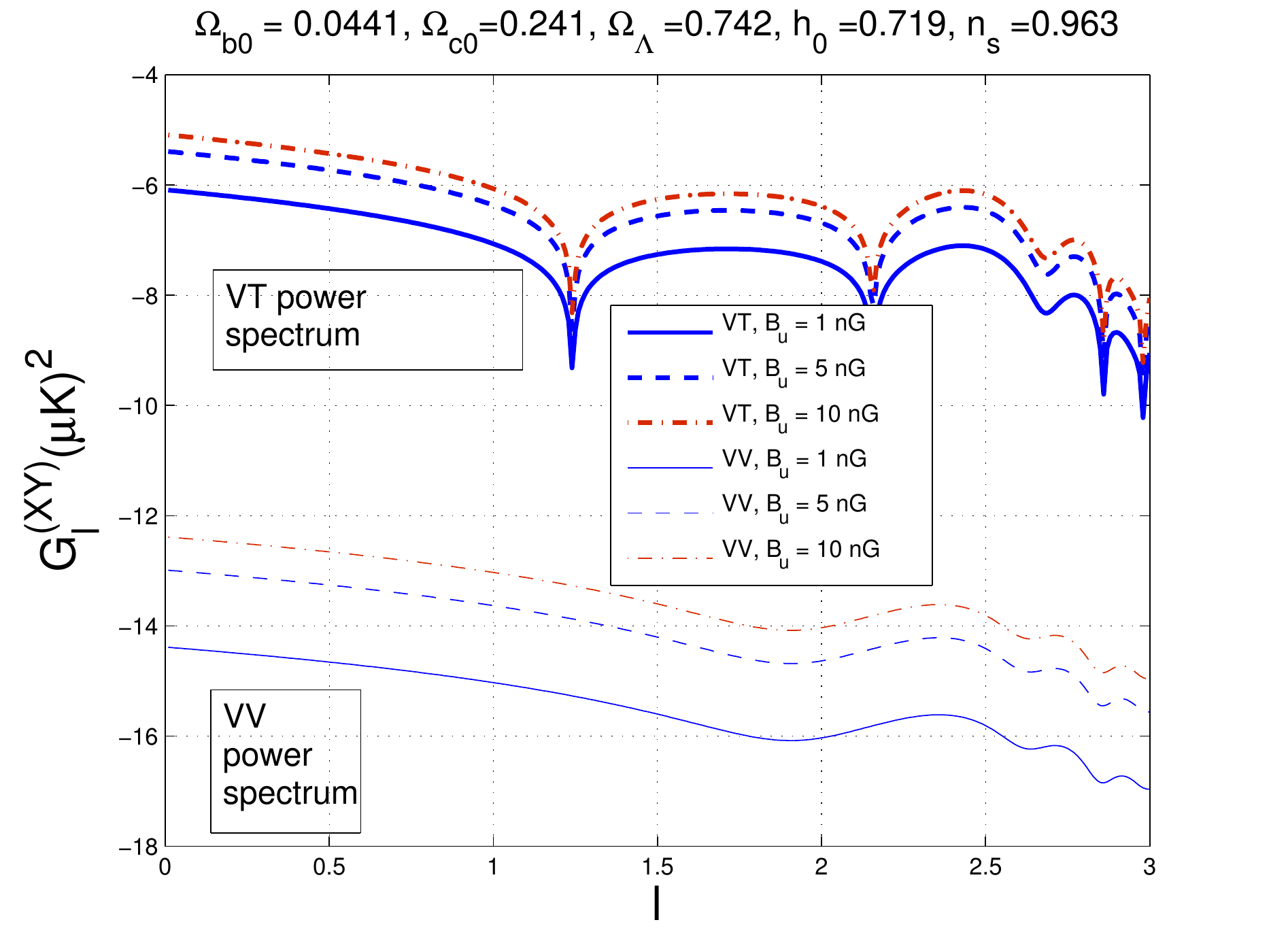}
\caption[a]{In the plot at the left the $VT$ and the $VV$ angular power spectra are reported 
for a fixed value of the magnetic field intensity $B_{u}$ (i.e. $1$ nG) but for different values of the 
comoving frequency. In the plot at the right the comoving frequency is fixed 
to $10$ GHz but the magnetic field strength increases. The thin lines denote 
the $VV$ correlations while the thick lines denote the $VT$ correlations. 
In the plots, on both axes, the common logarithm of the corresponding quantity is reported.}
\label{FIGCIRC1}      
\end{figure}

The degree of circular polarization directly induced by the magnetized 
plasma can be computed from Eqs. (\ref{deltaI}), (\ref{deltaP}) and (\ref{deltaV}). According to this 
perspective the initial $V$-mode polarization vanishes so that 
$\Delta_{\mathrm{V}1}=0$ and the primordial circular polarization can be neglected \cite{C1,C2}. 
The Cauchy data for the evolution of the brightness perturbations follow, in this case, from the standard 
adiabatic mode (possibly even magnetized). The induced circular polarization can be computed and 
the results of this analysis are illustrated in Fig. \ref{FIGCIRC1} in the case of the 
standard adiabatic mode. For the fiducial set of parameters of the Planck experiment
the quantitative differences are irrelevant so that we just adapted the results of Ref. \cite{C1}.
Note that, in Fig. \ref{FIGCIRC1}, we denoted ${\mathcal G}_{\ell}^{(XY)} = \ell(\ell +1) C_{\ell}^{(XY)}/(2\pi)$ 
where $X= V$ and $Y$ coincides either with $V$ (in the three curves at the bottom) or with 
$T$ (in the three curves at the top).
If the initial conditions are not adiabatic 
the $V$-mode polarization will have different physical features depending upon the 
specific entropic initial condition. 

The thin lines in both plots of Fig. \ref{FIGCIRC1} denote the $V$-mode autocorrelations while the thick lines denote 
the cross-correlation of the circular polarization anisotropies with the temperature inhomogeneities. 
The signal is larger for low multipoles and its shape reminds a bit of the temperature autocorrelations 
induced by the tensor modes of the geometry which reach their largest 
value for small $\ell$ and decline exponentially for $\ell > 90$.  
The $B$-mode autocorrelation induced by the tensor modes of the geometry is typically larger than the 
$V$-mode polarization. Recently the Spider collaboration 
reported a direct bound on the $V$-mode autocorrelation 
implying \cite{C3}
\begin{equation}
{\mathcal G}_{\ell}^{(VV)}= \frac{\ell(\ell+1) C_{\ell}^{(VV)}}{2\pi} < {\mathcal O}(100) \mu\mathrm{K}^2, \qquad 33< \ell < 307,
\label{spider}
\end{equation}
for a typical frequency of $150$ GHz. The bound of \cite{C3} is actually more accurate and the term 
${\mathcal O}(100)$ refers to constraints ranging from $141$ to $255$ $\mu\mathrm{K}^2$.
It is also useful, in some cases, to measure the circular polarization in terms of the 
square root of the $VV$ angular power spectrum. Clearly if a given angular 
power spectrum is in the range $10^{-6}\, \mathrm{\mu}K^2$, its square root 
is of the order of $10^{-3}$ mK. These are typically the sensitivities 
suggested by Fig. \ref{FIGCIRC1}.

\subsection{The magnetic field as a polarimeter}
If the initial radiation field is  circularly polarized prior to decoupling,
the problem is to deduce an upper limit on the initial degree of circular 
polarization. More specifically, if $\Delta_{\mathrm{V}1} \neq 0$ the line 
of sight solution of Eq. (\ref{deltaI}) implies that the power spectrum of 
the temperature correlations receives two separated contributions stemming, 
respectively, from the intensity of the radiation field (denoted by 
$\overline{a}^{(\mathrm{I})}_{\ell \, m}$) 
and from the circular polarization (denoted by $\overline{a}^{(\mathrm{V})}_{\ell \, m}$):
\begin{eqnarray}
&&  \overline{a}^{(\mathrm{I})}_{\ell \, m} = \frac{1}{(2\pi)^{3/2}} \int d^{3} k \int_{-1}^{1} \, d\mu \, \int_{0}^{2\pi} d\varphi  \,Y_{\ell\, m}^{*}(\mu,\varphi) \int_{0}^{\tau_{0}} 
\, e^{- i \mu x} \, e^{- \epsilon(\tau,\tau_{0})} \, {\mathcal N}_{\mathrm{I}}(k,\mu,\tau) \, d\tau,
\label{a2}\\
&&  \overline{a}^{(\mathrm{V})}_{\ell \, m} = \frac{1}{(2\pi)^{3/2}} \int d^{3} k \int_{-1}^{1} \, d\mu \, \int_{0}^{2\pi} d\varphi  \,Y_{\ell\, m}^{*}(\mu,\varphi) \int_{0}^{\tau_{0}} 
\, e^{- i \mu x} \, e^{- \epsilon(\tau,\tau_{0})} \, {\mathcal N}_{\mathrm{V}}(k,\mu,\tau) \, d\tau,
\label{a3}
\end{eqnarray}
where the two generalized sources ${\mathcal N}_{\mathrm{I}}(k,\mu,\tau)$ and ${\mathcal N}_{\mathrm{V}}(k,\mu,\tau)$ are given, respectively,  by:
\begin{eqnarray}
&& {\mathcal N}_{\mathrm{I}}(k,\mu,\tau) = \psi' - i k \mu \phi + \epsilon' \biggl[ \Delta_{\mathrm{I}0} + \mu v_{\mathrm{b}} - \frac{1}{2} P_{2}(\mu)S_{\mathrm{P}}\biggr],
\label{a4}\\
&& {\mathcal N}_{\mathrm{V}}(k,\mu,\tau) = - \frac{3}{2} \, i\, \epsilon' f_{\mathrm{e}}(\omega) ( 1 + \mu^2) 
\Delta_{\mathrm{V}1}.
\label{a5}
\end{eqnarray}
The temperature fluctuation will then be the sum of the intensity contribution and of circular polarization dipole:
\begin{equation}
\Delta_{\mathrm{T}}(\hat{n},\tau_{0}) = \sum_{\ell m} a^{(\mathrm{T})}_{\ell m} \, 
Y_{\ell m}(\hat{n}), \qquad a^{(\mathrm{T})}_{\ell\, m} = \overline{a}^{(\mathrm{I})}_{\ell \, m} + \overline{a}^{(\mathrm{V})}_{\ell \, m}.
\label{TT1}
\end{equation}
It is important to appreciate that $\overline{a}^{(\mathrm{V})}_{\ell \, m}$ denotes the $V$-mode contribution to 
the temperature correlation and should not be confused with what we will later call $a^{(\mathrm{V})}_{\ell \, m}$ (see below Eq. (\ref{VM1})). 
Prior to matter-radiation equality the dipole power 
spectrum is given by 
\begin{equation}
{\mathcal P}_{\mathrm{V}}(k) = {\mathcal A}_{\mathrm{V}} \biggl(\frac{k}{k_{\mathrm{p}}}\biggr)^{n_{\mathrm{v}} -1}, \qquad k_{\mathrm{p}} = 0.002\, \mathrm{Mpc}^{-1},
\label{PV}
\end{equation}
where, incidentally, $k_{\mathrm{p}}$ is the same pivot scale used to assign the adiabatic mode.
By demanding that the $V$-mode contribution to the 
temperature fluctuations be negligible in comparison with the intensity, an interesting bound 
on the circular polarization can be derived \cite{C2}:
\begin{eqnarray}
{\mathcal A}_{V} < {\mathcal N}_{\mathrm{TT}} \biggl(\frac{{\mathcal A}_{{\mathcal R}}}{2.43\times 10^{-9}}\biggr) \biggl(\frac{z_{*}+ 1}{1091.79}\biggr)^{-2} 
\biggl(\frac{D_{A}}{14116\, \mathrm{Mpc}}\biggr)^{n_{v} - n_{\mathrm{s}}} \biggl(\frac{B_{u}}{\mathrm{nG}}\biggr)^{-2} \biggl(\frac{\nu}{\mathrm{GHz}}\biggr)^{2}
\label{boundTT13}
\end{eqnarray}
where $D_{A}$ denotes the (comoving) angular diameter distance to last scattering while $z_{*}$ denotes the redshift 
to the last scattering;
the term ${\mathcal N}_{\mathrm{TT}}$ is given by:
\begin{eqnarray}
{\mathcal N}_{\mathrm{TT}} &= &1.156 \times 10^{6} \times \,(0.0354)^{n_{\mathrm{s}} - n_{v}} \, e^{- 2 \epsilon_{\mathrm{re}}} \, r(\ell, n_{v}, n_{\mathrm{s}}),
\label{boundTT14}\\
r(\ell, n_{v}, n_{\mathrm{s}}) &=&  \frac{4\,
\Gamma\biggl(\frac{3}{2} - \frac{n_{\mathrm{s}}}{2}\biggr) \, \Gamma\biggl(4 - \frac{n_{v}}{2} \biggr)}{
(n_{v}^2 - 12 n_{v} + 39) \Gamma\biggl(2 - \frac{n_{\mathrm{s}}}{2} \biggr) \Gamma\biggl(\frac{3}{2} - \frac{n_{v}}{2}\biggr)} 
\,   \ell^{n_{\mathrm{s}} - n_{v}}\,\biggl[ 1 + {\mathcal O}\biggl(\frac{1}{\ell}\biggr)\biggr].
\label{boundTT12}
\end{eqnarray}
where we assumed $-3 < n_{v} < 3$ and, moreover,  $2\leq \ell < 40$.  The numerical value of 
Eq. (\ref{boundTT12}), for different values of the spectral indices, varies btween ${\mathcal O}(0.1)$ (for $n_{v} > n_{\mathrm{s}}$) 
and ${\mathcal O}(10)$ (for $n_{v} < n_{\mathrm{s}}$). 
From Eq. (\ref{deltaP}), with the same logic used for the intensity, the contribution of the circular polarization
to the $E$-mode autocorrelation can be computed. This analysis results is a further bound on ${\mathcal A}_{V}$ reading 
\begin{eqnarray}
&& {\mathcal A}_{\mathrm{V}} < 5.42 \times 10^{-2}\, \sqrt{\frac{(n_{\mathrm{s}} +1)^{n_{\mathrm{s}} +1}}{(n_{v} -1)^{n_{v} -1}}} \, \biggl( \frac{4 e}{\ell_{\mathrm{D}}}\biggr)^{(n_{v} - n_{\mathrm{s}})/2 -1}
\nonumber\\
&& \times \biggl(\frac{{\mathcal A}_{{\mathcal R}}}{2.43\times 10^{-9}}\biggr) \biggl(\frac{z_{*}+ 1}{1091.79}\biggr)^{-2} 
\biggl(\frac{D_{A}(z_{*})}{14116\, \mathrm{Mpc}}\biggr)^{n_{v} - n_{\mathrm{s}}} \biggl(\frac{B_{u}}{\mathrm{nG}}\biggr)^{-2} \biggl(\frac{\nu}{\mathrm{GHz}}\biggr)^{2},
\label{EE26}
\end{eqnarray}
for $n_{v} >1$ and
\begin{eqnarray}
&& {\mathcal A}_{\mathrm{V}} <  5.42\times 10^{-2} \, (2 \ell_{\mathrm{V}})^{1 - n_{v}}\, 
(n_{\mathrm{s}} +1)^{(n_{\mathrm{s}} +1)/2} \, \biggl( \frac{4 e}{\ell_{\mathrm{D}}}\biggr)^{-( n_{\mathrm{s}} +1)/2}
\nonumber\\
&& \times \biggl(\frac{{\mathcal A}_{{\mathcal R}}}{2.43\times 10^{-9}}\biggr) \biggl(\frac{z_{*}+ 1}{1091.79}\biggr)^{-2} 
\biggl(\frac{D_{A}(z_{*})}{14116\, \mathrm{Mpc}}\biggr)^{n_{v} - n_{\mathrm{s}}} \biggl(\frac{B_{u}}{\mathrm{nG}}\biggr)^{-2} \biggl(\frac{\nu}{\mathrm{GHz}}\biggr)^{2},
\label{EE27}
\end{eqnarray}
for $n_{v} < 1$. In Eqs. (\ref{EE26}) and (ref{EE27}) $\ell_{\mathrm{V}} ={\mathcal O}(65)$ while $\ell_{\mathrm{D}}$ is the damping multipole appearing in the $E$-mode autocorrelations \cite{C2}.

\subsection{Limits on the $V$-mode autocorrelations}

The contribution of the circular polarization to the $TT$ and to the $EE$ correlations 
can be used for the derivation of two separate sets of bounds as suggested by 
Eqs. (\ref{boundTT13}) and (\ref{EE26})--(\ref{EE27}). 
 Following the discussion of Ref. \cite{C2} the following parametrization will be adopted for the direct limits on the $V$-mode power spectrum:
\begin{equation}
\sqrt{\frac{\ell (\ell +1)}{2\pi} C_{\ell}^{(\mathrm{VV})}} = \alpha \, T_{\gamma 0}, \qquad T_{\gamma 0} =2.725 \, \mathrm{K}.
\label{limit8}
\end{equation}
Different values of $\alpha$ will correspond to different observational limits either 
already obtained or potentially interesting for the present considerations. Using the line of sight integration, Eq. (\ref{deltaV}) implies
\begin{eqnarray}
&& a_{\ell \, m}^{(\mathrm{V})} = \frac{1}{(2\pi)^{3/2}} \int \, d\hat{n}\, Y_{\ell\, m}^{*}(\hat{n}) \, \int d^{3} k\, \Delta_{\mathrm{V}}(k,\mu,\tau_{0}),
\label{VM1}\\
&& \Delta_{\mathrm{V}}(k,\mu,\tau_{0}) =  -\frac{3}{4}\,i \, \mu \int_{0}^{\tau_{0}} d\tau \, {\mathcal K}(\tau) \, e^{- i \mu x} \Delta_{\mathrm{V}\,1}(k,\tau). 
\label{VM2}
\end{eqnarray}
The V-mode autocorrelation can be neatly computed in the sudden decoupling limit 
where  the coefficient $a_{\ell\, m}^{(\mathrm{V})}$ are:
\begin{equation}
a_{\ell\, m}^{(\mathrm{V})} = \frac{3\, (-i)^{\ell}}{4 \, (2 \pi)^{3/2}} \delta_{m0} \sqrt{\frac{4\pi}{2 \ell + 1}} \int d^{3} k \, \int_{0}^{\tau_{0}}
 [ \ell \,j_{\ell -1}(x) - (\ell + 1) \, j_{\ell +1}(x) ]\, \Delta_{\mathrm{V}\,1}(k,\tau) \, d\tau;
 \label{VM3}
 \end{equation}
  as usual, $ x = k (\tau_{0} - \tau)$. In the large-scale limit (i.e. in practice for $\ell < 40$) the angular power spectrum 
 of the $V$-mode polarization is given by:
 \begin{equation}
 C_{\ell}^{(\mathrm{VV})} =  \frac{9\pi}{4 \, ( 2 \ell+ 1)^2} \int_{0}^{\infty} \frac{d k}{k} \, {\mathcal P}_{V}(k) [\ell j_{\ell-1}(x) - (\ell +1) j_{\ell +1}(x) ]^2. 
 \label{VM4}
 \end{equation}
  \begin{figure}[!ht]
\centering
\includegraphics[height=6cm]{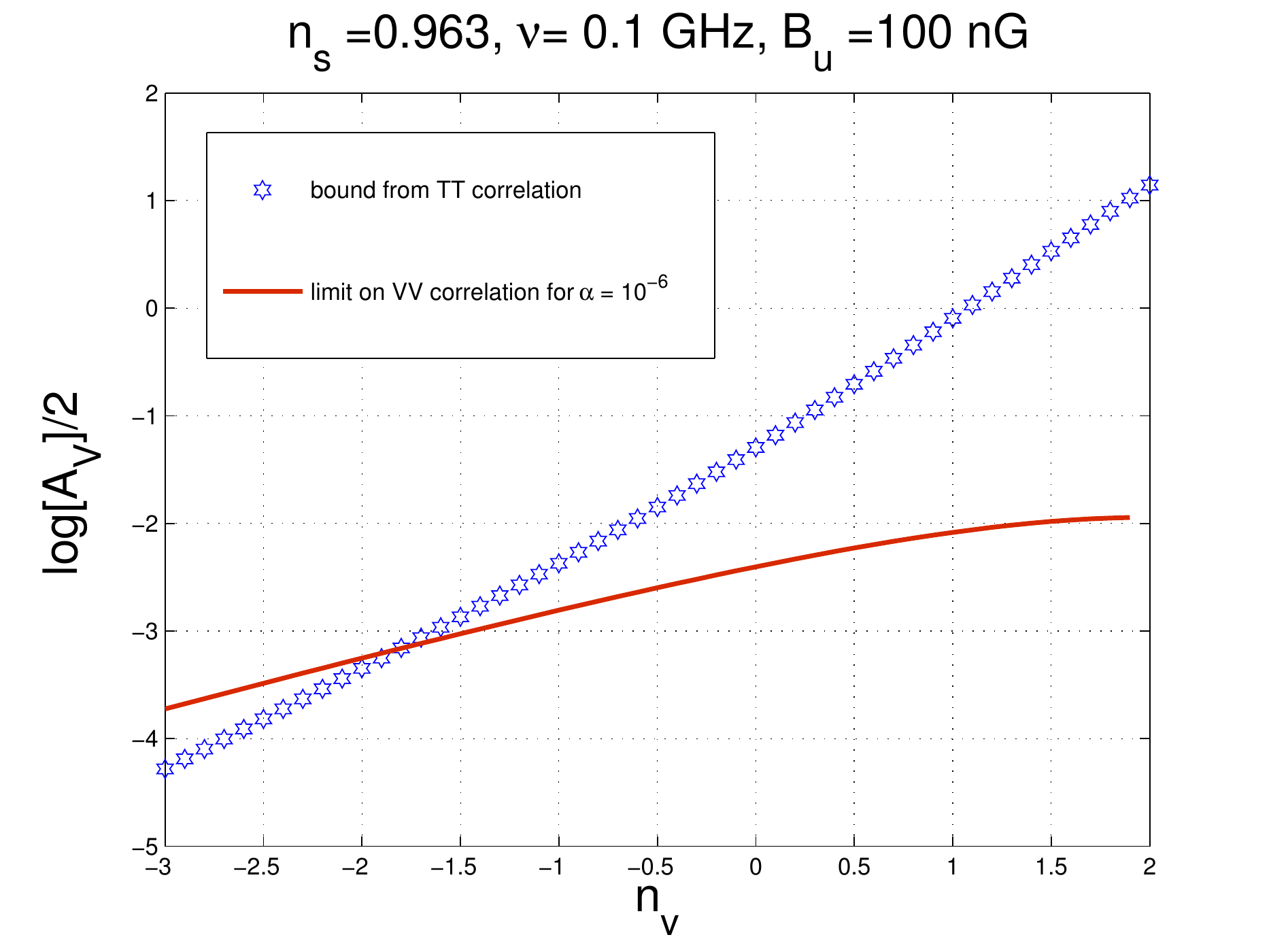}
\includegraphics[height=6cm]{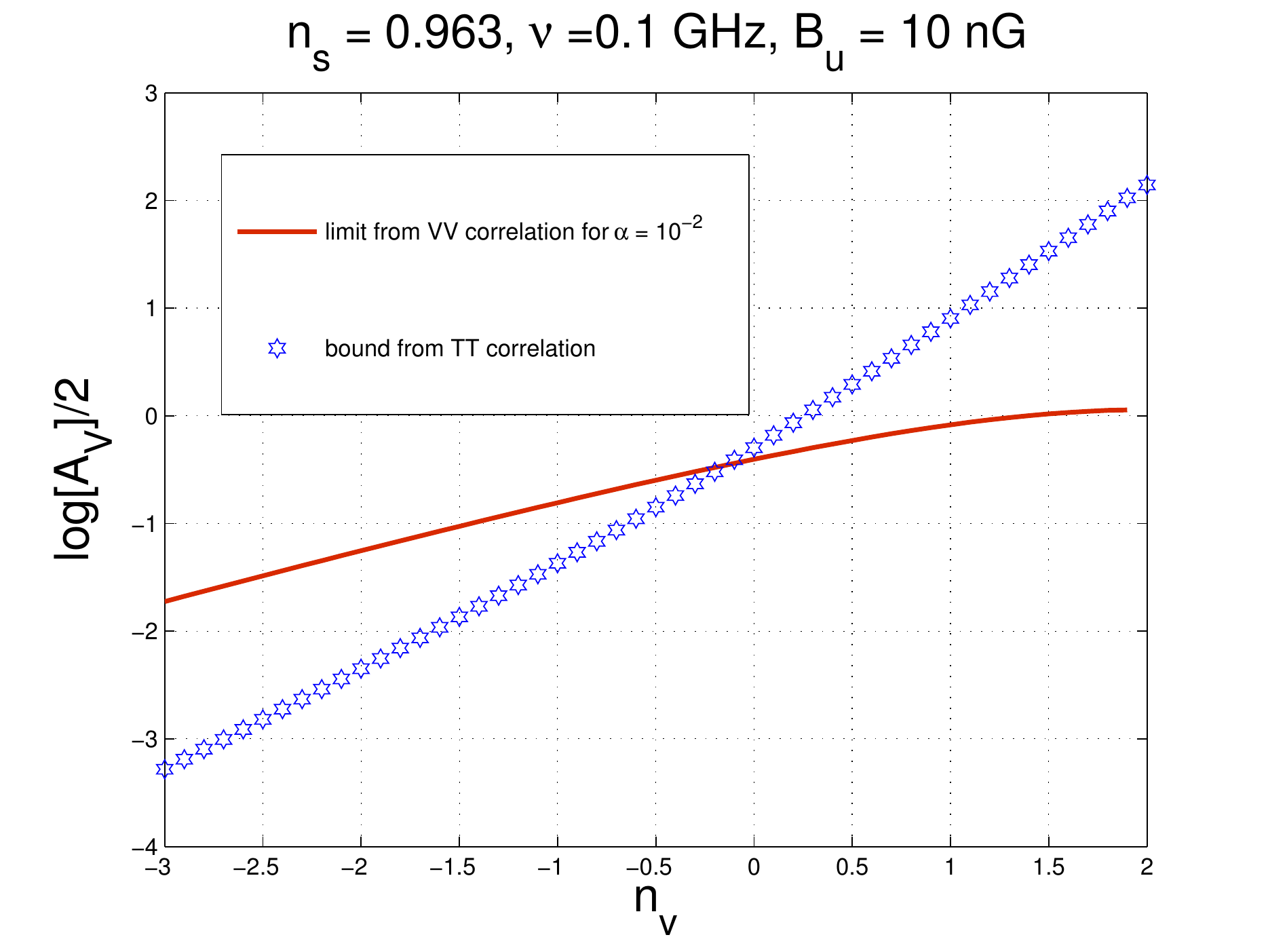}
\includegraphics[height=6cm]{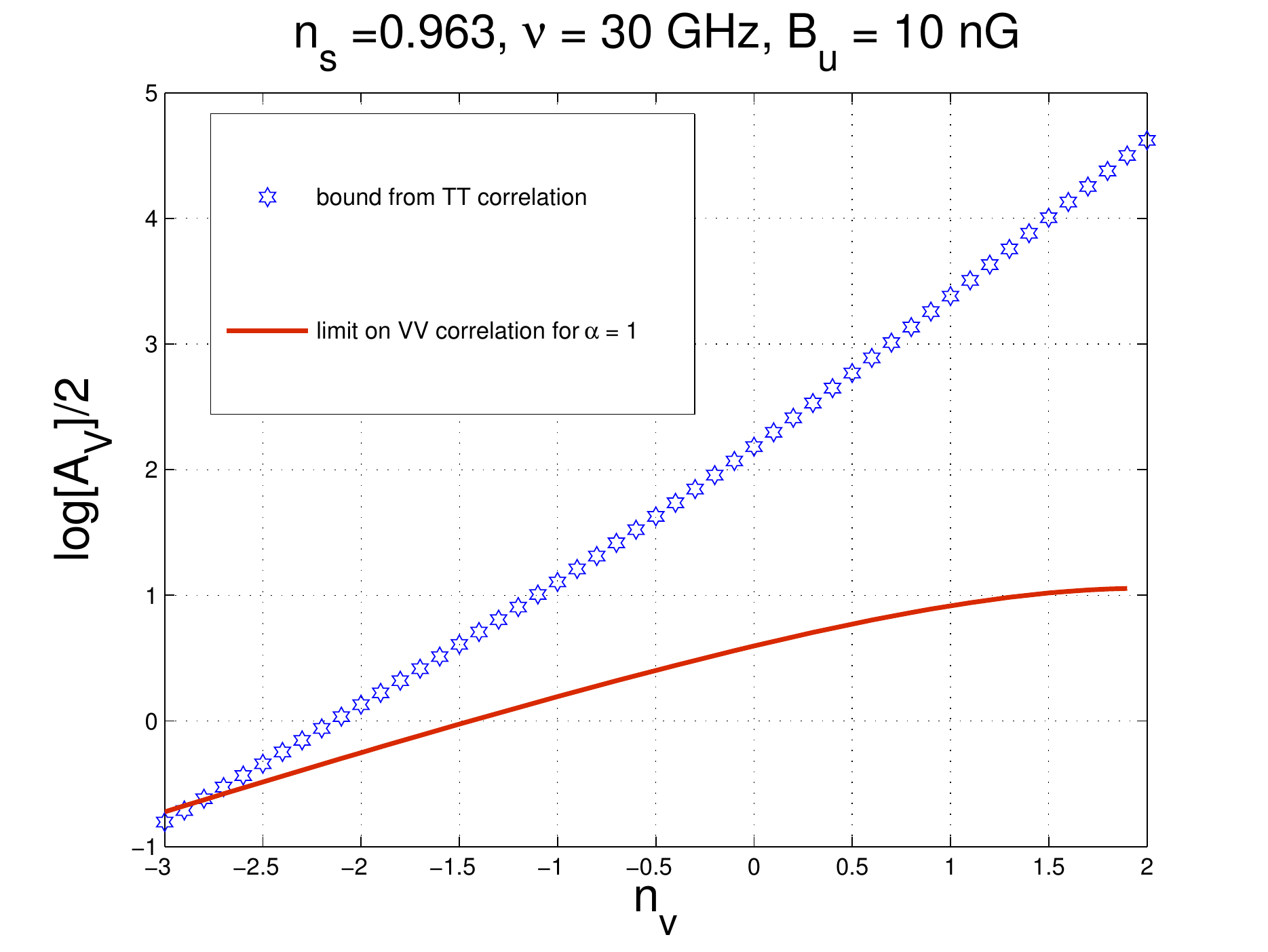}
\includegraphics[height=6cm]{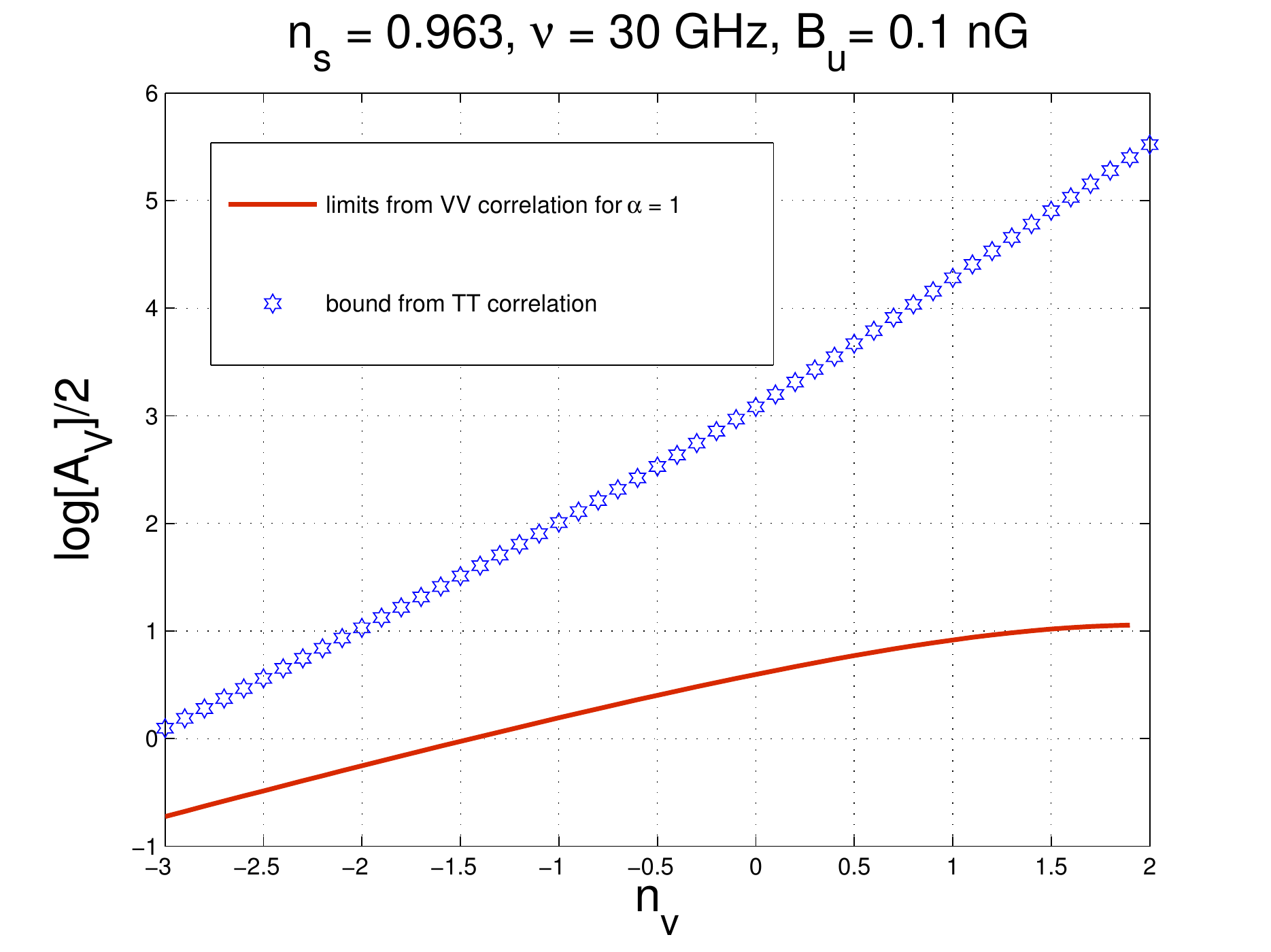}
\caption[a]{The bounds on the $V$-mode power spectrum are illustrated 
in terms of the amplitude and of the spectral index. The starred points correspond 
to the bounds arising from  Eqs. (\ref{boundTT13}) and (\ref{boundTT14}).
The full lines corresponds to Eq. (\ref{limit8}) with the values of $\alpha$ 
reported in each legend.}
\label{figure18}      
\end{figure}
The latter expression can be explicitly computed and the result is: 
\begin{eqnarray}
&& C_{\ell}^{(\mathrm{VV})} = \frac{9 \pi^2}{8} {\mathcal A}_{\mathrm{V}} \, \biggl(\frac{k_{0}}{k_{\mathrm{p}}}\biggr)^{n_{v} -1} {\mathcal V}(\ell, n_{v}),
\label{VM5}\\
&& {\mathcal V}(\ell,n_{v}) = \frac{\ell^2}{(2 \ell +1)^2} {\mathcal V}_{1}(\ell, n_{v}) + \frac{(\ell +1)^2}{(2\ell +1)^2} {\mathcal V}_{2}(\ell, n_{v}) 
- \frac{2 \ell (\ell +1)}{(2\ell +1)^2} {\mathcal V}_{3}(\ell, n_{v}),
\label{VM6}
\end{eqnarray}
where the functions ${\mathcal V}_{1}(\ell, n_{v})$, ${\mathcal V}_{2}(\ell, n_{v})$ and ${\mathcal V}_{3}(\ell, n_{v})$ are given as ratios of products of Gamma functions:
\begin{eqnarray}
{\mathcal V}_{1}(\ell,n_{v}) &=& \frac{1}{2 \sqrt{\pi}} \frac{\Gamma\biggl(\frac{3}{2} - \frac{n_{v}}{2}\biggr)
\Gamma\biggl(\ell - \frac{3}{2} + \frac{n_{v}}{2} \biggr)}{\Gamma\biggl(2 - \frac{n_{v}}{2}\biggr)\, \Gamma\biggl(\frac{3}{2} + \ell - \frac{n_{v}}{2} \biggr)},
\label{VM7}\\
 {\mathcal V}_{2}(\ell,n_{v}) &=&  \frac{1}{2 \sqrt{\pi}} \frac{\Gamma\biggl(\frac{3}{2} - \frac{n_{v}}{2}\biggr) \Gamma\biggl(\ell - \frac{1}{2} + \frac{n_{v}}{2}\biggr)}{\Gamma\biggl(2 - \frac{n_{v}}{2} \biggr) \Gamma\biggl(\frac{7}{2} +\ell - \frac{n_{v}}{2} \biggr)},
\label{VM8}\\
{\mathcal V}_{3}(\ell,n_{v}) &=&  \frac{(2- n_{v})}{4 \sqrt{\pi}} \frac{\Gamma\biggl(\frac{3}{2} - \frac{n_{v}}{2}\biggr) \Gamma\biggl(\ell - \frac{1}{2} + \frac{n_{v}}{2} \biggr)}{\Gamma\biggl(3 - \frac{n_{v}}{2} \biggr) \Gamma\biggl(\frac{5}{2} + \ell - \frac{n_{v}}{2} \biggl)}.
\label{VM9}
\end{eqnarray}
As previously done,  it is practical to deduce a simplified expression valid in the limit $\ell > 1$:
\begin{equation}
{\mathcal V}(\ell,n_{v}) = \frac{\ell^{n_{v} -3}}{2 \sqrt{\pi} ( 4 - n_{v})} \frac{\Gamma\biggl(\frac{3 - n_{v}}{2}\biggr)}{\Gamma\biggl(\frac{4 - n_{v}}{2}\biggr)}\biggl[ 1 + {\mathcal O}\biggl(\frac{1}{\ell}\biggr)\biggr].
\label{VM10}
\end{equation}
Since ${\mathcal A}_{\mathrm{V}}$ has been independently bounded from the analysis of the $TT$ and of the 
$EE$ angular power spectra, the $V$-mode angular power spectrum is also bounded. 
In Fig. \ref{figure18} the bounds stemming from the $V$-mode contribution to the $TT$
power spectrum are summarized for different values 
of the magnetic field intensity. In all four plots on the vertical axis we report 
the common logarithm of $\sqrt{{\mathcal A}_{\mathrm{V}}}$ 
while on the horizontal axis the corresponding spectral index is illustrated. 
\begin{figure}[!ht]
\centering
\includegraphics[height=6.cm]{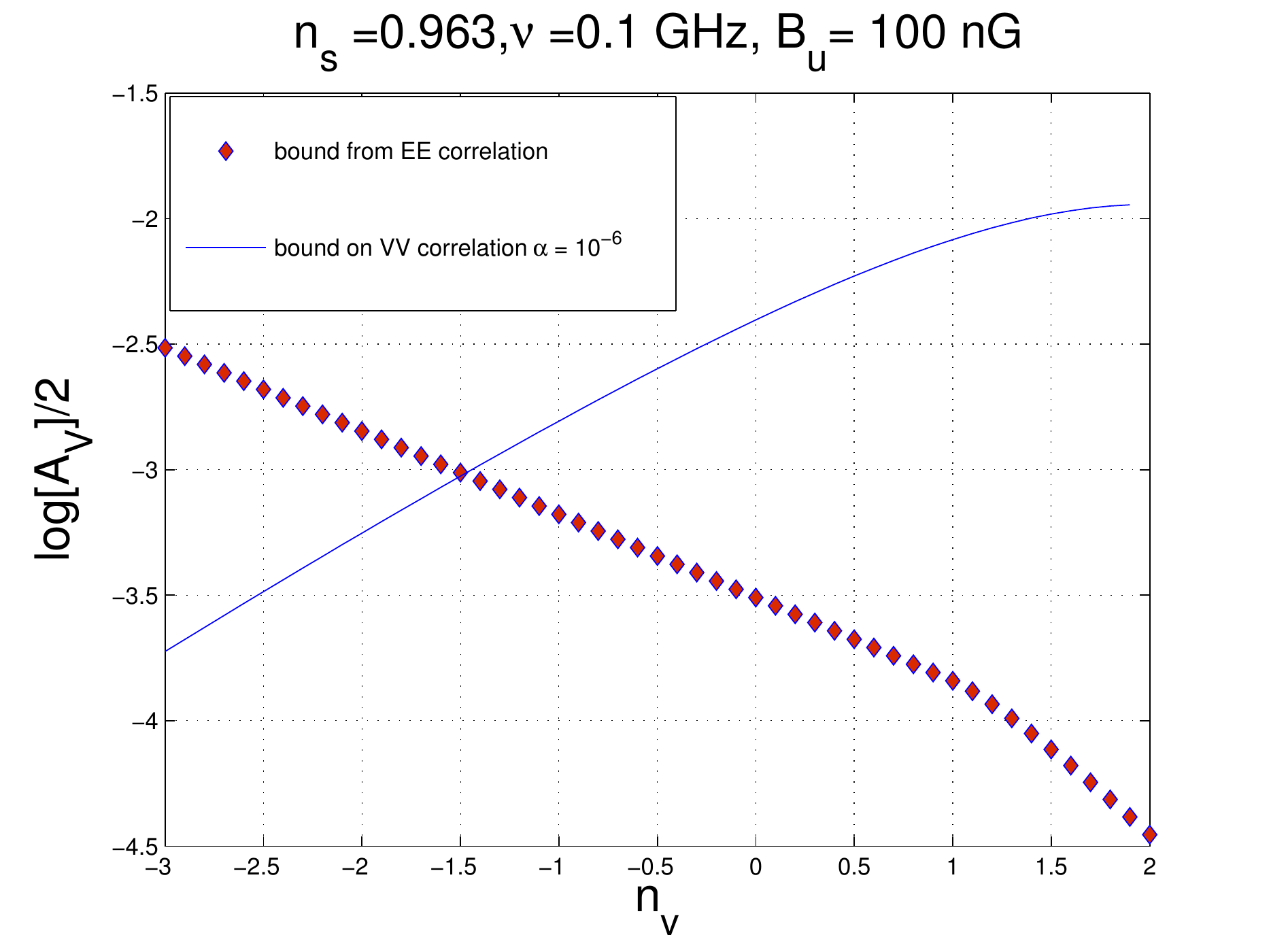}
\includegraphics[height=6.cm]{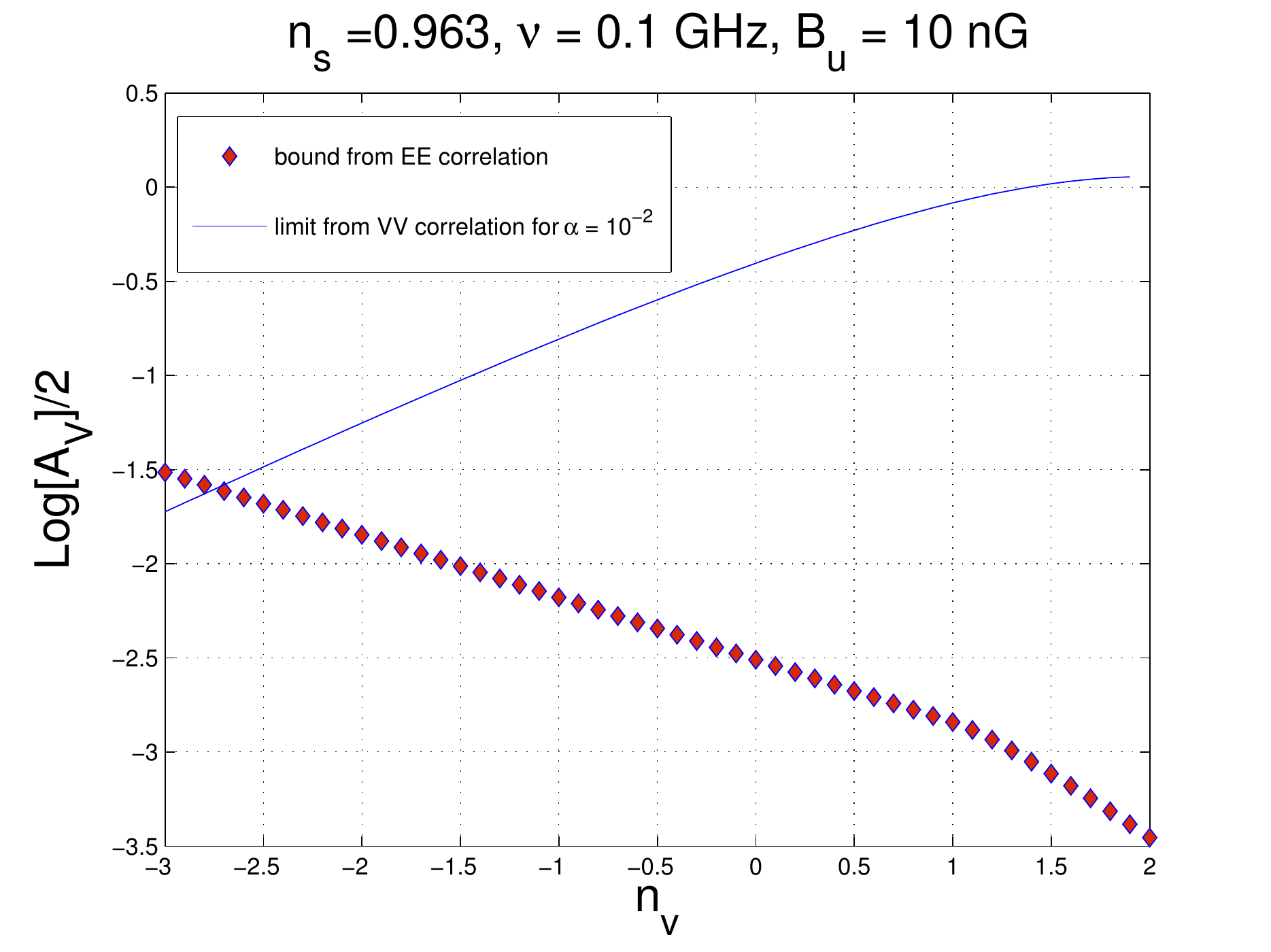}
\includegraphics[height=6.cm]{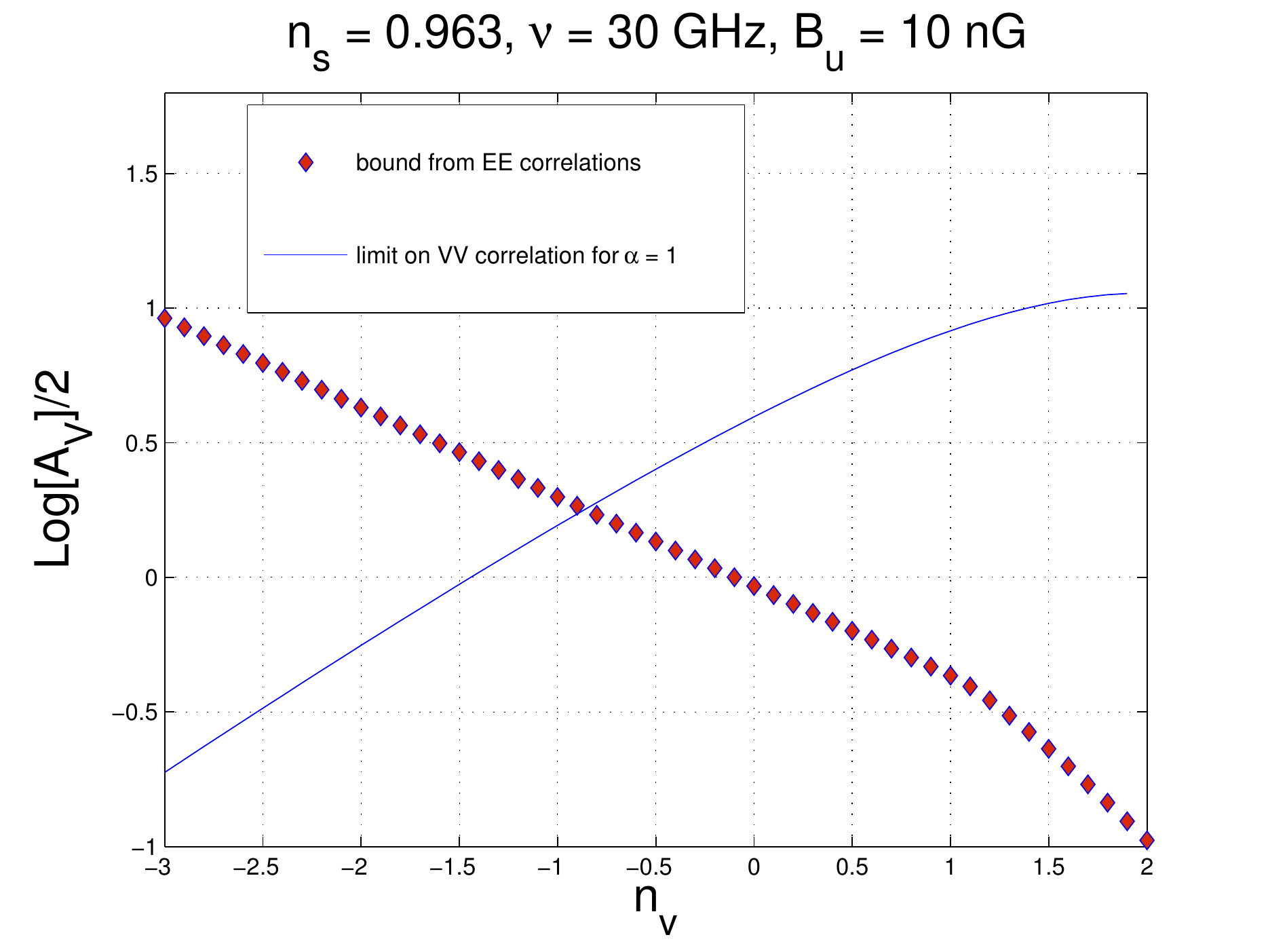}
\includegraphics[height=6cm]{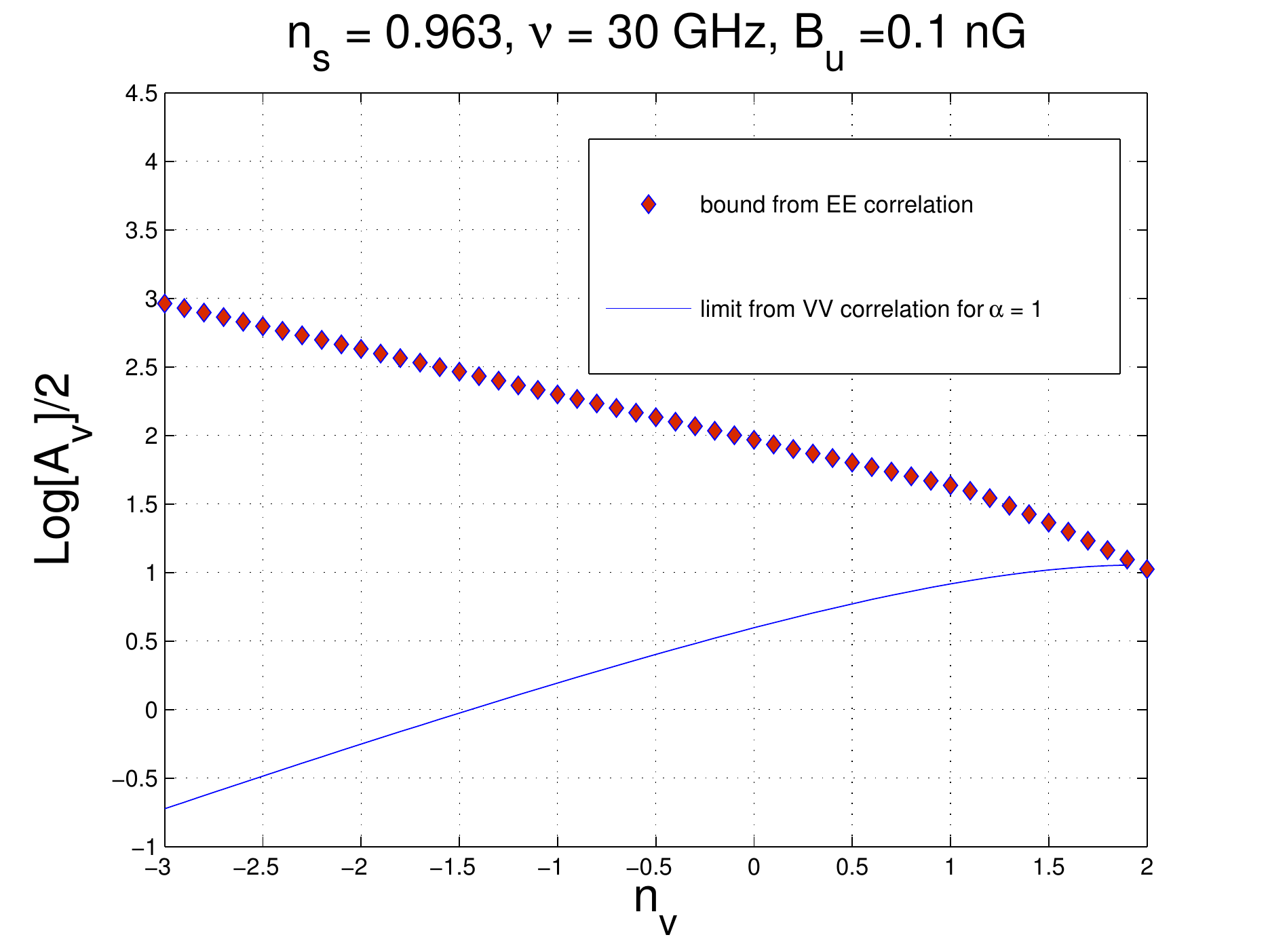}
\caption[a]{The bounds on the V-mode power spectrum are illustrated as they arise 
from the $EE$ correlations. The bounds derived in section \ref{sec4} are illustrated and compared 
with the potentially direct bounds parametrized, as in Fig. \ref{figure18}, in  terms of 
different values of $\alpha$ (see Eq. (\ref{limit8}).}
\label{figure20}      
\end{figure}
In Fig. \ref{figure18} with the full line we report the limit on 
$\sqrt{{\mathcal A}_{\mathrm{V}}}$ stemming from Eq. (\ref{limit8}) 
in terms of the corresponding value of $\alpha$. The various curves are 
 obtained by using, at the left hand side of Eq. (\ref{limit8}) 
the expression of Eq. (\ref{VM6}) appropriately averaged over the 
multipole range. Always in Fig. \ref{figure4} the starred points 
correspond to the bound on $\sqrt{{\mathcal A}_{\mathrm{V}}}$ 
derived in Eqs. (\ref{boundTT13}) and (\ref{boundTT14}). 

The results of Fig. \ref{figure18} suggest that 
for sufficiently large frequencies and for sufficiently small magnetic field 
intensity the bounds derived from the $TT$ correlations are 
not competitive with potential direct limits. This aspect can  be appreciated from the 
two bottom plots of Fig. \ref{figure4} where already a value 
$\alpha =1 $ would imply a more stringent limit 
on $\sqrt{{\mathcal A}_{\mathrm{V}}}$.  Notice that the allowed region is below the
full line (if the limit of Eq. (\ref{limit8}) is considered) or below the 
starred points if the limit of Eqs. (\ref{boundTT13}) and (\ref{boundTT14}) is 
enforced.  

In connection with Figs. \ref{figure18} and \ref{figure20} there are three possible situations.
The full line could always be above the starred line: this never happens 
in the case of Fig. \ref{figure18} but it would simply mean that any indirect limit 
is more stringent than the direct one.
If the full line is below the starred line the indirect limit from the $TT$ correlation 
is always compatible with the direct searches: this always happens if 
the magnetic field is sufficiently small (see, e.g. Fig. \ref{figure18} bottom left plot).
Finally the full line may cross the starred points: this is the most realistic 
situation in the light of the present and forthcoming direct limits on circular dichroism.

By looking at Fig. \ref{figure18} it is plausible that, depending upon the frequency of the experiment, 
magnetic fields $B_{\mathrm{u}}= {\mathcal O}(100\,\mathrm{nG})$ can be 
directly excluded for $\nu \simeq \mathrm{GHz}$ and with a sensitivity 
$\alpha \simeq 10^{-6}$ which would imply, in terms of Eq. (\ref{limit8}),
direct upper limits on the V-mode power spectrum ${\mathcal O}(\mu K)$ 
for $\ell < 40$. 

 In Fig. \ref{figure20} the same absolute bounds illustrated in Fig. \ref{figure18} are now compared with the bounds from the analysis of the $EE$ correlations which are numerically more significant, 
 especially for large spectral indices (i.e. $n_{v} >1$). As in the case of 
Fig. \ref{figure18} sufficiently small values of the magnetic field intensity 
make the indirect bounds rather loose in comparison with direct limits.
There are however numerical differences. 
From the top right and bottom left plots of Fig. \ref{figure20}
magnetic fields $B_{\mathrm{u}}= {\mathcal O}(10\,\mathrm{nG})$ can be 
directly excluded for $\nu \simeq \mathrm{GHz}$ and with a sensitivity 
$\alpha \simeq 10^{-6}$. The bounds stemming from the $EE$ correlations 
are therefore more stringent than the ones derived in the case of the $TT$ correlations. 

It is finally appropriate to mention that the frequency range assumed in the present 
discussion is in the GHz range because previous bounds, even if loose, were 
set over those frequencies. It is however tempting to speculate that, in a far future, 
microwave background measurements could be possible even below the GHz. In this case 
direct bounds will certainly  be more stringent but huge foregrounds 
might make this speculation forlorn (see, in this connection, \cite{S1,S2}). 

All in all, if a $V$-mode polarization (not correlated with the adiabatic mode) 
is present prior to matter-radiation equality both the $TT$ and the $EE$ 
power spectra are affected in a computable manner.
Specific constraints can then be inferred in terms of the amplitude and of the spectral 
index of the $V$-mode power spectrum. Improved direct experimental limits on the $VV$ 
correlations could be used for setting a limit on the magnetic field intensity.
For experimental devices operating in the 
GHz range, direct limits on the circular dichroism imply constraints  
on pre-decoupling magnetic fields in the $10$ nG range. Conversely, the current limits 
on large-scale magnetic fields derived from the distortions of the $TT$, $TE$ and $EE$
correlations (in the nG range) are compatible with current bounds on
the primordial dichroism.  Improved bounds 
on the $V$-mode polarization are not only interesting in their own right but they
might have rewarding phenomenological implications. Direct limits on the $V$-mode power spectrum 
in the range ${\mathcal O}(0.01\, \mathrm{mK})$ imply limits on  ${\mathcal A}_{\mathrm{V}}$
ranging from  ${\mathcal O}(10^{-8})$ to  ${\mathcal O}(10^{-4})$ depending on the value of 
the spectral index and for angular scales larger than ${\mathcal O}(1\, \mathrm{deg})$.

\newpage
\renewcommand{\theequation}{7.\arabic{equation}}
\setcounter{equation}{0}
\section{An ongoing trialogue}
\label{sec7}
As argued more than fifty years ago by Hoyle and Zeldovich, if the origin of  large-scale 
magnetism is primordial (as opposed to astrophysical), magnetic random fields
 evolving in the primeval plasma prior to the decoupling of radiation from matter 
 must necessarily affect the microwave background observables. 
During the past decade the magnetized temperature and polarization anisotropies 
have been analyzed in the context of a serendipitous trialogue 
involving plasma physics, general relativity and astrophysics. 
This ongoing effort led to a number of interesting progresses so that today 
the problem is correctly formulated, at least in principle if not always in practice.
Interesting constraints on magnetic random fields in the nG range 
have been established. The five WMAP releases and the Planck results provided 
a steady quantitive refinement of existing bounds 
(sometimes obtained with comparatively primitive methods). 
While the overall consistency of different approaches is 
rewarding from the theoretical viewpoint, 
it is fair to say that the primordial nature of large-scale 
magnetic fields has been neither confirmed nor ruled out. 

Various reference sets of Cauchy data 
of the Einstein-Boltzmann hierarchy are customarily arranged to 
scrutinize the microwave background observables in the context 
of the concordance paradigm and of its immediate extensions.
The very same strategy is a fortiori mandatory 
when the magnetic random fields are dynamical. 
Even if not all the aspects of this rich theoretical 
framework have been fully explored, the guiding logic developed 
in the past decade is that magnetic random fields cannot 
be generically excluded or discovered by only refining the tools of the 
data analyses: it is essential to understand what to look 
for in the data and what kind of initial conditions are more 
or less physical. Alternatively one should envisage specific 
tests that are independent on the initial data: 
this complementary tactic proved to be more difficult so far.

There are at least three general 
problems to be addressed in the near future if 
our course of action is to prove sound 
and effective. At the moment the magnetized initial conditions of 
the Einstein-Boltzmann hierarchy have been more or less classified 
but it is unclear which ones are more physical. While the adoption of 
the conventional adiabatic paradigm found a strong justification  
in the relative position of the first acoustic peak and of the first 
anticorrelation peak  of the cross-correlation between temperature 
and polarization, an analog model-independent test is not yet 
available in the case of magnetized perturbations. Similarly, 
while in the conventional case there are now good reasons 
to eliminate the entropic initial conditions from the Cauchy data 
of the Einstein-Boltzmann hierarchy, we do not have any specific 
rationale to exclude possible compensations effects coming from 
the interference of entropic modes and magnetized initial conditions.

Assuming a better understanding of the Cauchy data 
the second interesting area of investigation involves an improved 
theoretical scrutiny of the Faraday effect which is certainly 
one of the best model-independent tests for the 
primordial nature of large-scale magnetism. It could be 
that a $B$-mode polarization of tensor origin will be soon 
discovered and we shall therefore be in the situation of 
considering the interplay of magnetic and non-magnetic sources 
of $B$-mode polarization. These analyses seem particularly 
urgent in the light of the forthcoming full-sky surveys 
of Faraday rotations which could even reach frequencies 
of $25$ GHz.

It would be highly desirable to see a steady observational 
progress in the analysis of circular polarizations of the 
microwave background. Further scrutiny of these aspects 
is important in its own right since the direct analyses 
of circular polarizations are at the moment the most 
challenging in the remarkable agenda of the observational 
cosmologists. The study of circular dichroism is not more forlorn 
than other signals which are often invoked as conceptually 
important to consider but observationally difficult to assess. 
While the systematic effects plaguing the measurements of 
the $V$-mode power spectra differ from the case of linear polarizations,
whether or not they are less severe depends also upon 
the features of the instrument and on the specific frequency band.

It is finally plausible to expect that during the forthcoming score year a new 
channel for the observations of magnetic random fields will become hopefully 
available. Magnetic random fields 
may lead to a stochastic backgrounds of relic gravitational waves
 in the frequency interval ranging between few $\mu$Hz and $10$ kHz. 
 This intermediate range encompasses the operating windows of space-borne
 interferometers (hopefully available twenty years from now) and of 
 terrestrial detectors (already available but still insensitive to stochastic backgrounds of relic gravitons 
of cosmological origin). Hypermagnetic fields possibly 
present before and after the electroweak phase transition typically lead to a 
stochastic background which may even be $8$ orders of magnitude 
larger than the conventional inflationary contribution characterized by 
a spectral energy density in critical units 
${\mathcal O}(10^{-17})$. 

Given the encouraging progresses of the past decade it is fair to 
expect that the forthcoming years will be an exciting 
moment both for theory and for observations. This will be 
even more true if the forthcoming  flow of data will not only be 
regarded as a source of improved precision but also as 
a concrete inspiration for the scrutiny of novel and potentially 
unexpected paradigms.

\section*{Aknowledgements}
It is a pleasure to thank G. Sironi for sharing his wise remarks all 
along the last decade. The kind support and the encouragement of 
D. Pedrini is also acknowledged. Last but not least I wish to thank 
J. Vigen, T. Basaglia, A. Gentil-Beccot and S. Rohr of the CERN 
Scientific Information Service for their help and 
for their everlasting patience.
\newpage

\begin{appendix}
\renewcommand{\theequation}{A.\arabic{equation}}
\setcounter{equation}{0}
\section{Isotropic random fields}
\label{APPA}
\subsection{Scalar, vector and tensor random fields}
For coincident (conformal) times the two-point functions of isotropic scalar, vector and tensor random fields in real space only depend on the distance $r = |\vec{x} - \vec{y}|$ between the two spatially separated points:
\begin{eqnarray}
&& {\mathcal C}^{(s)}(r, \tau)= \langle {\mathcal R}(\vec{x},\tau) {\mathcal R}(\vec{y},\tau) \rangle, 
\label{IRF1}\\
&& {\mathcal C}^{(v)}_{ij}(r, \tau)=\langle B_{i}(\vec{x},\tau) B_{j}(\vec{y},\tau) \rangle,
\label{IRF2}\\
&& {\mathcal C}^{(t)}_{ijmn}(r, \tau)= \langle h_{ij}(\vec{x},\tau) h_{mn}(\vec{y},\tau) \rangle.
\label{IRF3}
\end{eqnarray}
In Eq. (\ref{IRF1}) ${\mathcal R}(\vec{x},\tau) $ denotes a generic scalar which can coincide, for instance, with the curvature perturbations on comoving orthogonal hypersurfaces introduced in Eqs. (\ref{NM2}), (\ref{NMR4}) or (\ref{NMODE5}). Similarly, in Eq. (\ref{IRF2})  $B_{i}(\vec{x},\tau) $ is valid for a three-dimensional vector but it also applies to  
the comoving electric and magnetic fields introduced  in Eqs.(\ref{S4}) and (\ref{S4a}). Finally, in Eq. (\ref{IRF3})  $h_{ij}(\vec{x},\tau)$ is a rank-two tensor in three-dimensional Euclidean space and it describes, for example, 
the tensor modes of the geometry or the anisotropic stress. 
According to the present conventions the Fourier transform of ${\mathcal R}(\vec{x},\tau)$ is defined as:
\begin{equation}
{\mathcal R}(\vec{x},\tau) = \frac{1}{(2\pi)^{3/2}} \int d^3 k {\mathcal R}(\vec{k},\tau) e^{- i \vec{k}\cdot\vec{x}}.
\label{IRFSC1}
\end{equation}
The power spectrum $P_{{\mathcal R}}(k)$ is simply given by
\begin{equation}
\langle {\mathcal R}(\vec{x},\tau) \,{\mathcal R}(\vec{x} + \vec{r},\tau) \rangle = \int d \ln{k} \,P_{{\mathcal R}}(k,\tau) \, j_{0}(k r), \qquad j_{0}(kr) = \frac{\sin{k r}}{k r},
\label{IRFSC2}
\end{equation}
where $j_{0}(kr)$ is the spherical Bessel function of zeroth-order \cite{abr1,abr2}. In Fourier space the two-point function 
of Eq. (\ref{IRFSC2}) is
\begin{equation} 
\langle {\mathcal R}(\vec{k},\tau) \,{\mathcal R}(\vec{p},\tau) \rangle = \frac{2 \pi^2}{k^3} P_{{\mathcal R}}(k,\tau)
\delta^{(3)}(\vec{k} + \vec{p}).
\label{IRFSC3}
\end{equation}
The power spectrum of Eq. (\ref{IRFSC3}) describes 
the large-scale inhomogeneities of the concordance paradigm and its explicit form has been introduced in Eq. (\ref{SPS}).
Equations (\ref{IRFSC2}) and (\ref{IRFSC3}) power spectrum $P_{{\mathcal R}}(k,\tau)$ (in Fourier space) has the  same dimensions of the correlation function (in real space).

The explicit expression of the two-point function of vector random fields given in Eq. (\ref{IRF2}) depends on three 
functions $M_{T}(r, \tau)$, $M_{L}(r,\tau)$ and $M_{G}(r,\tau)$ denoting, respectively, the {\em transverse}, the {\em longitudinal} and the {\em gyrotropic} components:
\begin{equation}
{\mathcal C}^{(v)}_{ij}(r, \tau) = M_{T}(r,\tau) p_{ij}(\hat{r}) + M_{L}(r,\tau) \hat{r}_{i} \hat{r}_{j} + M_{G}(r,\tau) \epsilon_{i j \ell} \hat{r}^{\ell},
\label{IRFVEC1}
\end{equation}
where $p_{ij}(\hat{r})= \delta_{ij} - \hat{r}_{i} \hat{r}_{j}$ is the transverse the projector and $\hat{r}^{i} = r^{i}/r$ is the unit vector. The gyrotropic contribution, proportional to the Levi-Civita totally antisymmetric symbol, is rotationally-invariant 
but not parity-invariant; if present it implies the 
existence of a non-vanishing magnetic gyrotropy\footnote{We prefer to use the terminology magnetic 
gyrotropy (instead of helicity) since the gyrotropy (unlike the helicity densities sometimes discussed in the literature) is gauge-invariant.} defined as $\vec{B}\cdot \vec{\nabla} \times \vec{B}$.  
Whenever the vector random fields are divergenceless, their transverse and the longitudinal components 
will be subjected to the following further condition:
\begin{equation}
 \frac{\partial M_{L}}{\partial r} + \frac{2}{r} (M_{L} - M_{T}) =0,
\label{IRFVEC2}
\end{equation}
which follows  by simply imposing that ${\mathcal C}^{(v)}_{ij}(r, \tau)$ be divergenceless.
It could also happen that the two-point function be traceless (i.e. $2 M_{T} + M_{L} =0$)
but this is not what happens in the case of the magnetic random fields. In complete analogy with the scalar case of Eq. (\ref{IRFSC1}) the Fourier transform of $B_{i}(\vec{x},\tau)$ is:
\begin{equation}
B_{i}(\vec{x},\tau) = \frac{1}{(2\pi)^{3/2}} \int d^{3} k \, B_{i}(\vec{k},\tau) \, e^{- i \vec{k}\cdot\vec{x}}.
\label{IRFVEC3}
\end{equation}
It follows from Eq. (\ref{IRFVEC3}) that the vector power spectra are given by:
\begin{equation} 
\langle B_{i}(\vec{k},\tau) \,B_{j}(\vec{p},\tau) \rangle = \frac{2 \pi^2}{k^3} 
\biggl[P_{B}(k,\tau) \, p_{ij}(\hat{k}) + {\mathcal P}_{G}(k,\tau) \epsilon_{i j \ell} \hat{k}^{\ell}\biggr]\delta^{(3)}(\vec{k} + \vec{p}).
\label{IRFVEC4}
\end{equation}
If $B_{i}(\vec{x},\tau)$ coincides with the magnetic field its dimensions will be of an inverse 
area (i.e. $L^{-2}$) while $B_{i}(\vec{k}, \tau)$ will obviously have 
dimensions of a length (i.e. $L$). This elementary remark shows in explicit terms that the magnetic power 
spectrum in Fourier space (i.e. Eq. (\ref{IRFVEC4})) has the same dimensions of the two point function in coordinate space 
(i.e.  Eq. (\ref{IRF2})). Since the two-point function of magnetic random fields has the dimensions of an energy density in real space, the corresponding power spectrum  will be measured in the same units (e.g. $\mathrm{Gauss}^2$ or $\mathrm{Tesla}^2$). 
In the case of the magnetic field, 
$M_{T}(r, \tau)$, $M_{L}(r,\tau)$ and $M_{G}(r,\tau)$ will be given by:
\begin{eqnarray}
M_{T}(r, \tau) &=& \int d\ln{k} \,P_{B}(k,\tau)\,\biggl[ \frac{\cos{k r}}{k^2 r^2} + \frac{k^2 r^2 -1}{k^3 r^3} \sin{k r}\biggr],
\label{IRFVEC5}\\
M_{L}(r, \tau) &=&2 \int d\ln{k} \,P_{B}(k,\tau) \,\frac{[ \sin{k r} - k r \cos{k r}]}{k^3 r^3} ,
\label{IRFVEC6}\\
M_{G}(r, \tau) &=& \int d\ln{k}\, P_{G}(k,\tau) \,\biggl[\frac{\cos{k r}}{k r} - \frac{\sin{k r}}{k^2 r^2}\biggr].
\label{IRFVEC7}
\end{eqnarray}
Note finally that the two vector polarizations $\hat{e}_{i}^{\alpha}(\hat{k})$ (with $\alpha=1,\,2$) of the divergenceless and traceless random vector fields obey $\sum_{\alpha=1,\,2} \hat{e}^{(\alpha)}_{i}(\hat{k}) \hat{e}^{(\alpha)}_{j}(\hat{k}) = p_{ij}(\hat{k})$.

The explicit form of the two-point function for 
tensor random fields given in Eq. (\ref{IRF3}) can be more explicitly written as:
\begin{eqnarray}
{\mathcal C}^{(t)}_{ijmn}(r, \tau) &=& N_{T}(r,\tau)[ p_{im}(\hat{r}) p_{jn}(\hat{r}) 
+ p_{in}(\hat{r}) p_{j m}(\hat{r}) - p_{ij}(\hat{r}) p_{mn}(\hat{r})] 
\nonumber\\
&+& N_{L}(r,\tau) \hat{r}_{i} \, \hat{r}_{j} \, \hat{r}_{m} \, \hat{r}_{n}
\nonumber\\
&+& N_{G}(r,\tau) [ \epsilon_{ij k} \epsilon_{m n \ell}\, \hat{r}^{k} \,\hat{r}^{\ell} 
+ \epsilon_{i m k} \epsilon_{j n \ell}\, \hat{r}^{k} \,\hat{r}^{\ell} + 
\epsilon_{i n k} \,\epsilon_{j m \ell} \,\hat{r}^{k}\, \hat{r}^{\ell} ].
\label{IRFTEN1}
\end{eqnarray}
Following the same conventions of Eqs. (\ref{IRFSC1}) and (\ref{IRFVEC3}) the Fourier transform of $h_{ij}(\vec{x},\tau)$ is defined as 
\begin{equation}
h_{ij}(\vec{x},\tau) = \frac{1}{(2\pi)^{3/2}} \int d^{3} k \, h_{i j}(\vec{k}, \tau) \, e^{- i \vec{k}\cdot \vec{x}}.
\label{IRFTEN2}
\end{equation}
Neglecting the terms that break explicitly parity and that are antisymmetric in $(ij)$ or $(mn)$, the two-point function traced over the indices and the correlation function in Fourier space are, respectively,
\begin{eqnarray}
\langle h_{ij}(\vec{x},\tau) h_{ij}(\vec{x}+\vec{r}, \tau) \rangle &=& \int d \ln{k} \, P_{\mathrm{T}}(k,\tau) j_{0}(k r),
 \label{IRFTEN3}\\
\langle h_{ij}(\vec{k},\tau) \,h_{mn}(\vec{p},\tau) \rangle &=& \frac{2 \pi^2}{k^3} P_{\mathrm{T}}(k,\tau) \,{\mathcal S}_{ijmn}(\hat{k}) \,\delta^{(3)}(\vec{k} + \vec{p}), 
\label{IRFTEN4}\\
{\mathcal S}_{ijmn}(\hat{k}) &=& \frac{1}{4} \biggl[p_{i m}(\hat{k}) p_{j n}(\hat{k}) + p_{i n}(\hat{k}) p_{j m}(\hat{k}) - p_{i j}(\hat{k}) p_{m n}(\hat{k})\biggr],  
\label{IRFTEN5}
\end{eqnarray}
 where ${\mathcal S}_{ijmn}(\hat{k})$ is traceless and divergenceless as implied by the requirement $h_{i}^{i}= \partial_{i} h^{ij}=0$:
 \begin{eqnarray}
\hat{k}^{i} {\mathcal S}_{ij mn} &=& \hat{k}^{j} {\mathcal S}_{ij mn}=
\hat{k}^{m} {\mathcal S}_{ij mn} = \hat{k}^{n} {\mathcal S}_{ij mn} =0,
\label{IRFTEN6}\\
{\mathcal S}_{i i m n}(\hat{k}) &=& {\mathcal S}_{i j m m}(\hat{k}) =0, \qquad {\mathcal S}_{i j i j}(\hat{k}) = 1.
\end{eqnarray}
The two tensor polarizations are defined in this paper as 
\begin{equation}
\hat{e}^{\oplus}_{ij}(\hat{k}) = \hat{m}_{i} \hat{m}_{j} - \hat{n}_{i} \hat{n}_{j}, \qquad 
\hat{e}^{\otimes}_{ij}(\hat{k}) = \hat{m}_{i} \hat{n}_{j} + \hat{n}_{i} \hat{m}_{j},
\label{IRFTEN8}
\end{equation}
where $\hat{m}$, $\hat{n}$ and $\hat{k}$ are a triplet of mutually orthogonal unit vectors. The sum over the polarizations leads, respectively, to 
\begin{equation}
 \sum_{\beta=\oplus,\, \otimes} \hat{e}^{(\beta)}_{ij}(\hat{k}) \hat{e}^{(\beta)}_{m n}(\hat{k}) = 
p_{i m}(\hat{k}) p_{j n}(\hat{k}) + p_{i n}(\hat{k}) p_{j m}(\hat{k}) - p_{i j}(\hat{k}) p_{m n}(\hat{k}) = 4 {\mathcal S}_{i j m n}(\hat{k}).
\label{IRFTEN9}
\end{equation}

\subsection{Vector identities and further power spectra}
The explicit components of the canonical energy-momentum tensor of Eq. (\ref{COV7})  are:
\begin{eqnarray}
T_{0}^{(EM)\,0} &=& \delta_{\mathrm{s}}\rho_{\mathrm{E}}(\vec{x},\tau) + \delta_{\mathrm{s}}\rho_{\mathrm{B}}(\vec{x},\tau),
\label{T00em}\\
T_{i}^{(EM)\,j}  &=& - \biggl[ \delta_{\mathrm{s}}p _{\mathrm{E}}(\vec{x},\tau) +  \delta_{\mathrm{s}}p_{\mathrm{B}}(\vec{x},\tau)\biggr] \delta_{i}^{j} +
\Pi^{(\mathrm{E})\, j}_{i}(\vec{x},\tau)  + \Pi^{(\mathrm{B})\, j}_{i}(\vec{x},\tau) ,
\label{Tijem}\\
T_{0}^{(EM)\,i}  &=& \frac{1}{4 \pi a^4} \biggl(\vec{E} \times \vec{B}\biggr)^{i},
\label{T0iem}
\end{eqnarray}
where $\vec{E}$ and $\vec{B}$ are the comoving electric 
and magnetic fields introduced in Eqs. (\ref{S1}), (\ref{S4}) and 
(\ref{S4a}). The energy densities already introduced in Eq. (\ref{GEOM1}) are preceded by $\delta_{\mathrm{s}}$ since they only affect the evolution of the scalar modes of the geometry. Conversely the magnetic and the electric anisotropic stresses $ \Pi^{(\mathrm{B})}_{i j}$ and 
$ \Pi^{(\mathrm{E})}_{i j}$ (already defined in Eq. (\ref{GEOM11})) 
not only contribute to the evolution of the scalar modes but also to the vector and tensor fluctuations of the geometry. In the case of the electric fields the following vector identity can be easily derived:
\begin{equation}
\frac{ \vec{\nabla}\cdot[ (\vec{\nabla}\times\vec{E}) \times \vec{E}]}{4 \pi a^{4} (\rho_{\gamma} + p_{\gamma})} = 
\nabla^2 \biggl[ \sigma_{E} - \frac{\Omega_{E}}{4} \biggr] - \frac{3}{16\pi \rho_{\gamma} a^4} (\vec{\nabla} \cdot \vec{E})^2
\label{VECID1}
\end{equation}
where, $\Omega_{\mathrm{E}}$ and $\sigma_{\mathrm{E}}$ have been introduced, respectively, 
in Eqs. (\ref{GEOM1a}) and (\ref{GEOM12}).
A similar vector identity  holds in the case of the magnetic field in the approximation where the total current is solenoidal namely
\begin{equation}
\frac{3}{4} \frac{\vec{\nabla}\cdot [ \vec{J} \times \vec{B}]}{a^4 \rho_{\gamma}} = \nabla^2 \sigma_{\mathrm{B}} - \frac{1}{4} \nabla^2 \Omega_{\mathrm{B}},
\label{VECID4}
\end{equation}
where, as in Eqs. (\ref{GEOM1a}) and (\ref{GEOM12}), we have referred the magnetic energy density 
and the corresponding anisotropic stress to the photon background.  
After simple algebra Eq. (\ref{VECID4}) can also be written as:
\begin{equation}
\nabla^2 \sigma_{\mathrm{B}} = \frac{3}{16\pi \rho_{\gamma} a^4} \partial_{i} B_{j} \partial^{j} B^{i} - 
\frac{1}{2} \nabla^2 \Omega_{\mathrm{B}}.
\label{VECID7}
\end{equation}
If the plasma is globally neutral also the electric field is solenoidal 
so that Eqs. (\ref{VECID1}) and (\ref{VECID4}) are symmetric.
In the slow description of the plasma modes summarized in Eq. (\ref{COLL5}) the total current, the magnetic fields and the electric field are all solenoidal.

Since $\Omega_{\mathrm{B}}(\vec{x},\tau)$ and $\sigma_{\mathrm{B}}(\vec{x},\tau)$ are scalars we can compute their associated power spectra. Recalling the conventions 
of Eqs. (\ref{IRFSC1}) and (\ref{IRFSC3}) in Fourier space their expression is: 
\begin{eqnarray}
\Omega_{\mathrm{B}}(\vec{q},\tau) &=& \frac{1}{(2\pi)^{3/2}} 
\frac{1}{8\pi a^4 \rho_{\gamma}} \int d^{3} k B_{i}(k,\tau) B^{i}(\vec{q} - \vec{k},\tau),
\label{VECID8}\\
\sigma_{\mathrm{B}}(\vec{q},\tau) &=& \frac{1}{(2\pi)^{3/2}} \frac{1}{16 \pi a^4 \rho_{\gamma}} \int 
d^{3} k \biggl[ \frac{3 ( q^{j} - k^{j}) k^{i}}{q^2} B_{j}(k,\tau)B_{i}(\vec{q} - \vec{k},\tau) 
\nonumber\\
&-& B_{i}(\vec{q} - \vec{k},\tau) B^{i}(\vec{k},\tau)\biggr].
\label{VECID9}
\end{eqnarray}
The correlation functions for $\Omega_{\mathrm{B}}(\vec{k},\tau)$ and $\sigma_{\mathrm{B}}(\vec{k},\tau)$ 
are then defined as 
\begin{eqnarray}
\langle \Omega_{\mathrm{B}}(\vec{q},\tau) \Omega_{\mathrm{B}}(\vec{p},\tau)\rangle &=& \frac{2\pi^2}{q^3} P_{\Omega}(q,\tau) \delta^{(3)}(\vec{q} + \vec{p}),
\nonumber\\
 \langle \sigma_{\mathrm{B}}(\vec{q},\tau) \sigma_{\mathrm{B}}(\vec{p},\tau)\rangle &=& \frac{2\pi^2}{q^3} P_{\sigma}(q,\tau) \delta^{(3)}(\vec{q} + \vec{p}).
\label{VECID10}
\end{eqnarray}
Defining, for the sake of simplicity, the following auxiliary scalar product  
\begin{equation}
\gamma(\vec{k}, \vec{q})= \frac{\vec{k} \cdot (\vec{q} - \vec{k})}{k |\vec{q} - \vec{k}|} = \frac{\hat{k} \cdot (\vec{q} - \vec{k})}{ |\vec{q} - \vec{k}|}, 
\end{equation}
the explicit expression of the two power spectra $P_{\Omega}(q,\tau)$ and $P_{\sigma}(q,\tau)$ is given in terms 
of the power spectra of the magnetic random fields as:
\begin{eqnarray}
P_{\Omega}(q,\tau) &=& \frac{q^{3}}{(2\pi)} \frac{1}{8\pi a^4 \rho_{\gamma}} \int 
d^{3} k \frac{P_{\mathrm{B}}(k,\tau)}{k^3} \frac{P_{\mathrm{B}}(|\vec{q} - \vec{k}|,\tau)}{|\vec{q} - \vec{k}|^3} \biggl[ 1 + 
\gamma^2(\vec{k}, \vec{q})\biggr],
\label{VECID11}\\
 P_{\sigma}(q,\tau) &=& \frac{q^3}{(2\pi) }\frac{1}{(16\pi a^4 \rho _{\gamma})^2} 
\int d^{3} k \frac{P_{\mathrm{B}}(k,\tau)}{k^3} \frac{P_{\mathrm{B}}(|\vec{q} - \vec{k}|,\tau)}{|\vec{q} - \vec{k}|^3}
\nonumber\\
&\times& \biggl\{1 + 
\gamma^2(\vec{k}, \vec{q}) +\frac{6 \,\vec{k}\cdot( \vec{q} - \vec{k})}{q^2} \biggl[ 1 - \gamma^2(\vec{k}, \vec{q}) \biggr]
\nonumber\\
&+& \frac{9 \,k^2 |\vec{q} - \vec{k}|^2}{q^4} \biggl[1 - \gamma^2(\vec{k}, \vec{q}) \biggr]^2 \biggr\}.
\label{VECID12}
\end{eqnarray}
The source terms for the evolution of the scalar modes of the geometry are visibly the most cumbersome since they 
involve the energy density, the pressure and the anisotropic stress. For the vector and for the tensor modes 
the source terms only involve the vector and the tensor components of the electromagnetic anisotropic stress. For instance, in the case of the magnetic fields we can always write:
\begin{eqnarray}
\Pi_{ij}^{(\mathrm{B})}(\vec{q},\tau) &=& \Pi_{ij}^{(scal,\mathrm{B})}(\vec{q},\tau) + \Pi_{ij}^{(vec,\mathrm{B})}(\vec{q},\tau) + \Pi_{ij}^{(ten,\mathrm{B})}(\vec{q},\tau),
\label{PItot}\\
\Pi_{ij}^{(scal,\mathrm{B})}(\vec{q},\tau) &=& \hat{q}_{i} \hat{q}_{j} \Pi_{ij}^{(\mathrm{B})}(\vec{q},\tau),
\label{PIS}\\
\Pi_{ij}^{(vec,\mathrm{B})}(\vec{q},\tau) &=&\biggl[ p_{i n}(\hat{q}) \, \hat{q}_{j} + p_{j n}(\hat{q}) \, \hat{q}_{i} \biggr] \hat{q}_{m} \Pi_{mn}^{(\mathrm{B})}(\vec{q},\tau),
\label{PIV}\\
\Pi_{ij}^{(tens,\mathrm{B})}(\vec{q},\tau) &=&\biggl[ p_{i m}(\hat{q}) \,p_{j n}(\hat{q})  + p_{j n}(\hat{q}) \, p_{i m}(\hat{q}) \biggr]  \Pi_{mn}^{(\mathrm{B})}(\vec{q},\tau),
\label{PIT}
\end{eqnarray}
where, by definition, the Fourier transform of the total magnetic anisotropic stress is given by:
\begin{equation}
 \Pi_{mn}^{(\mathrm{B})}(\vec{q},\tau) = \frac{1}{4 \pi a^4\, (2 \pi)^{3/2}} \int d^{3} k \biggl[ B_{m}(\vec{k}, \tau) B_{n}(\vec{q} - \vec{k},\tau) 
 - \frac{1}{3} B_{\ell}(\vec{k},\tau) B_{\ell}(\vec{q}-\vec{k},\tau) \delta_{mn} \biggr].
 \label{PIBB}
 \end{equation}
Equations (\ref{VECID8})--(\ref{VECID9}) as well as Eqs. (\ref{VECID10}), (\ref{VECID11}) and (\ref{VECID12}) 
can be rephrased in terms of the electric random fields at least as long as they 
are solenoidal, exactly as the magnetic random fields or the total current. 

\subsection{Parameters of the magnetized $\Lambda$CDM scenario}

Even if, according to Eq. (\ref{PSM1}),  the initial conditions of the Einstein-Boltzmann could be assigned directly in terms of the amplitude  $A_{\mathrm{B}}$ and of the spectral index, it is a common practice to trade $A_{\mathrm{B}}$ for the regularized magnetic energy density 
$B_{\mathrm{L}}^2$.  When the magnetic power spectra are blue (i.e. $n_{\mathrm{B}} > 1$) the energy density is predominantly concentrated over small scales; the opposite is true for the case of red spectra (i.e. $n_{\mathrm{B}} < 1$).  From Eqs. (\ref{IRFVEC1}) and (\ref{IRFVEC4}), $B_{\mathrm{L}}^2$ is the trace of the two-point function of the magnetic random fields:
\begin{equation}
B_{\mathrm{L}}^2= {\mathcal C}^{(v)}_{ii}(r, \tau)= \langle B_{i}(\vec{x},\tau) B^{i}(\vec{y},\tau) \rangle = 2 \int d\ln{k} P_{\mathrm{B}}(k,\tau)j_{0}(kr) W(k),
\label{REG1}
\end{equation}
where $W(k)$ is an appropriate window function which is not strictly necessary and could be simply replaced either
by an ultraviolet cut-off (in the case of blue spectra) or by an infra-red cut-off (in the case of red spectra).

When the spectrum is blue the energy density can be 
regularized over a typical comoving scale $L$ (which is related to the pivot wavenumber $k_{\mathrm{L}}$) by means of a Gaussian window function $W(k) = e^{-k^2L^2}$. 
Equation (\ref{REG1}) then implies:
\begin{equation}
B_{\mathrm{L}}^2(r) = (2\pi)^{1 - n_{\mathrm{B}}}  A_{\mathrm{B}} \Gamma\biggl(\frac{n_{\mathrm{B}} -1}{2}\biggr)  F_{11}\biggl(\frac{n_{\mathrm{B}}-1}{2},\frac{3}{2}, - \frac{r^2 k_{\mathrm{L}}^2}{16\pi^2}\biggr),
\label{REG2}
\end{equation}
where $F_{11}(a,b, z)$ is the Kummer function \cite{abr1,abr2}.
Since $\lim_{z\to 0} F_{11}(a,b, z) = 1$, 
\begin{equation}
B_{\mathrm{L}}^2 = \lim_{r\to0}  {\mathcal C}^{(v)}_{ii}(r, \tau) = A_{\mathrm{B}}^2 (2\pi)^{1 - n_{\mathrm{B}}} \Gamma\biggl(\frac{n_{\mathrm{B}} -1}{2}\biggr).
\label{REG3}
\end{equation}
When  $n_{\mathrm{B}} > 1$ we then have that $A_{\mathrm{B}} = 
(2\pi)^{n_{\mathrm{B}} -1} \, B_{\mathrm{L}}^2 /\Gamma[(n_{\mathrm{B}} -1)/2]$, as reported in Eq. (\ref{PSM2}).
In the radial integrals of Eqs. (\ref{VECID11}) and (\ref{VECID12}),  $ A_{\mathrm{B}}$ can be traded for
$B_{\mathrm{L}}^2$ so that $P_{\Omega}(k)$ and $P_{\sigma}(k)$ can be expressed, respectively, as \cite{MOD1,MOD7}:
\begin{equation}
P_{\Omega}(k) = \overline{\Omega}_{\mathrm{BL}}^2  \biggl(\frac{k}{k_{\mathrm{L}}}\biggr)^{2(n_{\mathrm{B}} -1)} {\mathcal F}(n_{\mathrm{B}}),\qquad P_{\sigma}(k) = \overline{\Omega}_{\mathrm{BL}}^2  \biggl(\frac{k}{k_{\mathrm{L}}}\biggr)^{2(n_{\mathrm{B}} -1)} {\mathcal G}(n_{\mathrm{B}}),
\label{REG4}
\end{equation}
where $\overline{\Omega}_{\mathrm{BL}}=B_{\mathrm{L}}^2/(8 \pi \overline{\rho}_{\gamma})$ and 
\begin{eqnarray}
{\mathcal F}(n_{\mathrm{B}}) &=&  \frac{(2\pi)^{2(n_{\mathrm{B}} -1)}}{\Gamma^2\biggl(\frac{n_{\mathrm{B}}-1}{2}\biggr)}\biggl[\frac{4( 7 - n_{\mathrm{B}})}{3 (n_{\mathrm{B}} -1) ( 5 - 2 n_{\mathrm{B}})} 
+  \frac{4}{(2 n_{\mathrm{B}} - 5)} \biggl( \frac{k}{k_{\mathrm{D}}}\biggr)^{5 - 2 n_{\mathrm{B}}} \biggr],
\label{REG6}\\
{\mathcal G}(n_{\mathrm{B}}) &=& \frac{(2\pi)^{2(n_{\mathrm{B}} -1)}}{\Gamma^2\biggl(\frac{n_{\mathrm{B}}-1}{2}\biggr)}\biggl[ \frac{ n_{\mathrm{B}} + 29}{15 ( 5 - 2 
n_{\mathrm{B}})( n_{\mathrm{B}} -1)} 
+ \frac{7}{5} \frac{1}{(2 n_{\mathrm{B}} - 5)} \biggl( \frac{k}{k_{\mathrm{D}}}\biggr)^{5 - 2 n_{\mathrm{B}}} \biggr].
\label{REG7}
\end{eqnarray}
When $1<n_{\mathrm{B}} < 5/2$, we can formally send the diffusion scale to infinity 
 (i.e. $k_{\mathrm{D}}\to \infty$) and the final result will still be convergent. Consequently
 the diffusion damping only enters the case when 
the spectral slopes are violet (i.e. $n_{\mathrm{B}} \gg 5/2$).
For red spectra (i.e. $n_{\mathrm{B}}<1$) the window function  can be chosen as a simple
step function $W(k)=\theta(k-k_0)$. In this case $A_{\mathrm{B}} =[ (1 -n_{\mathrm{B}})/2] (k_{0}/k_{\mathrm{L}})^{(1 - n_{\mathrm{B}})}B_{\mathrm{L}}^2$, where $H_{0} < k_{0}< k_{\mathrm{p}}$
(see also Eq. (\ref{PSM3})).
The power spectra $P_{\Omega}(k)$ 
and $P_{\sigma}(k)$ can be formally written exactly as in Eq. (\ref{REG4}) 
but with two slightly different pre-factors which shall be denoted by 
$\overline{{\mathcal F}}(n_{\mathrm{B}})$ and $\overline{\mathcal G}(n_{\mathrm{B}})$:
\begin{equation}
P_{\Omega}(k) = \overline{\Omega}_{\mathrm{BL}}^2  \biggl(\frac{k}{k_{0}}\biggr)^{2(n_{\mathrm{B}} -1)} \overline{{\mathcal F}}(n_{\mathrm{B}}),\qquad P_{\Omega}(k) = \overline{\Omega}_{\mathrm{BL}}^2  \biggl(\frac{k}{k_{0}}\biggr)^{2(n_{\mathrm{B}} -1)} \overline{{\mathcal G}}(n_{\mathrm{B}}),
\label{REG8}
\end{equation}
where 
\begin{eqnarray}
\overline{{\mathcal F}}(n_{\mathrm{B}}) &=&\frac{16}{3}(1-n_{\mathrm{B}})^2
\left[\frac{n_{\mathrm{B}}-7}{(n_{\mathrm{B}}-1)(2n_{\mathrm{B}}-5)}+\frac{2}{1-n_{\mathrm{B}}}\left(\frac{k_0}{k}\right)^{n_{\mathrm{B}}-1}\right],
\label{REG9}\\
\overline{{\mathcal G}}(n_{\mathrm{B}}) &=&(1-n_{\mathrm{B}})^2
\left[\frac{4n_{\mathrm{B}}+116}{15(5-2n_{\mathrm{B}})(n_{\mathrm{B}}-1)}+\frac{8}{3}\frac{1}{1-n_{\mathrm{B}}}
\left(\frac{k_0}{k}\right)^{n_{\mathrm{B}}-1}\right].
\label{REG10}
\end{eqnarray}

\renewcommand{\theequation}{B.\arabic{equation}}
\setcounter{equation}{0}
\section{Magnetized Thomson scattering}
\label{APPB}
The four distinct entries of the matrix $M(\Omega,\Omega^{\prime},\Omega^{\prime\prime})$ appearing in Eqs. (\ref{EV1}), (\ref{electric}) and (\ref{LIMM}) are given by:
\begin{eqnarray}
M_{\vartheta\vartheta}(\Omega,\Omega',\Omega^{\prime\prime}) &=& \frac{\zeta \Lambda_{1} - \Lambda_{3}}{2} \biggl[ \sqrt{1 - \mu^2} \sqrt{1 - \nu^2} + \mu \nu \cos{(\varphi -\alpha)} \cos{(\varphi' - \alpha)}\biggr]
\nonumber\\
&+& \frac{\zeta \Lambda_{1} + \Lambda_{3}}{2} \biggl\{\cos{2 \beta} \biggl[ \mu \nu \cos{(\varphi - \alpha)} \cos{(\varphi' - \alpha)} - \sqrt{ 1 - \mu^2} \sqrt{1 - \nu^2}\biggr]
\nonumber\\
&+&   \sin{2\beta} \biggl[ \mu 
\sqrt{1 - \nu^2} \cos{(\varphi - \alpha)} + \nu \sqrt{1 - \mu^2} \cos{(\varphi' -\alpha)}\biggr]\biggr\} 
\nonumber\\
&+&  \zeta \Lambda_{1} \mu \nu \sin{(\varphi -\alpha)} \sin{(\varphi' - \alpha)} 
+ i f_{\mathrm{e}} \zeta \Lambda_{2} \biggl\{ \sin{\beta} \biggl[ \mu \sqrt{1 - \nu^2} \sin{(\varphi - \alpha)} 
\nonumber\\
&-&  \nu \sqrt{1 - \mu^2} \sin{(\varphi' - \alpha)} \biggr] +
\mu \nu \cos{\beta} \sin{(\varphi - \varphi')}\biggr\},
\label{DIP4}\\
 M_{\vartheta\varphi}(\Omega,\Omega',\Omega^{\prime\prime}) &=& \frac{\Lambda_{3} - 
\Lambda_{1} \zeta}{2} \mu \sin{(\varphi' -\alpha)} \cos{(\varphi - \alpha)}
\nonumber\\ 
&-& \frac{\Lambda_{3} + \Lambda_{1} \zeta}{2} \biggl\{\mu \sin{(\varphi' -\alpha)} \cos{(\varphi - \alpha)} \cos{2\beta}
\nonumber\\
&+& \sqrt{1 - \mu^2} \sin{(\varphi' - \alpha)} \sin{2\beta} \biggr\} 
 + \zeta \Lambda_{1} \mu \sin{(\varphi - \alpha)} \cos{(\varphi' - \alpha)} 
\nonumber\\
&-& i f_{\mathrm{e}} \zeta \Lambda_{2} \biggl[ \mu \cos{\beta} \cos{(\varphi' - \varphi)}  + 
\sqrt{1 - \mu^2} \sin{\beta} \cos{(\varphi' - \alpha)}\biggr],
\label{DIP5}\\
M_{\varphi\vartheta}(\Omega,\Omega',\Omega^{\prime\prime}) &=&  - \frac{\zeta \Lambda_{1} + \Lambda_{3}}{2} \sqrt{1 - \nu^2} \, \sin{2 \beta} \sin{(\varphi - \alpha)}
\nonumber\\
&+& \frac{\nu}{4} (\zeta \Lambda_{1} + \Lambda_{3}) \biggl[ \sin{(\varphi + \varphi' - 2 \alpha)} - \sin{(\varphi' - \varphi)}\biggr] ( 1 - \cos{2\beta})
\nonumber\\
&+& i f_{\mathrm{e}} \Lambda_{2} \zeta \biggl[ \nu \cos{\beta} \cos{(\varphi' - \varphi)} + \sqrt{ 1 - \nu^2} \sin{\beta} \cos{(\varphi - \alpha)}\biggr],
\label{DIP6}\\
M_{\varphi\varphi}(\Omega,\Omega',\Omega^{\prime\prime}) &=& \frac{\zeta \Lambda_{1} + \Lambda_{3}}{4} \sin{2 \alpha} ( 1 - \cos{2 \beta}) \sin{(\varphi' + \varphi)}  - \Lambda_{3} \cos{(\varphi' - \varphi)}
\nonumber\\
&+&  \frac{\zeta \Lambda_{1} + \Lambda_{3}}{4} ( 1 + \cos{2\beta}) \biggl[ \cos{(\varphi' - \varphi)} - \cos{2 \alpha} \cos{(\varphi' + \varphi)}\biggr] 
\nonumber\\
&+& \frac{\zeta \Lambda_{1} + \Lambda_{3}}{2}\biggl[ \cos{(\varphi' - \varphi)} + \cos{ 2 \alpha} \cos{(\varphi' + \varphi)}\biggr]
\nonumber\\
&-&  i f_{\mathrm{e}} \zeta \Lambda_{2} \cos{\beta} \sin{(\varphi' - \varphi)},
\label{DIP7}
\end{eqnarray}
where, following the notations of section \ref{sec2}, we have $\nu= \cos{\vartheta^{\prime}}$, $\mu= \cos{\vartheta}$ and $\Omega^{\prime\prime}=(\alpha,\, \beta)$.
The components of the incident electric fields in the local frame $\hat{e}_{1}$, $\hat{e}_{2}$ and $\hat{e}_{3}$ 
can be related to the components off the electric field in the three Cartesian directions as 
\begin{eqnarray}
E_{1} &=& \cos{\alpha} \cos{\beta} E_{x}' + \sin{\alpha} \cos{\beta} E_{y}' - \sin{\beta} E_{z}Õ,\qquad E_{2} = -\sin{\alpha} E_{x}' +  \cos{\alpha} E_{y}' ,
\nonumber\\
E_{3} &=& \cos{\alpha} \sin{\beta} E_{x}' + \sin{\alpha} \sin{\beta} E_{y}' + \cos{\beta} E_{z}'.
\label{electric1}
\end{eqnarray}
The incident electric fields $E_{x}'$, $E_{y}'$ and $E_{z}'$ can be related, in turn, to their polar 
components as:
\begin{eqnarray}
E_{x}' &=& \cos{\vartheta'} \cos{\varphi'} E_{\vartheta}' - \sin{\varphi'} E_{\varphi}Õ,\qquad 
E_{y}' = \cos{\vartheta'} \sin{\varphi'} E_{\vartheta}' + \cos{\varphi'} E_{\varphi}',
\nonumber\\
E_{z}' &=& -\sin{\vartheta'} E_{\vartheta}',
\label{electric2}
\end{eqnarray}
where, as already spelled out in section \ref{sec2} the direction of propagation of 
the incident radiation $\hat{n}'$ coincides with $\hat{r}'$ and ($E_{\vartheta}'$,
$E_{\varphi}'$) are the components of the incident electric field in the spherical 
basis. 

\renewcommand{\theequation}{C.\arabic{equation}}
\setcounter{equation}{0}
\section{Synchronous gauge description}
\label{APPC}
The relation between the perturbed quantities in the longitudinal gauge (i.e. Eq. (\ref{STR4})) and the synchronous gauge of Eq. (\ref{SYN2}) is given by:
\begin{eqnarray}
&& \phi(k,\tau) = - \frac{1}{2k^2} \{[h(k,\tau) + 6 \xi(k,\tau)]'' + {\mathcal H} [h(k,\tau) + 6 \xi(k,\tau)]'\},
\nonumber\\
&& \psi(k,\tau) = - \xi(k,\tau) + \frac{{\mathcal H}}{2 k^2} [h(k,\tau) + 6 \xi(k,\tau)]',
\nonumber\\
&& \overline{\delta}(k,\tau) = \delta(k,\tau) + \frac{3 {\mathcal H}(w + 1)}{2k^2}[h(k,\tau) + 6 \xi(k,\tau)]',
\nonumber\\
&& \overline{\theta}(k,\tau) = \theta(k,\tau) - \frac{1}{2}[h(k,\tau) + 6\xi(k,\tau)]'.
\label{APB2}
\end{eqnarray}
The barred quantities (i.e. $\overline{\delta}$ and $\overline{\theta}$) 
are defined in the longitudinal gauge; $w$ is the barotropic 
index of the corresponding species.  The inverse transformations are instead given by:
\begin{eqnarray}
&& \xi(k,\tau) =  -\psi(k,\tau) - \frac{{\mathcal H}}{a} \int^{\tau} a(\tau') \phi(k,\tau') d\tau',
\nonumber\\
&& h(k,\tau) = 6 \psi(k,\tau) + 6  \frac{{\mathcal H}}{a} \int^{\tau} a(\tau') \phi(k,\tau') d\tau' - 2 k^2 
\int^{\tau} \frac{d\tau'}{a(\tau')} \int^{\tau'} a(\tau'') \phi(k,\tau'') d\tau'',
\nonumber\\
&& \delta(k,\tau) =  \overline{\delta}(k,\tau) + \frac{3{\mathcal H}(w + 1)}{a} \int^{\tau} a(\tau') \phi(k,\tau') d\tau',
\nonumber\\
&& \theta(k,\tau) = \overline{\theta}(k,\tau) - \frac{k^2}{a} \int^{\tau} a(\tau') \phi(k,\tau') d\tau'.
\label{APB3}
\end{eqnarray}
While $\Delta_{\mathrm{Q}}$ and $\Delta_{\mathrm{U}}$ are gauge-invariant, the brightness perturbation of the intensity transforms as:
\begin{equation}
\overline{\Delta}_{\mathrm{I}} = \Delta_{\mathrm{I}} - \frac{{\mathcal H}}{2 k^2} (h^{\prime} + 6 \xi^{\prime}) + \frac{i \mu}{2 k} ( h^{\prime} + 6 \xi^{\prime}),
\label{APB20}
\end{equation}
while its multipoles and the baryon velocity transform as:
\begin{eqnarray}
&& \overline{\Delta}_{\mathrm{I}1} = \Delta_{\mathrm{I}1} - \frac{1}{6k} (h' + 6 \xi^{\prime}),\qquad \overline{\Delta}_{\mathrm{I}0} =  \Delta_{\mathrm{I}0} + \frac{{\mathcal H}}{2 k^2}(h'  + 6 \xi').
\label{APB21}
\end{eqnarray}

To avoid ambiguities we shall now agree that all the remaining equations of this appendix \ref{APPC} hold in the synchronous gauge. The evolution equations of  the density contrast 
of CDM particles (i.e. $\delta_{\mathrm{c}}$) and of the corresponding peculiar velocity (i.e. $\theta_{\mathrm{c}}$) are:
\begin{equation}
\delta_{\mathrm{c}}^{\prime}= - \theta_{\mathrm{c}} + \frac{h'}{2},\qquad 
\theta_{\mathrm{c}}^{\prime} + {\mathcal H} \theta_{\mathrm{c}} =0.
\label{APB4}
\end{equation}
 The integrals appearing in Eq. (\ref{APB3}) for the expressions of $\theta$ and $h$ imply two integration 
constants which can be space dependent. Following the standard practice 
they are fixed by demanding that the CDM peculiar velocity vanishes (i.e. $\theta_{\mathrm{c}}=0$) and that $h$ 
has no constant mode. While different possibilities can be envisaged the synchronous description typically assumes the CDM rest frame. Defining, in analog terms, $\delta_{\nu}$ and  $\theta_{\nu}$ as the neutrino 
density contrast and as the three-divergence of the neutrino peculiar velocity, the corresponding evolution equations 
are:
\begin{eqnarray}
&& \delta_{\nu}' = -\frac{4}{3} \theta_{\nu} + \frac{2}{3} h',\qquad 
\theta_{\nu}' = - k^2 \sigma_{\nu}  + \frac{k^2}{4} \delta_{\nu},
\label{APB5}\\
&& \sigma_{\nu}' = \frac{4}{15} \theta_{\nu} - \frac{3}{10} k {\mathcal F}_{\nu 3} - \frac{2}{15} h' - \frac{4}{5} \xi',
\label{APB7}
\end{eqnarray}
where $\sigma_{\nu}$ is the neutrino anisotropic stress and ${\mathcal F}_{\nu 3}$ is the octupole 
of the (perturbed) phase space distribution. These two terms are automatically 
gauge-invariant and do not change from the longitudinal to the synchronous 
descriptions.

Recalling the baryon-photon ratio $R_{\mathrm{b}}$  of Eq. (\ref{Rbdef}),
the equations for the reduced baryon-photon system are given by:
\begin{equation}
\delta_{\mathrm{b}}' = - \theta_{\mathrm{b}} + \frac{h'}{2} + \frac{\vec{E}\cdot\vec{J}}{a^4 \rho_{\mathrm{b}}},\qquad
\theta_{\mathrm{b}}' + {\mathcal H} \theta_{\mathrm{b}} = \frac{\epsilon'}{R_{\mathrm{b}}}
(\theta_{\gamma} - \theta_{\mathrm{b}}) + \frac{k^2}{4 R_{\mathrm{b}}}\bigl[ \Omega_{\mathrm{B}} - 4 \sigma_{\mathrm{B}}\bigr],
\label{APB9}
\end{equation}
where $\epsilon^{\prime}$ is the differential optical depth (see Eq.(\ref{diffop})). 
The lowest two multipoles of the Boltzmann hierarchy of the photons, namely 
the density contrast (i.e. the monopole) and the three-divergence of the velocity field (related 
to the dipole of the intensity of the brightness perturbations) are:
\begin{equation}
\delta_{\gamma}' = - \frac{4}{3} \theta_{\gamma} + \frac{2}{3} h', \qquad 
\theta_{\gamma}' = -\frac{1}{4} \nabla^2 \delta_{\gamma} + \epsilon' (\theta_{\mathrm{b}} - \theta_{\gamma}).
\label{APB11}
\end{equation}
In the tight-coupling limit Eqs. (\ref{APB9}) and (\ref{APB11}) become 
\begin{eqnarray}
&& \theta_{\gamma\mathrm{b}}' + \frac{{\mathcal H} R_{\mathrm{b}}}{1 + R_{\mathrm{b}}} \theta_{\gamma\mathrm{b}} + \frac{\eta}{\rho_{\gamma} (R_{\mathrm{b}} + 1)} k^2 \theta_{\gamma\mathrm{b}}= 
\frac{k^2}{ 4 ( 1 + R_{\mathrm{b}})} \delta_{\gamma} + \frac{k^2 (\Omega_{\mathrm{B}} - 4 \sigma_{\mathrm{B}})}{4 ( 1 + R_{\mathrm{b}})},
\label{APB12}\\
&& \delta_{\gamma}' = \frac{2}{3} h'- \frac{4}{3} \theta_{\gamma\mathrm{b}},\qquad \delta_{\mathrm{b}}' = \frac{h'}{2} - \theta_{\gamma\mathrm{b}}.
\label{APB14}
\end{eqnarray}
The brightness perturbations of the radiation field are related to the inhomogeneities of the Stokes parameters.
In the synchronous coordinate system the evolution equations of the brightness perturbations are\footnote{Equations (\ref{APB17}), (\ref{APB18}) and (\ref{APB19}) refer to the scalar case; 
for the sake of conciseness the superscript specifying the 
scalar nature of the brightness perturbation has been omitted.}
\begin{eqnarray}
&&  \Delta_{\mathrm{I}}' + i k \mu \Delta_{\mathrm{I}} = - \biggl[ \xi' - \frac{\mu^2}{2}( h' + 6 \xi')\biggr] +
\epsilon' \biggl[  - \Delta_{\mathrm{I}} + \Delta_{\mathrm{I}0} + \mu v_{\mathrm{b}} - \frac{1}{2} P_{2}(\mu) S_{\mathrm{P}}\biggr],
\label{APB16}\\
&& \Delta_{\mathrm{Q}}' + i k \mu \Delta_{\mathrm{Q}} = \epsilon' \biggl[- \Delta_{\mathrm{Q}} + \frac{1}{2} ( 1 - P_{2}(\mu)) S_{\mathrm{P}}\biggr],
\label{APB17}\\
&& \Delta_{\mathrm{U}}' + i k \mu \Delta_{\mathrm{U}} = - \epsilon'  \Delta_{\mathrm{U}},
\label{APB18}\\
&& v_{\mathrm{b}}' + {\mathcal H} v_{\mathrm{b}} + \frac{\epsilon'}{R_{\mathrm{b}}} ( 3 i \Delta_{\mathrm{I}1} + v_{\mathrm{b}}) + 
i k \frac{\Omega_{\mathrm{B}} - 4 \sigma_{\mathrm{B}}}{4 R_{\mathrm{b}}}=0,
\label{APB19}
\end{eqnarray}
where $R_{\mathrm{b}}$ is the baryon to photon ratio and where we defined $v_{\mathrm{b}} = \theta_{\mathrm{b}}/(i k)$. Moreover, in Eqs. (\ref{APB16}) and (\ref{APB17}) we have that $S_{\mathrm{P}} = \Delta_{\mathrm{I}2} + \Delta_{\mathrm{Q}0} + \Delta_{\mathrm{Q}2}$ is automatically gauge-invariant.

By perturbing the Einstein equations, the Hamiltonian and the momentum constraints stemming 
from the $(00)$ and $(0i)$ components of Eq. (\ref{PertS1}) are 
\begin{equation}
 2k^2 \xi - {\mathcal H} h' = \ell_{P}^2 a^2 \biggl[\delta_{\mathrm{s}}\rho_{\mathrm{t}} + \delta_{\mathrm{s}}\rho_{\mathrm{B}}\biggr],\qquad k^2 \xi' = - \frac{\ell_{P}^2 a^2}{2} ( p_{\mathrm{t}} + \rho_{\mathrm{t}}) \theta_{\mathrm{t}}.
 \label{APB23}
 \end{equation}
 In Eq. (\ref{APB23}) $\delta_{\mathrm{s}}\rho_{\mathrm{t}}$ and $\theta_{\mathrm{t}}$ are, respectively,
 the density fluctuation of the plasma and the three-divergence of the total velocity field. The spatial components of the perturbed Einstein equations (i.e., respectively, $(i=j)$ and $(i\neq j)$)
  lead instead to:
 \begin{eqnarray}
 && h'' + 2 {\mathcal H} h' - 2 k^2 \xi = 3 \ell_{P}^2 a^2 [\delta p_{\mathrm{t}} + \delta_{\mathrm{s}} p_{\mathrm{B}}],
\label{APB26}\\
&& (h + 6 \xi)'' + 2 {\mathcal H} ( h + 6 \xi)' - 2 k^2 \xi = 3 \ell_{P}^2 
a^2 \Pi_{\mathrm{tot}},
\label{APB27}
\end{eqnarray}
where $\Pi_{\mathrm{tot}}= [ (p_{\nu} + \rho_{\nu}) \sigma_{\nu} + 
(p_{\gamma} + \rho_{\gamma}) \sigma_{\mathrm{B}}]$.
In terms of the synchronous degrees of freedom, the curvature perturbation on comoving orthogonal 
hypersurfaces (i.e. ${\mathcal R}$) and the curvature perturbation on uniform density hypersurfaces 
(i.e. $\zeta$) are defined, respectively, as
\begin{equation}
{\mathcal R} = \xi + \frac{{\mathcal H} \xi'}{{\mathcal H}^2 - {\mathcal H}'},\qquad
\zeta = \xi - \frac{{\mathcal H} (\delta_{\mathrm{s}} \rho_{\mathrm{t}} + \delta_{\mathrm{s}} \rho_{\mathrm{B}} + \delta_{\mathrm{s}} \rho_{\mathrm{E}} )}{\rho_{\mathrm{t}}'},
\label{APB28}
\end{equation}
where the definition of ${\mathcal R}$ has been already mentioned 
in Eq. (\ref{NMR4}) while the definition of $\zeta$ is the synchronous analog of  Eq. (\ref{NMODE11}).
By taking the difference of ${\mathcal R}$ and $\zeta$ and using the Hamiltonian constraint of Eq.
(\ref{APB23}), the following equation can be obtained:
\begin{equation}
 \zeta - {\mathcal R} = - \frac{2 k^2\xi - ( h + 6 \xi)'}{3 \ell_{P}^2 a^2 (p_{\mathrm{t}} + \rho_{\mathrm{t}})}.
\label{APB29}
\end{equation}
By combining the evolution of $\xi$, $h$ and ${\mathcal R}$ we can obtain, after some algebra, the following relation
\begin{equation}
 {\mathcal R}' = \Sigma_{R} - \frac{2 a^2 k^2 \xi}{\ell_{P}^2 {\mathcal H} z_{t}^2} + \frac{a^2 (h + 6 \xi)'}{8 \pi G z_{t}^2},
\label{APB30}
\end{equation}
which is the analog of Eq. (\ref{NMODE2}) already discussed in the longitudinal gauge.

\renewcommand{\theequation}{D.\arabic{equation}}
\setcounter{equation}{0}
\section{Power spectra of Faraday rotation}
\label{APPD}
If we expand Faraday rotation rate of Eq. (\ref{varphi}) in ordinary  spherical harmonics we can formally define the angular power 
spectrum of Faraday rotation as:
\begin{equation}
\mathrm{F}(\hat{n}) = \sum_{\ell\, m} f_{\ell\,m} Y_{\ell\,m}(\hat{n}), \qquad \langle f_{\ell \, m}^{*} f_{\ell' \, m'}\rangle = C_{\ell}^{(\mathrm{FF})} \delta_{\ell \,\ell'} \delta_{m \,m'}.
\label{F1}
\end{equation}
To get the explicit form of $C_{\ell}^{(\mathrm{FF})}$ in terms of the magnetic power spectrum 
we expand the magnetic field in vector spherical harmonics \cite{var,bie,blatt} (see also \cite{edmonds}). This technique is 
rather well established both in the study of electromagnetic processes as well as in nuclear physics. The vector analog of the Rayleigh expansion can be written, for the present ends, as:
\begin{equation}
\vec{B}(\vec{k}) e^{ - i k \mu \tau_{0}} =\sum_{\ell\, m} \sum_{\alpha} g_{\ell \, m}^{(\alpha)}(k,\mu) \vec{Y}_{\ell\, m}^{(\alpha)}(\hat{n}),
\label{F4}
\end{equation}
where, as usual $\mu = \hat{k}\cdot\hat{n}$. In Eq. (\ref{F4}) $\alpha$ is the polarization and 
$\vec{Y}_{\ell\, m}^{(\alpha)}$ are the vector harmonics \cite{var,bie}:
\begin{equation}
\vec{Y}_{\ell\, m}^{(1)}(\hat{n}) = \frac{ \vec{\nabla}_{\hat{n}} Y_{\ell \, m}(\hat{n})}{\sqrt{\ell (\ell + 1)}}, 
\qquad \vec{Y}_{\ell\, m}^{(0)}(\hat{n}) = -\frac{i(\hat{n} \times \vec{\nabla}_{\hat{n}} )Y_{\ell \, m}(\hat{n})}{\sqrt{\ell (\ell + 1)}},\qquad \vec{Y}^{(-1)}_{\ell\, m} = \hat{n} Y_{\ell\, m}(\hat{n}),
\label{F5}
\end{equation}
where $Y_{\ell\, m}(\hat{n})$ are the usual spherical harmonics. Since $\hat{n} \cdot \vec{Y}^{(1)}_{\ell\, m}(\hat{n}) = \hat{n}\cdot \vec{Y}^{(0)}_{\ell\, m}(\hat{n}) =0$ and 
 $\hat{k}\cdot \vec{B}(\vec{k})=0$, only one term of the sum over $\alpha$ survives in Eq. (\ref{F4}), i.e. the term $\alpha = -1$ and the result is: 
\begin{equation}
\hat{n} \cdot \vec{B}(\vec{k}) e^{- i k \mu \tau_{0}} = 4 \pi \sum_{\ell\, m} \sqrt{\ell (\ell +1)}  \frac{j_{\ell}(k\tau_{0})}{k\tau_{0}} \vec{B}(\vec{k}) \cdot \vec{Y}_{\ell \, m}^{(-1)\,\ast}(\hat{k}).
\label{F10}
\end{equation}
The angular power spectrum of Faraday rotation can be simply expressed as:
\begin{equation}
C_{\ell}^{(\mathrm{FF})} = 4\pi {\mathcal A}^2 \ell (\ell +1) \int \frac{d k}{k} P_{B}(k) \frac{j_{\ell}^2(k\tau_{0})}{k^2 \tau_{0}^2}.
\label{F13}
\end{equation}
The expression of the $B$-mode autocorrelation can then be written as:
\begin{eqnarray}
{\mathcal C}_{\ell}^{(\mathrm{BB})} &=& \sum_{\ell_{1},\,\,\ell_{2}}  {\mathcal Z}(\ell, \ell_1, \ell_{2}) C_{\ell_{2}}^{(\mathrm{EE})} C_{\ell_{1}}^{(\mathrm{F})}
\label{F13aa}\\
{\mathcal Z}(\ell, \ell_1, \ell_{2}) &=&
N_{\ell}^2 N_{\ell_{2}}^2 {\mathcal Q}(\ell,\ell_{1}, \ell_{2})^2 \frac{( 2\ell_{1} + 1) ( 2 \ell_{2} +1)}{4\pi ( 2 \ell + 1)}   [ {\mathcal C}^{\ell 0}_{\ell_{1} 0 \ell_{2} 0}]^2,
\label{CBBF}
\end{eqnarray}
which are the expressions mentioned in Eq. (\ref{F14a}). Note that in Eq. (\ref{CBBF}) ${\mathcal Q}(\ell,\ell_{1}, \ell_{2})$ is defined as:
\begin{eqnarray}
&&{\mathcal Q}(\ell,\ell_{1}, \ell_{2})= - \frac{1}{2}[L^2 + L_{1}^2 + L_{2}^2 
- 2 L_{1} L_{2} - 2 L_{1} L + 2 L_{1} - 2 L_{2} - 2 L],
\label{FFF1}\\
&& L = (\ell + 1)\ell,\qquad L_{1} = (\ell_{1} + 1)\ell_{1},\qquad
L_{2} = (\ell_{2} + 1)\ell_{2},
\label{FFF2}\\
&& N_{\ell}= \sqrt{\frac{2 (\ell -2)!}{(\ell + 2)!}},\qquad  N_{\ell_2}= \sqrt{\frac{2 (\ell_{2} -2)!}{(\ell_{2} + 2)!}}.
\label{FFF3}
\end{eqnarray}
Denoting with $n$ a positive integer, in Eq.  (\ref{CBBF})
${\mathcal C}^{\ell\, 0}_{\ell_{1}\, 0\, \ell_{2}\,0}$ vanishes whenever 
$\ell + \ell_{1} + \ell_{2} = (2 n + 1)$; conversely, 
\begin{equation}
 {\mathcal C}^{\ell\, 0}_{\ell_{1}\, 0\, \ell_{2}\,0}= \frac{(-1)^{n - \ell} \sqrt{2 \ell +1} n!}{( n -\ell_{1})! (n - \ell_{2})! (n -\ell)!} 
\sqrt{\frac{(2n - 2 \ell_{1})! ( 2 n - 2 \ell_{2})! ( 2n - 2\ell)!}{(2 n + 1)!}},
\nonumber
\end{equation}
provided $\ell + \ell_{1} + \ell_{2} = 2 n$. 
This form of the relevant Clebsch-Gordon coefficient
 is given by \cite{var}. The Clebsch-Gordon coefficient of the previous equation then vanishes unless  $|\ell_{1} - \ell_{2}| \leq \ell < \ell_{1} + \ell_{2}$ 
(triangle inequality) and unless $\ell_{1} + \ell_{2} + \ell$ is an even integer. In the two degenerate cases (i.e. $\ell = \ell_1 + \ell_{2}$ and $\ell = \ell_{1} - \ell_{2}$) 
the expressions become, respectively:
\begin{eqnarray}
&& {\mathcal C}^{\ell_{1} + \ell_{2}\,\,0}_{\ell_{1}\,0\,\ell_{2}\,0} = \frac{(\ell_{1} + \ell_{2})!}{\ell_{1}! \ell_{2}!} \sqrt{\frac{(2\ell_{1})! ( 2 \ell_{2})!}{( 2\ell_{1} + 2 \ell_{2})!}} , 
\nonumber\\
&&{\mathcal C}^{\ell_{1} - \ell_{2}\,\,0}_{\ell_{1}\,0\,\ell_{2}\,0} = (- 1)^{\ell_{2}} \frac{\ell_{1}!}{\ell_{2}! (\ell_{1} - \ell_{2})!}
\sqrt{\frac{( 2\ell_{1})! ( 2\ell_{1} - 2\ell_{2} +1}{(2\ell_{1} + 1)!}}.
\nonumber
\end{eqnarray}
\end{appendix}

\end{document}